\documentclass[fleqn,usenatbib]{mnras}

\usepackage{newtxtext,newtxmath}

\usepackage[T1]{fontenc}

\DeclareRobustCommand{\VAN}[3]{#2}
\let\VANthebibliography\thebibliography
\def\thebibliography{\DeclareRobustCommand{\VAN}[3]{##3}\VANthebibliography}

\usepackage{graphicx}	
\usepackage{amsmath}	
\usepackage{amssymb}	
\usepackage{subfigure}
\usepackage{threeparttable}
\usepackage{multirow}
\usepackage[linewidth=0.5pt]{mdframed} 

\def   \thCN        {$^{13}$CN}
\def   \HthCN       {$^{13}$CN}

\def   \twCN        {$^{12}$CN}
\def   \CftN        {C$^{15}$N}
\def   \HthCN       {$\rm H^{13}CN$}
\def   \HCfifN      {$\rm HC^{15}N$}
\def   \Tmb         {$T_{\rm mb}$}
\def   \Tex         {$T_{\rm ex}$}
\def   \Tcmb        {$T_{\rm CMB}$}

\def   \CtwCth      {$\rm ^{12}C/^{13}C$}
\def   \NftNfif     {$\rm ^{14}N/^{15}N$}

\def   \Rgc         {$R_{\rm gc}$}
\def   \Rtwth       {$R_{\rm ^{12}C/^{13}C}$}
\def   \Rftfif      {$R_{\rm ^{14}N/^{15}N}$}
\def   \Rsixeit     {$R_{\rm ^{16}O/^{18}O}$}
\def   \thC         {$\rm ^{13}C$}
\def   \ffN         {$\rm ^{15}N$}

\def\purple#1 {{\textcolor{purple}{#1}}\ }
\def\red#1 {\textcolor{red}{#1}}
\def\new#1 {{\bf #1 }}
\def\blue#1 {{\textcolor{blue}{#1}}\ }
\def\zy#1 {\textcolor{red}{zy: #1} }
\def\yc#1 {\textcolor{cyan}{yichen: #1}}

\graphicspath{{./}{Figures/}}

\title[CN isotopologues in the Galactic outer disk]{An improved method to measure $\rm ^{12}C/^{13}C$ and $\rm ^{14}N/^{15}N$ abundance ratios: revisiting CN isotopologues in the Galactic outer disk}

\author[Y.\,C.~Sun et al.]{\noindent
Yichen~Sun,$^{1,2}$
Zhi-Yu~Zhang,$^{1,2}$\thanks{E-mail:zzhang@nju.edu.cn}
Junzhi~Wang,$^{3}$\thanks{E-mail:junzhiwang@gxu.edu.cn}
Lingrui~Lin,$^{1,2}$
Padelis~P.~Papadopoulos,$^{4,5}$
\newauthor
Donatella~Romano,$^{6}$
Siyi~Feng,$^{7}$
Yan~Sun,$^{8}$
Bo~Zhang,$^{9}$
and Francesca~Matteucci$^{10,11,12}$
\\
\noindent
$^{1}$School of Astronomy and Space Science, Nanjing University, Nanjing 210093, the People’s Republic of China\\
$^{2}$ Key Laboratory of Modern Astronomy and Astrophysics (Nanjing University), Ministry of Education, Nanjing 210093, the People’s Republic of China\\
$^{3}$ School of Physical Science and Technology, Guangxi University, Nanning, the People’s Republic of China \\
$^{4}$ Department of Physics, Section of Astrophysics, Astronomy and Mechanics, Aristotle University of Thessaloniki, Thessaloniki, GR-54124, Greece \\
$^{5}$  Research Center for Astronomy, Academy of Athens, Soranou Efesiou 4, GR-11527, Athens, Greece \\
$^{6}$ Istituto nazionale di astrofisica, Astrophysics and Space Science Observatory
Via Gobetti 93/3, Bologna, IT 40129\\
$^{7}$ Department of Astronomy, Xiamen University, Zengcuo'an West Road, Xiamen, 361005, the People's Republic of China \\
$^{8}$ Purple Mountain Observatory and Key Laboratory of Radio Astronomy, Chinese Academy of Sciences, Nanjing 210008, the People’s Republic of China \\
$^{9}$ Shanghai Astronomical Observatory, Chinese Academy of Sciences, 80 Nandan Road, Shanghai
200030, the People’s Republic of China \\
$^{10}$ Sezione di Astronomia, Dipartimento di Fisica, Universit`a di Trieste, Via Tiepolo 11, I-34131 Trieste, Italy \\
$^{11}$ INAF, Osservatorio Astronomico di Trieste, Via Tiepolo 11, I-34131 Trieste, Italy \\
$^{12}$ INFN, Sezione di Trieste, Via Valerio 2, I-34127 Trieste, Italy
}

\date{Accepted 2023 November 18. Received 2023 November 10; in original form 2023 September 28}

\pubyear{2023}

\begin{document}
\label{firstpage}
\pagerange{\pageref{firstpage}--\pageref{lastpage}}
\maketitle

\begin{abstract}

The variations of elemental abundance and their ratios along the Galactocentric
radius result from the chemical evolution of the Milky Way disks. The
$\rm ^{12}C/^{13}C$ ratio in particular is often used as a proxy to determine
other isotopic ratios, such as $\rm ^{16}O/^{18}O$ and $\rm ^{14}N/^{15}N$.
Measurements of $\rm ^{12}CN$ and $\rm ^{13}CN$ (or $\rm C^{15}N$) -- with
their optical depths corrected via their hyper-fine structure lines
-- have traditionally been exploited to constrain the Galactocentric gradients
of the CNO isotopic ratios. Such methods typically make several simplifying
assumptions (e.g. a filling factor of unity, the Rayleigh–Jeans approximation,
and the neglect of the cosmic microwave background) while adopting
a single average gas phase. However, these simplifications introduce
significant biases to the measured \CtwCth\ and  \NftNfif. We demonstrate that
exploiting the optically thin satellite lines of \twCN\ constitutes a more
reliable new method to derive \CtwCth\ and \NftNfif\ from CN isotopologues. We
apply this satellite-line method to new IRAM 30-m observations of \twCN, \thCN,
and \CftN\ $N=1\to0$ towards 15 metal-poor molecular clouds in the Galactic
outer disk ($R_{\rm gc} > $ 12\,kpc), supplemented by data from the literature.
After updating their Galactocentric distances, we find that \CtwCth\ and \NftNfif\ gradients are in good agreement with those derived using independent optically thin molecular tracers, even in regions with the lowest metallicities.  We therefore recommend using optically thin tracers for Galactic and extragalactic CNO isotopic measurements, which avoids the biases associated with the traditional method.

\end{abstract}




\section{Introduction}

The isotopic abundance ratios of carbon, nitrogen, and oxygen, namely $\rm
^{12}C/^{13}C$ (\Rtwth), $\rm ^{16}O/^{18}O$ (\Rsixeit), and $\rm
^{14}N/^{15}N$ (\Rftfif), are crucial for constraining the evolutionary
chemical history of the Milky Way\citep{Wilson1994,Romano2022}. Systematic
variations in these ratios are caused by the different time scales according to
which different isotopes are ejected in the interstellar medium by stars of
different initial masses
\citep{Nomoto2013,Karakas2014,Cristallo2015,Limongi2018}. Furthermore, the
growth history of the Galactic thin disk has a significant impact on these
ratios, resulting in strong evolutionary features
\citep[e.g.,][]{Matteucci1991,Henkel1993,Romano2003,Romano2017,Romano2019}.  In
particular, \Rtwth, \Rsixeit, and \Rftfif\ generally increase with
Galactocentric distances (\Rgc)
\citep[e.g.,][]{Wilson1994,Wouterloot2008,Jacob2020,Zhang2020,Chen2021}, though
\Rftfif\ shows a decreasing trend in the Galactic outer disk\citep{Colzi2022}.

The production of metal elements and their isotopes depends on nucleosynthesis
and stellar evolutionary processes ~\citep{Burbidge1957,Meyer1994}.  For
example, the $\rm ^{12}C$ production is dominated by triple-$\alpha$ reactions
\citep{Timmes1995,Woosley1995}, while $\rm ^{13}C$ can be produced by the CNO-I
cycle, $\rm ^{12}C$-burning, or proton-capture nucleosynthesis
\citep{Meynet2002b,Botelho2020}.  The $\rm ^{14}N$ can be synthesized through
the cold CNO cycle in the H-burning zone of stars
\citep{Karakas2014,Romano2022},  $\rm ^{12}C$-burning in the low-metallicity
fast massive rotators, and proton-capture reactions in the hot convective
envelope of intermediate-mass stars
\citep{Marigo2001,Pettini2002,Meynet2002a,Limongi2018,Botelho2020}.  The
production of $\rm ^{15}N$ and $\rm ^{13}C$ can happen in hot-CNO cycles in
H-rich material accreting on white dwarfs
\citep{Audouze1973,Wiescher2010,Romano2022}. The Galactic $\rm ^{15}N$ is
contributed by novae \citep{Romano2017} and proton ingestion in the He shell of
massive stars \citep{Pignatari2015}.

Despite extensive study, the isotopic ratio gradients in the low-metallicity
outer disk of the Milky Way remain poorly constrained. To date, measurements of
only one target, WB89-391, at a Galactocentric distance \Rgc\ $\gtrsim$ 12 kpc,
have been made for the \Rtwth\ ratio using CN isotopologues \citep{Milam2005}.
This single measurement (\Rtwth\ $\sim$ 134) critically constrains the \Rftfif\
and \Rsixeit\ gradients derived from C-bearing isotopologues. To fully
constrain all CNO isotopic ratios in the outer Galactic disk, more data and
more precise measurements of \Rtwth\ are needed.

The \Rtwth\ ratios in the ISM are determined using pairs of molecular
isotopologues.  Measurements of $\rm ^{12}C^{18}O/^{13}C^{18}O$
\citep[e.g.,][]{Langer1990,Wouterloot1996,Giannetti2014} or $\rm ^{12}C^{34}S/^{13}C^{34}S$ \citep{Yan2023} may be limited to
nearby strong targets because of the weak emission of $\rm ^{13}C^{18}O$ or $\rm ^{13}C^{34}S$
expected in the metal-poor outer disk clouds.  Molecules such as $\rm
H_2CO/H_2^{13}CO$ \citep{Henkel1982,Henkel1985,Yan2019}, $\rm
CH^{+}/^{13}CH^{+}$ \citep{Ritchey2011}, and $\rm CH/^{13}CH$ \citep{Jacob2020}
can be good tracers to derive \Rtwth\ through their absorption lines. 
$\rm H_2CO$ isotopologues can be absorbed by the CMB but their absorption lines are still weak because of low abundance. 
The CH and $\rm CH^{+}$ absorption lines are in rare cases of strong background continuum and/or high column density.  
The optically-thick line pairs, such as \thCN\ and \twCN\, $N=1\to0$, which require
corrections to their optical depths that are mostly determined by fitting
hyper-fine-structure (HfS) lines~\citep{Savage2002,Milam2005}, could be adopted 
to measure \Rtwth\ in the outer disk regions. This method can also
derive \Rftfif\ without any assumptions of \Rtwth\ \citep{Adande2012}.

The current commonly used method for deriving the \Rtwth\ from CN isotopologues, with the HfS
fitting, makes several 
assumptions and approximations. In order to derive the column density \citep[i.e.,][]{Mangum2015}, it
assumes identical excitation temperature of \twCN\ and \thCN. However differential radiative trapping among spectral lines,
due to different optical depths (if these are significant), can yield different excitation temperatures between the spectral lines.
Also, a single gas phase is typically used (necessitated by the small number of transitions per species typically available, 
and lack of standard molecular cloud models), while a range of excitation conditions is present in molecular clouds impacting the
volume-average excitation temperatures of the CN isotopologues. In addition, most studies \citep[e.g.,][]{Savage2002,Milam2005} adopt the 
Rayleigh-Jeans (R-J) approximation, which becomes increasingly poor when $h\nu_{\rm ik}/k_{\rm B} T_{\rm ex} \lesssim 1$ starts approaching unity (and is of course inappropriate when $ h\nu_{\rm ik}/k_{\rm B} T_{\rm ex} \gtrsim 1$). In the cold ($ T_{\rm kin} \sim 15-20$\,K), often sub-thermal line excitation conditions ($T_{\rm ex}$$<$$T_{\rm kin}$) prevailing in the bulk of molecular clouds in the quiescent ISM of the Galaxy, the R-J approximation
is simply not good enough for abundance ratio studies conducted using the CN $N = 1 \to 0$ and $N=2 \to 1$ at 112\,GHz and 224\,GHz. 
In addition, the cosmic microwave background (CMB) must be taken into account given that for the low-temperature
molecular clouds where such isotopologue studies are conducted (and with the possible sub-thermal excitation of the lines utilized)
, the line Planck temperature $ J(\nu_{\rm ik}, T_{\rm ex})=h\nu _{\rm ik}/k_{\rm B}\,\left(e^{h\nu _{\rm ik}/k_B T_{\rm ex}} - 1\right)^{-1} $
can be so low that the  $J(\nu _{\rm ik}, T_{\rm CMB})$ is no longer negligible for the frequencies used, even for the low $ T_{\rm CMB, 0}=2.72$\,K)
in the local Universe \citep[see][for even more serious effects in the high-z Universe where $T_{\rm CMB}(z)=(1+z) T_{\rm CMB,0}$]{Zhang2016}.
All these assumptions/approximations have been used for such studies conducted in widely different ISM environments, 
ranging from the very cold and quiescent ISM in the Galactic outer disk \citep{Milam2005}, where they become questionable, up to the warm, dense ISM in starburst galaxies \citep[e.g.,][]{Henkel1993b,Henkel2014,Tang2019} where the aforementioned assumptions/approximations should be adequate.

In this work, we introduce an improved method to derive \CtwCth\ from CN isotopologues and compare it with current methods. We list the basic assumptions of the current methods of using CN isotopologues 
to derive \Rtwth\ and \Rftfif. We also present new \twCN, \thCN\ and \CftN\ $N=1\to0$ observations in a small sample of molecular clouds with \Rgc\ $>$ 12 kpc, which allows new constraints on the \Rtwth\ and \Rftfif\ outer gradients. Section~\ref{sec:observation} presents the observations and the data reduction. 
In Section~\ref{sec:traditional_methods}, we list the current methods of deriving \Rtwth\ and \Rftfif\ with CN isotopologues and show their underlying assumptions and drawbacks. In Section~\ref{sec:derive_C_N_ratio_from_satellite_line}, we introduce a new method to derive \Rtwth\ and \Rftfif. In Section~\ref{sec:results}, we present newly measured isotopic ratios and data derived from the improved traditional methods and the new method. In Section~\ref{sec:discussion}, we compare the Galactic \Rtwth\ and \Rftfif\ obtained by the different methods. We also compare CN isotopologues and the optically thin tracers $\rm ^{12}C^{18}O/^{13}C^{18}O$, and we discuss physical and chemical processes that may bias the abundances. We present the main conclusion in Section~\ref{sec:conclusions}.

\section{Observations and data reduction}
\label{sec:observation}

\subsection{Sample Selection} 

We select 15 molecular clouds from the literature 
\citep{Savage2002, Milam2005, Wouterloot2008, Sun2015, Li2016, Sun2017, Reid2019} 
based on the following criteria:  (a). located in the Galactic radii range of (11 -- 22) kpc; 
and (b). with strong $\rm ^{13}CO$ $J=1\to0$ detections of $\ge 5$K (in \Tmb) in \cite{Sun2015,Sun2017}. 
This allows a good chance of detecting CN isotopologues because of an expected strong \twCN\ $N=1\to0$ emission. We also include WB89~391 from \cite{Milam2005}, which provided the only \Rtwth\ data at $R_{\rm gc}> 12$ kpc from CN isotopologues before our study. 

\begin{figure}
\includegraphics[scale=0.38]{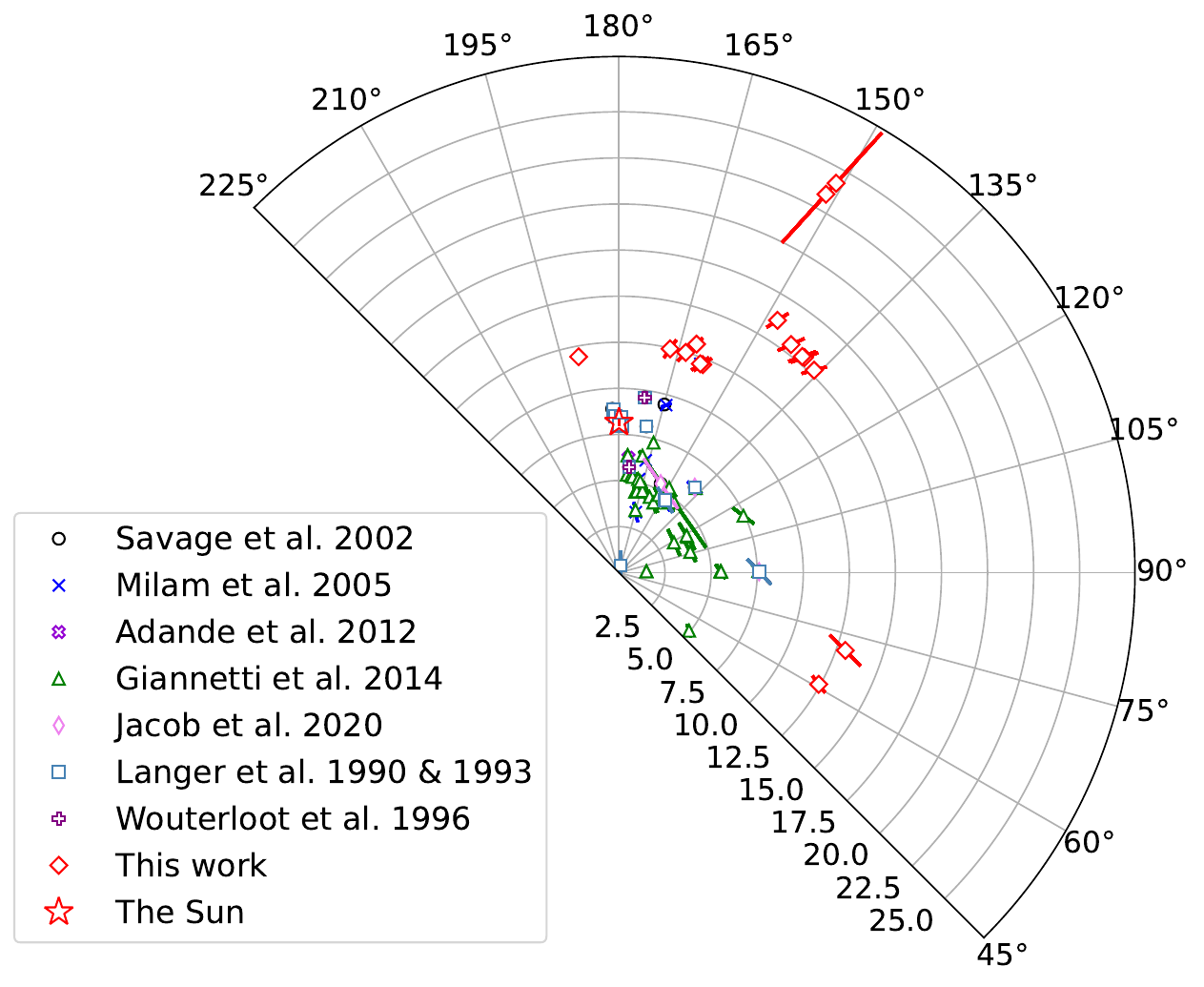}
\centering
\caption{Spatial distribution of sources in the Galactic plane. The pole is the Galactic center, and the red star shows the position of the Sun~\citep{Reid2019}. Rings indicate the Galactocentric distances ($R_{\rm gc}$). Red squares show our new observations. 
Black circles, blue multiplication signs, purple multiplication signs, green triangles, 
magenta thin diamonds, steel-blue rectangles, purple plus signs, and red diamonds, are positions of sources from~\citet{Savage2002},~\citet{Milam2005},~\citet{Adande2012},~\citet{Giannetti2014},~\citet{Jacob2020},~\citet{Langer1990, Langer1993}, and~\citet{Wouterloot1996}, respectively. } \label{fig:fig_posi}
\end{figure}

\subsubsection{Revision of distances} \label{sec:Rgc_revision}

We update the Galactocentric distances, for both our targets and sources in the
literature\citep{Savage2002,Milam2005,Adande2012,Giannetti2014,Jacob2020,Langer1990,Langer1993,Wouterloot1996}. 
For sources with direct trigonometric measurements \citep{Reid2014,Reid2019}, 
we adopt the measured values. For the others, we apply the up-to-date Galactic rotation 
curve model~\citep{Reid2019}, which has been well calibrated 
with accurate trigonometric measurements, to obtain the kinematic distance.  
Specifically, we employed results based on the Parallax-Based Distance Calculator~\citep{Reid2016}
\footnote{\url{http://bessel.vlbi-astrometry.org/node/378}}. 
The Galactocentric distances are then derived with the following equation:

\begin{equation}
	R_{\rm gc}=\sqrt{R_{\odot}^2+R_{\rm h}^2-2R_{\odot}R_{\rm h}{\rm cos}l},
\end{equation}

where $R_{\odot}=8.15\ \rm kpc$ is the distance of the Sun from the centre of
the Milky Way \citep{Reid2019}, $R_{\rm h}$ is the heliocentric
distance, $R_{\rm gc}$ is the Galactocentric distance, and $l$ is the Galactic
longitude of the sources.  

Figure~\ref{fig:fig_posi} shows the locations of these sources. 
The Probability density functions (PDFs) generated by the distance calculator 
are presented in Appendix~\ref{sec:dis_PDF}.
In Table~\ref{tab:target_information} we list the coordinates, velocity at the frame of 
Local Standard of Rest (LSR) ($V_{\rm LSR}$), estimated heliocentric distances, updated $R_{\rm gc}$, and observing time.

\subsection{IRAM 30-m observations}

Fourteen targets were observed with the 30-m telescope in Project 031-17 (PI: Zhi-Yu Zhang) 
from August 31 to September 05, 2017. 
An additional target, G211.59, was observed as part of Project 005-20 (PI: Junzhi Wang) 
during August 5--9, 2020. To ensure data quality, we only used observations with a 
precipitable water vapor (PWV) $<$ 10 mm and discarded data with PWV $\gtrsim$ 10 mm. 
In Project 031-17, the typical system temperature ($T_{\rm sys}$) was $\sim \rm (100-200)\, K$ 
and $\sim \rm (200-400)\, K$ at 3-mm and 2-mm bands, respectively. For G211.59, $T_{\rm sys}$ 
was $\sim \rm (150-200)\, K$ and $\sim \rm (400-650)\, K$ at 3-mm and 2-mm bands, respectively.

Both projects utilized the Eight Mixer Receiver (EMIR) as the front end, with
both E090 and E150 receivers under similar frequency setups. The backend is 
the Fast Fourier Transform Spectrometers working at 200 kHz resolution (FTS200), 
corresponding to  $\sim 0.53\ \rm km\cdot s^{-1}$ and $\sim 0.34\ \rm km\cdot s^{-1}$ at 3-mm and 2-mm, 
respectively. 
The frequency setup covers $^{13}\rm{CO}$ $J = 1 \to 0$ ($\nu_{\rm rest}=$ 110.201 GHz), $\rm ^{12}CN$ $N = 1 \to 0$ ($\nu_{\rm rest}=$ 113.490 GHz), $\rm ^{13}CN$ $N = 1 \to 0$ ($\nu_{\rm rest}=$ 108.780 GHz), and  $\rm C^{15}N$ $N = 1 \to 0$ ($\nu_{\rm rest}=$ 110.024 GHz). In addition, our setup at 2-mm covers the $\rm H^{13}CN$ $J=2\to1$ ($\nu_{\rm rest}=$ 172.678 GHz) and $\rm HC^{15}N$ $J=2\to1$ ($\nu_{\rm rest}=$ 172.108 GHz).

Planets, i.e., Saturn, Mars, and Venus, were used to perform the focus calibration,  
each after a prior pointing correction. In a few cases, we adopted PKS 2251+158, PKS 0316+413 and W3(OH) 
for the focus, when the planets were not available. These focus corrections were 
performed at the beginning of each observing slot and were repeated within 30 minutes 
after sunset or sunrise. We perform regular pointing calibration every 1--2 hours, 
with strong point continuum sources within the 15$\degr$ radius of targets, e.g., PKS 1749+096, 
PKS 0736+017,  PKS 0316+413, NGC7538, and W3(OH). The typical pointing error is $\sim$ 3$''$ (rms).

The observations were performed in two steps: we first performed an
On-The-Fly (OTF) mapping towards each target, to get the spatial distribution 
of $\rm ^{13}CO$ $J=1\to0$ emission. Then we performed a single-pointing 
deep integration towards the emission peak position on the $\rm ^{13}CO$ $J=1\to0$ map. 
During the OTF mapping, we scanned along both right ascension  
(R.A.) and declination (Dec.) directions, with a spatial scan interval of 9.0$''$.
Along the direction of each scan, it outputs a spectrum every 0.5 sec, which makes 
a 4.8 $''$ interval along the scan direction. Each OTF map covers an area of $\sim 2.4' \times 2.4'$.

We adopted a positional beam-switch mode and performed deep integration at 
the peak position of $\rm ^{13}CO$ $J=1\to0$ for the extended targets. The 
OFF positions were set 10$'$ (in Azimuth) away from the target. 
For targets with compact $\rm ^{13}CO$ $J=1\to0$ emission (spatial FWHM $< 1'$), we use 
the wobbler switch mode. The beam switching used had a frequency of 2\,Hz and a throw of
 120$''$ (in Azimuth) on either side of the target (to correct for any first-order 
beam-asymmetric beam effects between the two throw positions).

The beam sizes of the IRAM 30-m telescope are $\sim$ 22$''$ and 14$''$ at 110 GHz 
($\rm ^{13}CO$) and 170 GHz ($\rm H^{13}CN$), with main beam efficiencies 
($\eta_{\rm mb}$)  of $\sim 0.78$ and $\sim 0.69$\footnote{\url{https://publicwiki.iram.es/Iram30mEfficiencies}}, 
respectively.  We present the integration time of each target in 
Table~\ref{tab:target_information}. The noise levels of the 
final spectra are listed in Table~\ref{tab:target_base_rms}. Particularly, we list the transitions of CN isotopologues and the main beam efficiencies ($\eta_{\rm mb}$) of IRAM 30-m at the frequency of each transition in Table~\ref{tab:transition_and_main_beam_efficiency}.

\begin{table*}
\caption{Target information.}
\label{tab:target_information}
\begin{threeparttable}
\begin{tabular}{cccccccc}
\hline
\hline
    Sources & $l$ & $b$ & $V_{\rm{LSR}}$ & $R_{\rm{h}}$ &  $R_{\rm{gc}}$ & $t_{\rm obs}$ & references \\
            & (degree) & (degree) & ($\rm km\cdot s^{-1}$) & (kpc) & (kpc) & (hours) &  \\
\hline
G211.59   & 211.593 & 1.056  & 45.0   & $4.18\  \pm 0.18 $ & $11.92^{*}\ \pm 0.22 $ & 9.72 &  \citet{Reid2019}\\
G37.350   & 37.350  & 1.050  & $-$53.9  & $17.87\ \pm 0.49 $ & $12.42\ \pm 0.46 $ & 1.20 &  \citet{Sun2017} \\
G44.8     & 44.800  & 0.658  & $-$62.1  & $17.43\ \pm 1.10 $ & $12.98\ \pm 0.99 $ & 1.30 &  \citet{Sun2017}\\
IRAS0245  & 136.357 & 0.958  & $-$61.4  & $5.25\  \pm 0.57 $ & $12.49\ \pm 0.53 $ & 3.15 &  \citet{Li2016}\\
SUN15 14N & 109.292 & 2.083  & $-$101.1 & $10.67\ \pm 0.70 $ & $15.41\ \pm 0.62 $ & 0.55 &  \citet{Sun2015}\\
SUN15 18  & 109.792 & 2.717  & $-$99.3  & $10.58\ \pm 0.71 $ & $15.38\ \pm 0.63 $ & 0.84 &  \citet{Sun2015}\\
SUN15 21  & 114.342 & 0.783  & $-$100.9 & $10.25\ \pm 0.71 $ & $15.50\ \pm 0.63 $ & 0.73 &  \citet{Sun2015}\\
SUN15 34  & 122.775 & 2.525  & $-$100.7 & $10.24\ \pm 0.62 $ & $16.17\ \pm 0.58 $ & 0.55 &  \citet{Sun2015}\\
SUN15 56  & 137.758 & $-$0.983 & $-$103.6 & $17.53\ \pm 3.61 $ & $24.19\ \pm 3.52 $ & 1.63 &  \citet{Sun2015}\\
SUN15 57  & 137.775 & $-$1.067 & $-$102.1 & $16.72\ \pm 3.47 $ & $23.40\ \pm 3.38 $ & 0.94 &  \citet{Sun2015}\\
SUN15 7W  & 104.983 & 3.317  & $-$102.7 & $10.98\ \pm 0.62 $ & $15.26\ \pm 0.54 $ & 0.77 &  \citet{Sun2015}\\
WB89 380  & 124.644 & 2.540  & $-$86.2  & $5.53\  \pm 0.51 $ & $12.17\ \pm 0.45 $ & 6.08 &  \citet{Wouterloot2008}\\
WB89 391  & 125.802 & 3.048  & $-$86.0  & $5.45\  \pm 0.50 $ & $12.16\ \pm 0.44 $ & 4.38 &  \citet{Wouterloot2008}\\
WB89 437  & 135.277 & 2.800  & $-$71.6  & $5.65\  \pm 0.39 $ & $13.10^{*}\ \pm 0.38 $ & 5.67 &  \citet{Wouterloot2008}\\
WB89 501  & 145.199 & 2.987  & $-$58.1  & $4.87\  \pm 0.51 $ & $12.46\ \pm 0.49 $ & 2.07 &  \citet{Wouterloot2008}\\
\hline
\end{tabular}
\begin{tablenotes}
\footnotesize
\item Column 1: source names. Columns 2 and 3: the Galactic coordinates of sources. Column 4: $V_{\rm{LSR}}$ values gained by fitting a gauss profile for $^{13}$CO line at $J = 1 \to 0$ at 110201.354 MHz. Column 5: Heliocentric distances. Column 6: Galactocentric distances. Column 7: single-pointing observing time. Column 8: source references. *: The distance of G211.59 and WB89 437 are directly measured by $\rm H_2O$ masers in~\citet{Reid2014,Reid2019}. 
\end{tablenotes}
\end{threeparttable}
\end{table*}

\subsection{Data reduction} 
\label{sec:datareduction}

For data reduction, we used the Continuum and Line Analysis Single-dish Software
(CLASS) package from the Grenoble Image and Line Data Analysis Software
\citep[GILDAS,][]{Guilloteau2000}.  For each sideband, three independent
Fast Fourier Transform Spectroscopy (FTS) units cover a 4-GHz bandwidth, which
sometimes causes different continuum levels on the same spectrum (so-called the
platforming effect)\footnote{\url{https://www.iram.fr/GENERAL/calls/s21/30mCapabilities.pdf}}.
Therefore, we first split each spectrum into three frequency ranges
(corresponding to the three units) and treat them independently. Then we locate
the line-free channels and subtract a first-order baseline for each spectrum
with command \texttt{BASE}. For spectra affected by apparent standing waves,
which are less than 5\% of the total, a sinusoidal function is adopted to fit
and subtract the baseline.

We discarded spectra with high noise levels: $\sigma_{\rm obs} $$>$$ 1.2
\sigma_{\rm theoretical}$.  Spectra at the same position are then averaged with
the default weighting setup of \texttt{TIME}, by which the weight is
proportional to the integrated time, frequency, and $T_{\rm sys}^{-2}$. In
Table~\ref{tab:target_base_rms}, we list the typical root-mean-square (RMS) of
the antenna temperature in the  $T^\star_{\rm A}$ scale at 3-mm.   For the IRAM
30-m telescope\footnote{\url{https://safe.nrao.edu/wiki/pub/KPAF/KfpaPipelineReview/kramer_1997_cali_rep.pdf}},
the antenna temperature scale denotes one corrected only for atmospheric
absorption\footnote{Please note that in the fundamental paper on the
calibration of 
mm/submm radio telescopes by \cite{Kutner1981}, $T^\star_{\rm
A}$ designates a temperature scale corrected for atmospheric absorption and rearward beam spillover (see their Equation 14).}.

\begin{table*}
\caption{The RMS of 3-mm spectra in our targets.} \label{tab:target_base_rms}
\begin{tabular}{lccccccc}
  \hline
  \hline
    Sources & \multicolumn{3}{c}{E0UI} & \multicolumn{3}{c}{E0UO}  \\
\hline
            & Frequency Range & Channel Width & RMS &  Frequency Range & Channel Width & RMS  \\
            & (GHz)           & ($\rm km\cdot s^{-1}$) & (mK)  & (GHz)  & ($\rm km\cdot s^{-1}$) & (mK)\\
\hline
G211.59     & $103.318-105.826$ & $0.553-0.567$  & 3.0 & $109.557-113.608$ & $0.515-0.534$  & 4.1  \\
G37.350     & $106.850-110.900$ & $0.528-0.548$  & 7.9 & $110.580-114.629$ & $0.511-0.529$  & 13   \\
G44.8       & $106.850-110.900$ & $0.528-0.548$  & 8.1 & $110.580-114.629$ & $0.511-0.529$  & 13   \\
IRAS0245    & $106.850-110.914$ & $0.528-0.548$  & 4.8 & $110.580-114.643$ & $0.511-0.529$  & 7.3  \\
SUN15 14N   & $106.850-110.900$ & $0.528-0.548$  & 12  & $110.580-114.629$ & $0.511-0.529$  & 19   \\
SUN15 18    & $106.850-110.900$ & $0.528-0.548$  & 17  & $110.580-114.629$ & $0.511-0.529$  & 19   \\
SUN15 21    & $106.851-110.899$ & $0.528-0.548$  & 10  & $110.579-114.628$ & $0.511-0.529$  & 16   \\
SUN15 34    & $106.851-110.900$ & $0.528-0.548$  & 12  & $110.579-114.628$ & $0.511-0.529$  & 20   \\
SUN15 56    & $106.851-110.899$ & $0.528-0.548$  & 6.0 & $110.579-114.628$ & $0.511-0.529$  & 9.4  \\
SUN15 57    & $106.851-110.900$ & $0.528-0.548$  & 7.6 & $110.579-114.628$ & $0.511-0.529$  & 12   \\
SUN15 7W    & $106.851-110.899$ & $0.528-0.548$  & 14  & $110.579-114.628$ & $0.511-0.529$  & 23   \\
WB89 380    & $106.851-110.899$ & $0.528-0.548$  & 3.5 & $110.580-114.629$ & $0.511-0.529$  & 5.0  \\
WB89 391    & $106.851-110.900$ & $0.528-0.548$  & 4.0 & $110.579-114.628$ & $0.511-0.529$  & 6.2  \\
WB89 437    & $106.851-110.900$ & $0.528-0.548$  & 3.5 & $110.579-114.628$ & $0.511-0.529$  & 4.9  \\
WB89 501    & $106.851-110.900$ & $0.528-0.548$  & 5.6 & $110.579-114.629$ & $0.511-0.529$  & 8.6  \\
\hline
\end{tabular}
\end{table*}

\begin{table*}
\caption{Transitions of CN isotopologues and IRAM 30-m main beam efficiencies.}
\label{tab:transition_and_main_beam_efficiency}
\begin{threeparttable}
\begin{tabular}{cccccc}
\hline
\hline
 \multicolumn{2}{c}{ Isotopologue Transitions}    & Components$^{\rm a}$ & Frequency & Relative intensity$^{\rm b}$  & $\eta_{\rm mb}=B_{\rm eff}/F_{\rm eff}$  \\
              &  &   & (MHz) &  &     \\
\hline 
                &  \multirow{4}{*}{$J=1/2\to1/2$} & $F=1/2\to1/2$ &  113123.369  &  0.012   & 0.7819  \\
                &                                 & $F=1/2\to3/2$ &  113144.190  &  0.099   & 0.7819  \\
                &                                 & $F=3/2\to1/2$ &  113170.535  &  0.096   & 0.7819  \\   
$\rm ^{12}CN$   &                                 & $F=3/2\to3/2$ &  113191.325  &  0.125   & 0.7819  \\                
\cline{2-6}
$N=1\to0$       & \multirow{5}{*}{$J=3/2\to1/2$} & $F=3/2\to1/2$ &  113488.142  &  0.126   & 0.7816  \\
                &                                & $F=5/2\to3/2$ &  113490.985  &  0.334   & 0.7816  \\
                &                                & $F=1/2\to1/2$ &  113499.643  &  0.099   & 0.7816  \\
                &                                & $F=3/2\to3/2$ &  113508.934  &  0.097   & 0.7815  \\
                &                                & $F=1/2\to3/2$ &  113520.422  &  0.012   & 0.7815  \\
\hline     
$\rm ^{13}CN$   & $J=3/2\to1/2$   & $F=3\to2$     &  108780.201  &  0.195   & 0.7864  \\
$N=1\to0$       & $F_{13}=2\to1$  & $F=2\to1$     &  108782.374  &  0.103   & 0.7864  \\
\hline     
$\rm C^{15}N$   & \multirow{2}{*}{$J=3/2\to1/2$}   & $F=1\to0$     &  110023.540  &  0.165   & 0.7851  \\
$N=1\to0$       &                                  & $F=2\to1$     &  110024.590  &  0.417   & 0.7851  \\
\hline
\end{tabular}
\begin{tablenotes}
\footnotesize
\item a. For $\rm ^{13}CN$ and $\rm C^{15}N$ $N=1\to0$, we only list the two strongest components considered in our intensity estimation. \\
b. The intrinsic ratio between the intensity of individual line components and the total intensity of all the line components (based on CDMS/JPL). 
\end{tablenotes}
\end{threeparttable}
\end{table*}

\subsubsection{Line intensities}

We adopt the rest frequencies of molecular lines from NASA's Jet Propulsion
Laboratory (JPL)\footnote{\url{https://spec.jpl.nasa.gov/ftp/pub/catalog/catform.html}}.
We first fit a Gaussian profile to the $^{13}\rm{CO}$ $J=1\to0$ spectra, which
are all single peaked. Then, we use $\Delta v$= $8\times$
FWHM/$\rm{2\sqrt{2ln2}}$ as the velocity range for other emission lines, by
assuming that all lines of the same target have the same line width.  The
line-free channels are adopted as $\sim 4 \times \Delta v$ at both sides of
each line.  Then we obtain the velocity-integrated-intensity in the velocity
range of $\Delta v$, using the following equation:  

\begin{equation}
        I_{\rm line}=\int_{\Delta v}T_{\rm mb} dv, 
\end{equation}

\noindent
where $\Delta v$ is the Full Width at Zero Intensity (FWZI) of the emission line, which is set to be $8\times$ FWHM/$\rm 2\sqrt{2ln2}$. 
$T_{\rm mb} (=F_{\rm eff}/B_{\rm eff}\cdot T_{\rm A}^\star = T_{\rm A}^\star/\eta_{\rm mb})$ 
is the main beam temperature; and $T_{\rm A}^\star$, $F_{\rm eff}$, and $B_{\rm eff}$ are the 
antenna temperature, the forward efficiency, and the telescope beam efficiency, respectively.

We derive the thermal noise error following \citet{Greve2009}, with Eq.~\ref{eq:eq_area_sigma}, which accounts
for both the one associated with the velocity-integral of the line intensity over
its FWZI and the one associated with the subtracted baseline level (the
latter becoming significant only if a wide line leaves little baseline ``room'' 
within a spectral window).

\begin{equation}
\label{eq:eq_area_sigma}
\sigma_{I_{\rm line}}=\sigma_{T_{\rm{mb, chan}}}\Delta v_{\rm{res}}\sqrt{N_{\rm{line}}(1+\frac{N_{\rm{line}}}{N_{\rm{base}}})},
\end{equation}

\noindent
where $\sigma_{T_{\rm mb, chan}}$ is the channel noise level of the main beam
temperature, $N_{\rm line}$ is the number of channels covering the FWZI of the
line, $N_{\rm{base}}$ is the number of line-free channels as the baseline, and
$\Delta v_{\rm{res}}$ is the velocity resolution of the spectrum. The flux
calibration and beam efficiency uncertainties (typically $\sim $ 10--15\% in
such single-dish measurements), are not included in our final line ratio
uncertainties since all lines were measured simultaneously in our observations 
(the flux calibration and main beam efficiency factors are applied multiplicatively).

The two strongest satellite lines of $\rm ^{13}CN$ $N=1\to0$ ($\nu_{\rm rest}$
at 108.780 and 108.782 GHz) are blended, with a velocity separation of $\sim
6.0\ \rm km s^{-1}$.  We use the sum of their
velocity-integrated intensities because they are very likely optically thin
(see further discussion in Appendix \ref{Appendix:12C13Cderiv}). The
associated noise is obtained with Equation~\ref{eq:eq_area_sigma}.  

\subsubsection{Upper limits of non-detected line fluxes}
\label{sec:upperlimit_es}

We define a line detection feature with the following three criteria: 

\begin{itemize}
\item \rm  I, More than three contiguous channels have $T_{\rm mb} > 2 \sigma_{T_{\rm mb, chan}}$,   
\item \rm II, $T^{\rm peak}_{\rm mb} > 3 \sigma_{T_{\rm mb, chan}}$, and
\item \rm III, $I_{\rm line} \ge 3\sigma_{I_{\rm line}}\,\,\,,{\rm i.e., S/N \ge 3}$
\end{itemize}

For non-detected targets, we adopt $3\sigma_{I_{\rm line}}$ as the upper limit of the
velocity-integrated intensity.  For blended lines, such as the satellite lines
of \thCN, we estimate the upper limits of the summed fluxes.

\section{Past methods used in deriving \CtwCth\ and \NftNfif\ from CN isotopologues}
\label{sec:traditional_methods}

\subsection{Assumptions in the `traditional' models} \label{sec: assumptions}
 
First, we list the common assumptions in deriving \CtwCth\ and \NftNfif\ from
emission lines of CN isotopologues. Here we take a simple example to obtain
abundance ratios of $\rm{^{12}C/^{13}C}$ and  $\rm{^{14}N/^{15}N}$ from $^{12}$CN, $^{13}$CN, and $\rm C^{15}N$ with their $N = 1 \to 0$
transition lines. For higher $N$ levels, the method is essentially the same.  To
perform such derivations, several basic assumptions are needed for all models
(details are shown in Appendix~\ref{Appendix:12C13Cderiv}) :

\begin{enumerate} 
\item Column density ratios of isotopologues represent abundance ratios of isotopes, meaning that astrochemical effects are neglected.
\item In all regions, the populations at the energy levels that give rise to the HfS lines are assumed to have identical $T_{\rm ex}$ among them.
\item  Differences in the dipole moment matrix, the rotational partition function, the upper energy level, and the degeneracy of the upper energy between \twCN, \thCN, and \CftN\ are ignored.
\item The $N=1\to0$ transitions of \twCN, \thCN, and \CftN\ have identical rotational temperatures. 
\end{enumerate} 

With these assumptions, ratios between column densities of \twCN, \thCN, and \CftN\ (hereafter, we only consider the \twCN\ and \thCN\ pair, which is identical to the \twCN\ and \CftN\ pair. ) equal their respective optical depth ratios, for $N=1\to0$:

\begin{equation}
\label{eq:Nratio_eq_tauratio}
    \frac{N_{\rm ^{12}CN}}{N_{\rm ^{13}CN}} = \frac{\tau_{\rm ^{12}CN}}{\tau_{\rm ^{13}CN}}, 
\end{equation}

\noindent
where $N_{\rm ^{12}CN}$ and $N_{\rm ^{13}CN}$ are column densities of \twCN\ and \thCN, respectively.  $\tau_{\rm ^{12}CN}$ and $\tau_{\rm ^{13}CN}$ are the total optical depths of \twCN\ and \thCN\ $N=1\to0$, respectively. 

Ratios between the main beam temperatures of isotopologue lines would satisfy:

\begin{equation}
\label{eq:Tbratio_eq_taufunction}
      \frac{T_{\rm b, ^{12}CN}}{T_{\rm b, ^{13}CN}}=\frac{1-e^{-\tau_{\rm m, ^{12}CN}}}{1-e^{-\tau_{\rm m, ^{13}CN}}} = \frac{T_{\rm mb, ^{12}CN}}{T_{\rm mb, ^{13}CN}}
\end{equation}

\noindent
where $T_{\rm b, ^{12}CN}$ and $T_{\rm b, ^{13}CN}$ are the peak brightness
temperatures of the main components of \twCN\ and \thCN\ $N=1\to0$,
respectively. The main beam temperature of \twCN\ and \thCN\ $N=1\to0$ main
component is  $T_{\rm mb, ^{12}CN}$ and $T_{\rm mb, ^{13}CN}$, respectively.
The right-hand side of Equation \ref{eq:Tbratio_eq_taufunction} is satisfied if we assume the beam filling
factors $f$ of \twCN\ and \thCN\ $N=1\to0$ to be the same.  Should the sources
have been resolved by the 30-m beam, the same assumption would then have to be
made for the corresponding irreducible (source-structure)-beam
coupling factors $\rm \eta_{\rm c}$ \citep[e.g.][]{Kutner1981}. As $\tau_{\rm m,
^{12}CN}$ and $\tau_{\rm m, ^{13}CN}$ we indicate the optical depths of the main HfS
component lines, which  are defined as $J = 3/2 \to 1/2, F = 5/2 \to 3/2$ at
113.490985 GHz and  $J = 3/2 \to 1/2$, the sum of $F_2= 2 \to 1$ and $F= 3 \to
2 $ blended at 108.780201 GHz, for \twCN\ and \thCN, respectively.

\subsection{Formula to derive \CtwCth\ and \NftNfif\ in the literature}

The traditional equation to derive \CtwCth\ is presented in \citet[Equation~3]{Savage2002}:

\begin{equation}
\label{eq:Savage_and_Milam_formula_old}
    \frac{\rm{^{12}C}}{\rm{^{13}C}} = \frac{(3/5)\tau_{\rm m} T_{\rm ex, ^{12}CN} }{T^\star_{\rm R, ^{13}CN}/\eta_{\rm c, ^{13}CN}},
\end{equation}

\noindent

where $T^\star_{\rm R, ^{13}CN}$ is the line temperature measured from the observed spectra, 
$\eta_{\rm c, ^{13}CN}$ is the beam efficiency, $T_{\rm ex, ^{12}CN}$ is the excitation 
temperature of \twCN\ $N=1\to0$ main component; and the factor 5/3 is a conversion factor 
from the column density ratio of the main components between \twCN\ and \thCN\ $N=1\to0$ to 
all components of \twCN\ and \thCN\ in this transition. 

However, this Equation \ref{eq:Savage_and_Milam_formula_old} and the corresponding Equations 2 and 3 
in \cite{Savage2002}, where $T^{\rm \star}_{\rm R}/\eta _{\rm c}$ appears, contain two issues: 
First,  $\eta _{\rm c}$ is set as the antenna beam efficiency. This is incorrect since at the 
NRAO 12-m telescope the $ T^{\rm\star}_{\rm R}$ scale is already corrected for both atmosphere and
all telescope efficiency factors\footnote{\href{https://library.nrao.edu/public/memos/12/12U/12U_010.pdf}{User's Manual For The NRA0 12M Millimeter-Wave Telescope, J. Mangum, 01/18/00} }, while $\eta _{\rm c}$ stands for an {\it irreducible} (source-structure)-beam coupling factor, instead of a beam efficiency \citep[see][for details]{Kutner1981}.
Second, the background correction is only considered in the denominator. 
$T^{\rm\star}_{\rm R}/\eta_{\rm c} = T_{\rm R} - T_{\rm bg}$, where $T_{\rm R}$ and $T_{\rm bg}$ are the source radiation temperature and background emission (the CMB), respectively.   
The excitation temperature, \Tex\ in the nominator, on the other hand, does not subtract $T_{\rm bg}$. 
Unfortunately, the same problems exist also in \citet{Milam2005}.

If both $T^{\rm \star}_{\rm R}$ and $\eta _{\rm c}$ were set as their original definitions, i.e.,
$T^{\star}_{\rm R}$ is the observed source antenna temperature corrected for atmospheric attenuation, radiative loss, and rearward and forward scattering and spillover, and $\eta_{\rm c}$ as the efficiency at which the source couples to the telescope beam,  then Equation \ref{eq:Savage_and_Milam_formula_old} still stands, as long as the background temperature is negligible, i.e. the target is warm enough compared to the CMB. 
However, the coupling factor, $\eta_{\rm c}$, is unknown because the source size and geometry are unclear, unless we assume that the sources are big enough to cover the whole forward beam.

Besides the aforementioned assumptions and problems the traditional method 
contains also the following  unstated assumptions: 

\begin{itemize}
\item  A beam filling factor of $\sim $1 for both \thCN\ and \twCN\ $N=1\to0$ lines.
Equations 2 and 3 in \citet{Savage2002} can only be understood if the source geometric 
beam filling factor $f_{\rm s}$ is set to $\sim $1 (i.e. extended targets fully resolved 
in both \twCN\ and \thCN), and the intrinsic (source structure)-beam coupling factor $\eta _{\rm c}$ 
is also $\sim $1.
\item  The Rayleigh-Jeans approximation is adopted for expressing line radiation temperature,  
\item  Negligible contribution from the CMB emission,     
\item  The line optical depth is uniformly distributed within the beam size -- a flat spatial distribution. 
\end{itemize}

Most of these dense gas clumps are spatially compact within $<$ 1 pc scales \citep[e.g.][]{Wu2010,Tafalla2002}, 
especially for those targets from the outer Galactic disk. Most main-beam of single-dish telescopes could cover the emitting regions of \twCN\ and \thCN\ lines. 
Therefore, we update Equation~\ref{eq:Savage_and_Milam_formula_old} as follows, 
to accommodate the temperature definition by the IRAM 30-m telescope (with identical 
assumptions listed above):

\begin{equation}
\label{eq:Savage_and_Milam_formula}
    \frac{\rm{^{12}C}}{\rm{^{13}C}} = \frac{(3/5)\tau_{\rm m} T_{\rm ex, ^{12}CN} }{T^\star_{\rm A, ^{13}CN}/\eta_{\rm mb, ^{13}CN}},
\end{equation}

\noindent
where $T^{\rm \star}_{\rm A}$ is the corrected antenna temperature, or, the forward beam brightness 
temperature\citep{Wilson2013}, $\eta_{\rm mb, ^{13}CN} (= \frac{B_{\rm eff}}{F_{\rm eff}})$ is 
the main beam efficiency of \thCN\ $N=1\to0$, with $B_{\rm eff}$ and $F_{\rm eff}$ being the telescope beam efficiency
and forward efficiency, respectively;
$T_{\rm ex, ^{12}CN}$ is the excitation temperature of \twCN\ $N=1\to0$ main
component;  The factor 5/3 is a conversion factor from the column density ratio
of the main components between \twCN\ and \thCN\ $N=1\to0$ to all components of
\twCN\ and \thCN\ in this transition.

We re-organized this ``traditional'' method and present the detailed derivation
in Appendix~\ref{Appendix:12C13Cderiv}. Note that a similar formula has also been adopted
to derive \Rftfif\ \citep[e.g.,][]{Adande2012}.

\subsubsection{The hyper-fine structure of CN: fitting the optical depth}

For $^{12}$C$^{14}$N and $\rm ^{12}C^{15}N$,  the nuclear spin of $\rm ^{14}N$
and $\rm ^{15}N$ couples in the total angular momentum, which generates
hyper-fine structures ~\citep[here labeled by
$\boldsymbol{F}$;][]{Skatrud1983,Saleck1994}.  For more complex
$^{13}$C$^{14}$N, the angular momentum $\boldsymbol{J}$ first couples with the
nuclear spin of $^{13}$C atom to form an angular momentum
$\boldsymbol{F_{13}}$, which further couples with the nitrogen nuclear spin to
form the total angular momentum $\boldsymbol{F}$~\citep{Bogey1984}. 

For a $T_{\rm ex}$ common among the various HfS CN satellite lines (which could
be different from e.g., the rotational excitation temperature $T_{\rm rot}$,
and $T_{\rm kin}$), their corresponding optical depth ratios are fixed by the
ratios of the corresponding $S$ factors (line strengths) in the matrix element $\mu _{\rm ul}$
(see Equations 62, 75 in \cite{Mangum2015}, but also \citet{Skatrud1983}) that
enters the expression of the Einstein coefficients $A_{\rm ul}$ of the hyperfine
lines.  Should these lines be optically thin, the line optical depth ratios are
also the ratios of line strengths (assuming a common $T_{\rm ex}$ among the
satellite lines involved).

One can derive the optical depths of $\rm ^{12}CN$ $N=1\to0$ lines from main beam
temperature ratios between the nine components of the hyperfine structure lines
\citep[or, a subset of them, e.g.\, five components in ][]{Savage2002,Milam2005}, using:

\begin{equation}
\label{eq:HfS_fitting_assumption}
        \frac{T_{\rm mb, m, ^{12}CN}}{T_{\rm mb, sat, ^{12}CN}} = \frac{1-e^{-\tau_{\rm m, ^{12}CN}}}{1-e^{-\tau_{\rm sat, ^{12}CN}}} = \frac{1-e^{-\tau_{\rm m, ^{12}CN}}}{1-e^{-R_{\rm h}\cdot \tau_{\rm m, ^{12}CN}}},
\end{equation}

\noindent
where $T_{\rm mb, m, ^{12}CN}$ and $T_{\rm mb, sat, ^{12}CN}$ are the peak
main beam temperature of the main component and the satellite component of
\twCN\ $N=1\to0$, respectively.  We label the optical depths of the main
component and the satellite component as $\tau_{\rm m, ^{12}CN}$ and $\tau_{\rm
sat, ^{12}CN}$, and $R_{\rm h}$ is the intrinsic intensity ratio between the satellite
line and the main component. This of course assumes that all these lines share the same \Tex. This does not necessarily mean full local thermodynamic equilibrium (LTE). Only common
excitation among HfS lines would work out, as \Tex\, could be different from
$T_{\rm rot}$ or $T_{\rm kin}$).

We performed HfS fitting with the package developed by \cite{Estalella2017},
which shows better robustness and stability than the HfS fitting method
provided in CLASS/Gildas (For details, see
Appendix~\ref{appendix:compare_HfS}).  The fitted optical depths of the main
line of \twCN\ are shown in Table~\ref{tab:results_HfS}.

\subsection{The updated HfS method to derive \CtwCth\ and \NftNfif }
\label{sec:update_HfS_method}

In this section, we consider the effect of adopting the Planck-equivalent
radiation temperature scale, and a non-zero CMB temperature, respectively. We
then introduce our updated equation to derive \Rtwth\ and \Rftfif\  that
combines both reformulations.

\subsubsection{Corrections to the Planck's Equation}
\label{sec:effect_of_RJ_appro}

The R-J approximation gives deviation in the methods in \citet{Savage2002} and
\citet{Milam2005} to derive \Rtwth\ and the method in \citet{Adande2012} to
derive \Rftfif. The \Rtwth\ derived from the Planck's radiation temperature
will be smaller than that derived under the R-J approximation. As expected
(details in Appendix~\ref{Appendix:12C13Cderiv}), the decrease of \Rtwth\ after
revision will be larger when \Tex\ is smaller, in a non-linear fashion. At the
frequency of \twCN\ $N=1\to0$ main component (113.490 GHz), the decrease is
$\le 5$ \%, and $\sim 0.5$ for $T_{\rm ex} \gtrsim 53$ K and $T_{\rm ex}\sim
4.3$ K, respectively.

\subsubsection{Corrections to the CMB contribution} 
\label{sec:effect_of_Tcmb}
Considering the CMB temperature $T_{\rm CMB} = 2.73$ K as the background
temperature \citep[e.g., Equation 2.3 in][]{Zhang2016}, the derived \Rtwth\
will also be lower than those derived without CMB contribution (see in
Appendix~\ref{Appendix:12C13Cderiv}). In this case, the term $T_{\rm ex,
^{12}CN}$ should be replaced by $T_{\rm ex, ^{12}CN}-T_{\rm CMB}$ in
Eq.~\ref{eq:Savage_and_Milam_formula}.  After the CMB correction, the derived
\Rtwth\ would decrease by <5\% for a $T_{\rm ex}$$\gtrsim $54\,K (i.e.
negligible). However, for $T_{\rm ex}\sim $5.4\,K, and $\sim$ 3.0\,K, \Rtwth\
would decrease by $\sim $50\% and $\sim $90\%, respectively.  Some targets in
\cite{Milam2005} and \cite{Savage2002} show $T_{\rm ex}^{\rm ^{12}CN\, N=1\to0}$ $\sim$ (3--5) K. Therefore the CMB
corrections must be included.

\subsubsection{Combined Correction}

When the Planck equation and the CMB are both considered, $T_{\rm ex, ^{12}CN}$
in Equation \ref{eq:Savage_and_Milam_formula} is replaced by $J(T_{\rm ex,
^{12}CN})-J(T_{\rm CMB})$, where $J(T)$ is the Planck
temperature (i.e., the Rayleigh-Jeans equivalent temperature in \citet{Mangum2015}): $J(T)$=$ c^2/(2k_{\rm B}\nu^2) B(\nu, T)$, adopting the
Planck equation for $B(\nu, T)$ (see Appendix~\ref{Appendix:12C13Cderiv}
for the quantitative details).

Starting from Equations ~\ref{eq:Nratio_eq_tauratio} and
~\ref{eq:Tbratio_eq_taufunction}, we consider the complete expression of
the Planck's Equation (i.e., abandon the R-J approximation) and the radiative
transfer contribution from the CMB:

\begin{equation}
\label{eq:eq_12C13C_hfs}
  \begin{aligned}
	\frac{\rm{^{12}C}}{\rm{^{13}C}} & = 
         \frac{N_{\rm ^{12}CN}}{N_{\rm ^{13}CN}}\\
         & = -\frac{R^{\rm m/sum}_{13}}{R^{\rm m/sum}_{12}}\frac{\tau_{\rm m, ^{12}CN}}{{\rm ln}[1-(1-e^{-\tau_{\rm m, ^{12}CN}})\frac{T_{\rm mb,^{13}CN}}{T_{\rm mb,^{12}CN}}]}, \\
         & = -\frac{R^{\rm m/sum}_{13}}{R^{\rm m/sum}_{12}}\frac{\tau_{\rm m, ^{12}CN}}{{\rm ln}[1-(1-e^{-\tau_{\rm m, ^{12}CN}})\frac{T^\star_{\rm A,^{13}CN}/\eta_{\rm mb, ^{13}CN}}{T^\star_{\rm A,^{12}CN}/\eta_{\rm mb, ^{12}CN}}]}
  \end{aligned}
\end{equation}

\noindent
where $R^{\rm m/sum}_{12}$ and $R^{\rm m/sum}_{13}$ are line intensity ratios between 
the main HfS component and the sum of all HfS components, for $^{12}$CN and
$^{13}$CN $N=1\to0$, respectively. In Table \ref{tab:transition_and_main_beam_efficiency}, we list the relative intensities of all components of \twCN\ $N=1\to0$ and the two strongest components of \thCN\ $N=1\to0$.  $T^\star_{\rm A,^{12}CN}$ and $T^\star_{\rm A,^{13}CN}$ are the antenna temperatures of \twCN\ and \thCN\ $N=1\to0$ main components, respectively. The main beam efficiency at the frequency of \twCN\ and \thCN\ main components are $\eta_{\rm mb, ^{12}CN}$ and $\eta_{\rm mb, ^{13}CN}$, respectively, listed in Table~\ref{tab:transition_and_main_beam_efficiency}.

The \Rftfif\ can also be derived from the same formula by replacing the parameters of \thCN\ $N=1\to0$ with parameters of \CftN\ $N=1\to0$. The relative intensities of the two strongest components of \CftN\ $N=1\to0$ and the IRAM main beam efficiencies at the frequencies of these transitions are also listed in Table~\ref{tab:transition_and_main_beam_efficiency}.

\subsection{Converting Flux Ratios to $T_{\rm b}$ ratios}

We assume that all line components share the same Gaussian-like line profile and the optical depth broadening does not play a significant role. 
In this case, the ratios between the integrated line flux  $I_{\rm main}$/$I_{\rm sat}$  can represent the ratios between line brightness temperature $T_{\rm b, main}$/$T_{\rm b,sat}$ (assuming nearly identical line frequencies).

For targets without \thCN\ or \CftN\ $N=1\to0$ detections, we
estimate the upper limits of the total velocity-integrated intensity $I$ of 
the two strongest lines.

\subsection{Radiative transfer for multiple \Tex\ layers}
\label{sec:derivation_Two_Tex_layer_model}

\begin{figure}
\includegraphics[scale=0.42]{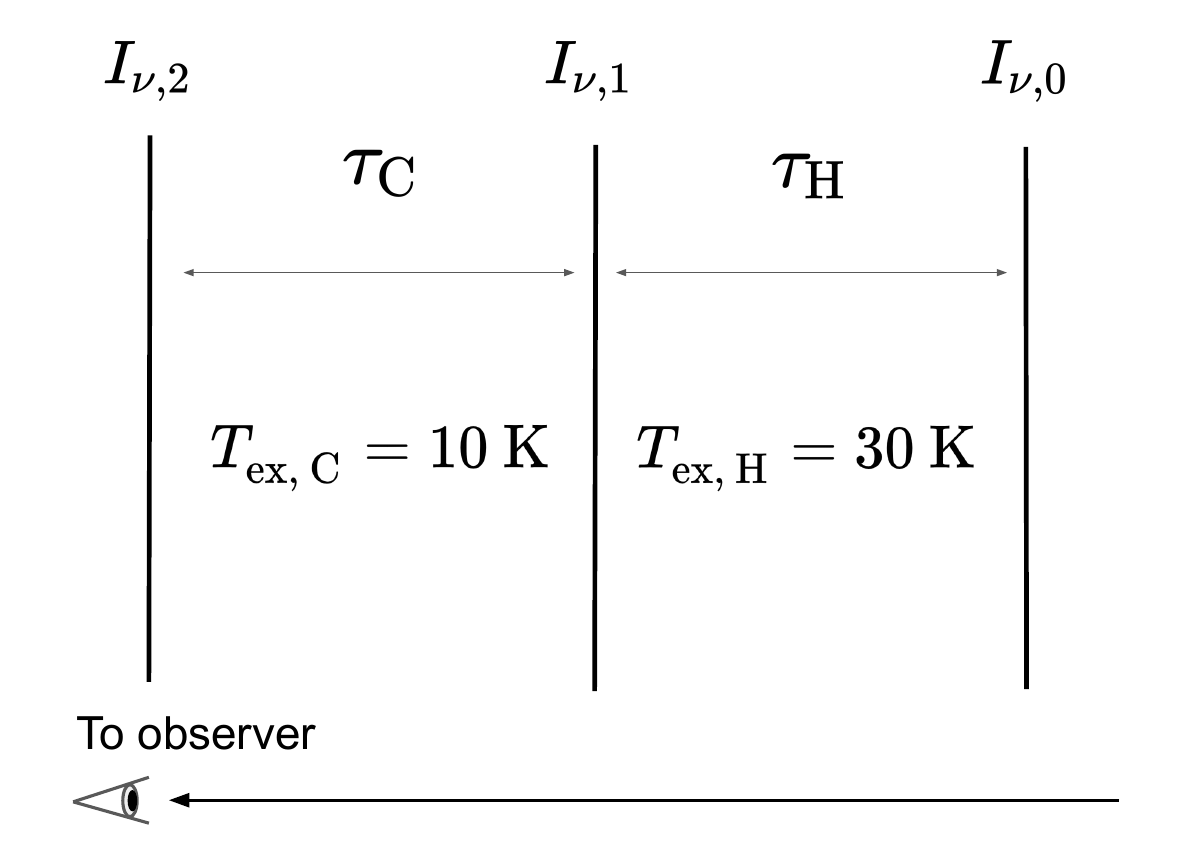}
\centering
	\caption{The toy model of molecular clouds with two \Tex\ layers. \label{fig:Two_layer_model}.} 
\end{figure}

In real molecular clouds, the excitation temperature often shows an inhomogeneous distribution 
inside molecular clouds, as multiple \Tex\ layers \citep[e.g., ][]{Zhou1993,Myers1996}. To test for the possible effects of such excitation temperature differentials
on the derivation of \Rtwth\ and \Rftfif, we set up a simple 
toy model with two different \Tex\ layers (Fig. ~\ref{fig:Two_layer_model}). In this model, 
the background layer has a high excitation temperature $T_{\rm ex, H}$ (H for hot) while 
the foreground layer has a low excitation temperature $T_{\rm ex, C}$ (C for cold). 
Detailed description and derivation are shown in Appendix~\ref{Appendix:Two_Tex_layer_model}.

Such a model indicates that the column density ratio and the brightness temperature ratio estimated from optical depth would systematically deviate from the intrinsic ratios with a simplified one-layer assumption. This is because the measured excitation temperature of \twCN\ and \thCN\ $N=1\to0$ will change with the optical depth with multiple \Tex\ layers. In our toy model, for $T_{\rm ex, H}= 30$ K, $T_{\rm ex, C}= 10$ K, $\tau_{\rm H}=\tau_{\rm C}=\tau$ and \Rtwth(intrinsic)=60, the intrinsic column density ratio and the intrinsic brightness temperature ratio will be $\sim$ 10\% and $\sim$ 15\% lower than the estimated ratios, respectively, when the optical depth of \twCN\ main component $\tau^{\rm main}_{\rm eff} \sim 1$ (i.e. $\sim$ 10\% and $\sim$ 15\% deviations from Eq. \ref{eq:Nratio_eq_tauratio} and Eq. \ref{eq:Tbratio_eq_taufunction}, respectively). When $\tau^{\rm main}_{\rm eff} \sim 1$, which is close to our measured $\tau$ of \twCN\ main components in our targets, such deviations will cause a $\sim$ 17\% decrease of the derived \Rtwth\ compared to the intrinsic ratio (red dashed line in Fig.~\ref{fig:two_model_12Cto13C_with_tau} and details in Appendix~\ref{Appendix:Two_Tex_layer_model}).

Multiple \Tex\ layers are significant when velocity spread is limited within linewidths of micro-turbulence and thermal motion, or when more than one fore/background cloud has identical $V_{\rm LSR}$. Observational evidence of self-absorption in $\rm ^{12}CO$~\citep{Enokiya2021}, $\rm HCO^{+}$~\citep{Richardson1986}, and even $\rm ^{13}CO$~\citep{Sandell2010}, which indicates multiple \Tex\ layers, are found in multiple sources of the CN sample \citep{Wouterloot1989,Savage2002}. 

Other issues can also influence the derivation of \Rtwth. For example, the excitation temperature of \twCN\ may be higher than that of \thCN\ because of large optical depths and radiative trapping thermalizing the lines of the most abundant isotopologue at densities $\rm n=n_{crit, thin} (1-e^{-\tau})/\tau < n_{crit, thin}$. Non-LTE will also affect the derivation by changing the relative intensities between \twCN\ line components. These are beyond the scope of this work and wait for future investigation.

\subsection{The method to derive \Rtwth\ in extragalactic targets}

\begin{figure}
    \includegraphics[scale=0.29]{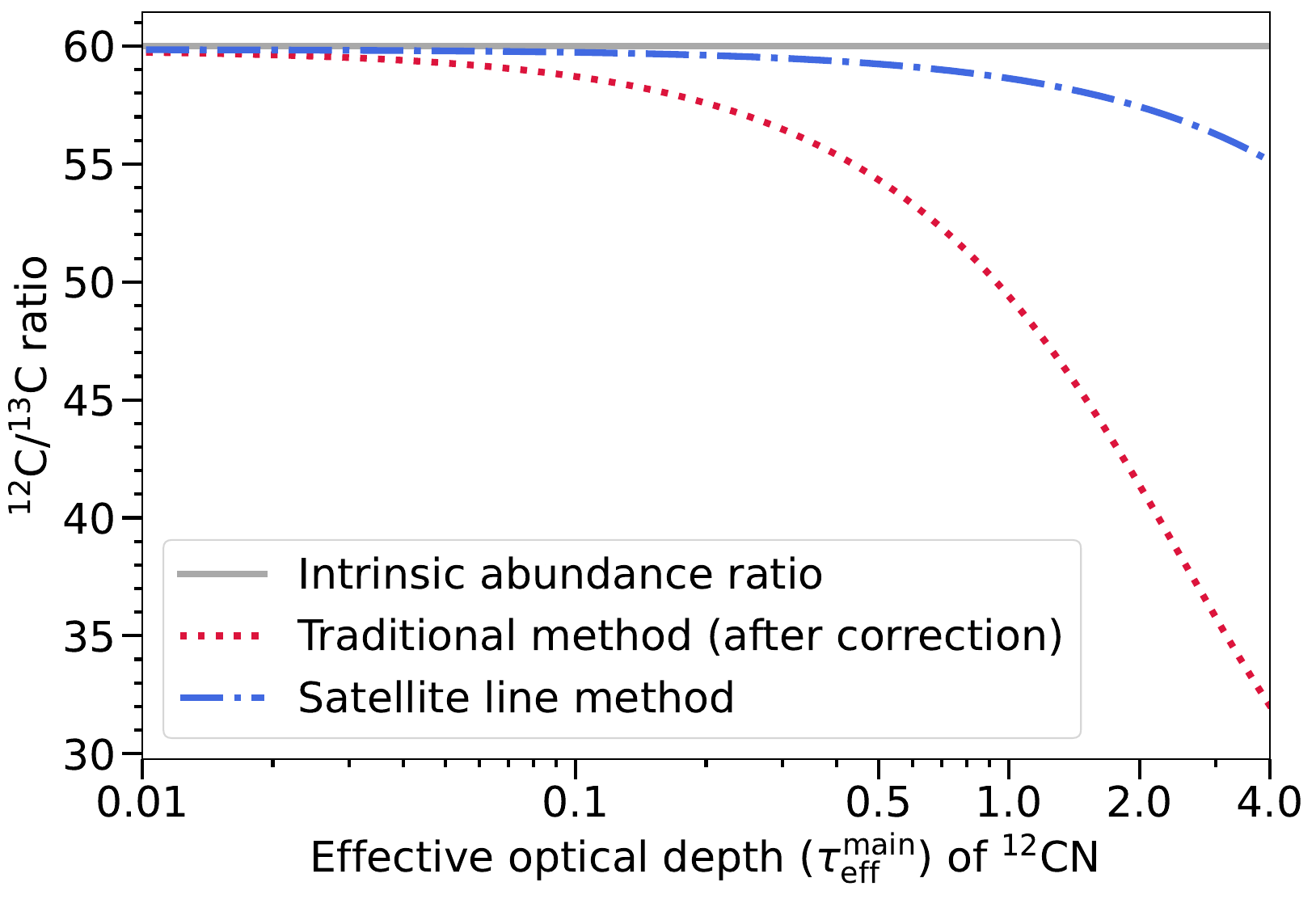}
\centering
	\caption{The theoretical values of derived \Rtwth\ compared with the intrinsic values of \Rtwth\ for the traditional HfS method and the satellite line method, varying with the optical depth of \twCN\ main component. We assume that $T_{\rm ex, H}=30$ K, $T_{\rm ex, C}=10$ K, $\tau_{\rm H}=\tau_{\rm C}=\tau$ and \Rtwth\  (intrinsic) = 60. \label{fig:two_model_12Cto13C_with_tau} } 
\end{figure}

The observations of CN isotopologues in the extragalactic star-burst galaxies provide \Rtwth\ derived from the following equation \citep{Henkel2014,Tang2019}:

\begin{equation}
    \frac{\rm ^{12}C}{\rm ^{13}C} = \frac{I_{\rm I, ^{12}CN}\cdot \frac{\tau_{\rm I}}{1-e^{-\tau_{\rm I}}}+I_{\rm II, ^{12}CN}\cdot \frac{\tau_{\rm II}}{1-e^{-\tau_{\rm II}}}}{1.082I_{\rm ^{13}CN}}
\end{equation}

Here, $I_{\rm I, ^{12}CN}$ and $I_{\rm II, ^{12}CN}$ are the total integrated intensities of \twCN\ $N=1\to0$ $J=3/2\to1/2$ and \twCN\ $N=1\to0$ $J=1/2\to1/2$, respectively. $\tau_{\rm I}$ and $\tau_{\rm II}$ are the total optical depth of the blended line components in $N=1\to0$ $J=3/2\to1/2$ and $J=1/2\to1/2$, respectively. 

This method will have the same deviation to \Rtwth\ in Section~\ref{sec:derivation_Two_Tex_layer_model} when the targets have complex excitation layers. Because of the large optical depth of \twCN\ and the small optical depth of \thCN, the effective \Tex\ of \twCN\ transitions will be smaller than that of \thCN\ in our toy model, which causes the underestimation of \Rtwth. In addition, the optical depth is derived from an equation similar to Eq.~\ref{eq:Tbratio_eq_taufunction} \citep[e.g., Eq. (1) in][]{Tang2019}. In our toy model, the $\tau_{\rm I}$ and $\tau_{\rm II}$ derived in this way may be overestimated. The \Tex\ layers will be more complex in reality than in our toy model, while differential excitation of
the lines due to radiative trapping will only add to such complexity. Thus the \Rtwth\ ratio estimated in this way still contains highly unclear uncertainties.

\section{New method: deriving \CtwCth\ and \NftNfif\ with CN satellite lines and isotopologues}
\label{sec:derive_C_N_ratio_from_satellite_line}

The satellite transitions of \twCN\ $N=1\to0$, including $J=1/2\to1/2, F=1/2\to1/2$ (113.123 GHz) and $J=3/2\to1/2, F=1/2\to3/2$ (113.520 GHz) are expected to be optically-thin because the intrinsic relative line strengths of \twCN\ $N=1\to0$ at these transitions are $\sim 3.6 \%$ of that of the main component, which means an optical depth $\tau_{\rm s, ^{12}CN} \sim 3.6 \%$ of $\tau_{\rm main,  ^{12}CN}$.

Therefore, the effect of different \Tex\ layers on the derivation to \Rtwth\ and \Rftfif\ should be small based on our analysis. If the optical depth of the \twCN\ $N=1\to0, J=3/2\to1/2, F=1/2\to3/2$ is less than 1, the theoretical deviation of the derived \Rtwth\ will be $\lesssim$ 0.5\%. This deviation is much smaller than the deviation in the traditional HfS method and can be ignored (see Fig~\ref{fig:two_model_12Cto13C_with_tau}).

The \Rtwth\ can be derived from the following equation:

\begin{equation}
\label{eq:12C_13C_from_satelliteline}
    \frac{\rm ^{12}C}{\rm ^{13}C}=\frac{R^{\rm I/sum}_{13}}{R^{\rm I/sum}_{12}}\frac{I_1+I_2}{I_{\rm ^{13}CN}} 
\end{equation}

\noindent
Here, $R^{\rm I/sum}_{12}$ is the integrated intensity ratio between the sum of the two satellite lines and the sum of all the nine lines of \twCN\ $N=1\to0$. $R^{\rm I/sum}_{13}$ is the ratio between the integrated intensities of the two strongest line components and that of all the components of \thCN\ $N=1\to0$. $I_1$ and $I_2$ are the integrated intensities of \twCN\ $N=1\to0$ $J=1/2\to1/2, F=1/2\to1/2$ and \twCN\ $N=1\to0$ $J=3/2\to1/2, F=1/2\to3/2$, respectively. $I_{\rm ^{13}CN}$ is the integrated intensity of the two strongest components of  \thCN\ $N=1\to0$. With the criteria in Section~\ref{sec:upperlimit_es}, if $I_1 + I_2$ is larger than the 3-$\sigma$ value of the corresponding integrated intensity, we treat it as a detection.

This method still assumes a common \Tex\ among the energy levels involved in the lines used \citep[i.e., the CTEX assumption,][their Section 12]{Mangum2015}. However, we sum up the integrated intensities of the two satellite lines to 
deduce the effect from hyperfine anomalies of \twCN, which has been observed in several studies \citep[e.g.,][]{Bachiller1997,Hily-Blant2010}.

Similarly, we can derive \Rftfif\ with the same method by replacing the relative intensity ratio and the integrated line intensity of \thCN\ $N=1\to0$ with those of \CftN\ $N=1\to0$.

\section{Results}
\label{sec:results}
With the improved HfS method (the traditional method after corrections) listed in Section~\ref{sec:update_HfS_method} and the new method (the satellite line method) in Section~\ref{sec:derive_C_N_ratio_from_satellite_line}, we obtained the Galactic \Rtwth\ and \Rftfif\ gradients and add our new \Rtwth\ and \Rftfif\ results in the Galactic outer disk. We also discuss the differences between these gradients from different methods.

\subsection{Line Detection and HfS fitting results}

Among the total 15 sources, 11 targets have detections (S/N $> 3$) of more than 
two $\rm ^{12}CN $ $N=1\to0$ satellite lines. 
One target, SUN15\,18, has a detection of the main component of \twCN\ $N=1\to0$.  
The other three targets only have non-detections of \twCN. 

Two sources, G211.59 and WB89\,380, show $\rm ^{13}CN$ $N=1\to0$ detections. 
G211.59 has a robust $\rm ^{13}CN$ $N=1\to0$ detection which has an S/N $\sim$ 8. 
The two blended main components of $\rm ^{13}CN$ in WB89 380 are weakly detected with a total signal-to-noise ratio of $\sim$ 4.5. 
We did not detect \thCN\, in WB89 391, at an $\sim 8\times$ better noise level than that 
 of ~\citet{Milam2005}. Further comparison will be shown in Section \ref{sec:WB89_391}.  
$\rm C^{15}N$ $N=1\to0$ was only detected in G211.59. 

\begin{figure*}
        \centering
		\includegraphics[scale=0.14]{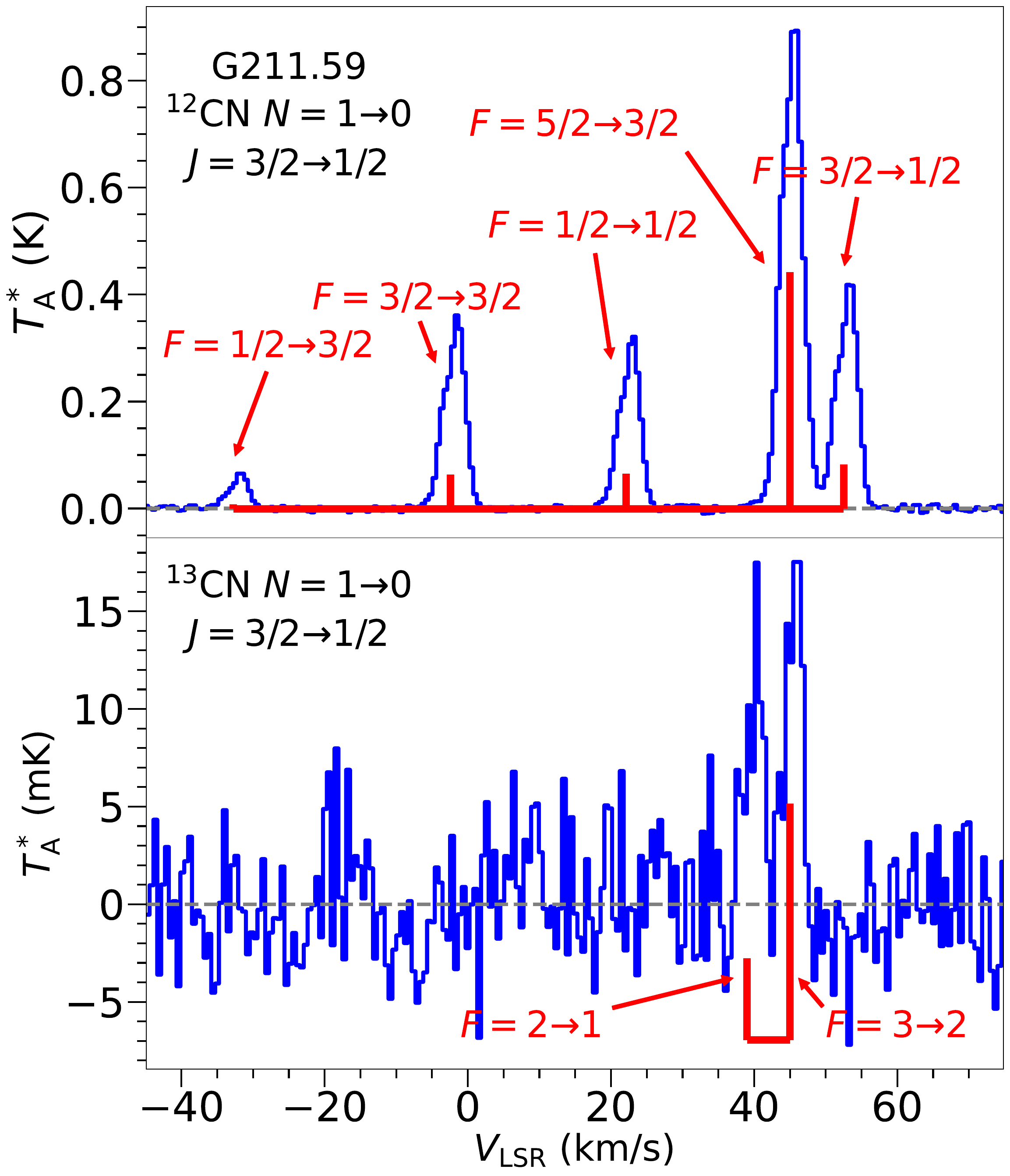}
		\includegraphics[scale=0.14]{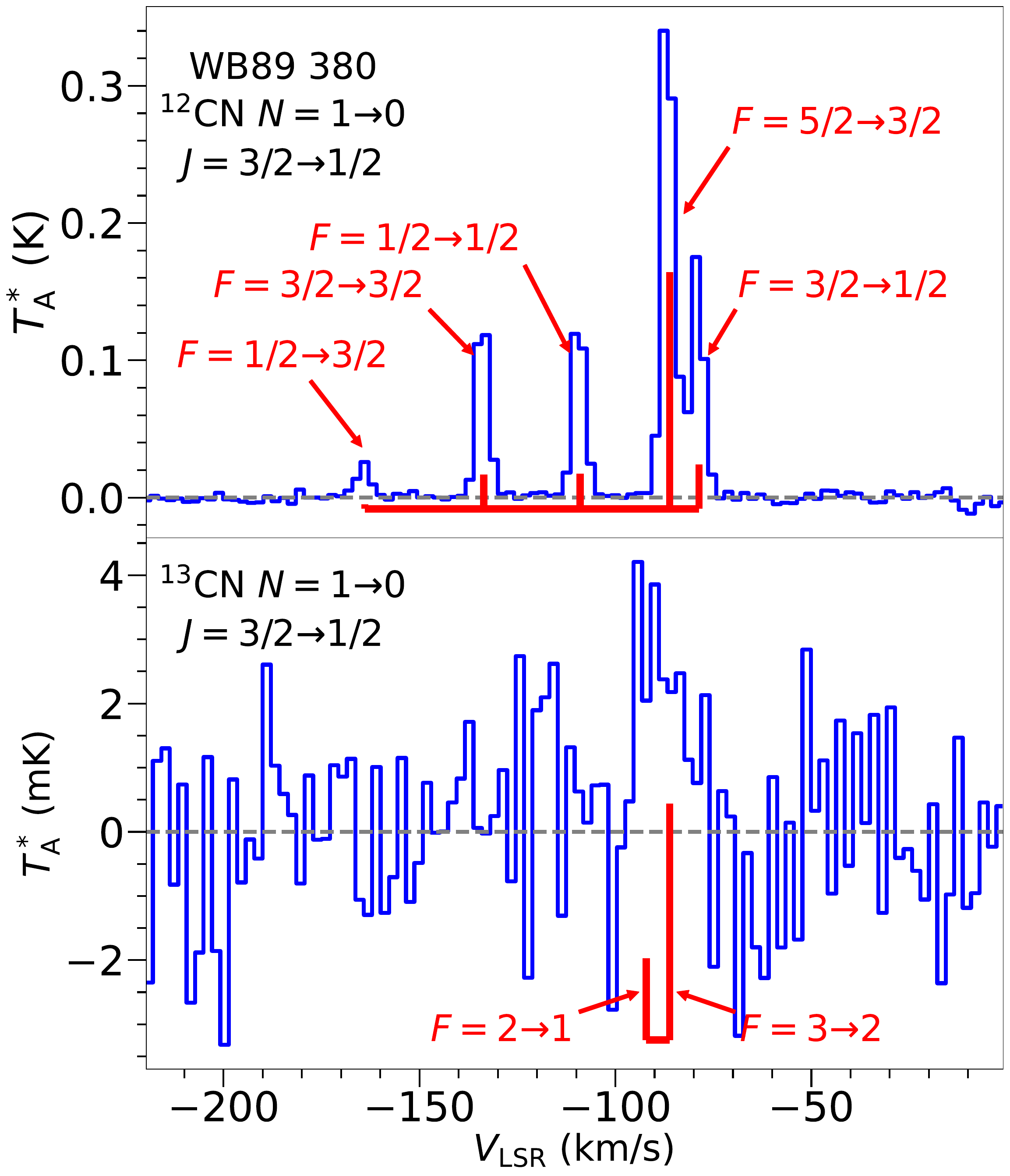}
		\includegraphics[scale=0.14]{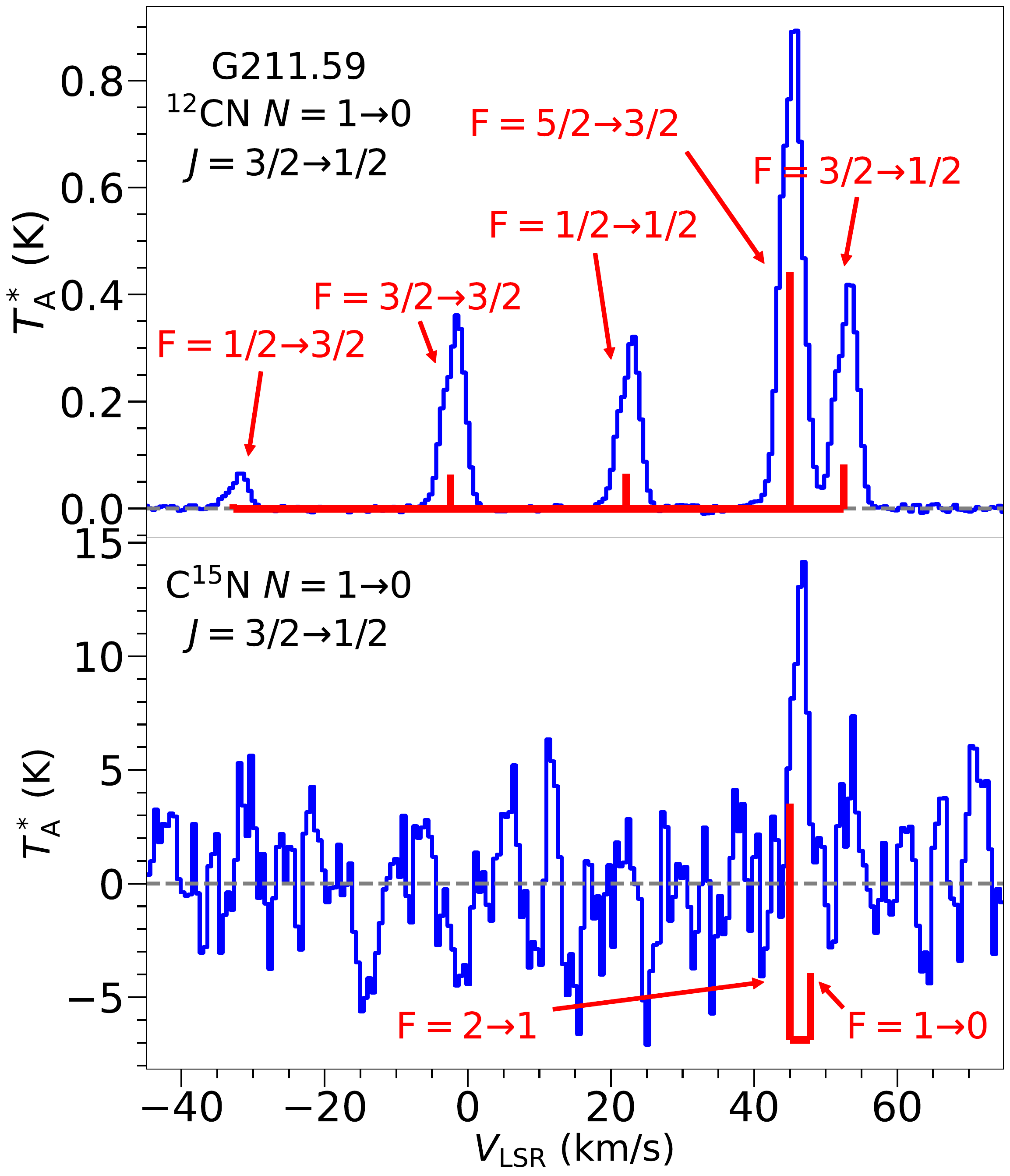}
\caption{Spectra of $\rm ^{12}CN$, $\rm ^{13}CN$ and $\rm C^{15}N$ . The upper panels show spectra of $\rm ^{12}CN$. The bottom panels are spectra of $\rm ^{13}CN$ in the left panel, the middle panel, and $\rm C^{15}N$ in the right panel. Blue histograms show the spectra. Red solid lines show the relative location of hyperfine-structure transitions, labeled by red arrows. The gray dashed lines show the baseline of each spectrum. Left: The spectra of G211.59 with no smoothness, with the velocity resolution at $\sim$ 0.53 $\rm km\cdot s^{-1}$ Middle: The spectra of WB89 380 after smoothing four channels into one, with the velocity resolution at $\sim$ 2.15 $\rm km\cdot s^{-1}$ Right: The spectra of G211.59, showing the $\rm C^{15}N$ detection with the velocity resolution at $\sim$ 0.53 $\rm km\cdot s^{-1}$.} \label{fig:detected_13CN_C15N_spectra}
\end{figure*}

We list the antenna temperatures of $\rm ^{12}CN$, $\rm ^{13}CN$ and  $\rm C^{15}N$ $N=1\to0$ in Table~\ref{tab:results_HfS}. The detected \twCN, \thCN\ and $\rm C^{15}N$ spectra are shown in Fig. ~\ref{fig:detected_13CN_C15N_spectra}. We show the spectra of \twCN\ and \thCN\ $N=1\to0$ non-detected transitions in Appendix~\ref{appendix:line_spectra}.

For sources with more than two $\rm ^{12}CN$ satellite line detections, our HfS fitting 
(Fig.~\ref{fig:HfS_fitting_spectra} in Appendix ~\ref{appendix:line_spectra}) shows that $\tau_{\rm 12, main} < 1$ in most of them (Table~\ref{tab:results_HfS}). 
Among them, three targets have large errors of $\tau_{\rm 12, main}$, so we do not include them in the following analysis.

\subsection{\CtwCth\ gradient from the HfS method}

\begin{figure*}
\includegraphics[scale=0.40]{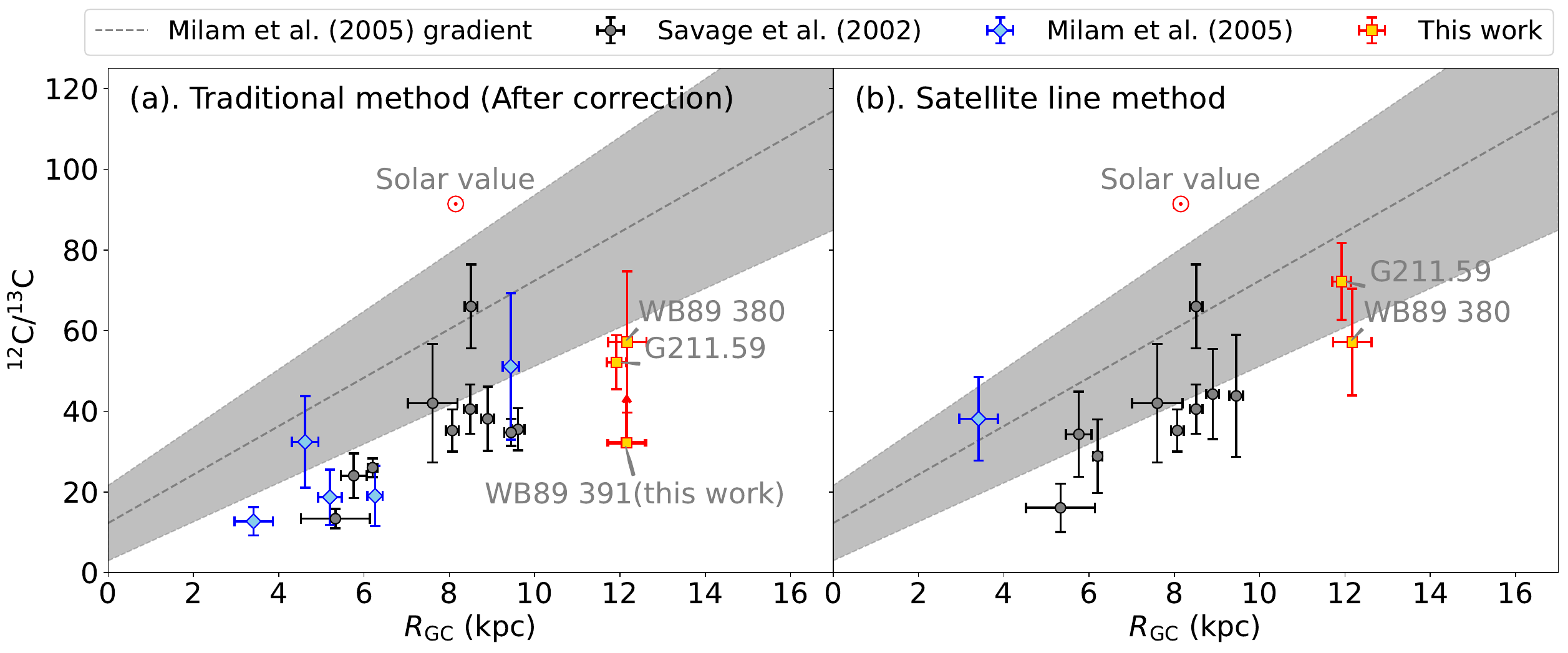}
\centering
	\caption{Ratio of $\rm{^{12}C/^{13}C}$ as a function of Galactocentric distance (\Rgc). (a): \Rtwth\ derived from the corrected HfS method. (b). \Rtwth\ derived from the optically-thin \twCN\ satellite line method. Black circles and blue diamonds are corrected values of results provided in~\citet{Savage2002} and~\citet{Milam2005}, respectively. Red squares are our new \Rtwth\ results of G211.59, WB89 380, and WB89 391. The grey dashed line with blocks shows the previous gradient provided in~\citet{Milam2005} with 1-$\sigma$ error, which is $\rm{^{12}C/^{13}C=6.01(1.19)\it{D}_{\rm{GC}}\rm{+12.28(9.33)}}$. \label{fig:12C_13C_grad}. The red circle with a central dot indicates the Solar value of \Rtwth\ $= 91.4\pm 1.3$ in this figure \citep{Ayres2013}, and in Figures ~\ref{fig:12C_13C_grad_diff_tracer} and~\ref{fig:12C_13C_show_revis}. }
\end{figure*}

With the antenna temperatures of $\rm ^{12}CN$ and $\rm ^{13}CN$ $N=1\to0$, the \Rtwth\ derived with Eq.~\ref{eq:eq_12C13C_hfs}  are listed in Table~\ref{tab:results_HfS}. We compare our newly measured \Rtwth\ with those from CN observations reported in \citet{Savage2002,Milam2005}. We update the Galactocentric distances of their targets and re-derive their \Rtwth\ with Eq.~\ref{eq:eq_12C13C_hfs}. In Fig.~\ref{fig:12C_13C_grad} (a), we show the Galactic \Rtwth\ gradients based on $\rm ^{12}CN/^{13}CN$ derived from Eq.~\ref{eq:eq_12C13C_hfs}, which is systematically lower than the gradient reported by \citet{Milam2005}.

\begin{table*}
\caption{$\rm{^{12}C/^{13}C}$ and $\rm ^{14}N/^{15}N$ ratios in our targets (The traditional HfS method).}
\label{tab:results_HfS}
\begin{threeparttable}
\begin{tabular}{ccccccc}
\hline
\hline
Sources              & $\tau_{\rm main}$    & $T_{\rm A, ^{12}CN}^\star$ & $T_{\rm A, ^{13}CN}^\star $ & $\rm{^{12}C/^{13}C}$ & $T_{\rm A, C^{15}N}^\star $ & $\rm ^{14}N/^{15}N$\\
                     &                      & ($10^{-1}\,$K)             & ($10^{-3}\,$K)              &                      & ($\rm{10^{-3}}\:$K)         &\\
\hline
G211.59              & 0.86 $\pm$ 0.03              & 7.85 $\pm$ 0.03                 & 13.1 $\pm$ 1.7      & 52.2 $\pm$ 6.7   & 8.7 $\pm$ 1.6     & 166 $\pm$ 32            \\
G37.350              & 0.6 $\pm$ 1.1                & 1.28 $\pm$ 0.09                   & $\le$ 16          & $\ge$ 6.1        & $\le$  16         & $\ge$  13               \\     
G44.8                & 0.71 $\pm$ 0.66              & 1.58 $\pm$ 0.09                   & $\le$ 13          & $\ge$ 9.4        & $\le$  16         & $\ge$  17               \\
IRAS0245             & 0.87 $\pm$ 0.23              & 3.29 $\pm$ 0.06                   & $\le$ 9.6         & $\ge$ 30         & $\le$  13         & $\ge$  46               \\
SUN15 14N            & -                            & $\le$ 0.6                       & $\le$ 3.6           & -                & $\le$  43         & -                       \\
SUN15 18$\rm ^{a}$   & -                            & 0.7 $\pm$ 0.1                   & $\le$ 3.1           & -                & $\le$  30         & -                       \\
SUN15 21$\rm ^{b}$   & 0.56 $\pm$ 11                & 1.15 $\pm$ 0.20                    & $\le$ 28         & $\ge$ 3.0        & $\le$  22         & $\ge$  8.4              \\
SUN15 34             & -                            & $\le$ 0.6                       & $\le$ 4.2           & -                & $\le$  48         & -                       \\
SUN15 56$\rm ^{b}$   & 1.33 $\pm$ 10                & 1.02 $\pm$ 0.08                   & $\le$ 17          & $\ge$ 6.0        & $\le$  21         & $\ge$  12               \\
SUN15 57$\rm ^{b}$   & 0.63 $\pm$ 11                & 1.04 $\pm$ 0.11                   & $\le$ 16          & $\ge$ 5.1        & $\le$  24         & $\ge$  8.1              \\
SUN15 7W             & -                            & $\le$ 0.6                       & $\le$ 4.3           & -                & $\le$  43         & -                       \\
WB89 380             & 0.58 $\pm$ 0.05              & 4.22 $\pm$ 0.03                 & 5.7$\pm$1.7         & 57$\pm$17        & $\le$  5.2        & $\ge$  134              \\
WB89 391             & 0.98 $\pm$ 0.12              & 3.62 $\pm$ 0.05                 & $\le$ 10            & $\ge$ 32         & $\le$  12         & $\ge$  59               \\
WB89 437             & 0.26 $\pm$ 0.07              & 4.97 $\pm$ 0.04                   & $\le$ 5.3         & $\ge$ 62         & $\le$  5.8        & $\ge$  122              \\
WB89 501             & 0.50 $\pm$ 0.40              & 2.02 $\pm$ 0.07                  & $\le$ 13           & $\ge$ 12         & $\le$  17         & $\ge$  19               \\
\hline
\end{tabular}

\begin{tablenotes}
\footnotesize
\item[a.] Failed to do HfS fitting for satellite lines of \twCN\ $N=1\to0$. 
\item[b.] Huge error bars of $\tau$ in these targets.
\end{tablenotes}

\end{threeparttable}
\end{table*}

\subsection{\NftNfif\ gradient from the HfS method}

\begin{figure*}
\includegraphics[scale=0.40]{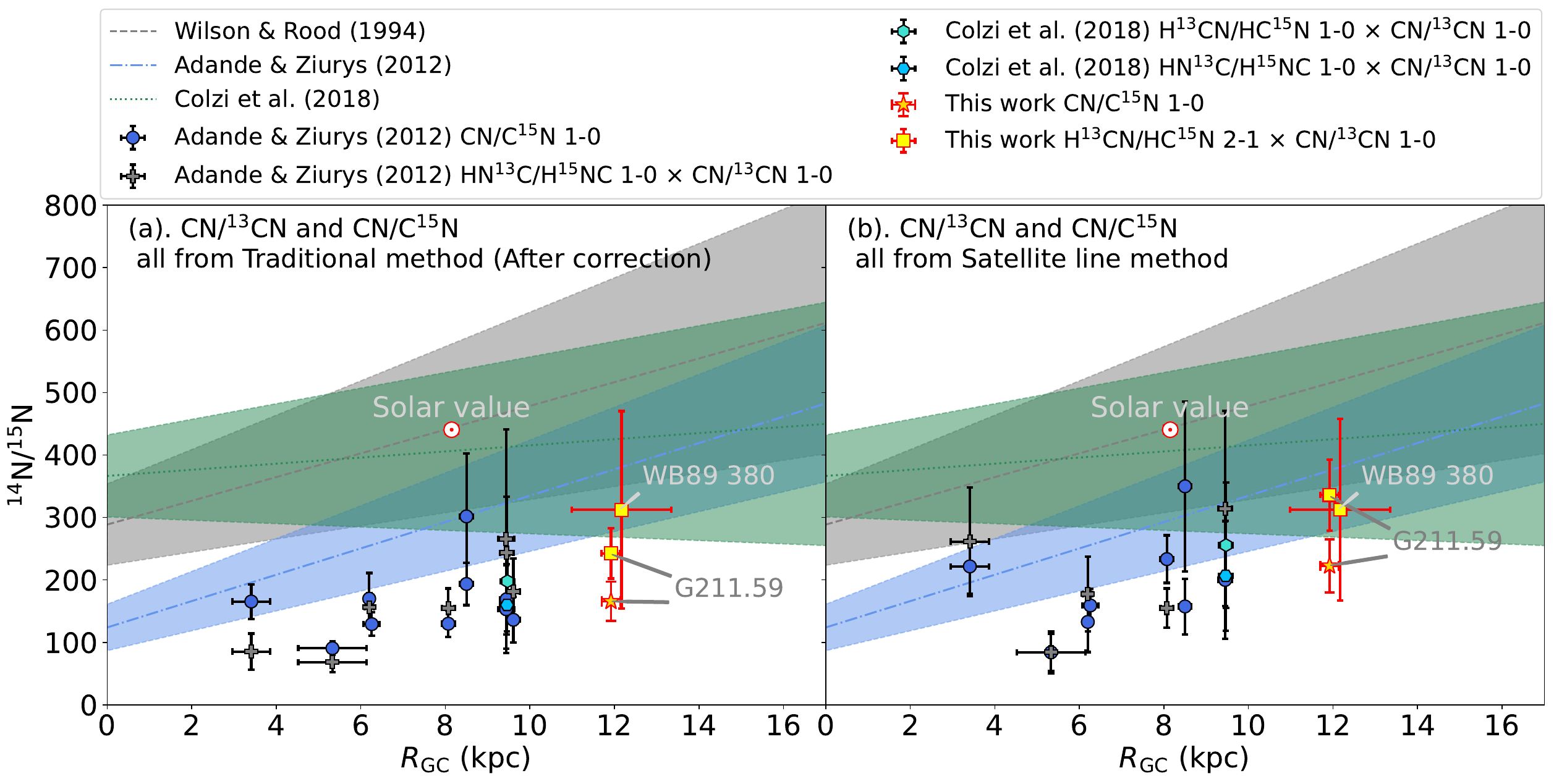}
\centering
\caption{\label{fig:14N_15N_grad}The Galactic $\rm{^{14}N/^{15}N}$ gradient. (a). The gradient with \Rftfif\ derived from the CN HfS method and from $\rm H^{13}CN/HC^{15}N$ multiplying with \Rtwth\ obtained by the HfS method. (b). The gradient with \Rftfif\ from CN satellite line method and from $\rm H^{13}CN/HC^{15}N$ multiplying with \Rtwth\ obtained by the satellite line method.  The red star is \Rftfif\ derived with CN $N=1\to0$ of G211.59. The red squares are \Rftfif\ from $\rm H^{13}CN/HC^{15}N$ multiplying \Rtwth\ in G211.59 and WB89 380. For comparison, we show the \Rftfif\ derived from $\rm CN/C^{15}N$~\citep[royal blue circles]{Adande2012}; $\rm HN^{13}C/H^{15}NC \times ^{12}C/^{13}C$ in ~\citet[grey crosses]{Adande2012} and~\citet[deep sky blue hexagons]{Colzi2018}; and $\rm H^{13}CN/HC^{15}N \times ^{12}C/^{13}C$ ~\citep[turquoise hexagons]{Colzi2018}. We also show the Galactic \Rftfif\ gradients provided in ~\citet[grey dashed line]{Wilson1994},~\citet[blue dashdot line]{Adande2012} and~\citet[green dotted line]{Colzi2018}. The corresponding color blocks show the 1-$\sigma$ ranges of these gradients. The red circle with a central dot shows the Solar value of \Rftfif $=440.5\pm5.8$ from the protosolar nebula \citep{Marty2011}.  
}
\end{figure*}

Fig. \ref{fig:14N_15N_grad} (a) displays our Galactic \Rftfif\ results from the HfS method. Besides \Rftfif\ derived from the CN isotopologues with the HfS method \citep{Adande2012}, we show the \Rftfif\ from other tracers together. We excluded targets for which \Rftfif\ was calculated by multiplying a fitted gradient with \Rtwth\ from the literature~\citep[e.g.,][]{Milam2005}. 

Our \Rftfif\ gradient is also consistently lower than gradients reported in previous studies~\citep[e.g.,][]{Wilson1994, Adande2012, Colzi2018}. We do revise $\rm ^{12}CN/C^{15}N$ in \citet{Adande2012}  as what we do also for $\rm ^{12}CN/^{13}CN$ in \citet{Savage2002} and \citet{Milam2005}.
In addition, Figure \ref{fig:14N_15N_grad} shows the \Rftfif\ derived from $\rm H^{13}CN/HC^{15}N$ multiplying \Rtwth\ in two of our targets with both \Rtwth\ and detected \HthCN\ and \HCfifN\ $J=2\to1$. More descriptions are in Appendix~\ref{Appendix:14N_15N_from_HCN}.

\subsection{$\rm ^{12}C/^{13}C$ and $\rm ^{14}N/^{15}N$ gradients from CN $N=1\to0$ satellite lines}

Among eleven targets with the detection of \twCN\ $N=1\to0$ main component, five of them have detected satellite lines. 
For targets in \citet{Savage2002} and \citet{Milam2005}, we use the peak temperature of $J=3/2\to1/2, F=1/2\to3/2$ provided in \citet{Savage2002} and \citet{Adande2012} to derive \Rtwth. In Table~\ref{tab:results_opthin}, we list the \Rtwth\ of our targets derived in the optically-thin condition. Fig~\ref{fig:12C_13C_grad} (b) shows the Galactic \Rtwth\ gradient from CN satellite lines. This gradient from optically thin satellite lines is systematically higher than the one derived from optical depth correction with HfS fitting. 

In Table~\ref{tab:results_opthin}, we also show the \Rftfif\ of our targets derived from optically-thin \twCN\ satellite lines and \thCN. Fig.~\ref{fig:14N_15N_grad} (b) shows the Galactic \Rftfif\ gradient where all the ratios from CN isotopologues have been revised to the optically-thin results. The \Rftfif\ derived from optically-thin CN satellite lines is higher than those from optical depth correction, which makes the Galactic \Rftfif\ gradient higher than the one in Fig.~\ref{fig:14N_15N_grad} (a).  In addition, we also show \Rftfif\ derived from $\rm H^{13}CN/HC^{15}N$ multiplying \Rtwth, where \Rtwth\ is obtained from \twCN\ optically-thin satellite lines, discussed in Section~\ref{sec:12Cto13C_impact_to_14Nto15N} and more details are in Appendix~\ref{Appendix:14N_15N_from_HCN}.

\begin{table*}
\caption{$\rm{^{12}C/^{13}C}$ and $\rm ^{14}N/^{15}N$ ratios in our targets (The optically-thin $\rm ^{12}CN$ satellite line method).}
\label{tab:results_opthin}
\begin{threeparttable}
\begin{tabular}{cccccc}
\hline
\hline
Sources              & $I_1+I_2$   & $I_{\rm ^{13}CN} $  & $\rm{^{12}C/^{13}C}$ & $I_{\rm C^{15}N} $ & $\rm{^{14}N/^{15}N}$ \\
                     & ($10^{-1}\, \rm K \cdot  km\ s^{-1}$)     & ($\rm 10^{-3}\,K \cdot  km\  s^{-1}$)   &  & ($\rm 10^{-3}\,K \cdot km\  s^{-1}$)  &  \\
\hline
G211.59   & 5.90 $\pm$ 0.17 & 101 $\pm$  13   & 72.2 $\pm$ 9.5  &  64 $\pm$  12  & 222 $\pm$ 43      \\
G37.350   & $\le$ 1.6       & $\le$ 87        & -               &  $\le$ 78      & -                 \\
G44.8     & $\le$ 1.9       & $\le$ 79        & -               &  $\le$ 84      & -                 \\
IRAS0245  & 0.76 $\pm$ 0.25 & $\le$ 35        & $\ge$ 27        &  $\le$ 44      & $\ge$ 42           \\
SUN15 21  & $\le$ 1.6       & $\le$ 91        & -               &  $\le$ 64      & -                 \\
SUN15 56  & $\le$ 0.81      & $\le$ 49        & -               &  $\le$ 48      & -                 \\
SUN15 57  & $\le$ 1.2       & $\le$ 49        & -               &  $\le$ 60      & -                 \\
WB89 380  & 2.61 $\pm$ 0.20 & 56  $\pm$  12   & 57 $\pm$ 13     &  $\le$ 37      & $\ge$ 174         \\
WB89 391  & 1.06 $\pm$ 0.21 & $\le$ 36        & $\ge$ 36        &  $\le$ 38      & $\ge$ 67          \\
WB89 437  & 1.57 $\pm$ 0.22 & $\le$ 33        & $\ge$ 59        &  $\le$ 33      & $\ge$ 116         \\
WB89 501  & $\le$ 0.96      & $\le$ 51        & -               &  $\le$ 63      & -                 \\
\hline
\end{tabular}


\end{threeparttable}
\end{table*}

\section{Discussion}
\label{sec:discussion}

\subsection{The revision to the \CtwCth\ gradient from the HfS method}
In this section, we discuss the Galactic \Rtwth\ and \Rftfif\ gradients along the Galactic disc that are derived by applying the corrections mentioned above and our new method from CN isotopologues.

\subsubsection{The updated Galactocentric distances}

We employ the updated Galactic rotation curve from \citet{Reid2019}. This choice leads to changes in the Galactocentric distances ($R_{\rm gc}$) of most sources in our sample compared to the values reported in \citet{Savage2002} and \citet{Milam2005}. 
For most targets in the Galactic inner (outer) disk, $R_{\rm gc}$ increases (decreases) after applying the new rotation curve in \citet{Reid2019}, with a typical difference of $\Delta R_{\rm gc} \sim 0-4$ kpc. One of our targets, WB89 437, has had its trigonometric parallax measured using VLBI in \citet{Reid2014},  which indicated a Galactocentric distance of $R_{\rm gc} = 13.10 \pm 0.38$ kpc. This value is consistent with the value of $R_{\rm gc} = 12.79 \pm 0.38$ kpc derived from the Parallax-Based Distance Calculator.

Most of our targets are located in the anti-center direction, where the distance determination is not affected by confusion at the tangent point curve. However, most $R_{\rm gc}$ values in the literature \citep[e.g.,][]{Brand1995, Savage2002, Milam2005, Wouterloot2008, Giannetti2014} were derived from the Galactic rotation curve measured more than twenty years ago \citep[e.g.,][]{Brand1986, Brand1988, Brand1993, McNamara2000}. This could introduce a strong bias in the derived abundance gradients. 

In contrast, the distances derived from the updated rotation curve from \citet{Reid2019} not only benefit from more accurate measurements of the trigonometric parallax with VLBI, but also  agree with the parallax-based distances of Galactic H{\sc ii} regions that rely on Gaia EDR3 data \citep{Mendez2022} for nearby targets. The updated Galactocentric distance estimates lead to a systematic reduction in the number of targets with $R_{\rm gc}$>10\,$\rm kpc$, yielding a steeper slope for the fitted gradient than the one in the literature where the previous, larger distances were used. For example, the Galactocentric distance of WB89~391 is now set to $\sim $12 kpc. This raises concerns about the previously reported high \Rtwth\ value for this source \citep{Milam2007thesis}  as this would seem reasonable for \Rgc\ at $\sim $16\,kpc, but not so for the updated distance of $\sim $12\,kpc
where a \CtwCth\ $\sim $134 seems rather high. After the update, the Galactocentric distance of the sample is  $R_{\rm gc}$<12\,kpc. Fig.~\ref{fig:12C_13C_show_revis} in Appendix~\ref{Appendix:effect_of_RJ_and_CMB} illustrates the changes in \Rtwth\ before and after our revision.

\subsubsection{Revised \Rtwth\ after the Planck Equation and CMB contribution corrections}

As expected, the \Rtwth\ values (for the transitions of \twCN, and \thCN\,  $N$=1-0) derived using Equation~\ref{eq:eq_12C13C_hfs} are systematically lower than those obtained using the R-J approximation and without considering the CMB temperature for targets in \citet{Savage2002,Milam2005}, see Fig.~\ref{fig:12C_13C_show_revis}. As we illustrated in Section~\ref{sec:effect_of_RJ_appro} and Section~\ref{sec:effect_of_Tcmb}, if the \Tex\ of \twCN\ is lower, the revision to \Rtwth\ will be larger. The \Tex\ of \twCN\ $N=1\to0$ of targets  in~\citet{Milam2005} is relatively lower (\Tex $\sim 3-7$ K) than those (\Tex $\sim 6-12$ K) from targets in \cite{Savage2002}, so the revisions of targets in \citet{Milam2005} are larger than those in \citet{Savage2002}. We note that this analysis assumes a common \Tex\ among the lines used. This CTEX assumption \citep[see also ][]{Mangum2015} is less constraining than LTE, and is likely to hold for the optically thin (or modestly optically
thick) lines from rare CN isotopologues (unlike the more abundant isotopologues of CO where radiative trapping of
more optically thick lines can yield different excitation temperatures among the isotopologues used).

\subsubsection{Constraint on the outer Galactic disk -- WB89 391}\label{sec:WB89_391}

WB89~391 is located in the outer Galactic disc and was previously observed by \citet{Milam2005,Milam2007thesis}. However, we did not detect the \thCN\ line for this target. As the original fluxes and their associated errors for the \twCN\ and \thCN\ lines were not reported in \cite{Milam2005,Milam2007thesis}, we could only make a rough estimate of the noise level based on the $N=1\rightarrow0$ spectra of \twCN\ and \thCN\ presented in \citet{Milam2007thesis}.

At the rest frequency of $\rm ^{13}CN$ $N=1\to0$ main component (108.78 GHz), the beam sizes of IRAM 30-m\footnote{https://publicwiki.iram.es/Iram30mEfficiencies} and ARO 12-m\footnote{described in~\citet{Milam2005}} are $\sim$ 22 $''$  and $\sim$ 59 $''$, respectively. Assuming a point source, the noise levels of $\rm ^{13}CN$ 
(in flux density) are $\sim 0.02$ Jy and $\sim 0.16$ Jy for our measurement and the measurement in~\citet{Milam2005}, respectively.  
Therefore, our noise level is approximately eight times deeper than that reported in~\citet{Milam2005}.

The typical sizes of dense molecular cores have been shown to be about 0.1\,pc \citep{Wu2010}, which is much smaller than our IRAM 30-m beamsize which corresponds to $\sim 1$\,pc. Thus, we assume a point source distribution for $\rm ^{12}CN$. In our study, we measured a $\rm ^{12}CN$ flux density of about 2.4 Jy, while in \citet{Milam2007thesis}, the flux density was about 2.3 Jy. The consistency between the two measurements suggests that this source is a compact target and the pointing directions of the observations were not severely offset from each other.

We also revisit the $\rm ^{12}CN/^{13}CN$ ratio of WB89\,391, using Equation~\ref{eq:eq_12C13C_hfs} with the Planck 
expression and the CMB correction, based on the optical depth and the peak temperature reported in \citet{Milam2005} and \citet{Milam2007thesis}. 
Our newly derived \Rtwth\ for WB89~391 is $\sim$5.3 $\pm$ 1.7, which is about twenty-five times lower than the previously reported ratio of $\sim$134.

Our IRAM 30-m non-detection of WB89~391 sets a 3-$\sigma$ lower limit of $>$\, 36 for the $\rm ^{12}CN/^{13}CN$ ratio, 
a result also supported by our recent NOEMA observation (Sun et al. in prep). Therefore, we recommend that 
the $\rm ^{13}CN$ detection report in  \citet{Milam2005} needs to be reconsidered for future analyses of the \Rtwth\ gradient.

\subsection{The impact of \CtwCth\ gradient to the \NftNfif\ one}
\label{sec:12Cto13C_impact_to_14Nto15N}
The \Rtwth\ gradient is commonly used to derive many other isotopic ratios. For example, by multiplying \Rtwth\ \citep[e.g.,][]{Wouterloot2008} with $\rm ^{13}CO/C^{18}O$ and $\rm H^{13}CN/HC^{15}N$ \citep[e.g.,][]{Colzi2018}, or $\rm HN^{13}C/H^{15}NC$ \citep[e.g.,][]{Adande2012,Colzi2018}, it is possible to obtain \Rsixeit\ and \Rftfif, respectively.

The $\rm ^{12}CN/^{13}CN$ ratios derived from optically thin satellite lines are systematically higher than those derived from optical depth correction. It will lead to systematically higher \Rftfif\ when they are derived from multiplying \Rtwth\ in optically-thin condition, compared with those multiplying the \Rtwth\ from $\tau$-correction method.  

In particular, we derive \Rftfif\ of G211.59 from two methods: (a). Using $\rm H^{13}CN/HC^{15}N$ multiplying \Rtwth\ (b). Directly using \twCN\ and \CftN\ $N=1\to0$. Using the $\rm H^{13}CN$ and $\rm HC^{15}N$ data from G211.59, we find that $\rm H^{13}C^{14}N/H^{12}C^{15}N = 4.7 \pm 2.5$. Using method (a), if we adopt that \Rtwth\ $= 52.2\pm 6.7$ derived from the HfS method, we will get \Rftfif\ $\sim 242 \pm 40$. Otherwise, if we adopt that \Rtwth\ $= 72.2 \pm 9.5$ derived from the optically-thin satellite line method, we will have \Rftfif\ $= 336 \pm 57$, which is much higher than the first value of \Rftfif. Using method (b), we get \Rftfif\ $= 166\pm 32$ from the HfS method and \Rftfif\ $= 222 \pm 43$ from the satellite line method. The \Rftfif\ from optical-thin satellite lines is also higher than the one from the HfS method. There is a discrepancy between the \Rftfif\ values obtained using the method that directly uses CN isotopes and the method that relies on \Rtwth\ as a conversion factor. This discrepancy persists regardless of whether we employ the HfS method or the optical-thin method, and it may be attributed to astrochemistry effects.

Recently, new \Rftfif\ measurements on the Galactic outer disk were reported by \cite{Colzi2022}. They obtained the \Rftfif\ value by computing the abundance ratios of $\rm H^{13}CN/HC^{15}N$ and $\rm HN^{13}C/H^{15}NC$, which were then multiplied by the \Rtwth\ gradient value provided in~\citet{Milam2005}.  This work does not include corrections to astro-chemical effects on HCN isotopologues, and the $\rm ^{14}N/^{15}N$ ratios derived from their analysis are systematically higher than ours. Particularly, the Galactic chemical evolution model in \citet{Colzi2022} shows lower \Rftfif\ values in the Outer disk (\Rftfif\ $\sim $250-400 at \Rgc\ $\sim$ 12\,kpc) than those derived from their observations, but model predictions remain consistent with \Rftfif\ derived in this work. 

However, WB89~391 induces a strong bias to the \Rtwth\ gradient in \citet{Milam2005} which then propagates to the \Rftfif\, gradient, making \Rftfif\ ratios highly overestimated in the Galactic outer disk.  If instead, the $\rm H^{13}CN/HC^{15}N$ and $\rm HN^{13}C/H^{15}NC$ ratios were multiplied by the \Rtwth\ values obtained from optically-thin CN satellite lines in our work, the resulting \Rftfif\ would be more consistent with our measurement of \Rftfif\ derived from $\rm ^{12}CN/C^{15}N$, with a small discrepancy possibly due to astrochemical effects.

\subsection{Comparing CN with optically thin isotopic ratio tracers}
\label{sec:discuss_diff_tracers}

Among the current tracers to derive \Rtwth, the abundance ratio of CO isotopologues (i.e., $\rm ^{12}C^{18}O/^{13}C^{18}O$) have been adopted in plenty of Galactic targets, which gives a reliable sample size. In addition, $\rm C^{18}O$ and $\rm ^{13}C^{18}O$ lines are expected to be optically thin, which will significantly reduce the effect of different excitation layers (similar to CN isotopologues, see Section \ref{sec:derivation_Two_Tex_layer_model}).  

\begin{figure*}
    \includegraphics[scale=0.39]{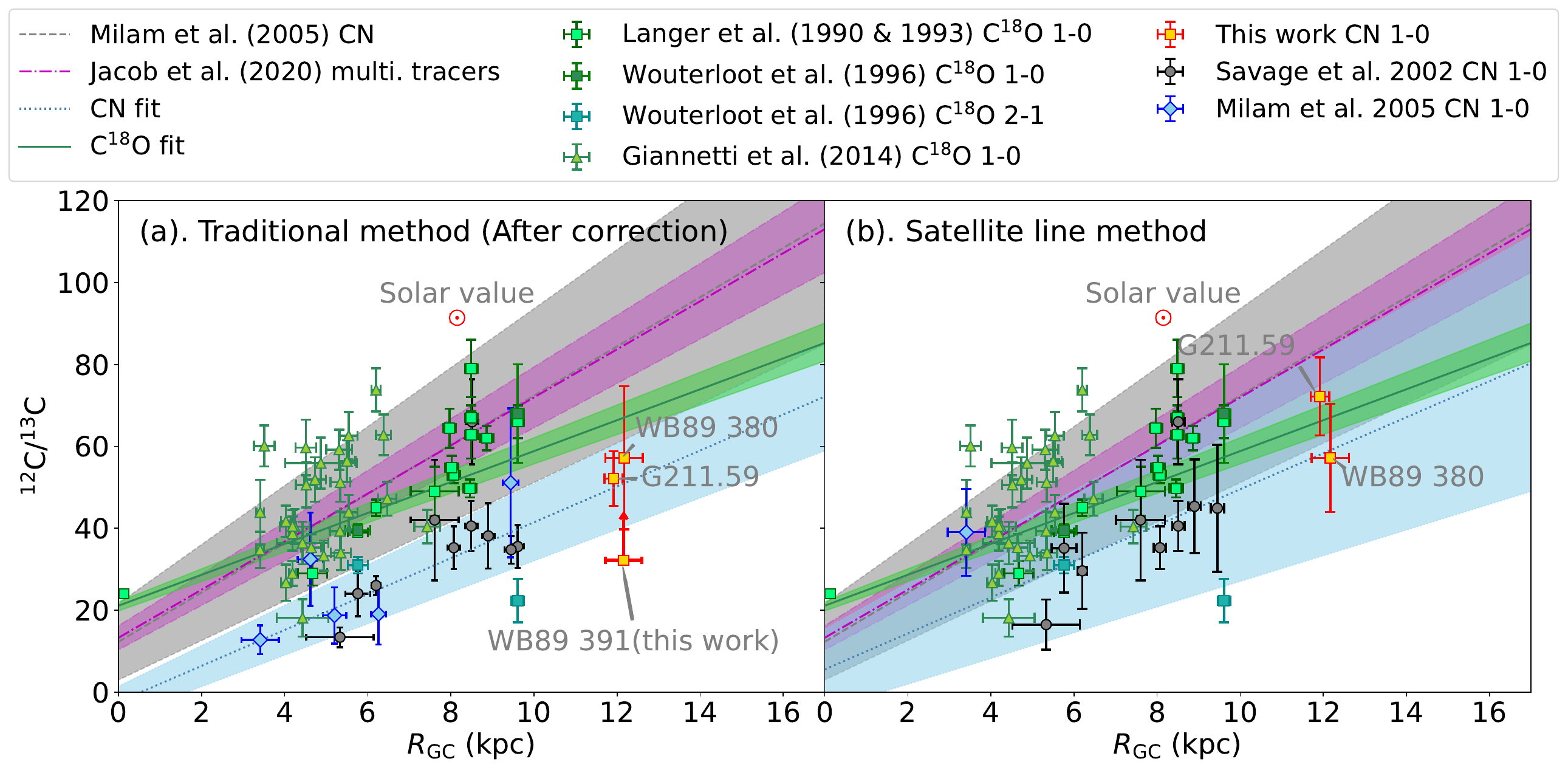}
\centering
\caption{The  \Rtwth\  ratios obtained from different tracers. Left: The ratios derived from $\rm ^{12}CN/^{13}CN$ with the traditional method. Right: The ratios derived from $\rm ^{12}CN/^{13}CN$ with the optically-thin satellite line method. The grey dots, blue diamonds, and red squares are the $\rm ^{12}CN/^{13}CN$ of targets in \citet{Savage2002}, \citet{Milam2005}, and in this work, respectively. The spring-green squares, sea-green squares, and yellow-green triangles are $\rm C^{18}O/^{13}C^{18}O$ at their $N=1\to0$ transitions of targets in \citet{Langer1990,Langer1993}, \citet{Wouterloot1996} and \citet{Giannetti2014}, respectively. The light-sea-green squares are $\rm C^{18}O/^{13}C^{18}O$ at their $N=2\to1$ transitions in \citet{Wouterloot1996}. The grey dashed line shows the previous $\rm ^{12}CN/^{13}CN$ gradient provided in~\citet{Milam2005}. The magenta dash-dotted line refers to the previous Galactic $\rm ^{12}C/^{13}C$ gradient fitted with ratios derived from multiple tracers, provided in \citet{Jacob2020}. The green solid line and the blue dotted line refer to the linear fitting curves of \Rtwth\ from $\rm ^{12}C^{18}O/^{13}C^{18}O$ and $\rm ^{12}CN/^{13}CN$, respectively. The color blocks show the 1-$\sigma$ error range of corresponding linear-fitting curves. \label{fig:12C_13C_grad_diff_tracer}} 
\end{figure*}

We select \Rtwth\ derived from $\rm ^{12}C^{18}O/^{13}C^{18}O$ with enough signal-to-noise ratios provided in the literature \citep{Langer1990,Langer1993,Wouterloot1996,Giannetti2014} and combine them with our \Rtwth\ derived from $\rm ^{12}CN/^{13}CN$, shown in Fig. ~\ref{fig:12C_13C_grad_diff_tracer}. The Galactocentric distances of all targets are derived with the same Galactic rotation curve model, following \cite{Reid2019}.

Fig.~\ref{fig:12C_13C_grad_diff_tracer} (a) overlays \Rtwth\ of Galactic outer disk clouds derived with the traditional method 
(the CN $\tau $-correction method, \Rtwth\ re-calculated with data from \citet{Savage2002} and \citet{Milam2005}),  
and \Rtwth\ derived from $\rm ^{12}C^{18}O/^{13}C^{18}O$ collected in the literature. The \Rtwth\ ratios derived from $\rm ^{12}CN/^{13}CN$ 
(the traditional method) appear systematically lower than that derived from the $\rm C^{18}O/^{13}C^{18}O$ method.

Fig.~\ref{fig:12C_13C_grad_diff_tracer} (b) shows the same comparison 
as Fig.~\ref{fig:12C_13C_grad_diff_tracer} (a), but now \Rtwth\ is derived with the 
Satellite-line method. 
The satellite line method seems to produce \Rtwth\ better matching 
the \Rtwth\ derived from $\rm C^{18}O/^{13}C^{18}O$, compared with those from the traditional method. Some \Rtwth\ ratios have already been derived in the optically-thin condition in \citet{Savage2002}, which do not change between Fig. \ref{fig:12C_13C_grad_diff_tracer} (a) and Fig. \ref{fig:12C_13C_grad_diff_tracer} (b). These \Rtwth\ ratios are consistent with \Rtwth\ from $\rm C^{18}O/^{13}C^{18}O$.  

In Fig.~\ref{fig:12C_13C_grad_diff_tracer}, we also show two previous linear fitted gradients. The previous $\rm ^{12}CN/^{13}CN$ gradient provided in \citet{Milam2005} is highly overestimated because of the R-J approximation and the neglect of the CMB temperature. The magenta one shows the fitted Galactic $\rm ^{12}C/^{13}C$ gradient with previous data points from multiple tracers in the literature, concluded in \citet{Jacob2020}. This gradient is also influenced by the previously overestimated $\rm ^{12}CN/^{13}CN$ ratios. In the outer disk, both the $\rm ^{12}C^{18}O/^{13}C^{18}O$ and $\rm ^{12}CN/^{13}CN$ are expected to be lower than what the two previous gradients predict.

While there is significant scatter from each individual tracer, a general trend of increasing \Rtwth\ from the Galactic center 
to the outer disk is apparent. The results based on the CO isotopologues also show a dependence on the chosen $J-$transition, i.e., \Rtwth\ derived from the $J=1\to0$ transition does not match with those obtained from the $J=2-\to1$ lines \citep{Wouterloot1996}, likely due to differential excitation from clouds \citep{Jacob2020}. 
The two new data points from our survey, G211.59, and WB89\,380, match both increasing gradients fitted from CN and CO isotopologues. Recently, \citet{Yan2023} provide $\rm ^{12}C^{34}S/^{13}C^{34}S$ ratios  also show an increasing \Rtwth\ gradient with \Rtwth $\sim 75$ at \Rgc\ $\sim$ 10 kpc. Therefore, we would expect that the even further outer disk region (\Rgc $>$ 15 kpc) may have even higher \Rtwth\ ratios if the low-metallicity fast rotators are not dominated in the chemical evolution in this region. However, more observations are required to really constrain the chemical evolution in the further outer disk. In Appendix~\ref{Appendix:linear_fitting_fuction}, we show the linear fitted functions of the Galactic \Rtwth\ and \Rftfif\ gradients based on measurements from optically-thin conditions shown in Fig.~\ref{fig:12C_13C_grad_diff_tracer} (b) and Fig.~\ref{fig:14N_15N_grad} (b), respectively. However, it is highly risky to use \Rtwth\ or \Rftfif\ from the linear fitting gradients instead of the direct measurement in individual targets (more details in Appendix~\ref{Appendix:linear_fitting_fuction}).

\subsection{Astrochemical effects} \label{sec:discuss_astrochemical}

Although the satellite method can reduce the deviation of abundance ratio measurements (Fig.~\ref{fig:12C_13C_grad_diff_tracer} (b)), astrochemical effects may still bias the 
measured molecular abundances. This can introduce additional discrepancies between \Rtwth\ of the same target derived from different tracers, and that between different targets (thus in different chemical environments) but derived from the same tracer. 

{\bf $\bullet$ UV self-shielding:} When exposed to UV radiation, more abundant $\rm ^{12}C$-bearing molecules can self-shield themselves in the inner clouds, making them less vulnerable to destruction by UV photons. On the other hand, the less abundant $\rm ^{13}C$-bearing isotopologues would be more easily destroyed even in the dense interior of clouds. This self-shielding effect can lead to an overestimation of \Rtwth\, in emission line pairs such as $\rm C^{18}O$ and $\rm ^{13}C^{18}O$, or $\rm CN$ and $\rm ^{13}CN$~\citep{vanDishoeck1988,Rollig2013}.  Given the high localization of 
strong Photon-Dominated Regions (PDRs) around O stars in molecular clouds ($\rm L\leq $0.1\,pc), such FUV-induced fractionization effects cannot affect
 globally-averaged isotopologue ratios over galaxy-sized molecular gas reservoirs, but they can affect local ISM regions like those observed in our study.

{\bf $\bullet$ Depletion:} In dense and well-shielded regions, the temperature of dust can drop below the CO condensation temperature, typically around 17 K \citep[e.g.,][]{Nakagawa1980}. As a result, CO isotopologues freeze out onto the surface of dust grains \citep[e.g.,][]{Willacy1998,Savva2003,Giannetti2014,Feng2020}. Observations suggest that the depletion factors of $\rm ^{12}C^{18}O$ and $\rm ^{13}C^{18}O$ vary with density or temperature, but no systematic differences have been found \citep{Savva2003,Giannetti2014}.

{\bf $\bullet$ Chemical fractionation:} In addition, the difference between the zero-point energy of $\rm ^{12}CO$ and $\rm ^{13}CO$ 
causes the carbon isotopic fractionation reactions \citep[e.g.\, ][]{Watson1976,Langer1984}:
\begin{equation}
^{13}{\rm C}^{+} +\hspace{0.1cm}^{12}{\rm CO} \leftrightarrow \hspace{0.1cm}^{13}{\rm CO} + \hspace{0.1cm}^{12}{\rm C}^{+} + 35 \hspace{0.1cm}{\rm K}
\end{equation}
The CN isotopologues have a similar effect~\citep{Kaiser1991,Roueff2015}:
\begin{equation}
^{13}{\rm C}^{+} +\hspace{0.1cm}^{12}{\rm CN} \leftrightarrow \hspace{0.1cm}^{13}{\rm CN} + \hspace{0.1cm}^{12}{\rm C}^{+} + 31.2 \hspace{0.1cm}{\rm K}
\end{equation}

Both gas-phase and grain-phase chemical networks have been adopted to model such fractionation effects on CO and CN molecules~\citep[e.g.\,][]{Smith1980,Roueff2015,Colzi2020,Loison2020}. The experiments in \citet{Smith1980} show $\rm ^{13}C$ and $\rm ^{18}O$ fractionation with a temperature $\lesssim$ 80 K. \citet{Ritchey2011} suggest that $\rm ^{12}CO/^{13}CO$ and $\rm ^{12}CN/^{13}CN$ evolution have inverse trends in the fractionation model in diffuse gas. Gas-phase models in \citet{Roueff2015} show that the diversion of $\rm ^{12}CN/^{13}CN$ from original \CtwCth\ is not significant with a temperature $\sim 10$ K. These models show stable $\rm C^{14}N/C^{15}N$ with an evolution time larger than $10^{6}$ years. 
This is consistent with the fractionation work containing gas and dust \citep{Colzi2018}, where the $\rm ^{12}CN/^{13}CN$ becomes stable after an evolution time $\sim 10^{7}$ years.  Recent results in \citet{Colzi2022} further find that nitrogen fractionation partly contributes to the scatter in the \Rftfif\ measurements, while higher angular resolution observations are needed to disentangle local effects from nucleosynthesis contribution.

 A detailed astrochemical study of the molecules used in such Galactic
isotope ratio surveys, subjected to the varying far-UV radiation, cosmic ray, and pressure environments found in the sources used for such surveys 
will be particularly valuable. Should any $R_{\rm gc}$-systematic astrochemistry effects be found,
such a study could yield valuable corrections to the (isotopologue)$\rightarrow$(isotope) abundance ratio conversion
and reduce the scatter in Figure \ref{fig:12C_13C_grad_diff_tracer}.

\subsection{Comparing with abundance ratios from stars}

The \Rtwth\ ratio can also be measured from stars.  \citet{Botelho2020} found local \Rtwth\  values of $\sim70-100$ in 63 Solar twins, higher than those derived from 
the interstellar medium (\Rtwth $\sim 40-60$ with \Rgc \ $\sim $ 8 kpc). Recently, \citet{Zhang2021} found an \Rtwth\ ratio of $97^{+25}_{-18}$ in an isolated brown dwarf with an age of $\sim125$ Myr. This value is similar to the ratio found in the Sun \citep{Ayres2013} and higher than the previously measured \Rtwth\ ratio in the Solar neighborhood \citep{Milam2005}.

This is reasonable since abundance ratios measured from stars reflect the abundance of their parental ISM typically several billion years ago, which may be different from those in the local ISM because of stellar moving and the time evolution of \Rtwth. Recent studies based on Gaia and LAMOST data reveal that young ($<$100 Myr) stellar clusters in the Solar neighbourhood exhibit a variety of metallicity in their member stars, which indicates strong inhomogeneous mixing or fast star formation \citep{Fu2022}. These young clusters span a wide range of age and non-circular orbits, indicating multiple spatial origins~\citep{Fu2022}.

\subsection{Homogeneous mixing in the Solar neighbourhood}

Optical spectroscopic observations of N and O towards Galactic H{\sc ii} regions indicate well-mixed gas in the Galactic plane\citep[e.g.,][]{Esteban2022}, at least in the second and the third quadrants~\citep{Arellano2020}.  Fig.~\ref{fig:fig_posi_compared_with_HIIregion} in Appendix~\ref{appendix:HII_region_figure} shows the location of those H{\sc ii} regions and molecular clouds with \Rtwth\ in this work. Most targets are located in the same quadrants as those H{\sc ii} regions, indicating similar abundance ratios at the same Galactic~radii.

However, these studies only consider the major isotopes of carbon and nitrogen, and the behavior of the rare isotopes \thC\ and \ffN\ may not be the same as that of the major isotopes. In fact, the production and enrichment of these rare isotopes often involve different mechanisms from those of the major isotopes, such as novae \citep[e.g.,][]{Romano2003,Cristallo2009,Romano2022}.  Furthermore, molecular tracers can be affected by additional chemical biases that may increase the scatter in measured abundances compared to those obtained from stellar tracers.

\subsection{Future prospects}

Our analysis of the different methods used to derive \Rtwth\ and \Rftfif\ indicates that the radiative transfer conditions may play an important role in affecting the basic assumptions of those methods. In particular, deriving isotopic ratios from lines of two molecular transitions with significantly different optical depth values (e.g., \twCN\ and \thCN\ $N=1\to0$ main components, with $\tau $$\sim $1 and $\tau $$\sim $0.01, respectively) after performing optical depth corrections is highly non-trivial because of the expected non-uniform excitation conditions within the sources.

For the Galactic targets, parts of the \Rtwth\ and \Rftfif\ have been derived from $\rm ^{12}CN/^{13}CN/C^{15}N$ \citep{Savage2002,Milam2005,Adande2012} or $\rm ^{14}NH_3/^{15}NH_3$ \citep{Chen2021}, with optical depth correction. After correcting the results using the full Planck function approximation and including the \Tcmb, the results are still subjected to the uncertainties of the underlying line excitation structure of the sources.
It is always preferable to use the optically-thin line components, such as $\rm C^{18}O/^{13}C^{18}O$ or the optical-thin satellite lines of \twCN\ and $\rm ^{14}NH_3$ with their isotopologues to derive \Rtwth\ and \Rftfif\ in Galactic targets.

For extragalactic sources the \Rtwth\ can be derived from the $\rm ^{12}CN/^{13}CN$ with optical depth correction~\citep{Henkel1993,Henkel2014,Tang2019} but this method has the same issues as those used in the Milky Way, and it is hard to separate the optically-thin satellite line from the blended \twCN\ line components. It is better to use isotopologues with optically thin transitions (e.g., $\rm C^{18}O/^{13}C^{18}O$) for deriving isotopic ratios, but this requires longer integration time with high sensitivity instruments. According to our 2/15 detection rate of $\rm ^{13}CN$ with an integration time $\sim 0.5-10$ hours of our targets (see in Table~\ref{tab:target_information}, \ref{tab:results_HfS} and \ref{tab:results_opthin}), we suggest that an observing time of more than six hours may be required to detect $\rm ^{13}C^{18}O$ in the targets of the outer disk with a sensitivity similar to that of IRAM 30-m. 

On the other hand, all the current methods are based on LTE assumptions. Non-LTE will not only cause different excitation between isotopologues but also change the relative intensities of hyperfine structures~\citep[e.g., ][]{Henkel1998}. These issues will be examined by the non-LTE models in the future.

\section{Conclusions} \label{sec:conclusions}

We examine the assumptions and shortcomings of three different methods used to derive \Rtwth\ from $\rm ^{12}CN/^{13}CN$ namely: (a) the one using HfS fitting to do optical-depth correction of \twCN, adopted in the literature, (b) the improved $\tau$-correction method incorporating the Planck radiation temperature and the CMB contribution, (c) the method deriving \Rtwth\ from the ratio of optically-thin \twCN\ satellite lines and \thCN\ lines. We also point out the similar issues of the methods of deriving \Rftfif\ from \twCN\ and \CftN.  We observed $\rm ^{12}CN$,  $\rm ^{13}CN$, and $\rm C^{15}N$ $N=1\to0$ towards 15 molecular clouds on the outer Galactic disk ($R_{\rm gc} \sim \rm 12\ kpc - 24\ kpc$). The Galactic \Rtwth\ and \Rftfif\ gradients obtained from different methods are shown by combining \Rtwth\ and \Rftfif\ derived from our new observations and those in the literature revised by our improved methods. Our results are as follows: 

(i). The current method for deriving $\rm ^{12}CN/^{13}CN$ in the literature adopts the Rayleigh-Jeans approximation and neglects the CMB, which then highly overestimates \Rtwth\ when the excitation temperature of \twCN\ lines is $<$10\,K. The improved $\tau$-correction method using the Planck equation and considering the CMB avoids these systematic overestimates, however, it still adopts the assumption of uniform excitation conditions. We show the latter still yields the under-estimate of \Rtwth\ using a simple 2-phase excitation model.  Scaling the intensity ratio of optically thin \twCN\ satellite lines and \thCN\ to the column density ratio of $\rm ^{12}CN/^{13}CN$ avoids the shortcomings of the other two methods and is a better method for deriving reliable \Rtwth\ from CN isotopologues. 

(ii). Our method requires long integration time. For most targets, we can only set lower limits, but for the objects with the longer integration time (G211.59 and WB89 380), we measure $\rm ^{12}C/^{13}C \sim$ 60. WB89~391, the target located at the outermost Galactic radius in the literature, shows no detection of $\rm ^{13}CN$ from our IRAM 30-m data, which is inconsistent with the previous result reported in \cite{Milam2005}. With $\sim$ 8$\times$ better sensitivity, our new data sets a lower limit of \Rtwth $\gtrsim 36$  in WB89~391 in the optically thin condition. We also give \Rftfif\ $\sim$ 220 from $\rm ^{12}CN/C^{15}N$ for one target and \Rftfif\ $\sim$ 300 from $\rm H^{13}CN/HC^{15}N$ with \Rtwth\ for two targets, at $R_{\rm gc} \sim 12$ kpc. These ratios are much lower than \Rftfif\ derived from $\rm H^{13}CN/HC^{15}N$ or $\rm HN^{13}C/H^{15}NC$ multiplying \Rtwth\ gradient in \citet{Milam2005}, in better agreement with predictions of chemical evolution models.

(iii). The updated Galactic gradient of \Rtwth\ (derived from $\rm ^{12}CN/^{13}CN$ with $\tau $-correction) yields systematically lower values than previous results. The updated \Rtwth\ gradient from CN optically-thin satellite lines is systematically steeper than the one from optical-depth correction. Such changes will regulate the Galactic \Rftfif\ deriving from other ratios (e.g., $\rm H^{13}CN/HC^{15}N$) multiplying \Rtwth. In addition, the \Rtwth\ obtained from $\rm ^{12}CN/^{13}CN$ in the optically thin condition is more consistent with \Rtwth\ from $\rm ^{12}C^{18}O/^{13}C^{18}O$ than the one derived from the $\tau $-corrected method.

\section*{Acknowledgements}
ZYZ and YCS acknowledge Prof. Rob Ivison and Dr. Xiaoting Fu for helpful discussions about this work. We also acknowledge the very helpful comments of our reviewer. This work is based on observations carried out under project numbers 031-17 and 005-20 with the IRAM 30-m telescope. IRAM is supported by INSU/CNRS (France), MPG (Germany), and IGN (Spain).
This work is supported by the National Natural Science Foundation of China (NSFC) under grants No. 12173016, 12041305, 12173067, and 12103024, the fellowship of China Postdoctoral Science Foundation 2021M691531, the Program for Innovative Talents, Entrepreneurs in Jiangsu, and the science research grants from the China Manned Space Project with NO.CMS-CSST-2021-A08 and NO.CMS-CSST-2021-A07.  DR acknowledges the Italian National Institute for Astrophysics for funding the project "An in-depth theoretical study of CNO element evolution in galaxies"  through Theory Grant Fu. Ob. 1.05.12.06.08.

\section*{Data Availability}

The \CtwCth\ and \NftNfif\ ratios from our new observations are listed in Table~\ref{tab:results_HfS} and Table~\ref{tab:results_opthin}. The revised isotopic ratios from the targets in the literature can be derived with data provided in the corresponding works~\citep{Savage2002,Milam2005,Adande2012,Colzi2018,Langer1990,Langer1993,Wouterloot1996,Giannetti2014}, via Equations in this paper. The original IRAM 30-m data underlying this work will be shared on reasonable request to the corresponding author.




\bibliographystyle{mnras}
\bibliography{reference_v2}




\appendix

\section{The PDFs of estimated distances for our targets}
\label{sec:dis_PDF}
\begin{figure*}
        \centering
		    \includegraphics[scale=0.3]{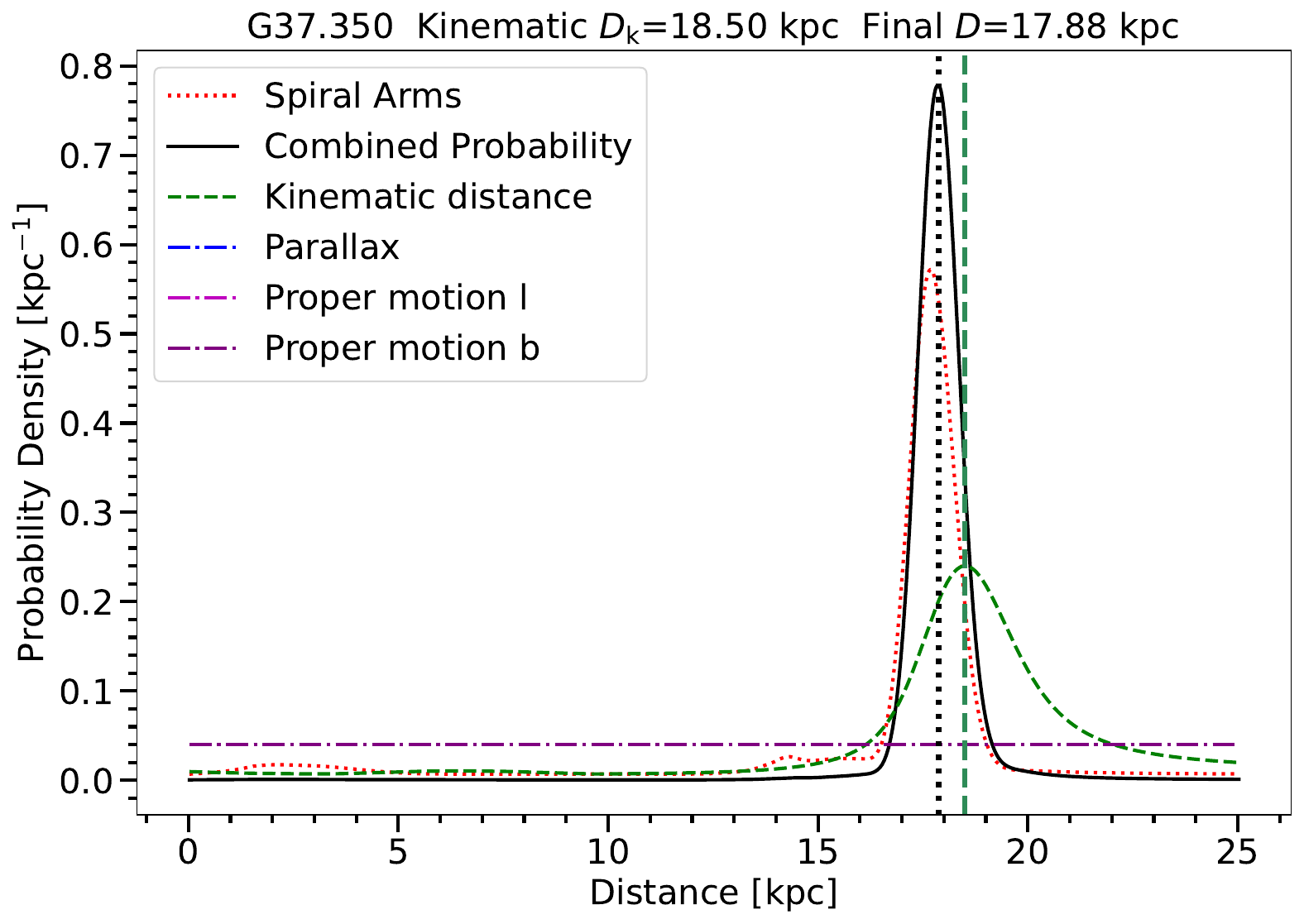}
                \includegraphics[scale=0.3]{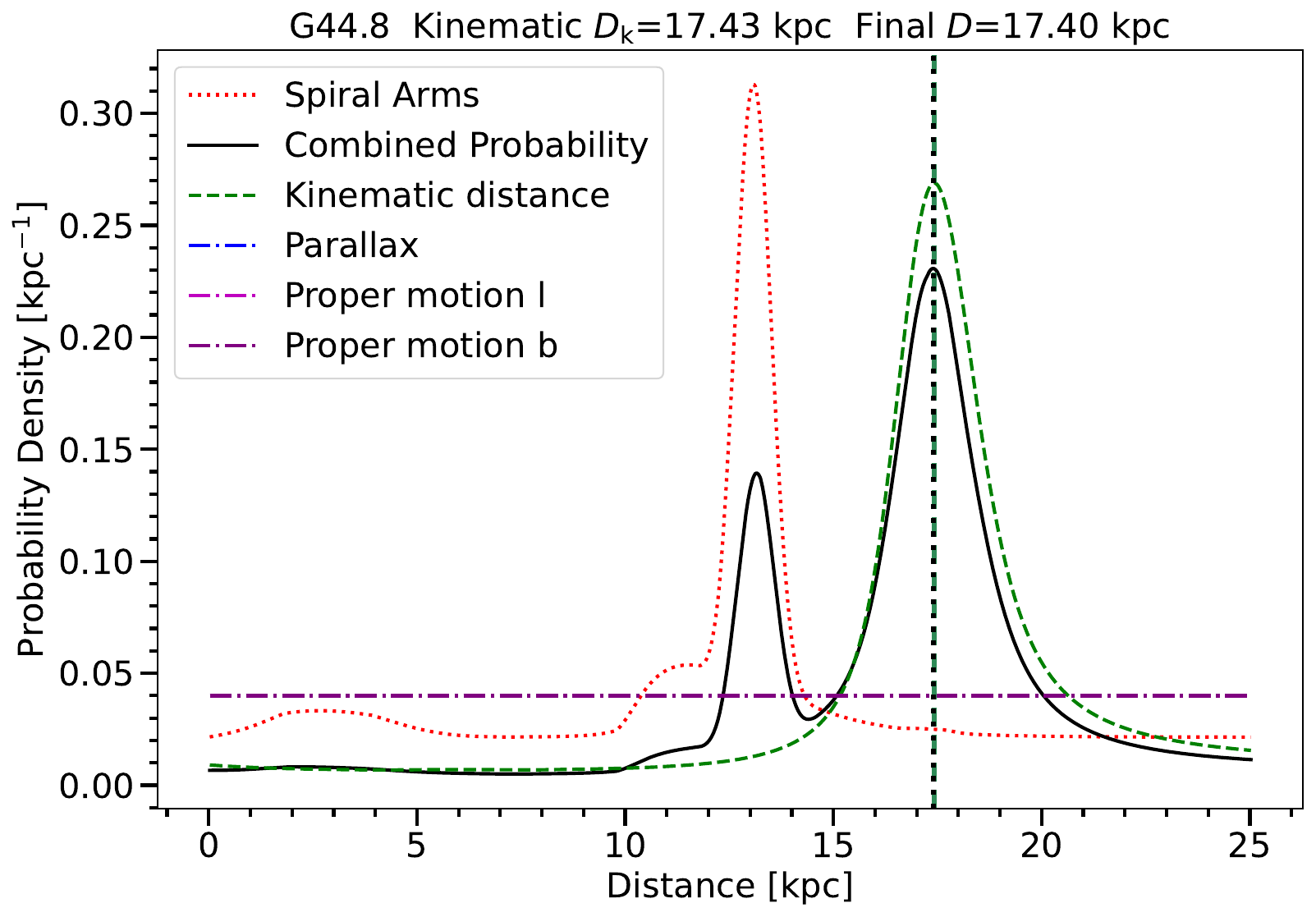}
                \includegraphics[scale=0.3]{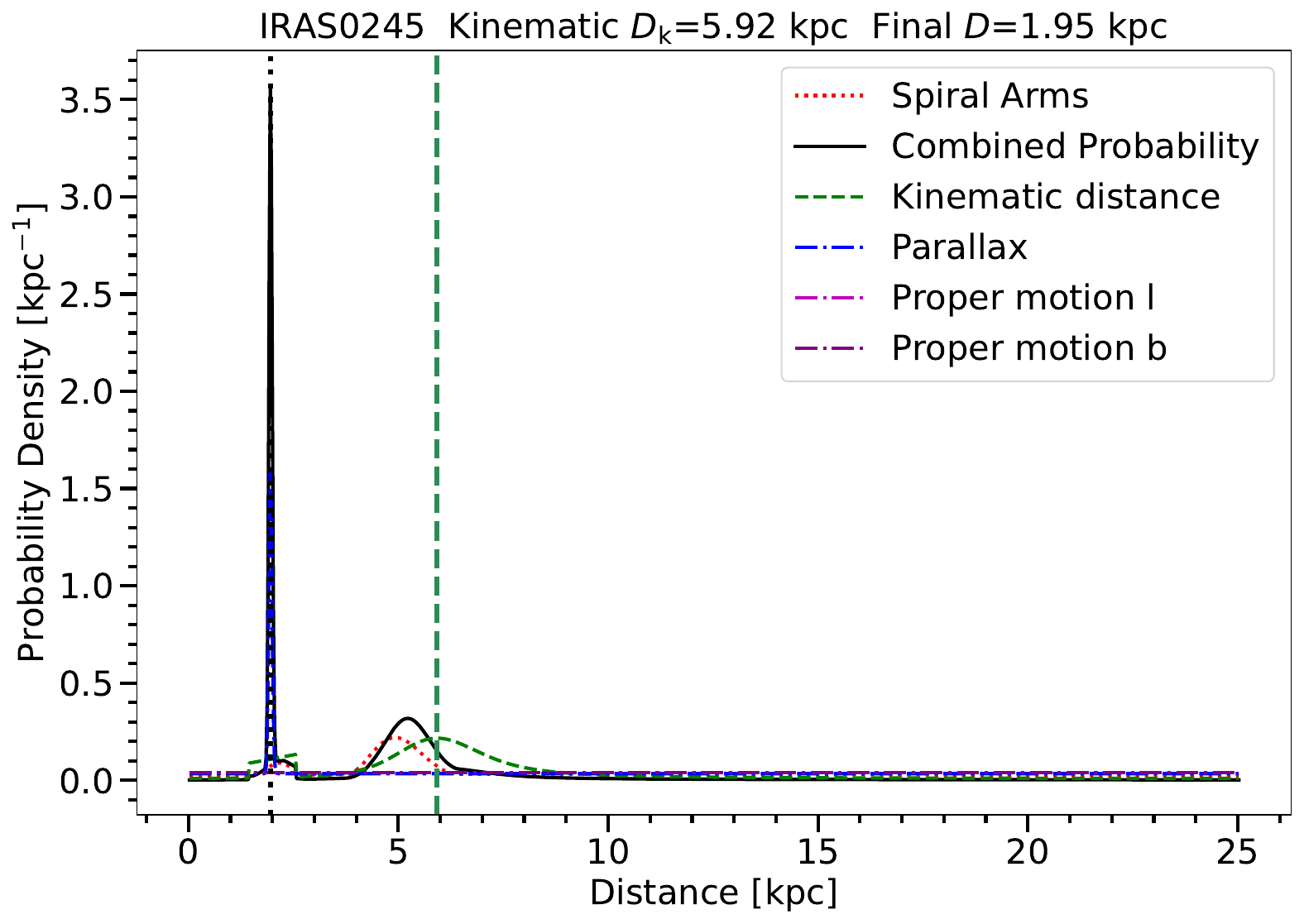}
                \includegraphics[scale=0.3]{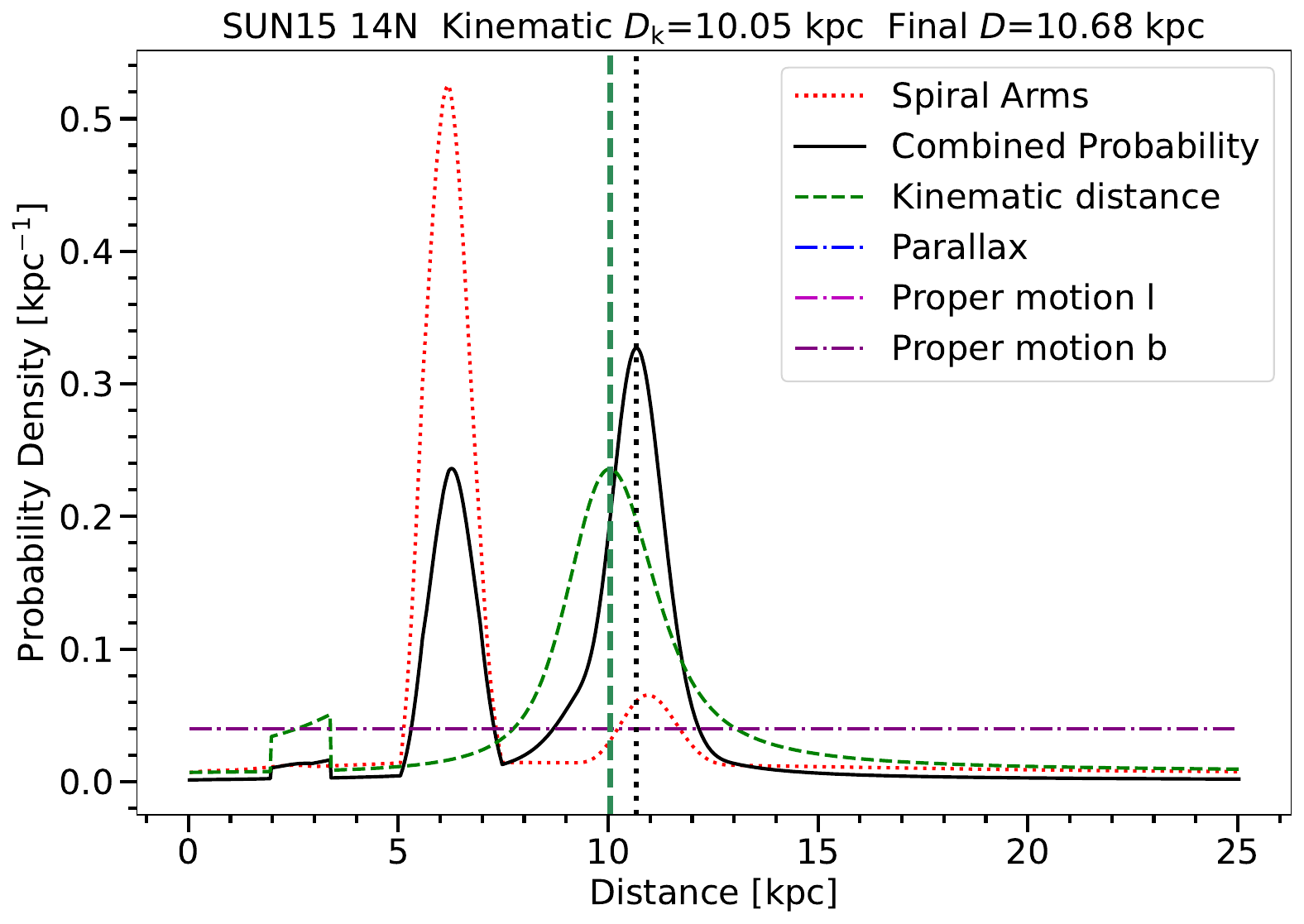}
                \includegraphics[scale=0.3]{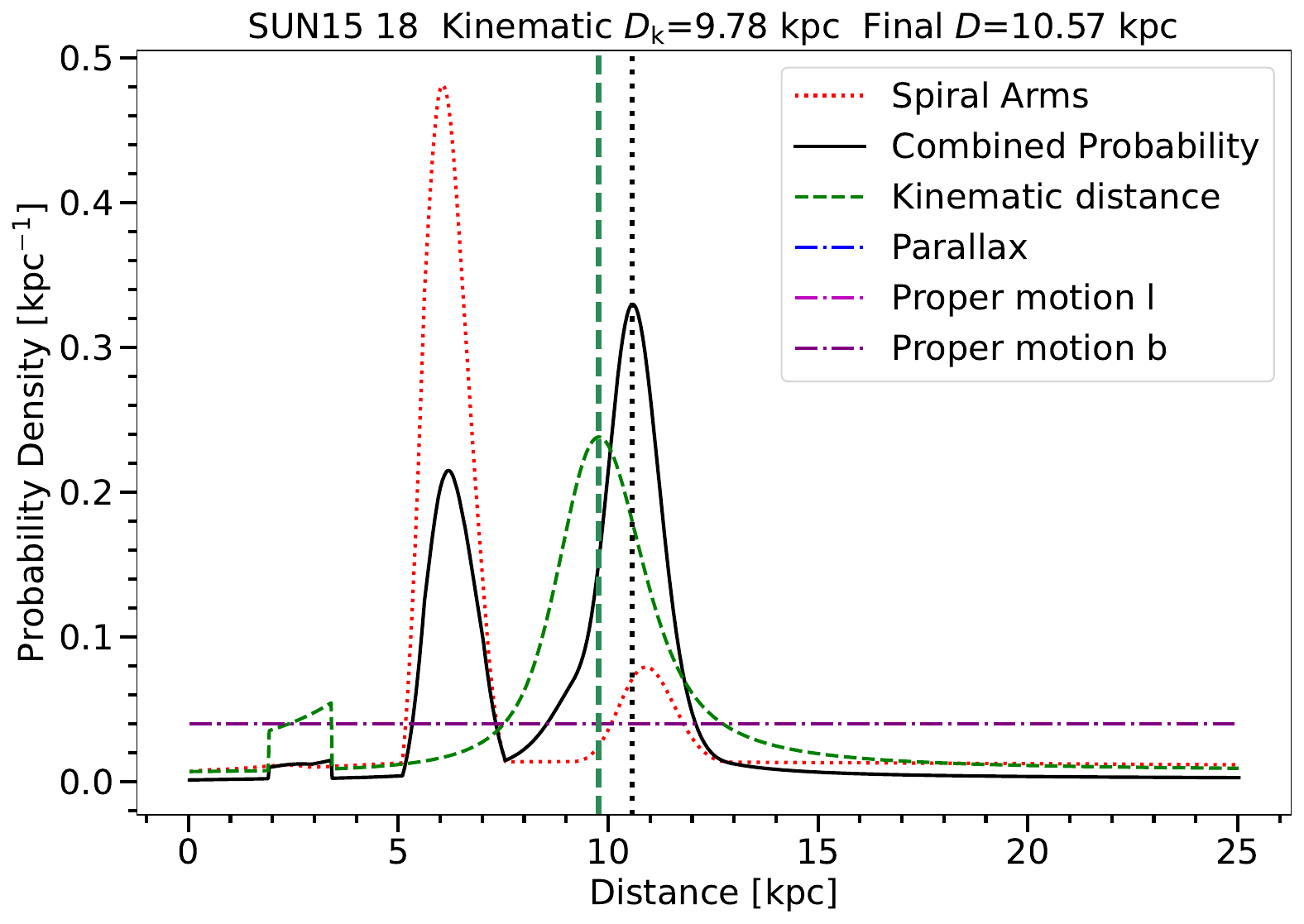}
		    \includegraphics[scale=0.3]
                {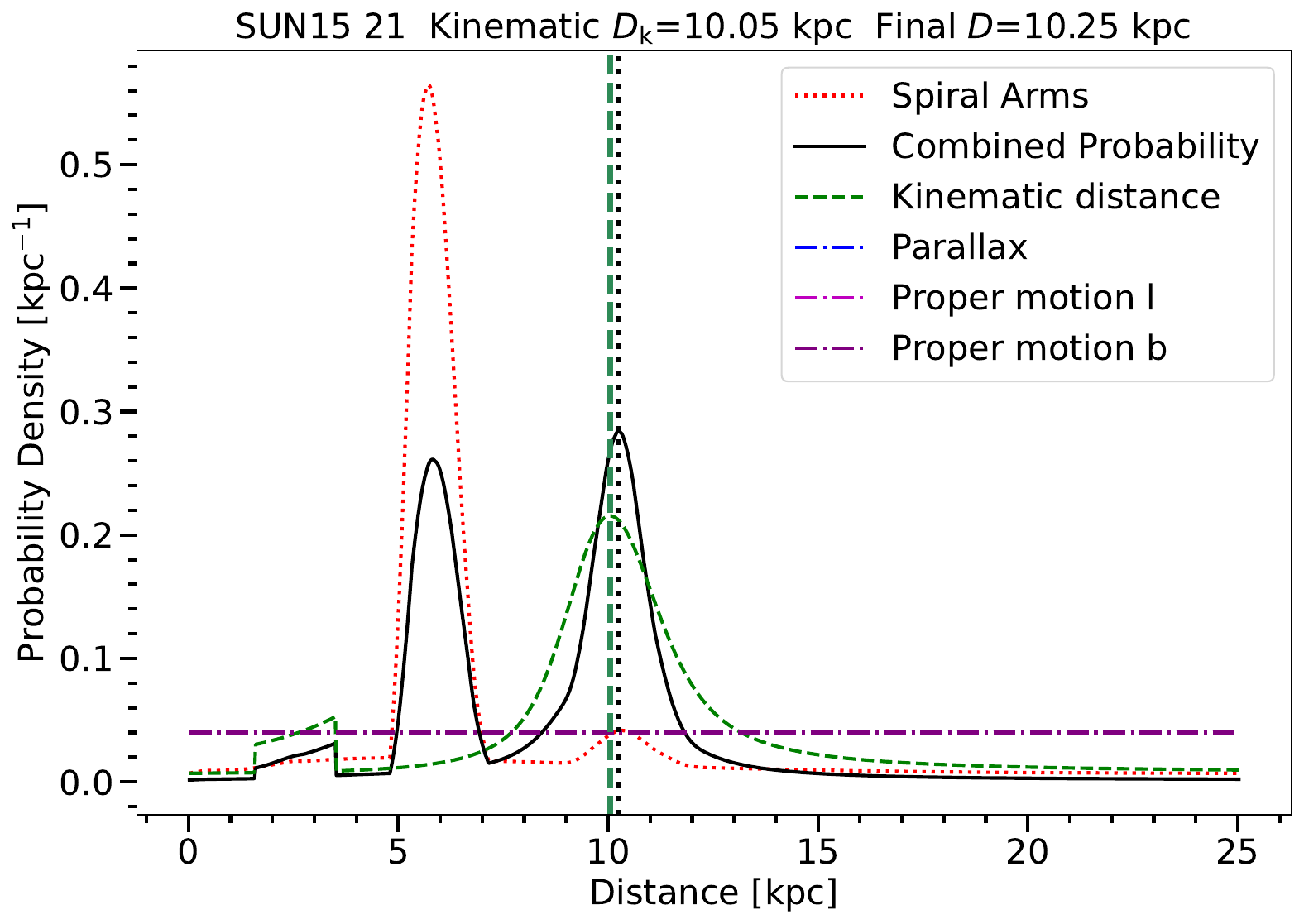}
        \caption{\label{fig:source_dis_PDFs}The probability distribution functions of the target distances. The red dotted curve, green dashed curve, blue dash-dot curve, magenta dash-dot curve, and purple dash-dot curve refer to the distance PDFs from Spiral arm distribution, kinematic methods, parallax, $l$ and $b$, respectively. The black solid curve is the combined PDF from all these PDFs. The green dashed vertical line shows the distance estimated only by kinematic PDF and the back dotted vertical line shows the final estimated distance. }
\end{figure*}

\begin{figure*}
        \centering
                \includegraphics[scale=0.3]{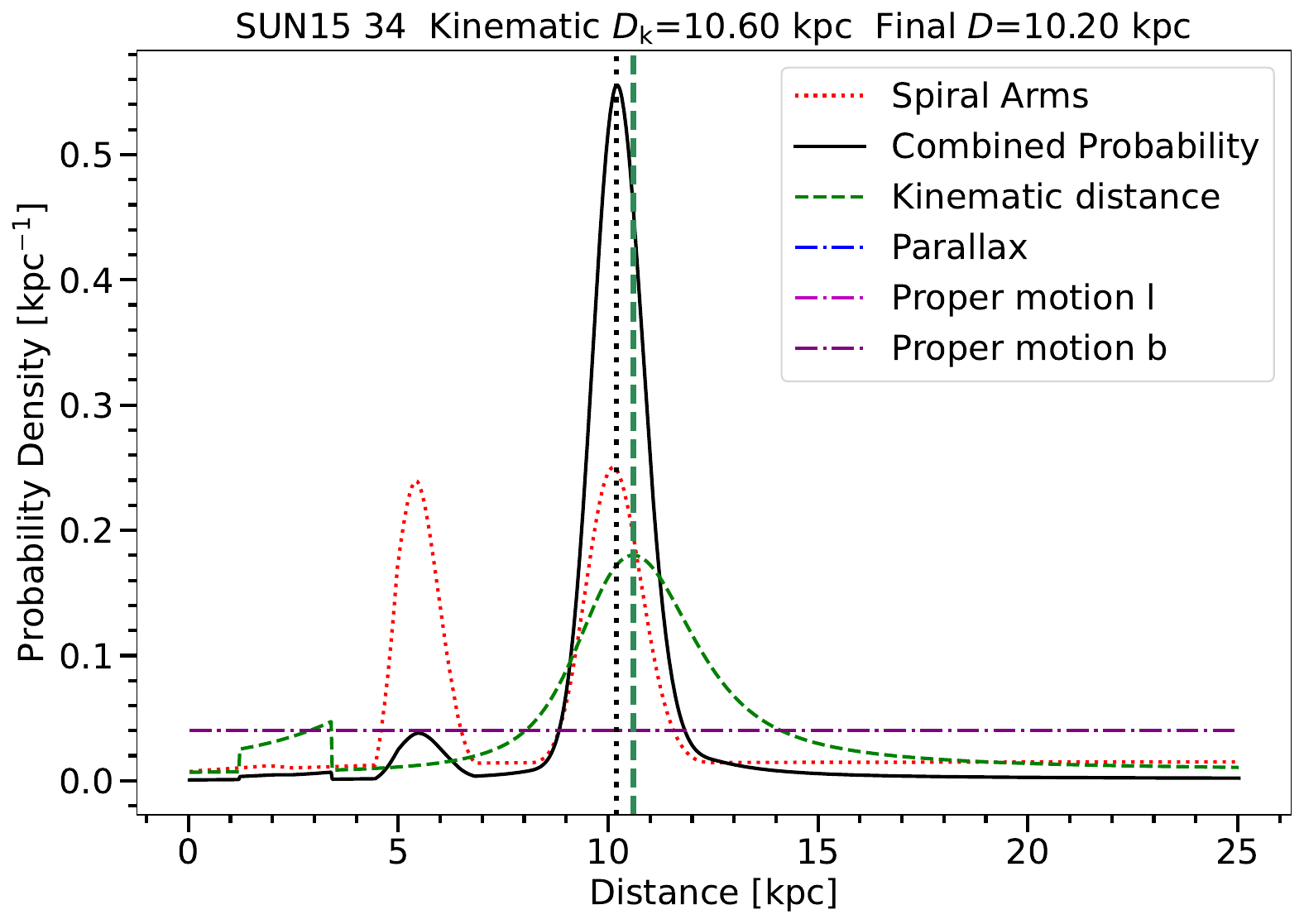}
                \includegraphics[scale=0.3]{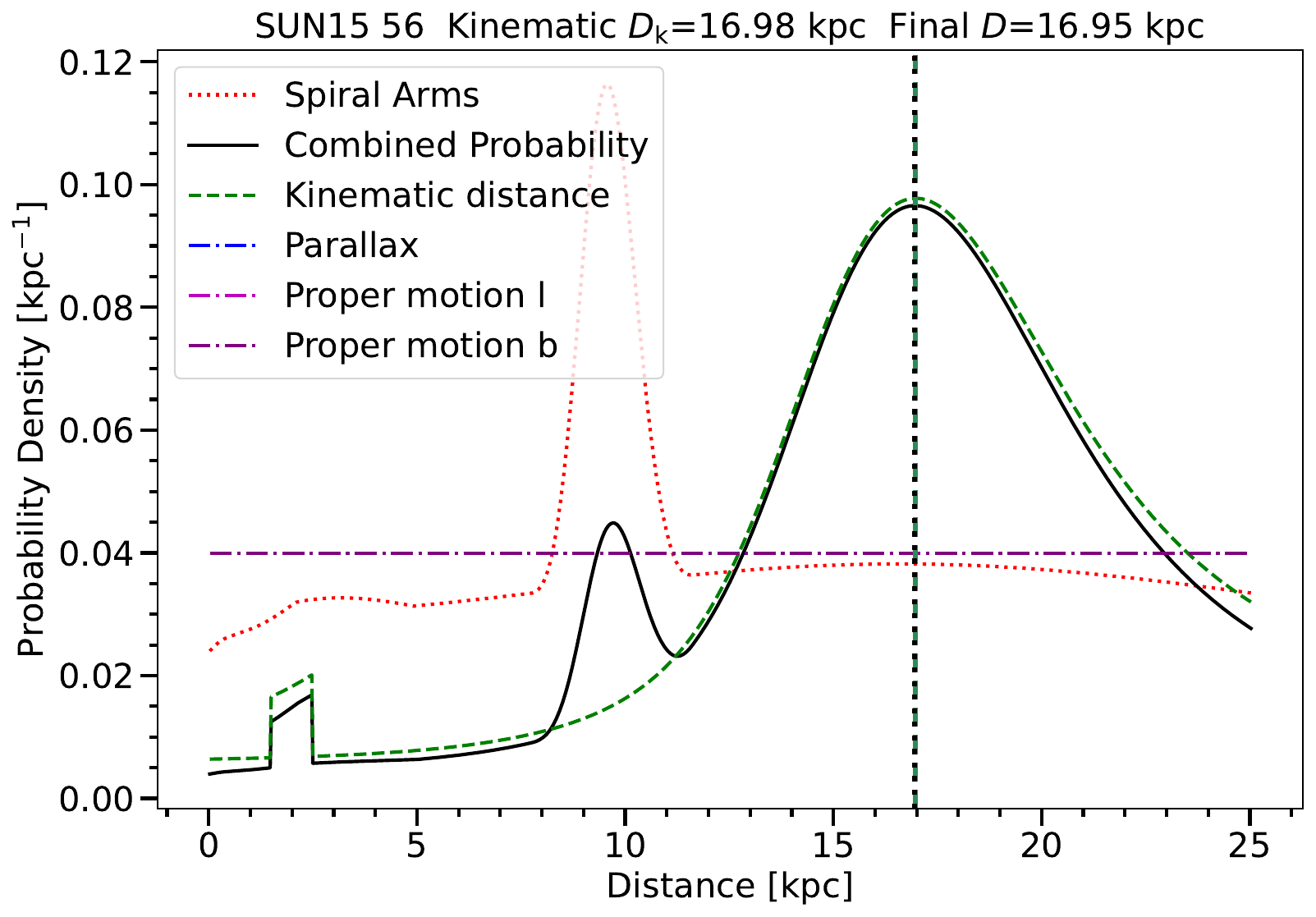}
                \includegraphics[scale=0.3]{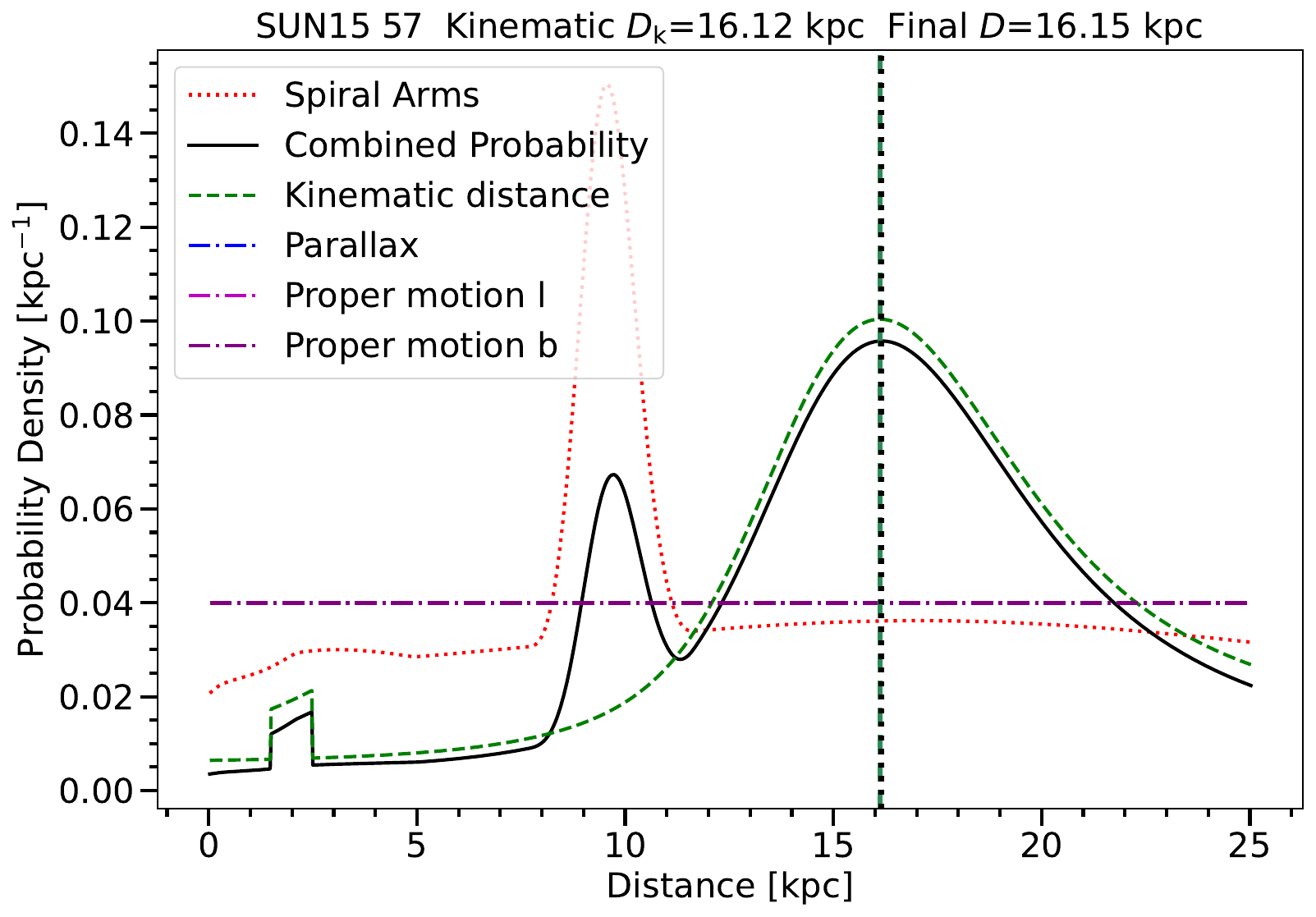}
                \includegraphics[scale=0.3]{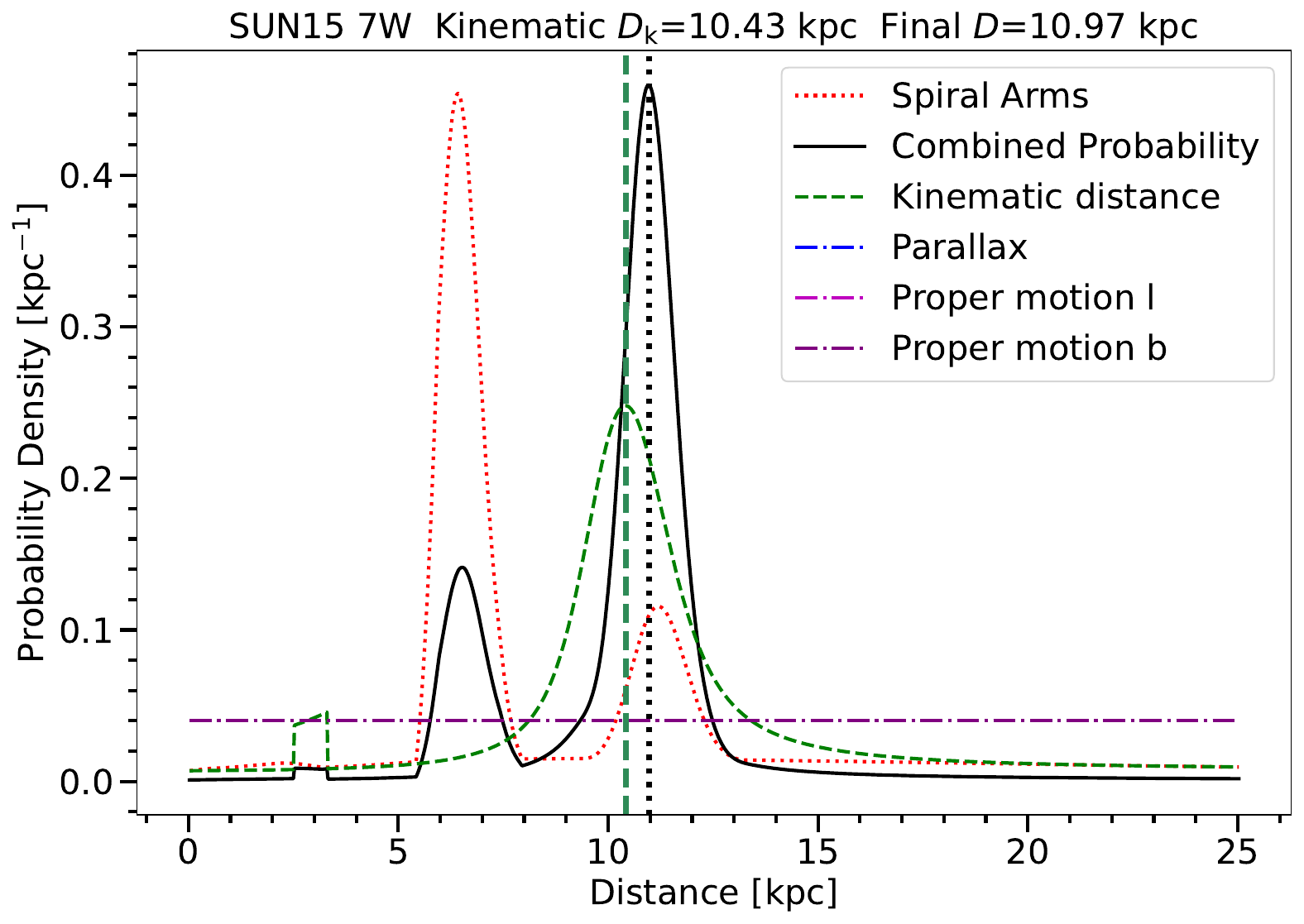}
                \includegraphics[scale=0.3]{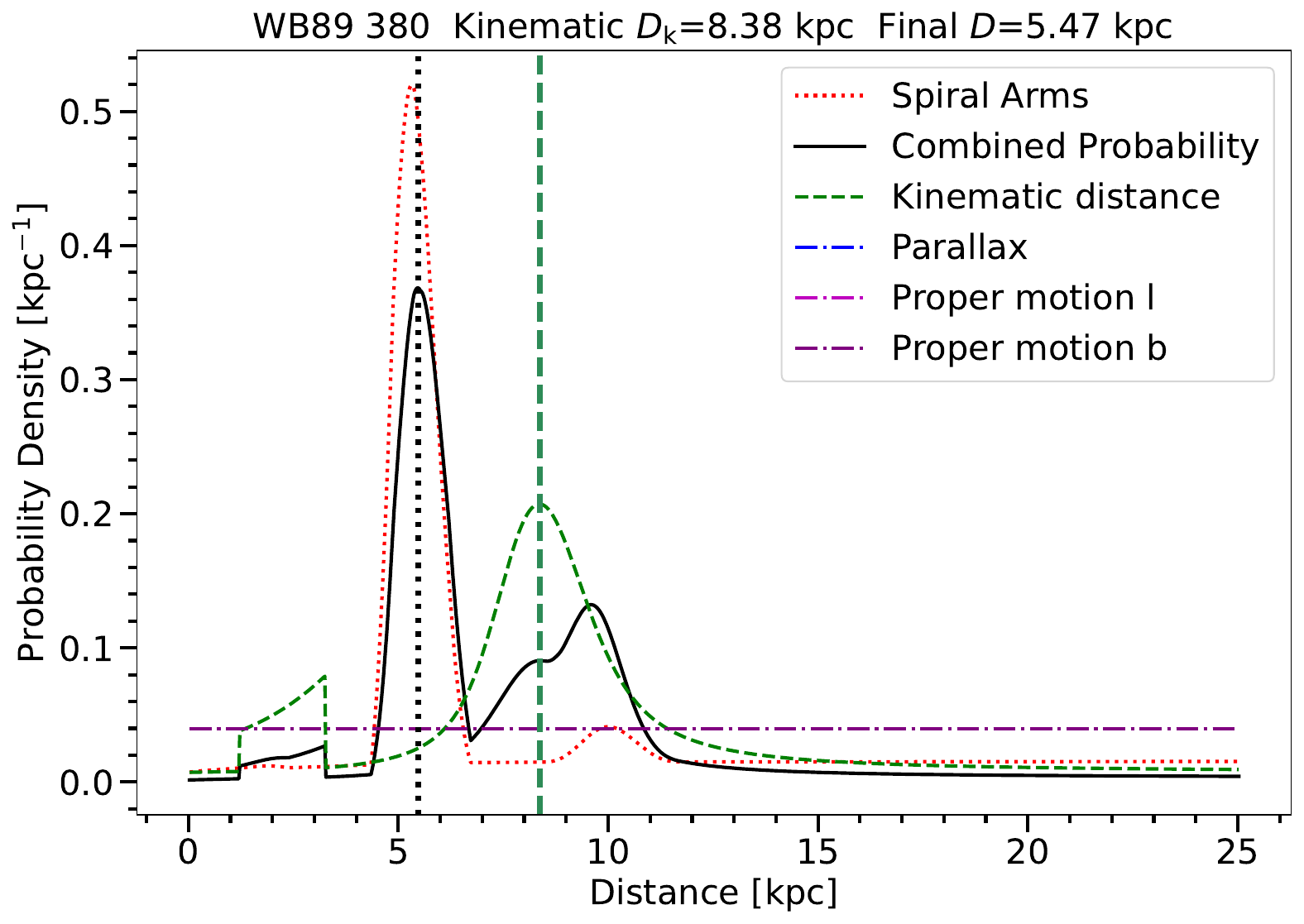}
                \includegraphics[scale=0.3]{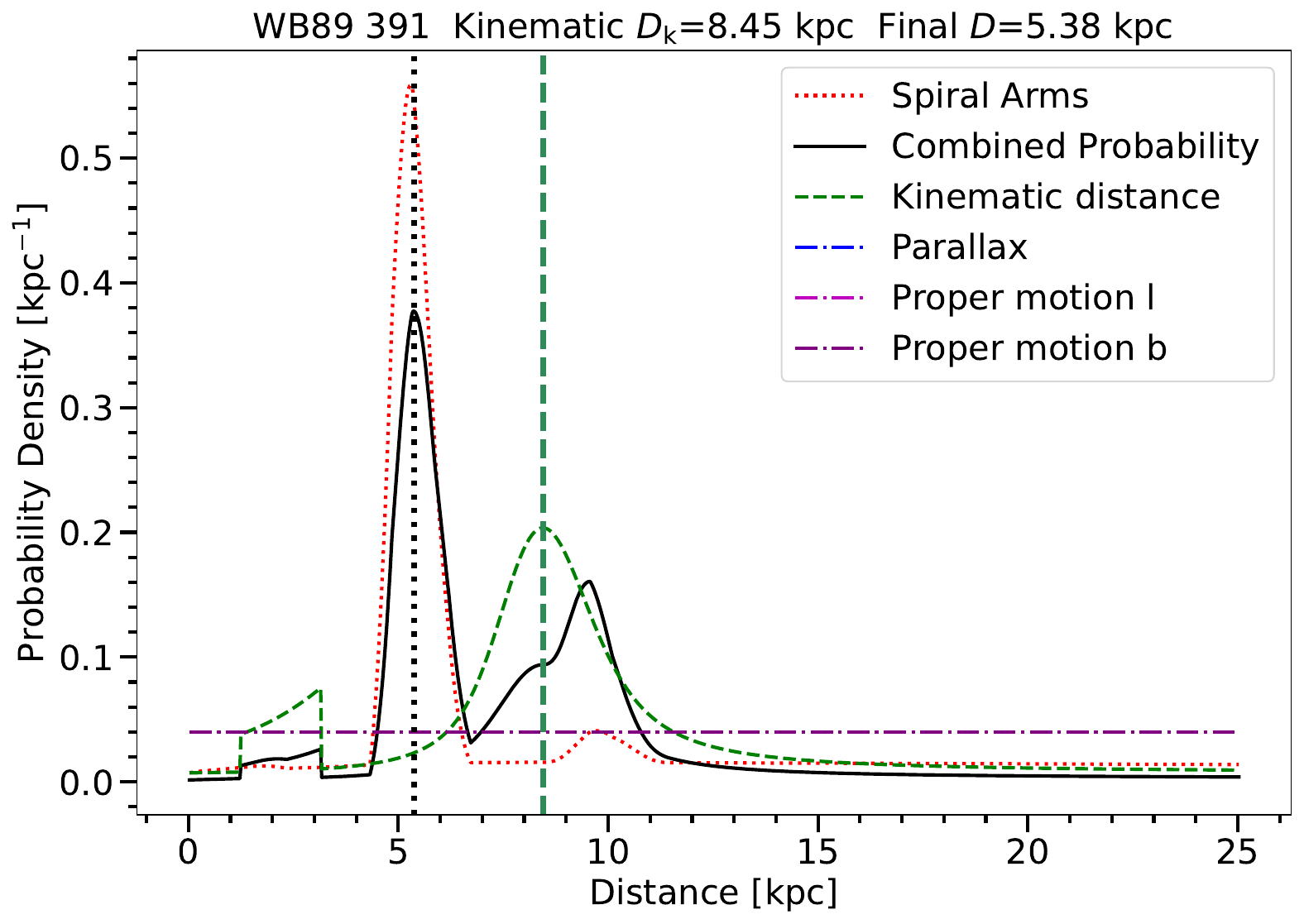}
                \includegraphics[scale=0.3]{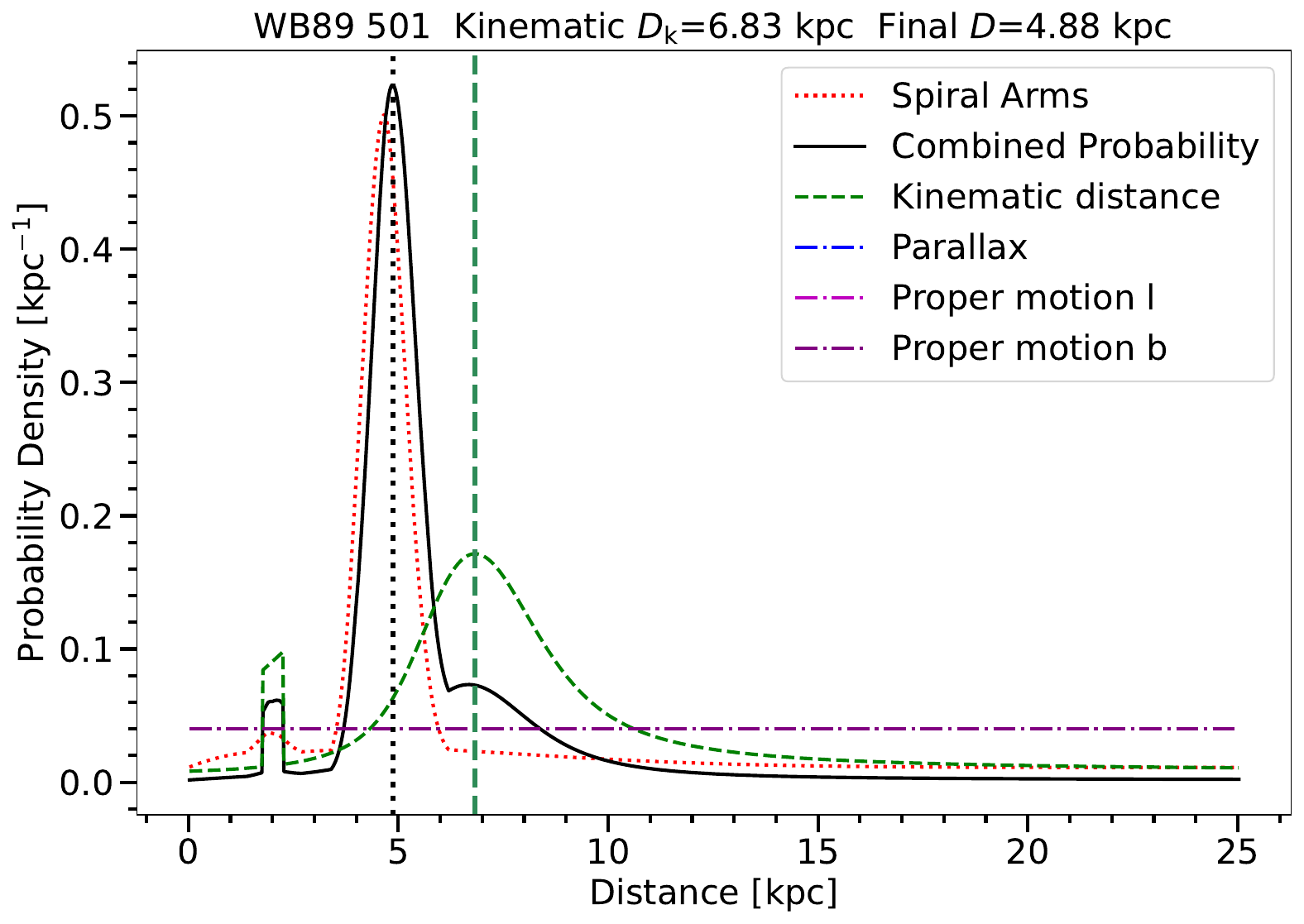}
        \caption{\label{fig:conti_source_dis_PDFs}The probability distribution functions of the target distances (Continued).}
\end{figure*}

We show the PDFs of the distances generated by the Parallax-Based Distance Calculator~\citep{Reid2016} for our targets in Fig.~\ref{fig:source_dis_PDFs}, and \ref{fig:conti_source_dis_PDFs}. Two of our targets, G211.59 and WB89 437, have direct parallax measurements of the water masers \citep{Reid2019}. So we use the parallax measurements to derive the Galactocentric distances for the two sources and do not show the PDFs of these two sources here. 

\section{The derivation of a more complete the \Rtwth\ formula}
\label{Appendix:12C13Cderiv}

\subsection{The derivation of \Rtwth\ formula in the literature}
\label{Appendix:12C13Cderiv_for_previous}
In this section, we illustrate our derivation of \Rtwth. We assume that the element abundance ratio \Rtwth\ equals the column density ratio between $\rm ^{12}CN$ and $\rm ^{13}CN$ (i.e. ignore the astrochemistry effects). The column density can be described as~\citep{Mangum2015}:

\begin{equation}
\label{eq:eqB1}
	N_{\rm tot}=\frac{3h}{8\pi^3|\mu_{\rm lu}|^2}\frac{Q_{\rm tot}}{g_{\rm u}}\frac{{\rm exp}(\frac{E_{\rm u}}{kT_{\rm ex}})}{{\rm exp}(\frac{h\nu}{kT_{\rm ex}})-1}\int \tau_{\nu} dv
\end{equation}

Here, $N_{\rm tot}$ means the total molecular column density of all energy levels. $h$ is the Planck constant and $\mu_{\rm lu}$ represents the dipole moment matrix element. $Q_{\rm tot}$ represents the rotational partition function. $E_{\rm u}$ and $g_{\rm u}$ represent the energy level and the degeneracy of the upper-level u, respectively. 
In this work, we assume the molecular structures of $\rm ^{12}CN$ and $\rm ^{13}CN$ are similar so the difference between $\mu_{\rm lu}$, $Q_{\rm tot}$, $E_{\rm u}$ and $g_{\rm u}$ can be ignored. We also assume a common excitation temperature $ T_{\rm ex}$\footnote{While strictly speaking
LTE is when $ T_{\rm ex}$=$T_{\rm  kin}$, it is unlikely that all the rotational energy levels would have a common
excitation achieved without the aid of frequent collisions.}. 
Then the column density ratio of $^{12}$CN and $^{13}$CN $N = 1 \to 0$ can be described as:

\begin{equation}
\label{eq:eqB2}
	\frac{N_{\rm tot, ^{12}CN}}{N_{\rm tot, ^{13}CN}} = \frac{\int \tau_{\nu, \rm ^{12}CN} dv}{\int \tau_{\nu, \rm ^{13}CN} dv}=\frac{\tau_{12}}{\tau_{13}}
\end{equation}

Here $\tau_{12}$ and $\tau_{13}$ represent the optical depth of $\rm ^{12}CN$ and $\rm ^{13}CN$ $N=1\to0$, respectively. We can derive the column density ratio of the main components of $\rm ^{12}CN$ and $\rm ^{13}CN$ $N=1\to 0$ and then convert this ratio to the column density ratio of all the energy levels. We can set $R_{12}$ and $R_{13}$ as conversion factors from the main component to all states of $\rm ^{12}CN$ and $\rm ^{13}CN$, respectively. That means:

\begin{equation}
	\frac{N_{\rm main, ^{12}CN}}{N_{\rm tot, ^{12}CN}}=R_{12},\quad \frac{N_{\rm main, ^{13}CN}}{N_{\rm tot, ^{13}CN}}=R_{13}
\end{equation}

\noindent We will also have:
\begin{equation}
	\frac{N_{\rm main, ^{12}CN}}{N_{\rm main, ^{13}CN}}=\frac{R_{12}\tau_{12}}{R_{13}\tau_{13}}=\frac{\tau_{\rm m, 12}}{\tau_{\rm m, 13}}
\end{equation}
The $\tau_{\rm m, 12}$ and $\tau_{\rm m, 13}$ are the optical depth of the main component of $\rm ^{12}CN$ and $\rm ^{13}CN$, respectively. 

On the other hand, the equation of radiative transfer for a uniform-excitation source is \citep[e.g.,][]{Mangum2015}:

\begin{equation}
\label{eq:eqB5}
	T_{\rm R}=f[J_{\nu}(T_{\rm ex})-J_{\nu}(T_{\rm bg})][1-e^{-\tau_{\nu}}]
\end{equation}
Here $T_{\rm R}$ is the beam-averaged source radiation temperature with $f$ the source beam-filling factor and $T_{\rm bg}$ the 
background temperature here assumed to be the CMB at 2.73\,K.  The beam filling factor: f=$\Omega_{\rm s}/\Omega_{\rm mb}$, with $\rm \Omega_{\rm s}$=$\int _{\omega _{\rm s}} P_{\rm n}(\omega) \Psi_{\rm s}(\omega) d\omega _s$, where $P_{\rm n}$ is the beam pattern function, $\Psi_{\rm s}$  the normalised
source brightness distribution function \citep[see][Equation 3]{Kutner1981} and $\omega_{\rm s}$ the solid angle range subtended by the 
source, while $\Omega_{\rm mb}$=$\int_{\omega_{\rm d}} P_{\rm n} d\omega $, with $\omega_{\rm d} $ the beam forward-pattern (typically Gaussian
for the IRAM 30-m telescope at its operating frequencies, but not necessarily so for high-frequency sub-mm telescopes).

The $J_{\nu}(T)$ expression gives the Rayleigh-Jeans equivalent temperature, but after using the
 Planck function for the black body radiation:
\begin{equation}
	T_{\rm b}=\frac{c^2}{2 k_{\rm B} \nu^2} B_{\rm \nu}(T)=J_{\rm \nu}(T)\equiv \frac{\frac{h\nu}{k}}{e^{\frac{h\nu}{kT}}-1}
\end{equation}

\noindent
If the line is optically thin, Equation~\ref{eq:eqB5} becomes 

\begin{equation}
\label{eq:eqB7}
	T_{\rm R}=f[J_{\nu}(T_{\rm ex})-J_{\nu}(T_{\rm bg})]\tau_{\nu}
\end{equation}
Insert Equation~\ref{eq:eqB7} into~\ref{eq:eqB1} and simplify the~\ref{eq:eqB1} as:
\begin{equation}
       N_{\rm tot}={\rm A}({\rm mol},T_{\rm ex})\int \tau_{\nu} dv
\end{equation}
Here the $ {A}({\rm mol}, T_{\rm ex})$ refers to all the parameters in Equation~\ref{eq:eqB1} to the left of the integral sign. Then we get:

\begin{equation}
	N_{\rm tot}={\rm A}({\rm mol},T_{\rm ex})\int \frac{T_{\rm R}}{f[J_{\nu}(T_{\rm ex})-J_{\nu}(T_{\rm bg})]} dv
\end{equation}

\noindent
as the source-averaged column density, with $\langle N_{\rm tot} \rangle _{\rm beam}$=$fN_{\rm tot}$ being the beam-averaged
column density. Here it is important to note that while line ratio excitation analysis can constrain the $T_{\rm ex}$ values, this
cannot be done for the source beam filling factor unless independent source size estimates can be had (e.g. via past interferometric
observations of the same molecules or concomitant line and/or continuum emitters (e.g. dust emission). Thus typically in the 
literature $\langle N_{\rm tot} \rangle _{\rm beam}$ is typically reported.

If the line has some modest optical depths, $N_{\rm tot}$ can still be derived after multiplying a $\tau $-correcting factor $\frac{\tau}{1-e^{-\tau}}$ \citep{Goldsmith1999,Mangum2015}, yielding:

\begin{equation}
\label{eq:eqB10}
	N_{\rm tot}=\frac{\tau}{1-e^{-\tau}}{\rm A}({\rm mol},T_{\rm ex})\int \frac{T_{\rm R}}{f[J_{\nu}(T_{\rm ex})-J_{\nu}(T_{\rm bg})]} dv
\end{equation}

\noindent
, but such corrections become unreliable for large line optical depths ($\tau$$\geq $2-3).
For the small velocity range of the lines in this work ($\rm FWHM $$\sim$ $1.5-4$\,$\rm km\, s^{-1}$), the optical depth correction factor, $ A({\rm mol}, T_{\rm ex})$, and $J(\nu, T_{\rm ex})$ factors can 
all be considered constant across the line width. Thus we can compute the velocity-averaged
column density across the entire line profile (which yields the highest possible S/N ratio
for spectral line observations) from Equation \ref{eq:eqB10} as:

\begin{equation}
N_{\rm tot, V}=\frac{\tau}{1-e^{-\tau}}\frac{{\rm A}({\rm mol},T_{\rm ex})}{f[J_{\nu}(T_{\rm ex})-J_{\nu}(T_{\rm bg})]}\left[\frac{1}{\Delta V}\int T_{\rm R}dv\right]
\end{equation}

\noindent
For the \Rtwth\ derivation, with $\rm ^{12}CN$ $N=1\to0$ optically thick and the $\rm ^{13}CN$ $N=1\to0$ optically thin (thus a 
$\tau$-correction factor only applies to $^{12}$CN), and$ A({\rm mol}, T_{\rm ex})$ and the source beam-filling factors 
$f$ identical between the two lines, we can write:

\begin{equation}
\label{eq:eqB12}
	\frac{N_{\rm main, ^{12}CN}}{N_{\rm main, ^{13}CN}}=\frac{\tau}{1-e^{-\tau}}\frac{\int T_{\rm R, ^{12}CN}dv}{\int T_{\rm R, ^{13}CN}dv}
\end{equation}

\noindent
Now we consider the formula to calculate \Rtwth\ used by~\citet{Savage2002} and ~\citet{Milam2005}. The formula they used is:
\begin{equation}
\label{eq:eqB13}
	\frac{\rm ^{12}C}{\rm ^{13}C} = \frac{(3/5)\tau_{\rm main}T_{\rm ex, ^{12}CN}}{T^\star_{\rm R, ^{13}CN}/\eta_{\rm c, ^{13}CN}}
\end{equation}

Here they erroneously set $\eta_{\rm c}$ as the beam efficiency, despite the $T^{\star}_{\rm R}$ temperature scale of the 12-m telescope
being already corrected for all telescope efficiency factors (and of course atmospheric absorption). Here, $\tau_{\rm main}$ is the optical depth of the main component of $\rm ^{12}CN$ and $T^\star_{\rm R, ^{13}CN}$ is the measured radiation temperature of the $\rm ^{13}CN$ main component and 3/5 is the conversion factor from the ratio of main components to the ratio of all levels. 

To deduce Equation~\ref{eq:eqB13}, we need to obtain a formula with $T_{\rm ex, ^{12}CN}$, $\tau_{\rm main}$ and $T^\star_{\rm R, ^{13}CN}$. We replace $T^\star_{\rm R, ^{12}CN}$ by inserting Equation~\ref{eq:eqB7} into Equation~\ref{eq:eqB12}:

\begin{equation}
	\frac{N_{\rm main, ^{12}CN}}{N_{\rm main, ^{13}CN}}=\frac{\tau}{1-e^{-\tau}}\frac{\int f[J_{\nu}(T_{\rm ex})-J_{\nu}(T_{\rm bg})][1-e^{-\tau_{\nu}}] dv}{\int T_{\rm R, ^{13}CN}dv}
\end{equation}

\noindent
Assuming the optical depth is nearly constant across the narrow line profile, we can further write:

\begin{equation}
\begin{aligned}
    \frac{N_{\rm main, ^{12}CN}}{N_{\rm main, ^{13}CN}} & = \frac{\tau}{1-e^{-\tau}}\frac{f[J(T_{\rm ex})-J(T_{\rm bg})](1-e^{-\tau}) }{T_{\rm R, ^{13}CN}} \\
    & =\frac{\tau f[J(T_{\rm ex})-J(T_{\rm bg})] }{T_{\rm R, ^{13}CN}}
\end{aligned}
\end{equation}

\noindent
If we regard $T^\star_{\rm R, ^{13}CN}$ as the observed corrected source antenna temperature \footnote{Corrected for atmospheric attenuation, radiative loss, rearward and forward scattering, and spillover, according to the NRAO 12-m manual.} of the \thCN\ main component as reported
by the (former) NRAO 12m-telescope we will have:

\begin{equation}
\label{eq:eqB16}
	\frac{N_{\rm main, ^{12}CN}}{N_{\rm main, ^{13}CN}}=\frac{\tau [J(T_{\rm ex})-J(T_{\rm bg})] }{T^\star_{\rm R, ^{13}CN}/f_{\rm ^{13}CN}}
\end{equation}

\noindent
Assuming the $^{13}$CN  beam filling factor $f=1$, and converting the column density at one transition to the total molecular gas column density, we obtain:

\begin{equation}
\label{eq:eqB17}
\begin{aligned}
    \frac{N_{\rm tot, ^{12}CN}}{N_{\rm tot, ^{13}CN}} & =\frac{R_{13}\tau [J(T_{\rm ex})-J(T_{\rm bg})] }{R_{12}T^
    \star_{\rm R, ^{13}CN}/(\eta_{\rm c, ^{13}CN})} \\ 
     & \approx \frac{(3/5)\tau [J(T_{\rm ex})-J(T_{\rm bg})] }{T^\star_{\rm R, ^{13}CN}}
 \end{aligned}
\end{equation}

\noindent
Using the Rayleigh-Jeans approximation ($ J(T)$$ \approx T$) and ignoring CMB temperature $T_{\rm bg}$, yields the same equation as Equation~\ref{eq:eqB13}, but without the erroneous $\eta_{\rm c}$ factor.\footnote{Here is a good opportunity to mention that an $\eta_{\rm c}$ factor with $ T^{\star }_{\rm R}$=$\eta_{\rm c} T_{\rm R} $ is used to denote a remaining non-trivial (source-structure)-beam coupling factor
that exists even for resolved sources \citep{Kutner1981}.  In such cases (where $\omega _{\rm d}$<$\omega _{\rm s} $) the source beam filling factor
$f$, is longer $f \propto 1/\Omega _{\rm mb}$, but it converges to a value $\sim \eta_{\rm c}$ than can still be $<$1 (assuming only the Gaussian part of the forward beam pattern coupling to the extended source). This factor cannot be corrected without strong assumptions about the source structure underlying the main beam pattern (an impossibility for molecular clouds), and thus it is assumed as $\sim $1 in line studies of resolved molecular clouds when an absolute value of $f$ ($\sim \eta_{\rm c}$) is needed. For line ratio studies conducted for resolved molecular clouds, the less 
constraining assumption of a common $\eta_{\rm c}$ factor among the various lines used is only needed. \\ On the question of how we actually know
that $\rm \eta_{\rm c} <1$, even for well-resolved molecular clouds (unless one observes especially compact regions such as Bok globules with high
angular resolution), the answer is given from both theory and observations. The fractal-like structures
of molecular clouds, with their supersonic gas motions, contain compact velocity-coherent clumps that are extremely small \citep[$\sim $$10^{-3}$\,pc, e.g.,][]{Falgarone1998}. These will certainly fail to completely fill in the angular area of any typical spectral line observations of molecular clouds, 
thus keeping the respective $\eta_{\rm c}<1$,  even if the cloud structures themselves appear well-resolved by the beams used. Multi-$J$ CO observations 
that include the easily thermalized optically thick CO $J=1\to0$ line, when modeled via various radiative transfer codes yield also the expected emergent radiation temperature of the $J=1\to0$ line $T_{\rm R, J=1\to0}$. For cold clouds in the Galaxy this is typically $\sim$ ($10-15$)\,K, similar to the
$T_{\rm kin}$ and $T_{\rm dust}$ of such clouds, yet the typical {\it brightness temperature} observed for well-resolved such clouds are
$T_{\rm R} \sim$ ($5-6$)\,K (i.e. indicating an $\eta_{\rm c}$ $\sim $1/3-1/2 factor).}

\subsection{The effects of the R-J approximation and the  $ T_{\rm CMB}$ emission}
\label{Appendix:effect_of_RJ_and_CMB}

In order to quantify the effect of the R-J approximation we derive the excitation temperature \Tex\ of $\rm ^{12}CN$ $N=1\to0$ for our targets, according to the following equation:

\begin{equation}
\label{eq:eqB18}
	T^\star_{\rm A}=\eta_{\rm mb}f[J_{\nu}(T_{\rm ex})-J_{\nu}(T_{\rm bg})][1-e^{-\tau_{\nu}}] = \eta_{\rm mb}T_{\rm mb}
\end{equation}

\noindent
We assume that $T_{\rm bg}$=$T_{\rm CMB}$=2.73\,K, with $T^\star_{\rm A}$ and $\tau_{\nu}$ of \twCN\ $N=1\to0$ obtained from HfS fitting. The \Tex\ of \twCN\ is then derived from Eq.~\ref{eq:eqB18}. 
The results are listed in Table~\ref{tab:Target_Tex}. The mapping data of $\rm ^{13}CO$ is used to estimate the filling factor $f$, with the assumption that the shape of targets is round with an area including all pixels having $T_{\rm mb, ^{13}CO} > 0.5 \ T_{\rm mb, ^{13}CO}^{\rm peak}$.

\begin{table}
\caption{$T_{\rm ex}$ of $\rm ^{12}CN$ $N=1\to0$ in our targets.}
\label{tab:Target_Tex}
\begin{threeparttable}
\begin{tabular}{cc}
\hline
\hline
Sources              &  $T_{\rm ex, ^{12}CN}$ \\
                      & (K)                              \\
\hline
G211.59              &  4.61 $\pm$ 0.08$\rm ^{a}$        \\
G37.350              & 3.5 $\pm$ 1.5                     \\
G44.8                & 3.7 $\pm$ 1.0                     \\
IRAS0245             & 3.6 $\pm$ 0.3                     \\
SUN15 14N            &  -                                \\
SUN15 18             &  -                                \\
SUN15 21$\rm ^{b}$   &  3 $\pm$ 10                       \\
SUN15 34             &  -                                \\
SUN15 56$\rm ^{b}$   &  2.9 $\pm$ 1.4                    \\
SUN15 57$\rm ^{b}$   &  3.1 $\pm$ 7.7                    \\
SUN15 7W             &  -                                \\
WB89 380             &  4.34 $\pm$ 0.18                  \\
WB89 391             &  4.16 $\pm$ 0.20                  \\
WB89 437             &   6.7 $\pm$ 1.1                   \\
WB89 501             &   4.0 $\pm$ 1.0                   \\
\hline
\end{tabular}

\begin{tablenotes}
\footnotesize
\item[a.]  A filling factor f=1 (and $\eta _c$=1) is assumed due to the unresolved mapping data, thus yielding a {\it minimum} $T_{\rm ex}$ value.
\item[b.] Huge errors are shown for SUN15 21, SUN15 56, and SUN15 57, which means that the ratio limit results for these three sources are not credible.
\end{tablenotes}

\end{threeparttable}
\end{table}

\begin{figure*}
\includegraphics[scale=0.42]{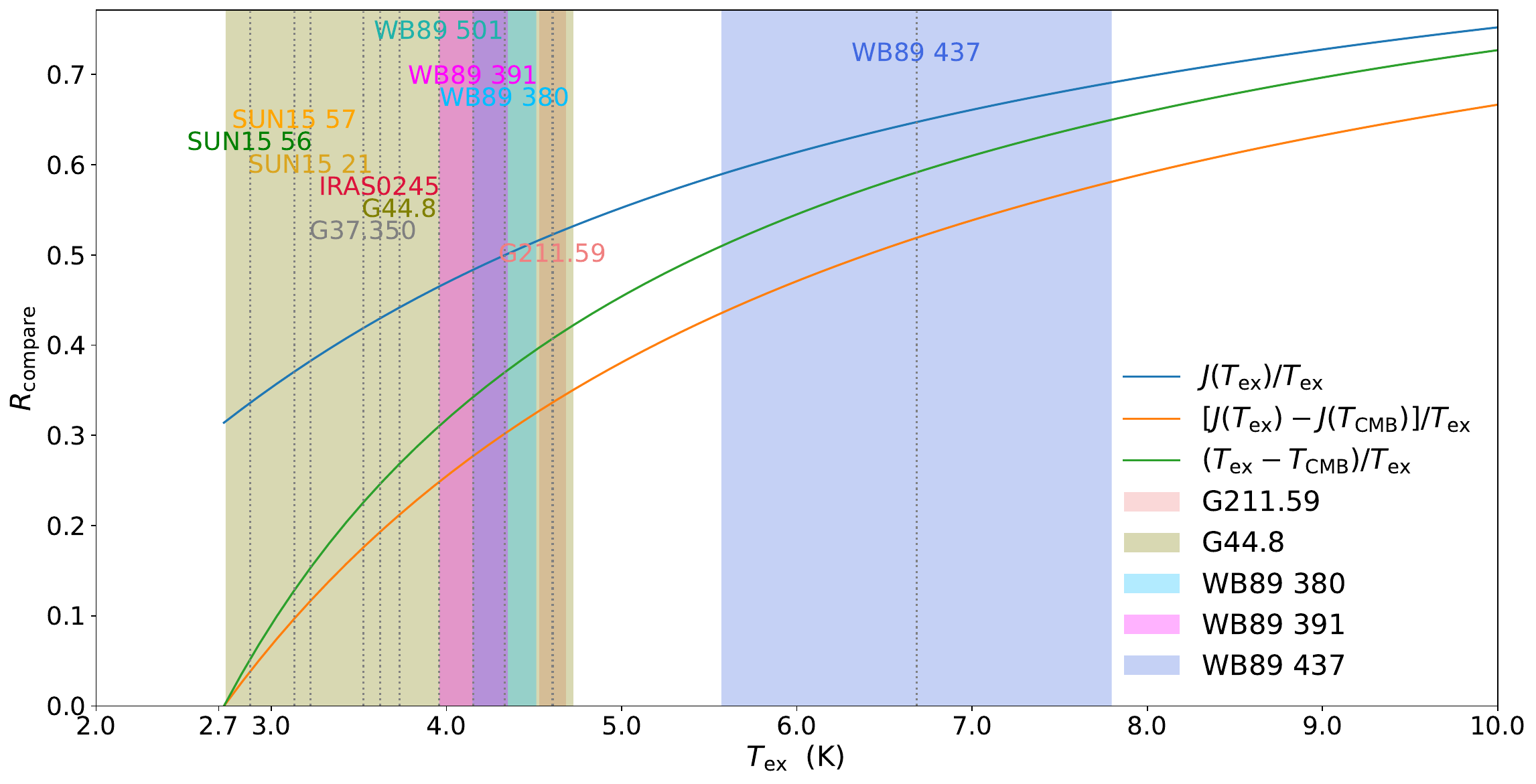}
\centering
	\caption{\label{fig:figA1} The effects of Rayleigh-Jeans approximation and background temperature. The solid lines show the change of $R_{\rm compare}$ with the excitation temperature $T_{\rm ex}$, in the condition a (blue), condition b (green), and condition c (orange). The dashed lines show the \Tex of $\rm ^{12}CN $ $N=1\to0$ in our sources and colorful blocks show the error range of $T_{\rm ex}$.}
\end{figure*}

\begin{figure*}
    \includegraphics[scale=0.5]{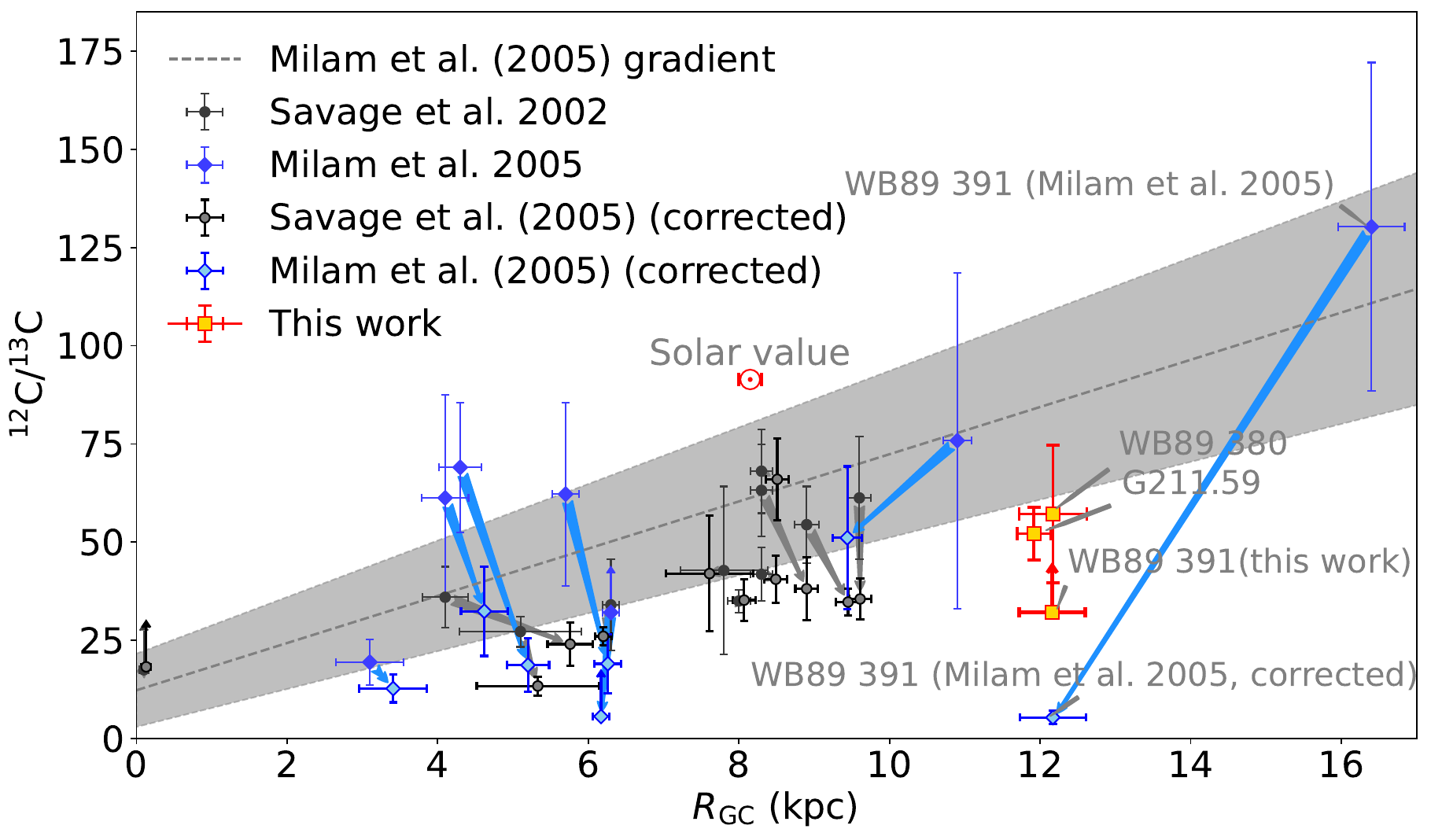}
\centering
\caption{Galactic \Rtwth\ gradients derived from CN, before
and after our corrections. The grey dots, blue diamonds, and red squares refer to
results from \citet{Savage2002}, \citet{Milam2005}, and this work, respectively.
The symbols filled with darker colors refer to previous values in the literature and those filled with lighter colors are the corrected values. The arrows show the change.} \label{fig:12C_13C_show_revis}
\end{figure*}

We define the ratios between the revised \Rtwth\ and the previous \Rtwth, labeled as $R_{\rm compare}$. Then we consider three conditions:

(a) Using the Planck equation and without $T_{\rm CMB}$:
\begin{equation}
        R_{\rm compare, a}= \frac{R_{\rm 12C13C, Planck\ eq, no\ CMB}}{R_{\rm 12C13C, R-J\ approx, no\ CMB}} 
        = \frac{J(T_{\rm ex})}{T_{\rm ex}}
\end{equation}

(b) With R-J approximation and with $T_{\rm CMB}$:
\begin{equation}
        R_{\rm compare, b}= \frac{R_{\rm 12C13C, R-J\ approx, CMB}}{R_{\rm 12C13C, R-J\ approx, no\ CMB}} 
        = \frac{T_{\rm ex}-T_{\rm CMB}}{T_{\rm ex}}
\end{equation}

(c) Using the Planck equation and with $T_{\rm CMB}$:

\begin{equation}
\begin{aligned}
        R_{\rm compare, c}= \frac{R_{\rm 12C13C, Planck\ eq, CMB}}{R_{\rm 12C13C, R-J\ approx, no\ CMB}} & \\
        = \frac{J(T_{\rm ex})-J(T_{\rm CMB})}{T_{\rm ex}}
\end{aligned}
\end{equation}

\noindent
Fig. ~\ref{fig:figA1} shows the change of $R_{\rm compare}$ with \Tex. For all the conditions $R_{\rm compare}$ will decrease when \Tex decreases. In condition (a),  $R_{\rm compare}$ > 0.95, which means the decrease of \Rtwth\ after revision is $\le 5$ \%, for $T_{\rm ex} \gtrsim 53$ K. $R_{\rm compare}$ becomes $\sim 0.5$ for $T_{\rm ex}\sim 4.3$ K. In condition (b), $R_{\rm compare}$ > 0.95 and $\sim 0.5$ for $T_{\rm ex}\gtrsim 54$ K and $\sim 6.3$ K, and  $\sim$ 3.1 K, respectively. For case (c), used in this work, it is $R_{\rm compare}$>0.95, $\sim 0.5$, and $\sim$0.1, for $T_{\rm ex}$$\gtrsim $70\,K, $\sim $6.3\,K, and  $\sim$ 3.1\,K, respectively. We conclude that if \Tex$\leq $10\,K, the Planck equation and the CMB  must be considered to derive \Rtwth\ from $\rm ^{12}CN/^{13}CN$.

In Fig.~\ref{fig:12C_13C_show_revis}, we show how the effects of the R-J approximation and the CMB temperature will revise the previous Galactic $\rm ^{12}CN/^{13}CN$ gradient derived from the HfS method. After the Galactocentric distance revision, the targets have a change of \Rgc\ at $\sim 0-4$ kpc, and the \Rgc\ are limited to $\lesssim$ 12 kpc. The values of \Rtwth\ systematically decrease after the revision. Especially, the previous out-most constraint, \Rtwth\ in WB89 391, has been in an unphysical region after the revision, which means the previous measurement of this target is highly doubtful.

\subsection{The derivation of a $\tau$-corrected \Rtwth\ formula}
\label{Appendix:improved_HfS_method}

\begin{table}
\caption{$T_{\rm ex}$ of $\rm ^{12}CN$ $N=1\to0$ in targets of \citet{Savage2002} and \citet{Milam2005}.}
\label{tab:Target_Tex_Savage_Milam}
\begin{threeparttable}
\begin{tabular}{ccc}
\hline
\hline
Sources & $T_{\rm ex, ^{12}CN, pre.}$ & $T_{\rm ex, ^{12}CN, revis.}$ \\
    & (K) & (K) \\
\hline
W3(OH) & 6.00  $\pm$  0.50 & 5.41 $\pm$ 0.41 \\
G34.3 & 6.00  $\pm$  0.20 & 5.41 $\pm$ 0.16 \\
W51M & 12.4  $\pm$  2.0 & 10.7 $\pm$ 1.6 \\
NGC 7538 & 8.10  $\pm$  0.50 & 7.13 $\pm$ 0.41 \\
NGC 2024 & 7.30  $\pm$  0.10 & 6.48 $\pm$ 0.08 \\
W33 & 8.90  $\pm$  0.40 & 7.79 $\pm$ 0.33 \\
G29.9 & 5.10  $\pm$  0.72 & 4.67 $\pm$ 0.59 \\
G19.6 & 3.39  $\pm$  0.08 & 3.27 $\pm$ 0.07 \\
W31 & 6.95  $\pm$  0.85 & 6.19 $\pm$ 0.70 \\
G35.2 & 4.49  $\pm$  0.40 & 4.17 $\pm$ 0.33 \\
S156 & 6.6  $\pm$  1.6 & 5.9 $\pm$ 1.3 \\
WB89 391 & 2.90  $\pm$  0.03 & 2.87 $\pm$ 0.02 \\
\hline
\end{tabular}


\end{threeparttable}
\end{table}

Equation~\ref{eq:eqB17} contains the assumption that the filling factor $f=1$. However, in this work, we take a more simple consideration, namely
we assume \Tex\ is the same for both $\rm ^{12}CN$ and $\rm ^{13}CN$ $N=1\to0$ main components. Because the corresponding frequencies are so similar
for these two components the now frequency-dependant expressions $J(T_{\rm ex})-J(T_{\rm bg})$ will also be very similar for the two lines (and
of course the main beam efficiency factors $\eta_{\rm mb}$ as well). We assume the spatial distributions of these isotopologues are similar so that  $f$ (or $\eta_{\rm c}$ for resolved-source observations) is the same. According to Eq.~\ref{eq:eqB18}, we then have:

\begin{equation}
\label{eq:eqB22}
	R_{\rm b}=\frac{T_{\rm b, ^{12}CN}}{T_{\rm b, ^{13}CN}}=\frac{f_{\rm ^{12}CN}(1-e^{-\tau_{\rm m, 12}})}{f_{\rm ^{13}CN}(1-e^{-\tau_{\rm m, 13}})}=\frac{1-e^{-\tau_{\rm m, 12}}}{1-e^{-\tau_{\rm m,  13}}}=\frac{T_{\rm mb, ^{12}CN}}{T_{\rm mb, ^{13}CN}}
\end{equation}

Here $T_{\rm b, ^{12}CN}$ and $T_{\rm b, ^{13}CN}$ are the peaks of the brightness temperature of the main components of $\rm ^{12}CN$ and $\rm ^{13}CN$. We set $R_{\rm b}$=$T_{\rm b, ^{12}CN}/T_{\rm b, ^{13}CN}$, which is the brightness temperature ratio of the main components of $\rm ^{12}CN$ and $\rm ^{13}CN$. With the assumption that the filling factor of \twCN\ ($f_{\rm ^{12}CN}$) equals the filling factor of \thCN\ ($f_{\rm ^{12}CN}$), we have $R_{\rm b}$=$T_{\rm mb, ^{12}CN}/T_{\rm mb, ^{13}CN}$. $\tau_{\rm m, 12}$ and $\tau_{\rm m, 13}$ are the optical depths of the \twCN\ and \thCN\ main components, respectively. We also~set:

\begin{equation}
\label{eq:eqB23}
	R_{\rm 1213}=\frac{N_{\rm main, ^{12}CN}}{N_{\rm main, ^{13}CN}}=\frac{\tau_{\rm m, 12}}{\tau_{\rm m, 13}}
\end{equation}

with $R_{\rm 1213}$=$N_{\rm main, ^{12}CN}/N_{\rm main, ^{13}CN}$ being the column density ratio between the main components of \twCN\ and \thCN\ $N=1\to0$. Then we have $\tau_{\rm m, 13}=\tau_{\rm m, 12}/R_{\rm 1213}$. Inserting this into~\ref{eq:eqB22} yields:
\begin{equation}
\label{eq:eqB21}
	R_{\rm b}=\frac{1-e^{-\tau_{\rm m, 12}}}{1-e^{-\tau_{\rm m, 12}/R_{\rm 1213}}},
\end{equation}

\noindent
which, solving for $R_{\rm 1213}$, it simply yields:
\begin{equation}
	R_{\rm 1213}=-\frac{\tau_{\rm m, 12}}{{\rm ln}[1-(1-e^{-{\tau_{\rm m, 12}}})/R_{\rm b}]}
\end{equation}
That is:
\begin{equation}
	\frac{N_{\rm main, ^{12}CN}}{N_{\rm main, ^{13}CN}}=-\frac{\tau_{\rm m, 12}}{{\rm ln}[1-\frac{T_{\rm mb, ^{13}CN}}{T_{\rm mb, ^{12}CN}}(1-e^{-{\tau_{\rm m, 12}}})]}
\end{equation}
We then convert the column density ratio of the main components to that of all levels of $\rm ^{12}CN$ and $\rm ^{13}CN$ $N=1\to0$. That is:

\begin{equation}
\label{eq:eqB27}
\begin{aligned}
    \frac{\rm ^{12}C}{\rm ^{13}C} & = \frac{N_{\rm tot, ^{12}CN}}{N_{\rm tot, ^{13}CN}} \\
    & =-\frac{R_{13}}{R_{12}}\cdot \frac{\tau_{\rm m, 12}}{{\rm ln}[1-\frac{T_{\rm mb, ^{13}CN}}{T_{\rm mb, ^{12}CN}}(1-e^{-{\tau_{\rm m, 12}}})]} \\
    & =-\frac{R_{13}}{R_{12}}\cdot \frac{\tau_{\rm m, 12}}{{\rm ln}[1-\frac{T^\star_{\rm A, ^{13}CN}/\eta_{\rm mb, ^{13}CN}}{T^\star_{\rm A, ^{12}CN}/\eta_{\rm mb, ^{12}CN}}(1-e^{-{\tau_{\rm m, 12}}})]}
\end{aligned}
\end{equation}

Equation~\ref{eq:eqB27} is the one used in this work to derive \Rtwth\ with optical-depth correction. There is no need to consider particular $T_{\rm ex}$ and filling factor $f$ (as long as they are common for the lines/species used) once we convert the measured antenna temperature $T^\star_{\rm A}$ to the main beam temperature $T_{\rm mb}$ with the main beam efficiencies $\eta_{\rm mb}$ at the frequency of  \twCN\ and \thCN\ main components. This derivation takes into account the CMB and abandons the simple Rayleigh-Jeans approximation. It is thus more appropriate than  Equation~\ref{eq:eqB13}  for the low $T_{\rm ex}$ often found in these sources (and Eq.~\ref{eq:eqB27} is unaffected by the error in Eq.~\ref{eq:eqB13} regarding the $\eta_{\rm c}$ factor). 

\begin{figure}
\includegraphics[scale=0.4]{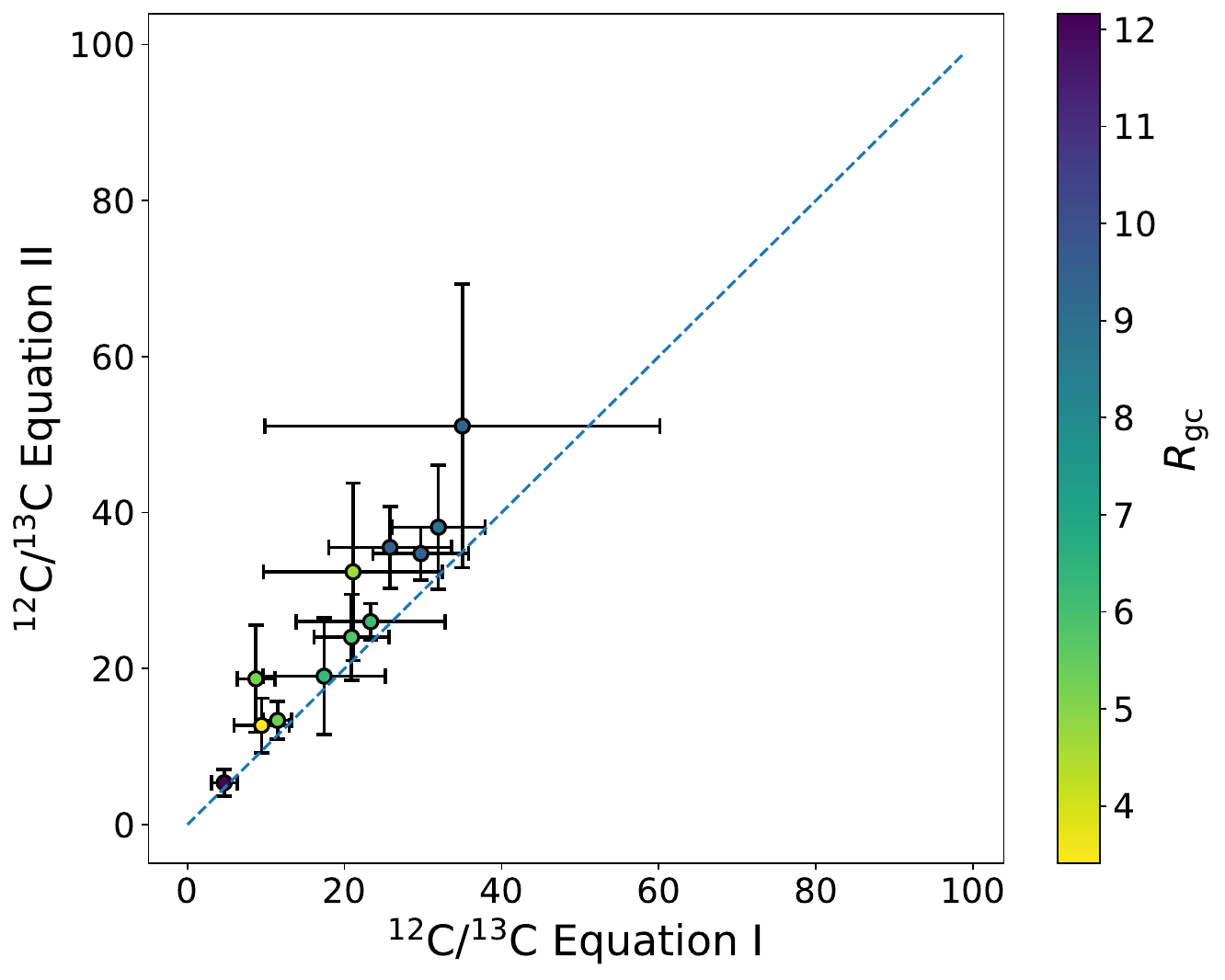}
\centering
	\caption{The comparison with $\rm{^{12}C/^{13}C}$ results derived by Eq.~\ref{eq:eqB17} (Equation I) and Eq.~\ref{eq:eqB27} (Equation II). The blue dashed line refers to where the x-value equals the y-value. The color map shows the Galactocentric distances of the targets. \label{fig:12C_13C_two_methods}} 
\end{figure}

We now compare \Rtwth\ derived from Eq.~\ref{eq:eqB17} (labeled as Equation I) and Eq.~\ref{eq:eqB27} (labeled as Equation II) in Fig.~\ref{fig:12C_13C_two_methods}. We use the \Tex, $\tau$, $T^\star_{\rm A, ^{12}CN}$ and $T^\star_{\rm A, ^{13}CN}$ of targets in \citet{Savage2002,Milam2005} and derive \Rtwth\ in these targets with the two equations. 

Because of the incorrect treatment of the irreducible beam coupling factor $\eta_{\rm c}$ in \citet{Savage2002} and \citet{Milam2005} (Mentioned in Section~\ref{sec:traditional_methods}), the excitation temperature they derived should be revised as follows:

\begin{equation}
    T_{\rm ex, revis.}=(T_{\rm ex, pre.}-T_{\rm CMB})*\eta_{\rm mb, 12-m}+T_{\rm CMB}
\end{equation}

Here $T_{\rm ex, pre.}$ and $T_{\rm ex, revis.}$ are the previous excitation temperature in \citet{Savage2002} and \citet{Milam2005}, and the revised values of excitation temperature. $\eta_{\rm mb, 12-m}=0.82$ is the main beam efficiency of NRAO 12-m telescope at the rest frequency of \twCN\ $N=1\to0$ main component, which is incorrectly used in Eq.(2) and (3) in \citet{Savage2002} with a misunderstanding symbol $\eta_{\rm c}$. The revised \Tex\ are listed in Table. \ref{tab:Target_Tex_Savage_Milam}.

In Fig. \ref{fig:12C_13C_two_methods}, the \CtwCth\ ratios derived from Eq.~\ref{eq:eqB17} and Eq.~\ref{eq:eq_12C13C_hfs} are mostly consistent within the errorbar. However, the ratios from Eq.\ref{eq:eqB17} are slightly but systematically lower than those from Eq.~\ref{eq:eq_12C13C_hfs}, which is highly likely because the actual filling factor $f\le 1$. This comparison means the revision of the previous \Rtwth\ derivation is essential no matter whether we consider the $T_{\rm ex}$ and the filling factor $f$ or not.

Another possibility is that in \citet{Savage2002} and \citet{Milam2005} they derive \Tex\ correctly by considering $\eta_{\rm c}$ as the irreducible beam coupling factor, but express it wrongly as beam efficiency. In this condition, the ratios from Eq. \ref{eq:eqB17} and \ref{eq:eq_12C13C_hfs} will be even more consistent than what we show in Fig.~\ref{fig:12C_13C_two_methods}.

\subsection{Different excitation conditions for $\rm ^{12}CN$ and $\rm ^{13}CN$}

\subsubsection{A two \Tex\ layer toy model}
\label{Appendix:Two_Tex_layer_model}

In Fig.~\ref{fig:Two_layer_model} we show a toy model cloud with two layers having different \Tex\ for the same molecule/transition (e.g., \twCN). From the right to the left, there is the background, a layer with a higher \Tex, a layer with a lower \Tex, and the observer. Indicatively, we set the high excitation temperature $T_{\rm ex, H}$=30\,K and the lower excitation temperature $T_{\rm ex, C}$=10\,K. The optical depth of the high \Tex\ layer and the low \Tex\ layer is $\tau_{\rm H}$ and $\tau_{\rm C}$, respectively. The intensity at a certain frequency on the background, on the interface of the two layers, and the intensity observed by us are labeled as $I_{\rm \nu,0}$, $I_{\rm \nu,1}$ and $I_{\rm \nu,2}$, respectively. 

We use this toy model in order to explore, in a simple manner, the possible effects of underlying excitation variations
(i.e. a non-uniform \Tex)  on the extracted abundance ratios Assuming uniform \Tex\ conditions within 
each layer the emergent line brightnesses will be \citep[e.g.,][]{Mangum2015}:

\begin{equation}
    I_{\rm \nu, 1}=I_{\rm \nu, 0}e^{-\tau_{\rm H}} + B_{\nu}(T_{\rm ex, H})(1-e^{-\tau_{\rm H}})    
\end{equation}

\begin{equation}
    I_{\rm \nu, 2}=I_{\nu, 1}e^{-\tau_{\rm C}} + B_{\nu}(T_{\rm ex, C})(1-e^{-\tau_{\rm C}})
\end{equation}

\noindent
where $B_{\nu}$ is the Planck function. Here, the spectral line optical depths
 are strong functions of frequency: $\tau_{\rm H, C}$=$\tau _{0}\, \phi (\nu-\nu_{0})$, 
with $\phi (\nu-\nu_{0})$  a normalized emission line profile, strongly peaked
at some central frequency, and a width determined by the gas motions within each layer.
 The assumption of a common line profile  function 
 $\phi (\nu -\nu _{\circ})=\phi _{\rm C}(\nu-\nu_{0})=\phi _{\rm H}(\nu-\nu_{0})$ for both gas layers
 is equivalent to the assumption of micro-turbulent and/or thermal gas velocity fields dominating
  both gas layers.  Indeed only under this assumption, the radiative transfer formalism
 that follows is applicable\footnote{For the macroturbulent/supersonic gas velocity fields of
 molecular clouds,  the velocity gradients are so steep that $\phi _{\rm C}(\nu-\nu_{0})\neq \phi _{\rm H}(\nu-\nu_{0})$ even for
 neighbouring gas cells. This is why line absorption occurs only locally within such cells, 
 and line optical depths are not added across a line of sight (the LVG approximation used
 to solve radiative transfer in molecular clouds is based
 on this, with $\tau = f(\vec{V})$ being a very strong function of the velocity field.)}
 (with front layer gas able to absorb line radiation emanating from
 the back layer). Thus we can combine the two previous equations to find:

\begin{equation}
    I_{\rm \nu, 2}= [I_{\rm \nu, 0}e^{-\tau_{\rm H}} + B_{\nu}(T_{\rm ex, H})(1-e^{-\tau_{\rm H}})]e^{-\tau_{\rm C}} + B_{\nu}(T_{\rm ex, C})(1-e^{-\tau_{\rm C}})
\end{equation}

\noindent
In spectroscopic observations, we often measure the line intensity against a non-zero continuum radiation field. This
may emanate from the source itself (e.g., dust continuum, synchrotron) and/or from a background source (e.g., the CMB).
 This (line)-(continuum) differential measurement is obtained spectrally by subtracting the continuum 
 as it is measured at neighbouring frequencies that are not line dominated.  If the source itself does
 not emit any continuum, the subtraction of the OFF-source measurement (as done in single-dish observations)
 also yields such a (line)-(continuum) differential (with the OFF-line part of such spectrum $\sim $0).
The intensity measured then is:

\begin{equation}
\begin{aligned}
    \Delta I_{\nu}  & = I_{\rm \nu, 2} - I_{\rm \nu, 0} \\
                    & = [I_{\rm \nu, 0}e^{-\tau_{\rm H}} + B_{\nu}(T_{\rm ex, H})(1-e^{-\tau_{\rm H}})]e^{-\tau_{\rm C}} \\
                    & + B_{\nu}(T_{\rm ex, C})(1-e^{-\tau_{\rm C}})-I_{\rm \nu, 0} \\
                    & = B_{\nu}(T_{\rm ex, H})(1-e^{-\tau_{\rm H}})e^{-\tau_{\rm C}} + B_{\nu}(T_{\rm ex, C})(1-e^{-\tau_{\rm C}}) \\
                    & - I_{\rm \nu, 0}[1-e^{-(\tau_{\rm H}+\tau_{\rm C})}] 
\end{aligned} 
\end{equation}

\noindent
Setting the background emission to be a  blackbody means $I_{\rm \nu, 0}=B_{\nu}(T_{\rm bg})$, and we will have:

\begin{equation}
\begin{aligned}
\label{eq:two_layer_radiavtive_transfer}
    \Delta I_{\nu}  & = B_{\nu}(T_{\rm ex, H})(1-e^{-\tau_{\rm H}})e^{-\tau_{\rm C}} + B_{\nu}(T_{\rm ex, C})(1-e^{-\tau_{\rm C}}) \\
            & - B_{\nu}(T_{\rm bg})[1-e^{-(\tau_{\rm H}+\tau_{\rm C})}]  
\end{aligned}
\end{equation}

\noindent
where $T_{\rm bg}$ is the background source temperature which here we set to be the CMB temperature ($T_{\rm bg}=T_{\rm CMB}=2.73$ K). Eq.~\ref{eq:two_layer_radiavtive_transfer} is the same as Eq. 2 in \citet{Myers1996}, showing that the two-layer \Tex\ model benefits understanding the line profiles of a collapsing cloud \citep[e.g., ][]{Zhou1993,Myers1996}.

An observer that assumes  a uniform \Tex\ single-layer source to analyze its line emission from the two layers, he/she will then assume an effective optical depth and an effective excitation temperature as
 $\tau_{\rm eff}$ and $T_{\rm ex, eff}$, defined from:

\begin{equation}
\begin{aligned}
\label{eq:two_layer_effective_radiavtive_transfer}
        \Delta I_{\nu} & = [B_{\nu}(T_{\rm ex, eff})-B_{\nu}(T_{\rm bg})](1-e^{-\tau_{\rm eff}}) \\
                       & = B_{\nu}(T_{\rm ex, eff})(1-e^{-\tau_{\rm eff}})-B_{\nu}(T_{\rm bg})(1-e^{-\tau_{\rm eff}})
\end{aligned}
\end{equation}

\noindent
From Eq.~\ref{eq:two_layer_radiavtive_transfer} and Eq.~\ref{eq:two_layer_effective_radiavtive_transfer}, we then have: 

\begin{equation}
    \tau_{\rm eff}=\tau_{\rm H}+\tau_{\rm C}
\end{equation}

\noindent
and

\begin{equation}
\begin{aligned}
     B_{\nu}(T_{\rm ex, eff})\cdot [1-e^{-(\tau_{\rm H}+\tau_{\rm C})}] & = B_{\nu}(T_{\rm ex, H})\cdot [e^{-\tau_{\rm C}}-e^{-(\tau_{\rm H}+\tau_{\rm C})}] \\
     & + B_{\nu}(T_{\rm ex, C})\cdot [e^{-(\tau_{\rm H}+\tau_{\rm C})}-e^{-\tau_{\rm H}}]
\end{aligned}
\end{equation}

\noindent
reordering the Equation above, we obtain: 

\begin{equation}
    B_{\nu}(T_{\rm ex, eff}) = B_{\nu}(T_{\rm ex, H})\cdot a+B_{\nu}(T_{\rm ex, C})\cdot b
\end{equation}

\begin{equation}
    a=\frac{e^{\tau_{\rm H}}-1}{e^{\tau_{\rm H}+\tau_{\rm C}}-1},\quad b=\frac{e^{\tau_{\rm H}+\tau_{\rm C}}-e^{\tau_{\rm H}}}{e^{\tau_{\rm H}+\tau_{\rm C}}-1}
\end{equation}

We can find that $a+b\equiv 1$, it means that the effective $B_{\nu}(T_{\rm ex, eff})$ is actually a weighted-averaged value of $B_{\nu}(T_{\rm ex, H})$ and $B_{\nu}(T_{\rm ex, C})$, with two weighting factors $a$ and $b$. The weighting factors $a$ and $b$ are both functions of $\tau_{\rm H}$ and $\tau_{\rm C}$, which means the effective \Tex of \twCN\ and \thCN\ emission lines should be different because of their different optical depth.

In Fig.~\ref{fig:test_a_b_epsi1_epsi2} (a), we show the change of $a$ and $b$ with optical depth $\tau$, assuming $\tau_{\rm H}=\tau_{\rm C}=\tau$. The $a$ and $b$ have monotonous and opposite trends to change with $\tau$. The effective temperature of \twCN\ main component will be much lower than the effective temperature of the optically thin \thCN\ main component.

If the linewidth is dominated by macro-turbulence or bulk motion, instead of thermal/micro-turbulence, the two-layer \Tex\ model may not play a dominant role. This might 
also affect the case in this work, in which the observed linewidths are $\sim1.5 - 4\ {\rm km\cdot s^{-1}}$.  The radiation between gas layers may not be well-coupled. 
However, such radiative coupling effects would still bias the isotopic ratios for all methods that need optical-depth corrections.

\subsection{The deviation from original assumptions}
Deriving \Rtwth\ from CN isotopologues has the basic assumptions including Eq.~\ref{eq:eqB22} and Eq.~\ref{eq:eqB23}. However, considering the different effective \Tex\ between \twCN\ and \thCN\ $N=1\to0$, and back to Eq.~\ref{eq:eqB1} and Eq.~\ref{eq:eqB5}, we have:

\begin{equation}
\label{eq:eqB39}
\begin{aligned}
    R_{\rm 1213} & =\frac{N_{\rm main, ^{12}CN}}{N_{\rm main, ^{13}CN}} \\
              & =[\frac{{\rm exp}(\frac{E_{\rm u, ^{12}CN}}{kT_{\rm ex, ^{12}CN}})}{{\rm exp}(\frac{h\nu}{kT_{\rm ex, ^{12}CN}})-1}]/[\frac{{\rm exp}(\frac{E_{\rm u, ^{13}CN}}{kT_{\rm ex, ^{13}CN}})}{{\rm exp}(\frac{h\nu}{kT_{\rm ex, ^{13}CN}})-1}] \times \frac{\tau_{\rm m, 12}}{\tau_{\rm m, 13}} \\
              & =\epsilon_{1} \frac{\tau_{\rm m, 12}}{\tau_{\rm m, 13}}
\end{aligned}
\end{equation}

\begin{equation}
\label{eq:eqB40}
\begin{aligned}
    R_{\rm b} & = \frac{T_{\rm b, ^{12}CN}}{T_{\rm b, ^{13}CN}} = \frac{T_{\rm mb, ^{12}CN}}{T_{\rm mb, ^{13}CN}} \\
                & = \frac{J_{\nu}(T_{\rm ex, ^{12}CN})-J_{\nu}(T_{\rm bg})}{J_{\nu}(T_{\rm ex, ^{13}CN})-J_{\nu}(T_{\rm bg})}\times \frac{1-e^{-\tau_{\rm m, 12}}}{1-e^{-\tau_{\rm m, 13}}} \\
                & = \epsilon_{2} \frac{1-e^{-\tau_{\rm m, 12}}}{1-e^{-\tau_{\rm m, 13}}}
\end{aligned}
\end{equation}

We have assumed that the filling factors $f_{\rm ^{12}CN}=f_{\rm ^{13}CN}$ in Eq.~\ref{eq:eqB40}, so that the brightness temperature ratio equals the main beam temperature ratio. Here we simplify Equations \ref{eq:eqB39} and \ref{eq:eqB40} with $\epsilon_1$ and $\epsilon_2$:

\begin{equation}
    \epsilon_1=[\frac{{\rm exp}(\frac{E_{\rm u, ^{12}CN}}{kT_{\rm ex, ^{12}CN}})}{{\rm exp}(\frac{h\nu}{kT_{\rm ex, ^{12}CN}})-1}]/[\frac{{\rm exp}(\frac{E_{\rm u, ^{13}CN}}{kT_{\rm ex, ^{13}CN}})}{{\rm exp}(\frac{h\nu}{kT_{\rm ex, ^{13}CN}})-1}]
\end{equation}

\begin{equation}
    \epsilon_2=\frac{J_{\nu}(T_{\rm ex, ^{12}CN})-J_{\nu}(T_{\rm bg})}{J_{\nu}(T_{\rm ex, ^{13}CN})-J_{\nu}(T_{\rm bg})}
\end{equation}

With our previous assumption that $T_{\rm ex, ^{12}CN}=T_{\rm ex, ^{13}CN}$, the estimated $R_{\rm 1213}$ and $R_{\rm b}$ are from Eq.~\ref{eq:eqB22} and Eq.~\ref{eq:eqB23}. In this way, $\epsilon_1=R^{\rm intrinsic}_{\rm 1213}/R^{\rm estimated}_{\rm 1213}$ and $\epsilon_2=R^{\rm intrinsic}_{\rm b}/R^{\rm estimated}_{\rm b}$. We still ignore the difference between other molecular parameters in Eq.~\ref{eq:eqB1} of $\rm ^{12}CN$ and $\rm ^{13}CN$, except for $E_{\rm u}$ and $T_{\rm ex}$. 

Assuming that the column density ratio of \twCN\ and \thCN\ $N=1\to0$ equals 60, and $\tau_{\rm H}=\tau_{\rm C}=\tau$, we quantify the deviation from the basic assumptions in Eq. \ref{eq:Nratio_eq_tauratio} and Eq. \ref{eq:Tbratio_eq_taufunction} with the factors $\epsilon_1$ and $\epsilon_2$ varying with the effective optical depth of \twCN\ $N=1\to0$ main component $\tau^{\rm main}_{\rm eff}$.

\begin{figure*}
	\centering
            \includegraphics[scale=0.3]{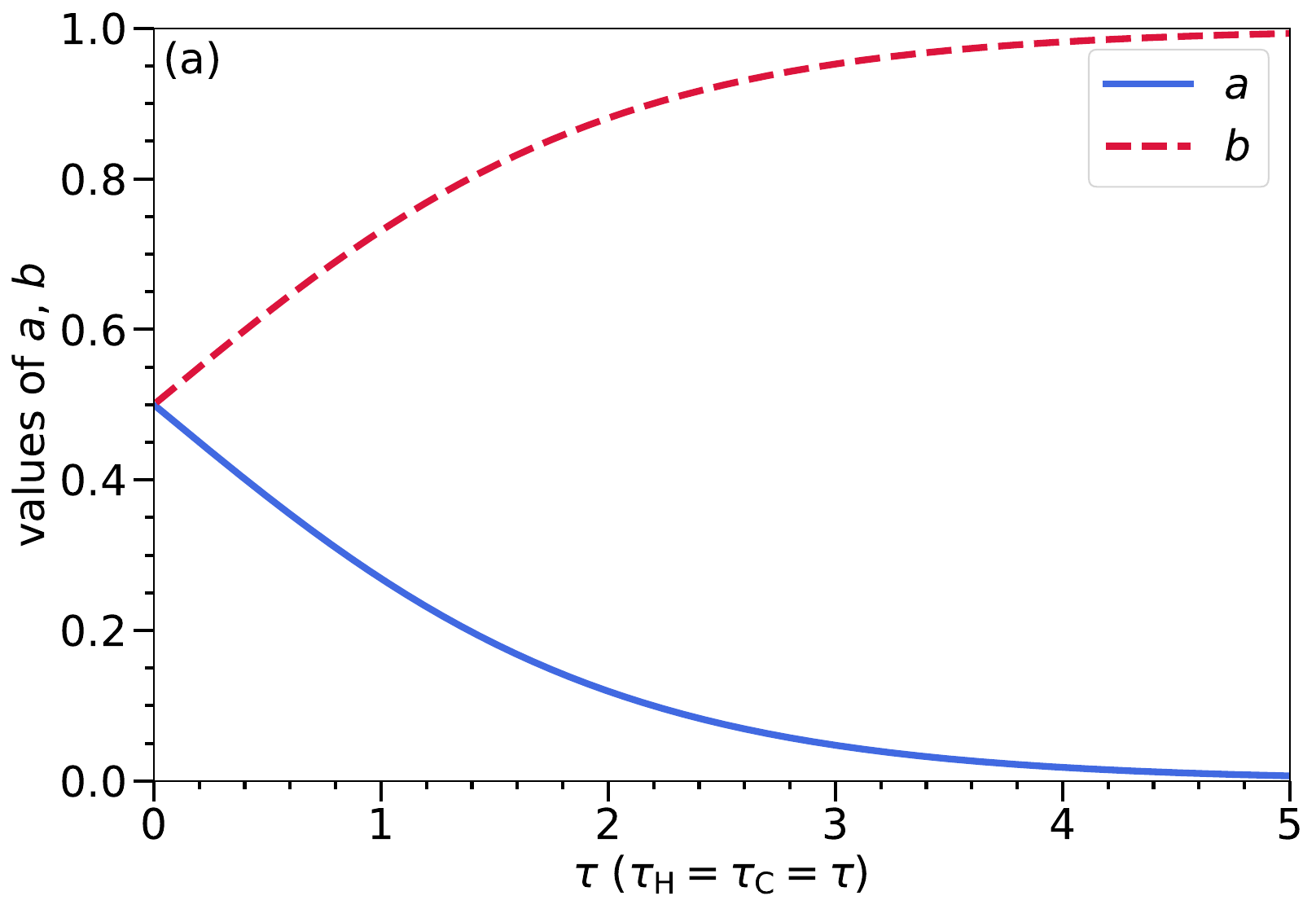}
            \includegraphics[scale=0.3]{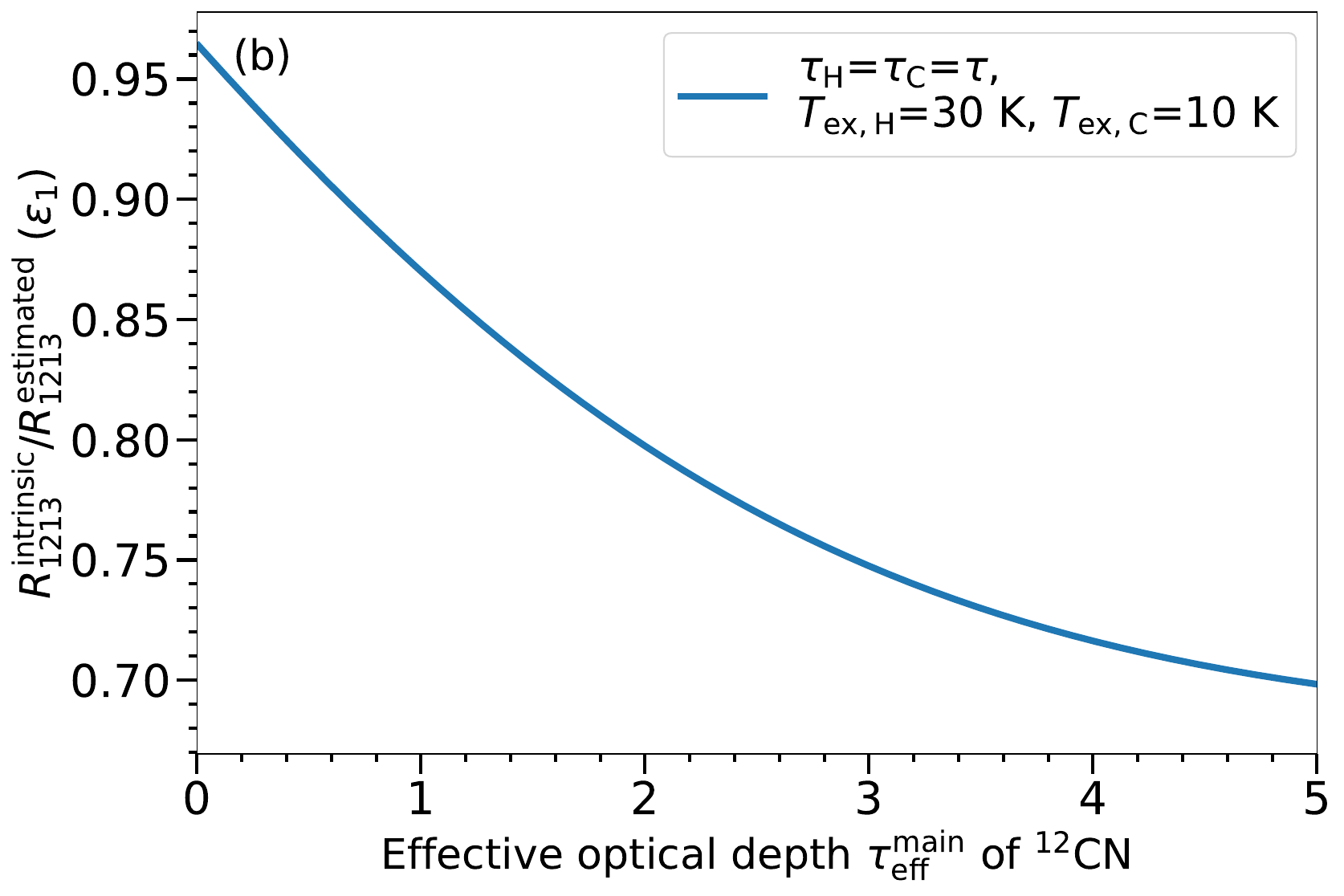}
            \includegraphics[scale=0.3]{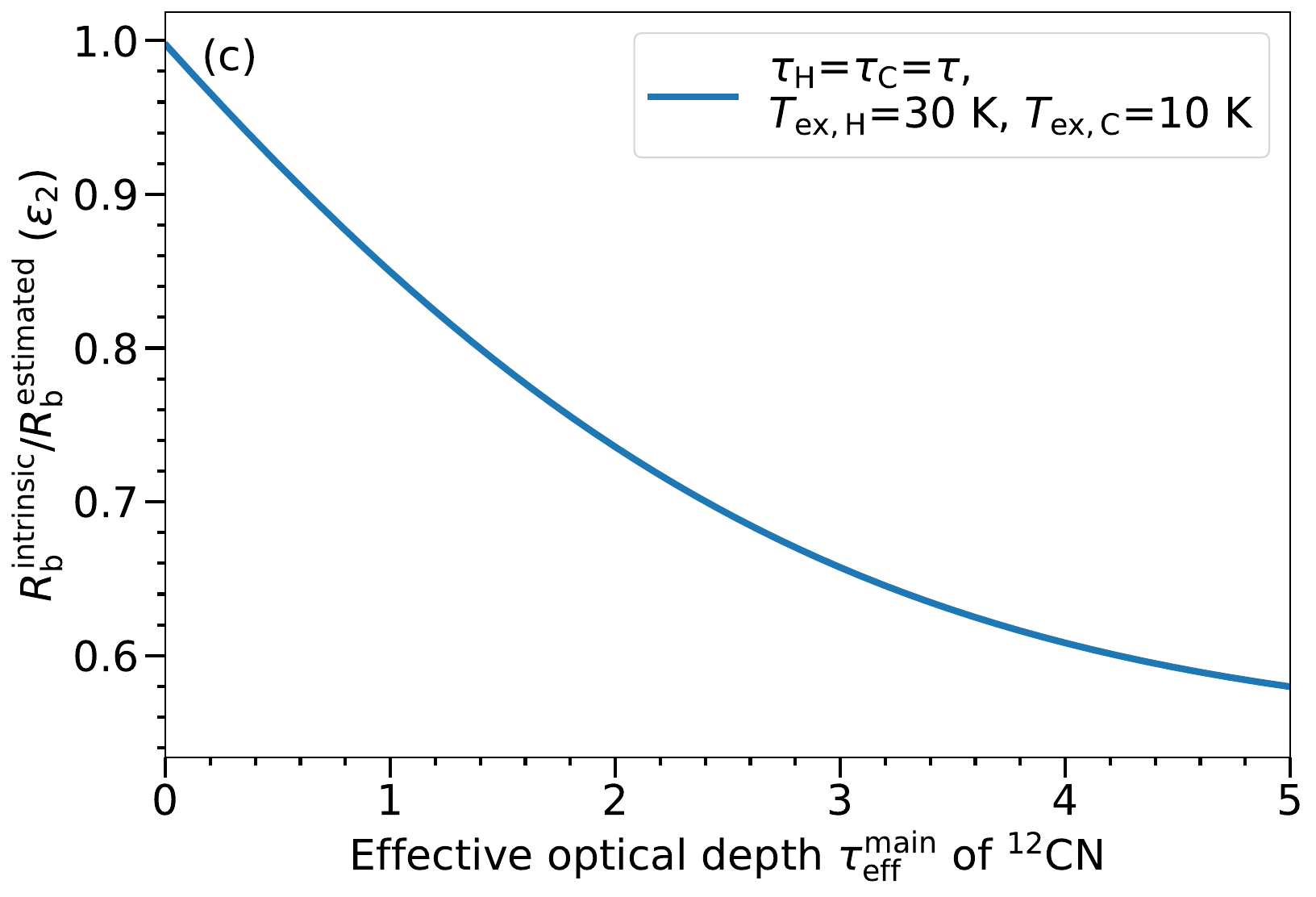}    
	\caption{ (a). The variation of the weighting factor $a$ (blue solid line) and $b$ (red dashed line) with optical depth. (b). The ratio ($\epsilon_1$) between the intrinsic column density ratio and the estimated column density ratio, varying with the effective optical depth of \twCN\ main component ($\tau^{\rm main}_{\rm eff}=\tau_{\rm H}+\tau_{\rm C}$). (c). The ratio ($\epsilon_2$) between the intrinsic brightness temperature ratio and the estimated brightness temperature ratio, varying with $\tau^{\rm main}_{\rm eff}$. \label{fig:test_a_b_epsi1_epsi2}}
\end{figure*}

In Fig. \ref{fig:test_a_b_epsi1_epsi2} (b) and (c), we show the change of $\epsilon_1$ and $\epsilon_2$ as a function of $\tau^{\rm eff}_{\rm main}$.  In our toy model with $T_{\rm ex, H}=30$ K and $T_{\rm ex, C}=10$ K, the intrinsic column density ratio will be smaller than the estimated column density ratio. If the optical depth is larger, the discrepancy will be larger. It means using Eq.~\ref{eq:Nratio_eq_tauratio} and Eq.~\ref{eq:Tbratio_eq_taufunction} will overestimate the column density ratio and the brightness temperature ratio between \twCN\ and \thCN\ $N=1\to0$, respectively. When $\tau^{\rm main}_{\rm eff} = \tau_{\rm H}+\tau_{\rm C} \sim 1$, which is similar to the measured optical depth of \twCN\ $N=1\to0$ main component in our targets, the intrinsic column density ratio will be $\sim$ 15 \% lower than the estimated ratio, and the intrinsic brightness temperature ratio will be 10 \% lower than the estimated one.

In Fig,~\ref{fig:two_model_12Cto13C_with_tau}, we quantify the total effects of these two discrepancies on the derived \Rtwth. Based on our toy model assumptions and adopting Eq.~\ref{eq:Nratio_eq_tauratio} and Eq.~\ref{eq:Tbratio_eq_taufunction}, using the HfS fitting method with the \twCN\ main component will let the derived \Rtwth\ $\sim$ 17 \% lower than the intrinsic \Rtwth\ when the optical depth of the \twCN\ main component $\sim$ 1. However, if we use the \twCN\ optical-thin satellite line (e.g., $J=3/2\to1/2$, $F=1/2\to3/2$ at 113.520 GHz), the derived \Rtwth\ will be closed to the intrinsic column density ratio of \twCN\ and \thCN.

\subsubsection{More general conditions}

The real excitation conditions of a molecular cloud may be much more complex than our toy model. The detailed deviation of the effective optical depth and effective excitation temperature of \twCN\ and \thCN\ is beyond this work. Here we free the values of $T_{\rm ex, ^{12}CN}$ and $T_{\rm ex, ^{13}CN}$ and analyze the deviation on the \Rtwth\ ratios if the excitation temperature of \twCN\ and \thCN\ $N=1\to0$ are different.

Considering the $\epsilon1$ and $\epsilon2$, Equation~\ref{eq:eqB27} should become:

\begin{equation}
\label{eq:eqB43}
	\frac{\rm ^{12}C}{\rm ^{13}C}=-\frac{R_{13}}{R_{12}}\cdot \frac{\epsilon_1\tau_{\rm m, 12}}{{\rm ln}[1-\epsilon_2\frac{T_{\rm mb, ^{13}CN}}{T_{\rm mb, ^{12}CN}}(1-e^{-{\tau_{\rm m, 12}}})]}
\end{equation}

\begin{figure*}
	\centering
            \includegraphics[scale=0.25]{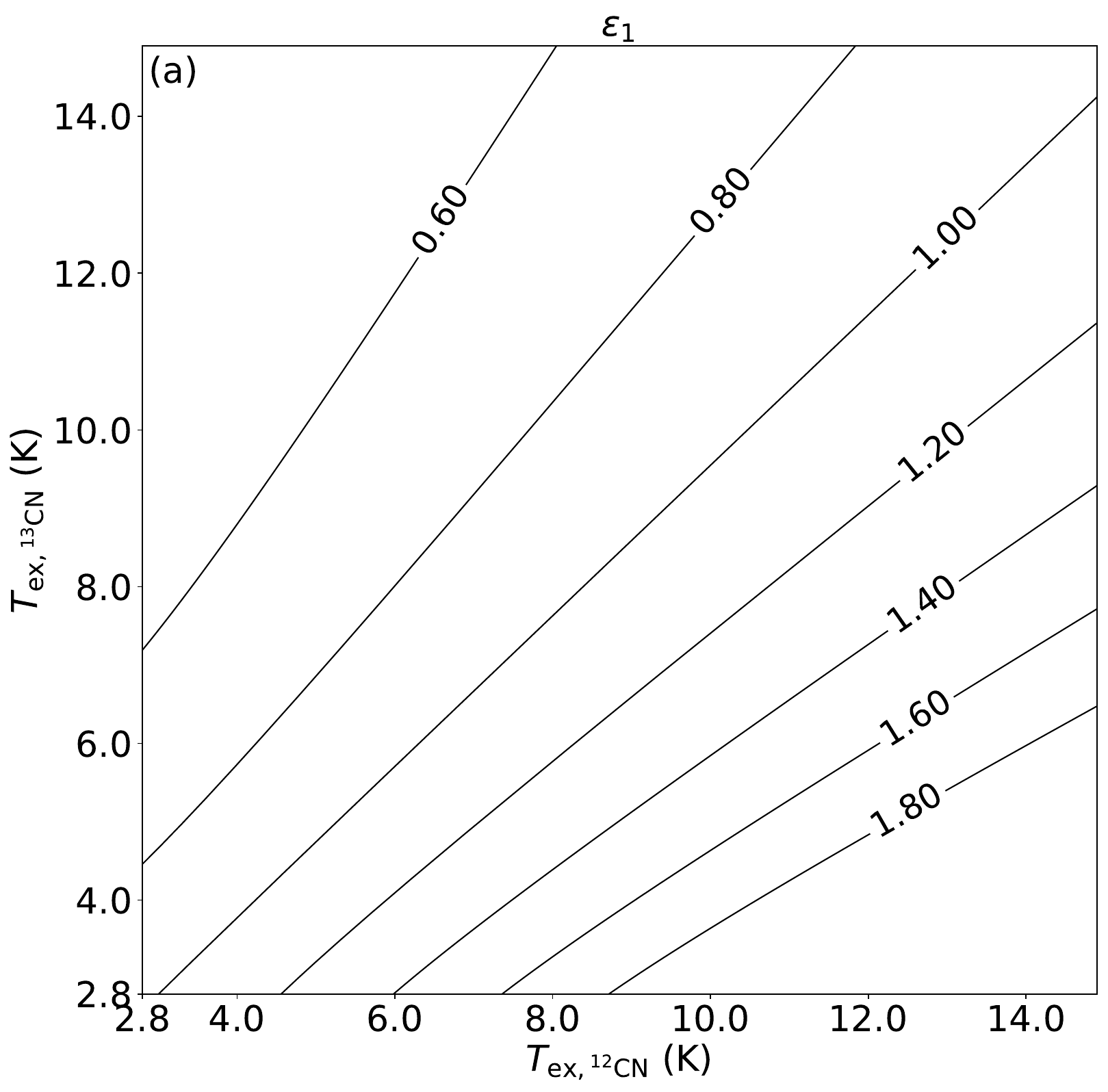}
            \includegraphics[scale=0.25]{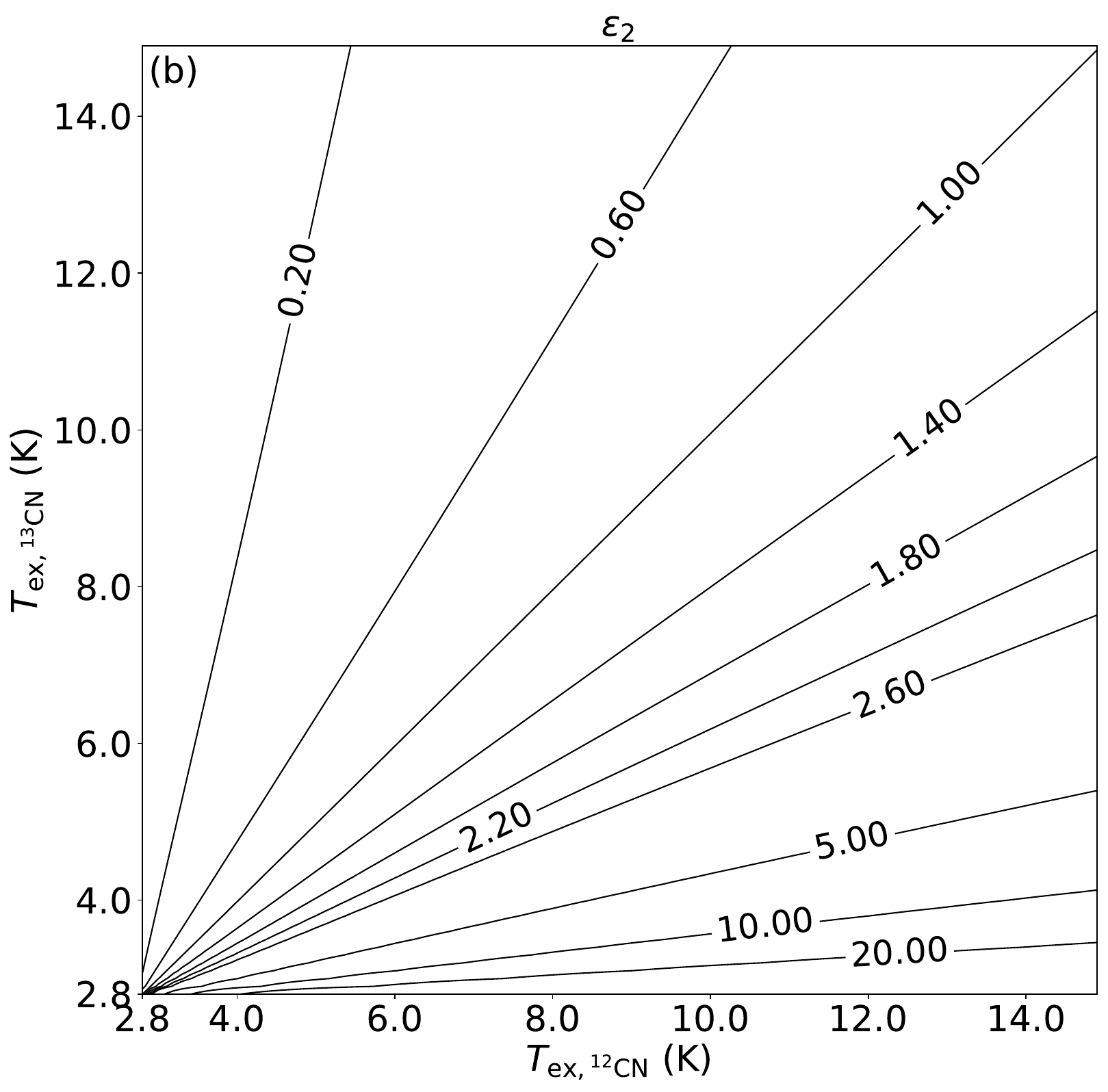}
            \includegraphics[scale=0.25]{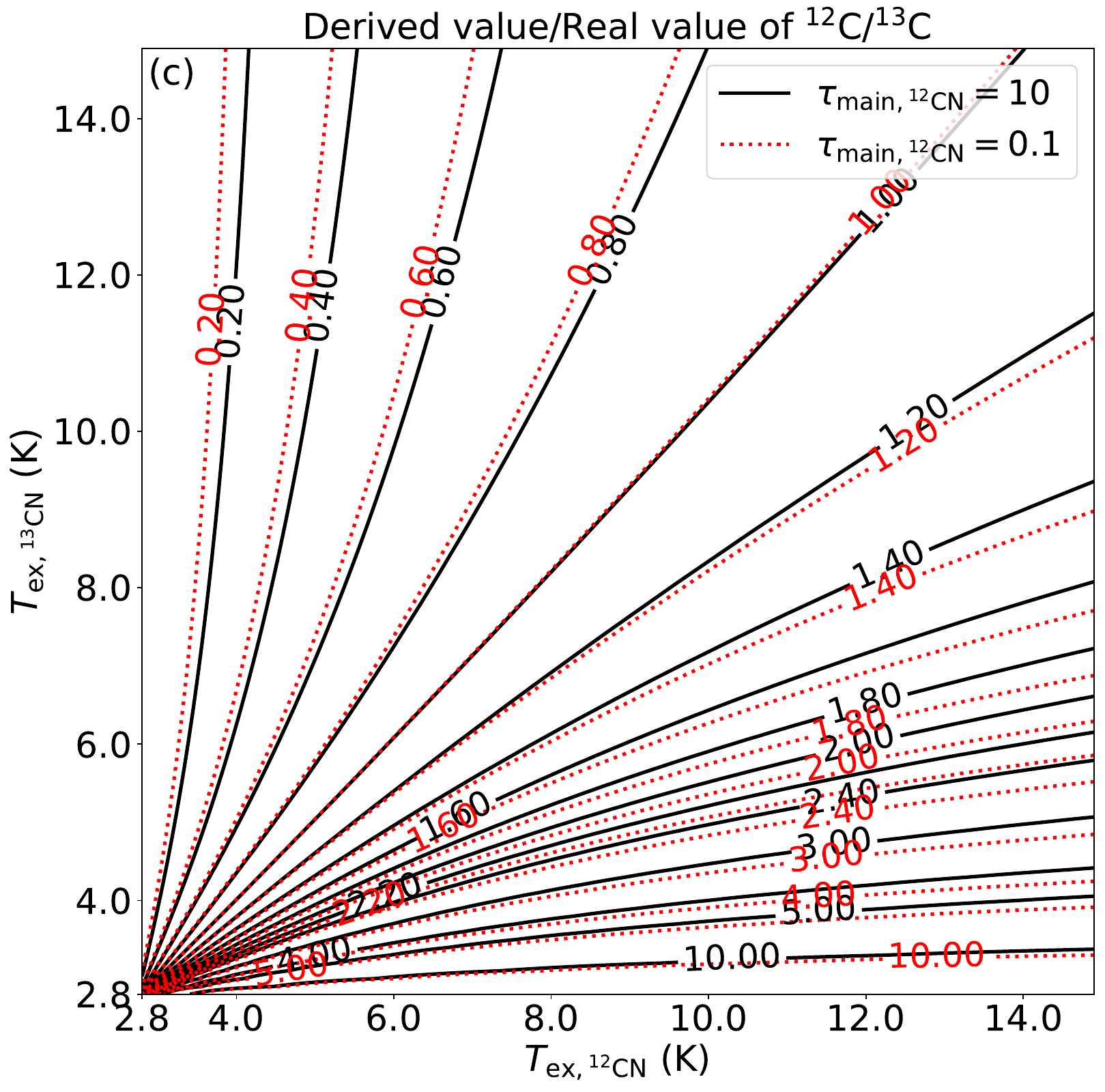}
	\caption{\label{fig:figA2} (a). the variation of $\epsilon_1$ with \Tex\ of $\rm ^{12}CN$ and $\rm ^{13}CN$. (b): the variation of $\epsilon_2$ with \Tex\ of $\rm ^{12}CN$ and $\rm ^{13}CN$. The contours show the values of $\epsilon_1$ and $\epsilon_2$ on the upper two figures. (c): assuming the optical depth of the main component of $\rm ^{12}CN$ ($\tau_{\rm m, ^{12}CN} = 10$ and $\tau_{\rm m, ^{12}CN} = 0.1$, respectively), the ratios between the derived \Rtwth\ from Equation~\ref{eq:eqB27} and \Rtwth\ from Equation~\ref{eq:eqB43}. The contours show the value of these ratios. }
\end{figure*}

In Figure~\ref{fig:figA2}, we quantify the effect on our derived \Rtwth\ when \Tex\ of $\rm ^{12}CN$ and $\rm ^{13}CN$ $N=1\to0$ are different.  Especially, when $T_{\rm ex, ^{12}CN}$ is much higher than $T_{\rm ex, ^{13}CN}$, $\epsilon_2$ will be high and cause a large deviation. The total effect is also sensitive to the difference of $T_{\rm ex, ^{12}CN}$ and $T_{\rm ex, ^{13}CN}$. The derived \Rtwth\ is not equal to the intrinsic abundance ratio when $T_{\rm ex, ^{12}CN}$ equals $T_{\rm ex, ^{13}CN}$, because the upper energy levels $E_{\rm u}$ of $\rm ^{12}CN$ and $\rm ^{13}CN$ have a slight difference. We ignore it in our \Rtwth\ derivation. \\

\section{Comparison of two HfS fitting procedures}
\label{appendix:compare_HfS}
In this section, we test the performances of two procedures that can be used for the HfS fitting. We generate $\rm ^{12}CN$ $N=1\to0$ spectra artificially. Then we use the HfS procedure in CLASS and the procedure developed by~\citet{Estalella2017} to fit these spectra and test the accuracy and scatter of the results.

We assume the line widths of \twCN\ $N=1\to0$ components are the same. We also assume that both the intrinsic line and the optical depth $\tau$ have Gaussian profiles. If $\tau=0$, the profiles of the line components can be described as:

\begin{equation}
	T_{\rm A}^\star(v)=Ae^{-\frac{(v-v_0)^2}{2\sigma^2}}
\end{equation}
Here $\sigma = {\rm FWHM}/2\sqrt{\rm 2ln2 }$ and we assume that $\sigma = 2\ {\rm km~s^{-1}} $. $A$ refers to the intrinsic intensity of a line component, so the total profile can be described as:

\begin{equation}
	T_{\rm A}^\star(v)=\sum_{k=1}^nA_ke^{-\frac{(v_{\rm k}-v_{0,k})^2}{2\sigma^2}}
\end{equation}
Here $A_k$ indicates the intensities of different line components. We assume $A=5$ K for the main component.

For the line components with the optical depth, we have the following equation:
\begin{equation}
\label{eq:eqC3}
	\frac{T^\star_{\rm A, hf}}{T^\star_{\rm A, m}}=\frac{1-e^{-\tau_{\rm hf}}}{1-e^{-\tau_{\rm m}}}	
\end{equation}
Here $\tau_{\rm hf}$ and $T^\star_{\rm A, hf}$ are the central optical depth and the peak antenna temperature of a certain HfS line component, respectively, while $\tau_{\rm m}$ and $T^\star_{\rm A, m}$ refer to the same properties of the main component. Because the difference between the main beam efficiency at the frequency of the main component and the satellite lines is very small (< 0.1\% for IRAM 30-m), we replace the $T_{\rm mb}$ in Eq.~\ref{eq:HfS_fitting_assumption} with the antenna temperature $T^\star_{\rm A}$.

We have $\tau_{\rm hf} = R_{\rm hf}*\tau_{\rm m}$. Here, $R_{\rm hf}$ is the column density ratio between an HfS line component and the main component. So line profiles can be described as follows:
\begin{equation}
\begin{aligned}
	T_{\rm A}^\star(v) & = \sum_{k=1}^nA(1-e^{-\tau_{\rm hf}\frac{(v_{\rm k}-v_{0,k})^2}{2\sigma^2}}) \\ 
       & =\sum_{k=1}^nA(1-e^{-R_{\rm hf, k}\tau_{\rm m}\frac{(v_{\rm k}-v_{0,k})^2}{2\sigma^2}})
\end{aligned}
\end{equation}

The basic equation HfS fitting procedures used is ~\ref{eq:eqC3}. With the known $R_{\rm hf}$ and the observed antenna temperature $T^\star_{\rm A, hf}$ of multiple HfS line components, the optical depth of the main component ($\tau_{\rm m}$) and $A$ can be derived by the fitting of line profiles. Then we can derive the excitation temperature $T_{\rm ex}$ with Equation~\ref{eq:eqB18}:

\begin{equation}
	T_{\rm ex}=T_{\rm A}(1-e^{-\tau})/(\eta_{\rm mb}f)+T_{\rm bg}
\end{equation}
Here we consider \Tex\ in the Rayleigh-Jeans approximation because the procedure in CLASS output $T_{\rm ex}$ under R-J approximation. We use this equation to derive $T_{\rm ex}$ from the procedure of \citet{Estalella2017} in our test.

\begin{figure*}
        \centering
		\includegraphics[scale=0.55]{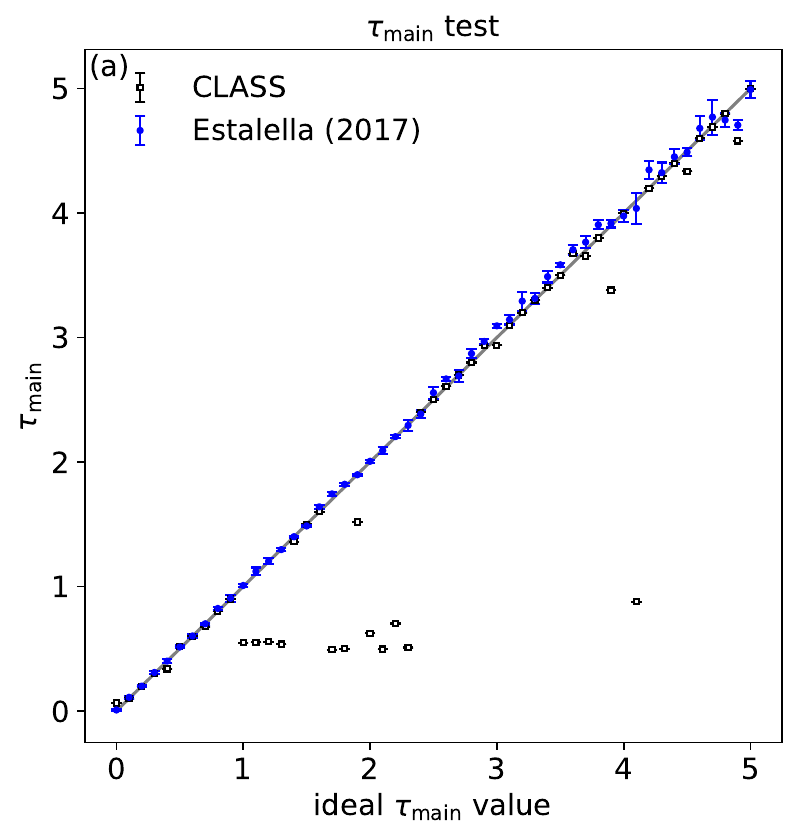}
            \includegraphics[scale=0.55]{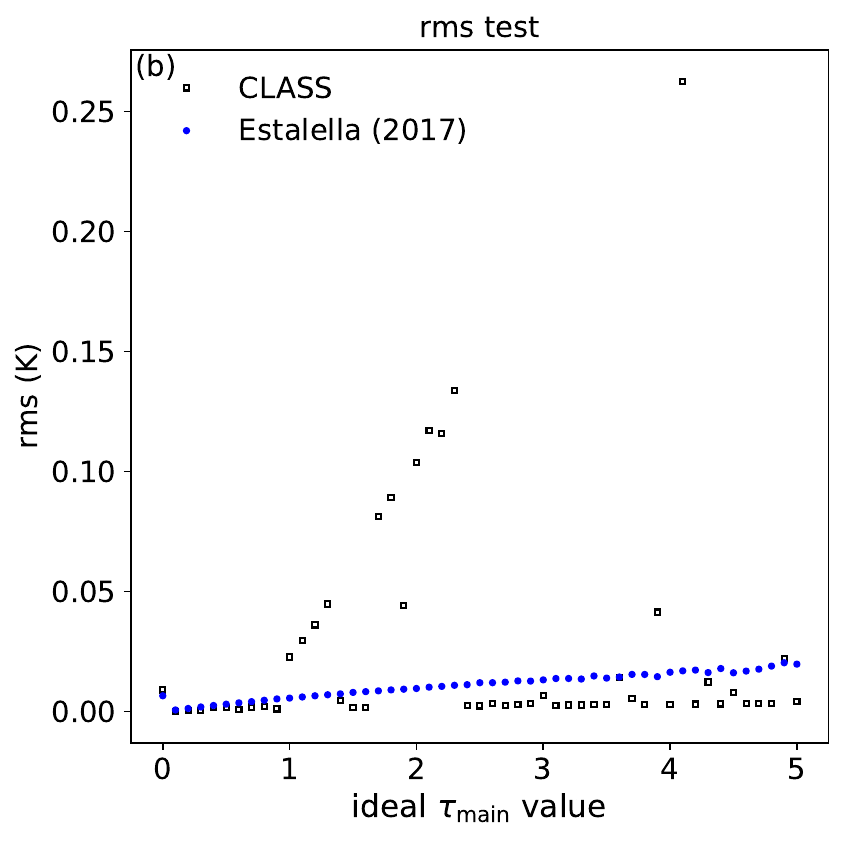}
            \includegraphics[scale=0.55]{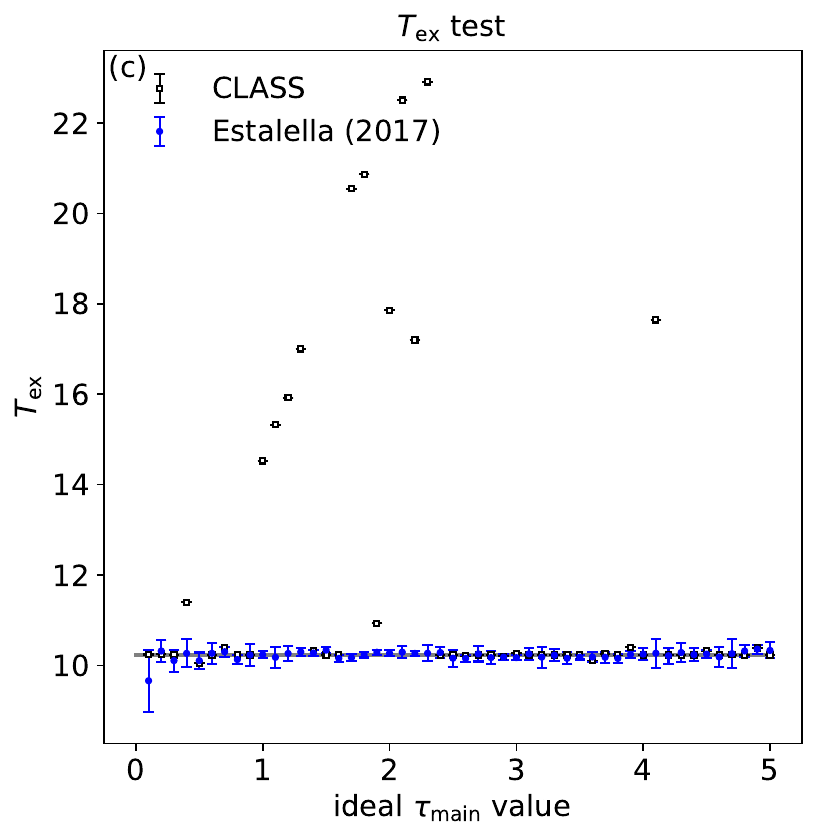}
            \includegraphics[scale=0.55]{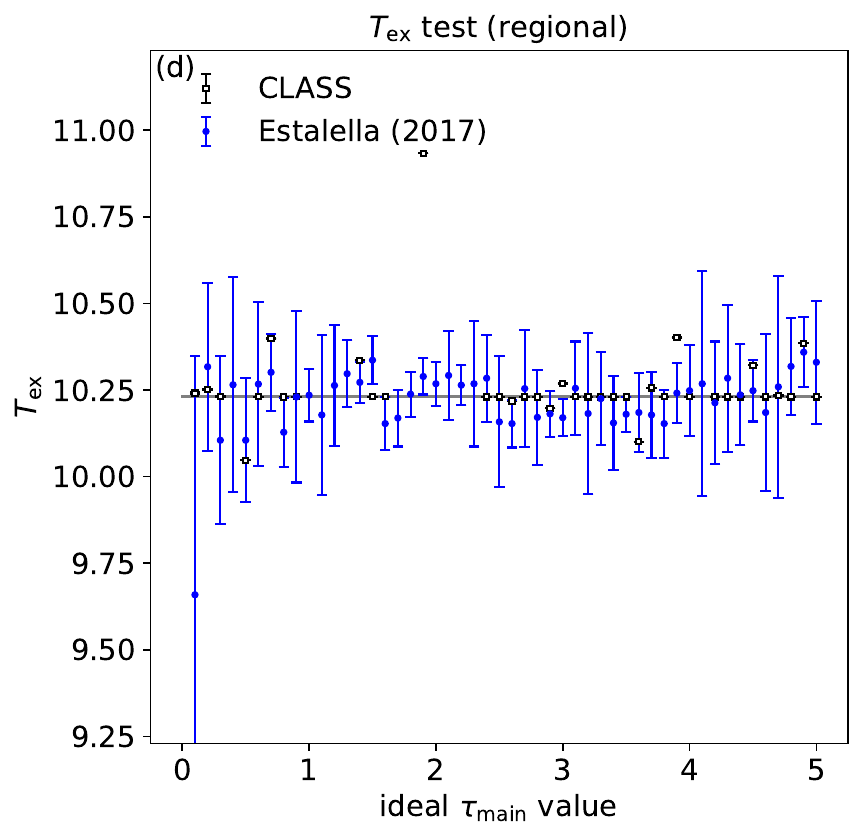}
	\caption{\label{fig:HfS_test}Results for the two HfS fitting procedures. The black hollow squares and solid blue circles show the results from the HfS fitting procedure in CLASS and the procedure developed by ~\citet{Estalella2017}, respectively. The grey line indicates where the fitted values equal ideal values. (a). The comparison between the ideal $\tau_{\rm m}$ value and the fitting values of the two procedures. The grey line refers to the ideal $\tau_{\rm m}$ we generate. (b). The comparison of the RMS of the fitting to the $\rm ^{12}CN$ line profile by these two procedures. (c). $T_{\rm ex}$ derived from the two procedures. (d). Same meaning as (c) but zoom in on the y-axis.}
\label{fig:figB1}
\end{figure*}

We generated a list of values ranging from 0 to 5 to be the list of $\tau_{\rm m}$, with an interval at 0.1, to generate ideal $\rm ^{12}CN$ spectra. There are 500 spectra, which is enough for the test. Figure~\ref{fig:HfS_test} shows the fitting results of the two procedures and the comparison with ideal values. The procedure from~\citet{Estalella2017} has a slightly larger RMS compared with the CLASS procedure, and the accuracy of the $T_{\rm ex}$ is not as good as in CLASS. However, the stability of the procedure from~\citet{Estalella2017} is much better than the procedure in CLASS while the results from CLASS include some points that differ a lot from the ideal values. To avoid the risk that the fitting to the real spectra is too bad, we choose the procedure from~\citet{Estalella2017} to fit our observed $\rm ^{12}CN$ $N=1\to0$ spectra.

\section{Spectra of lines}
\label{appendix:line_spectra}
\begin{figure*}
        \centering
		    \includegraphics[scale=0.16]{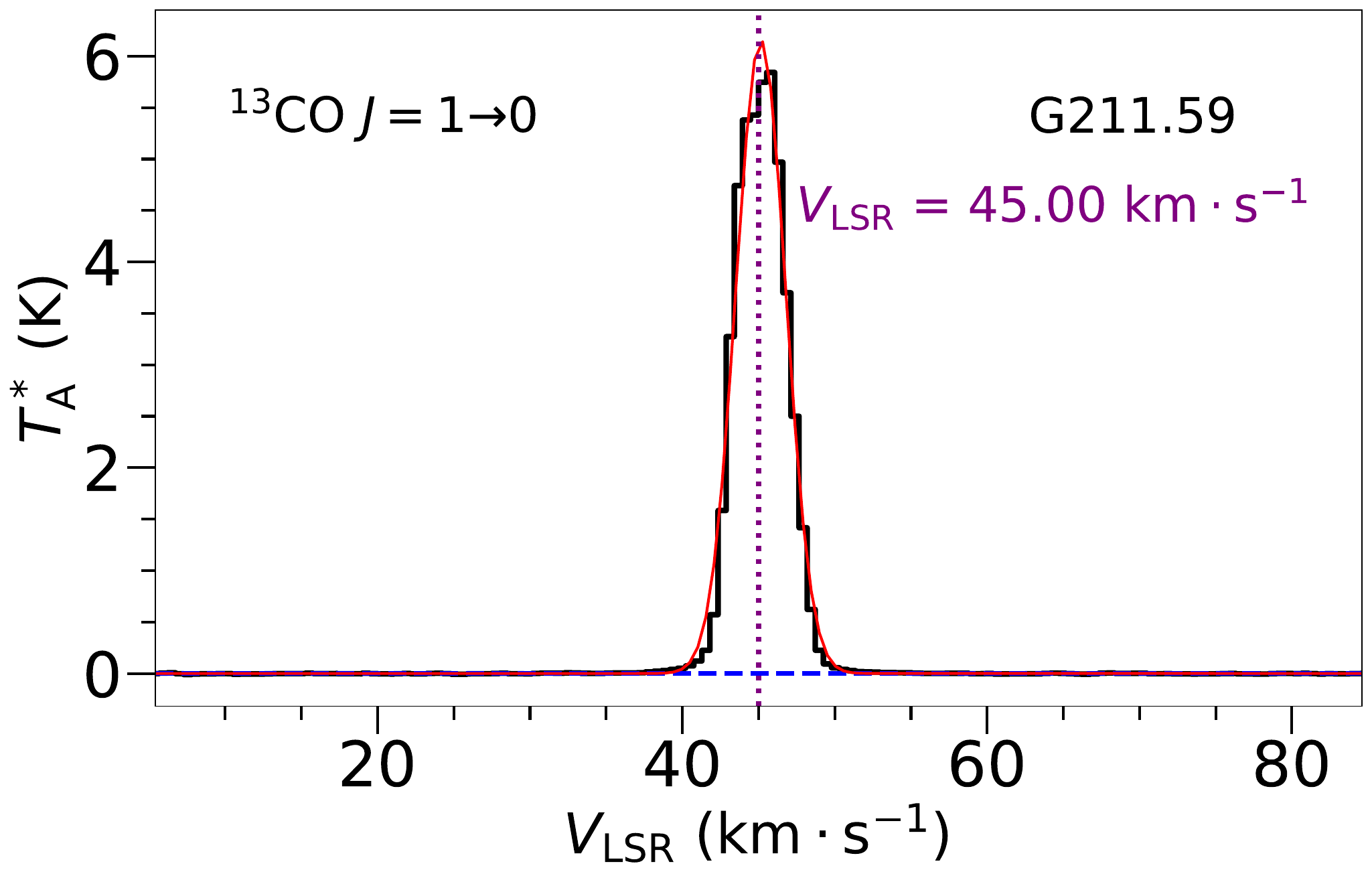}
                \includegraphics[scale=0.16]{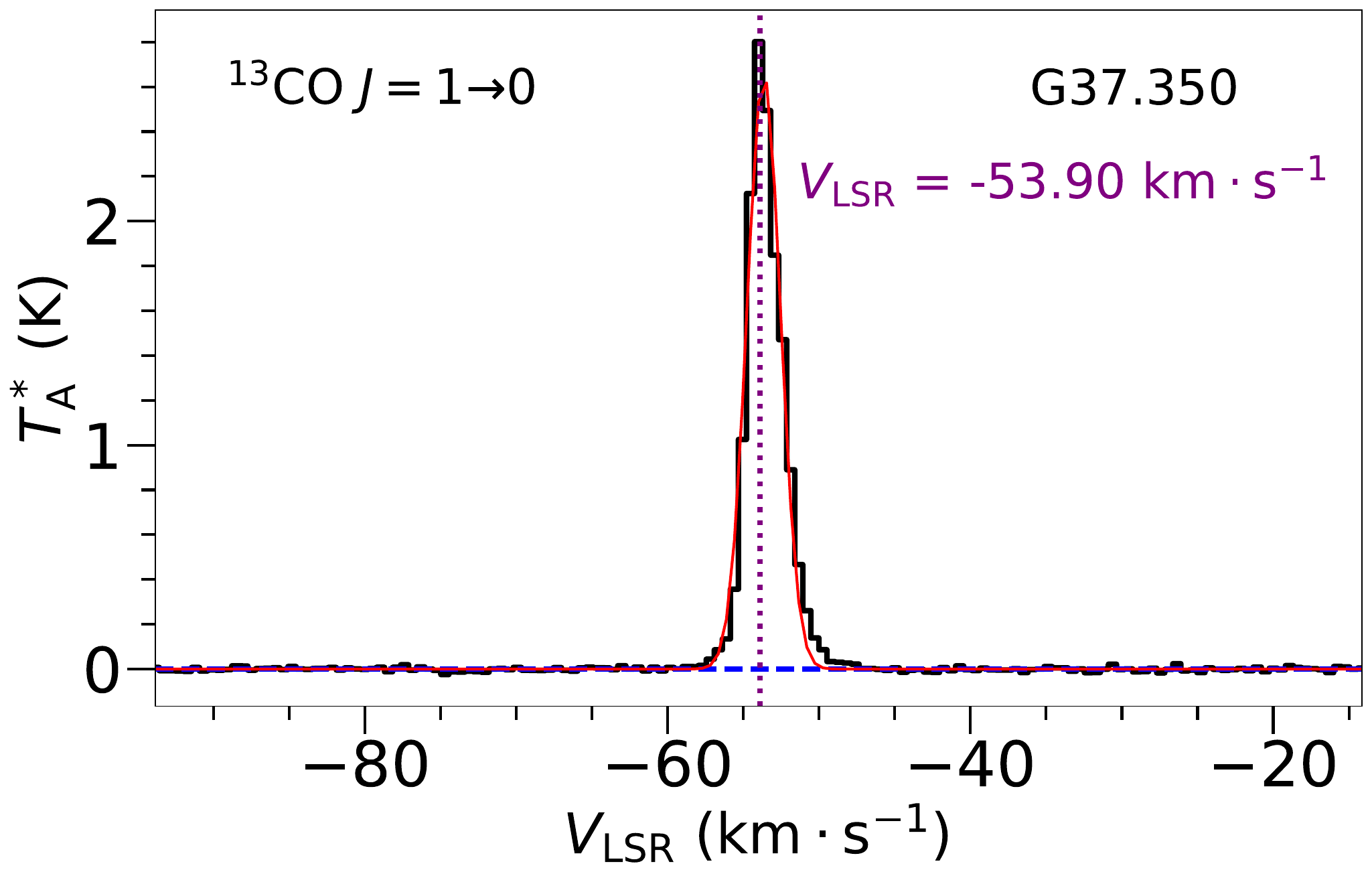}
                \includegraphics[scale=0.16]{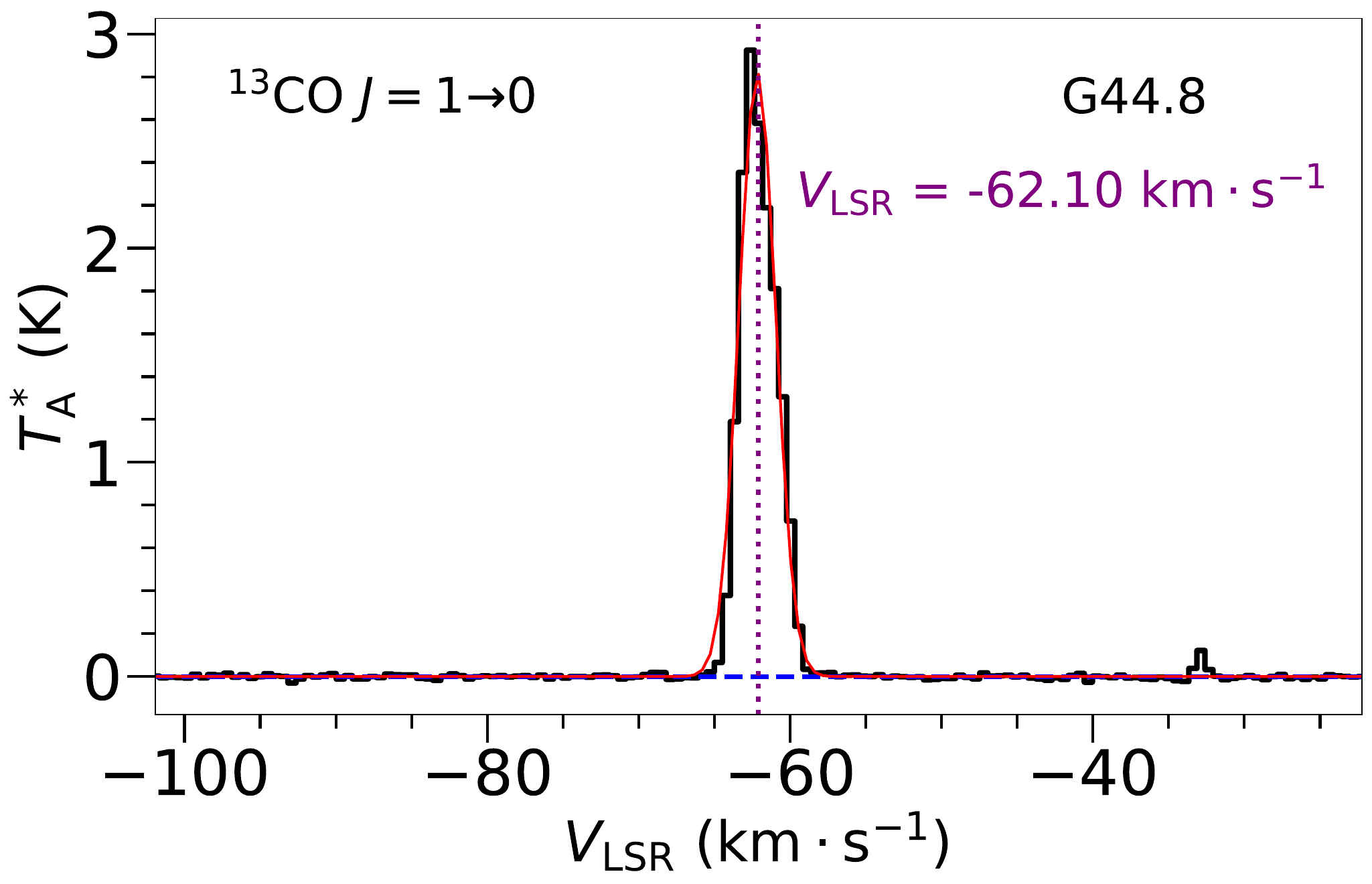}
                \includegraphics[scale=0.16]{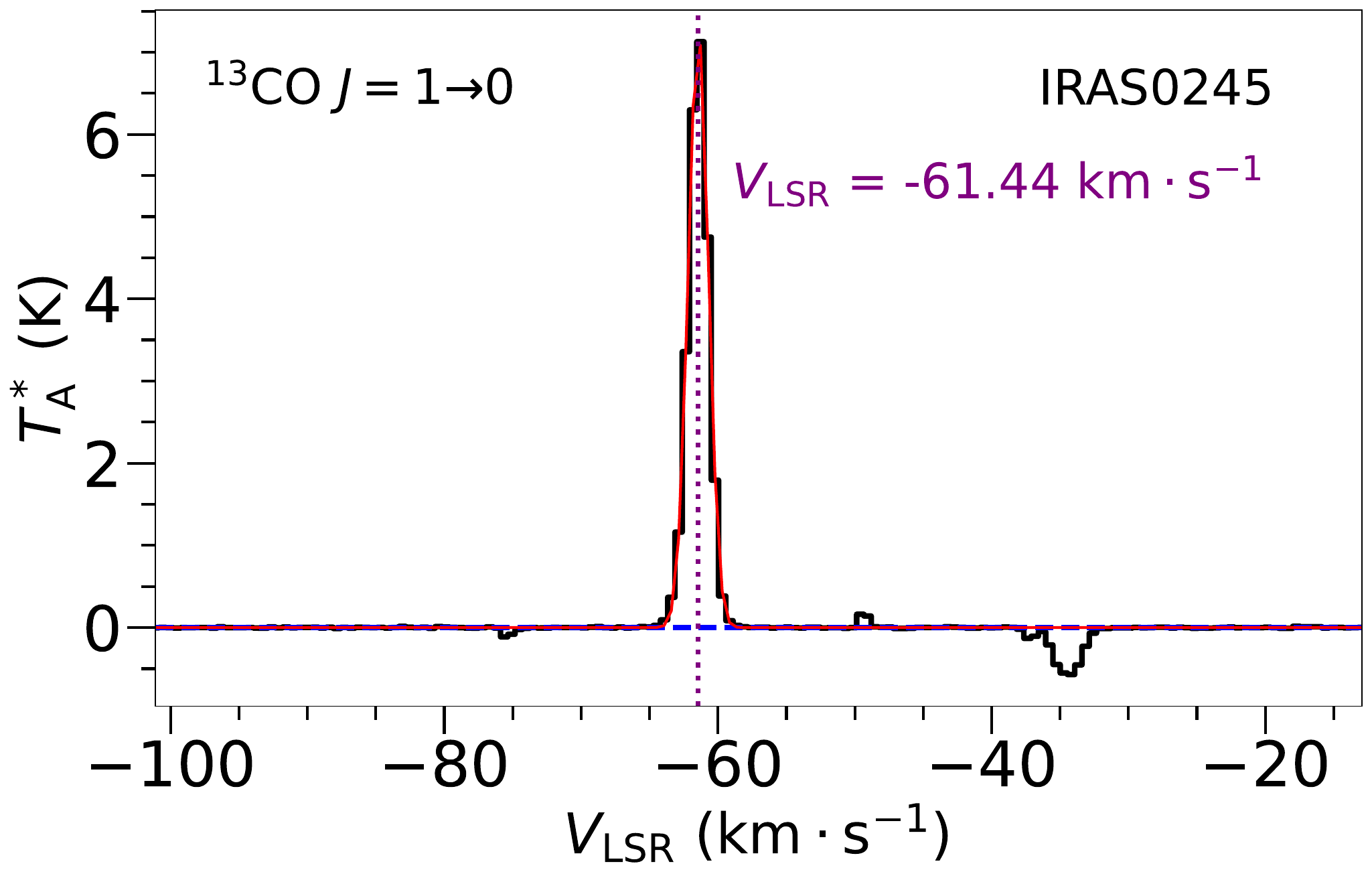}
                \includegraphics[scale=0.16]{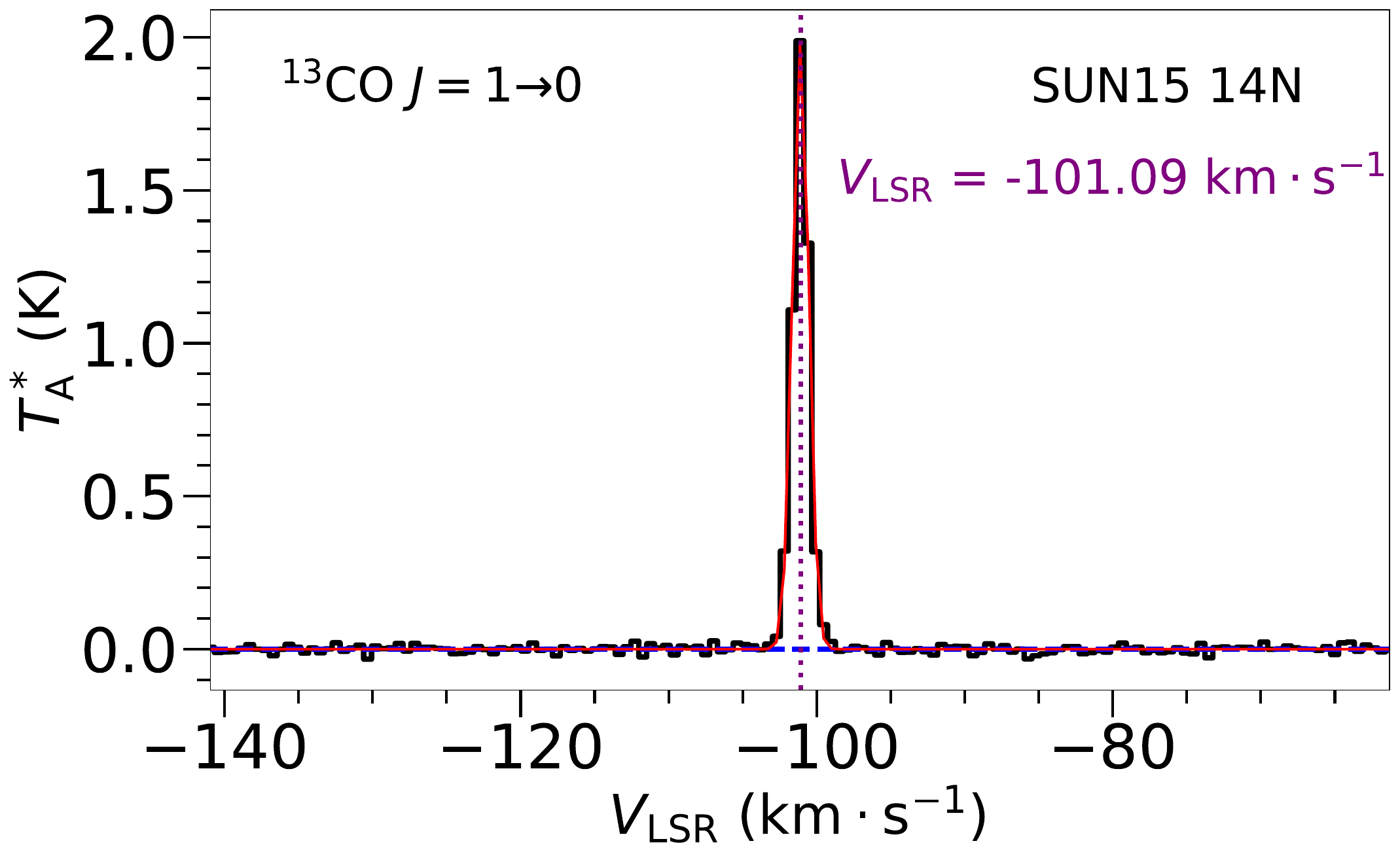}
		    \includegraphics[scale=0.16]{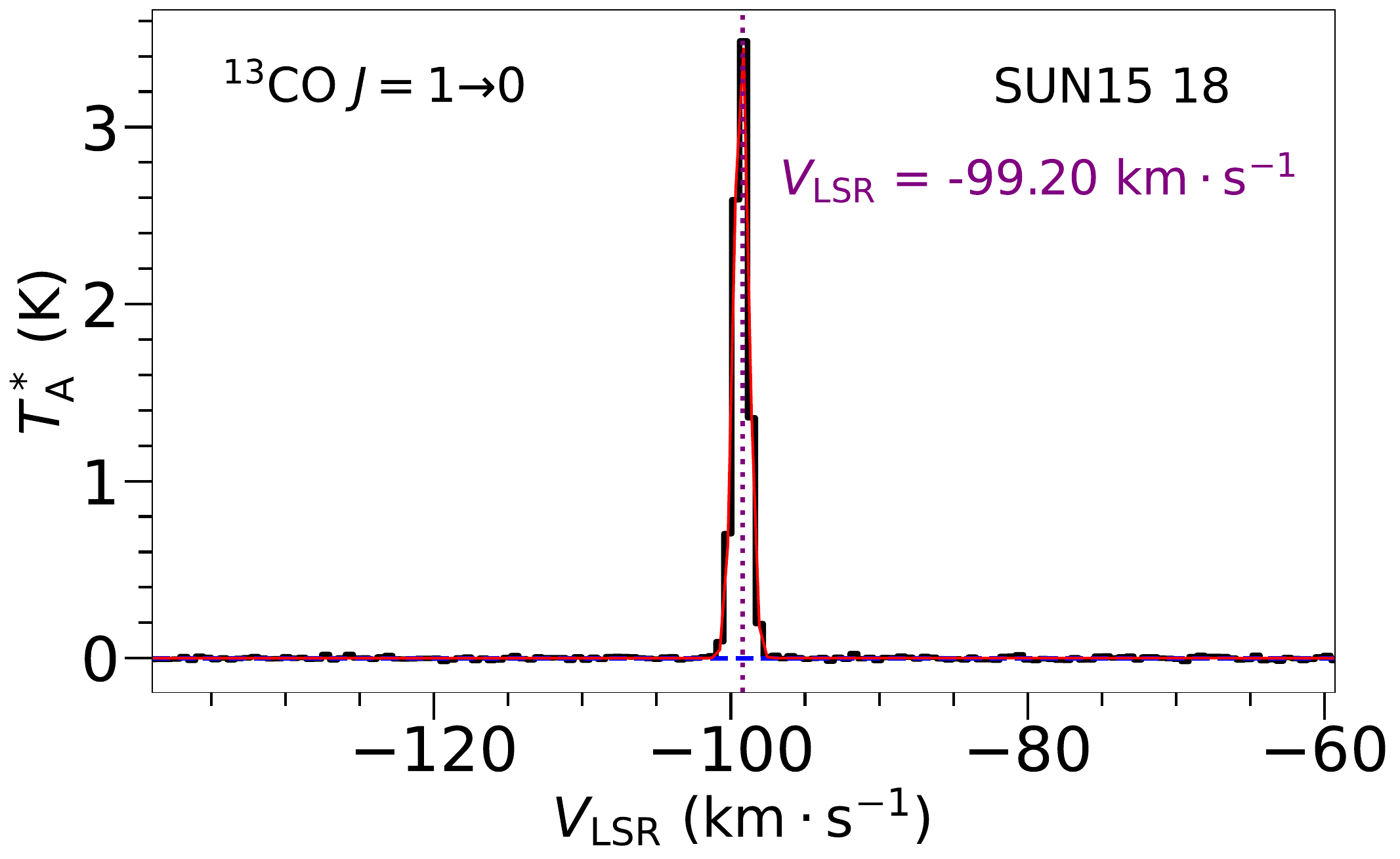}
                \includegraphics[scale=0.16]{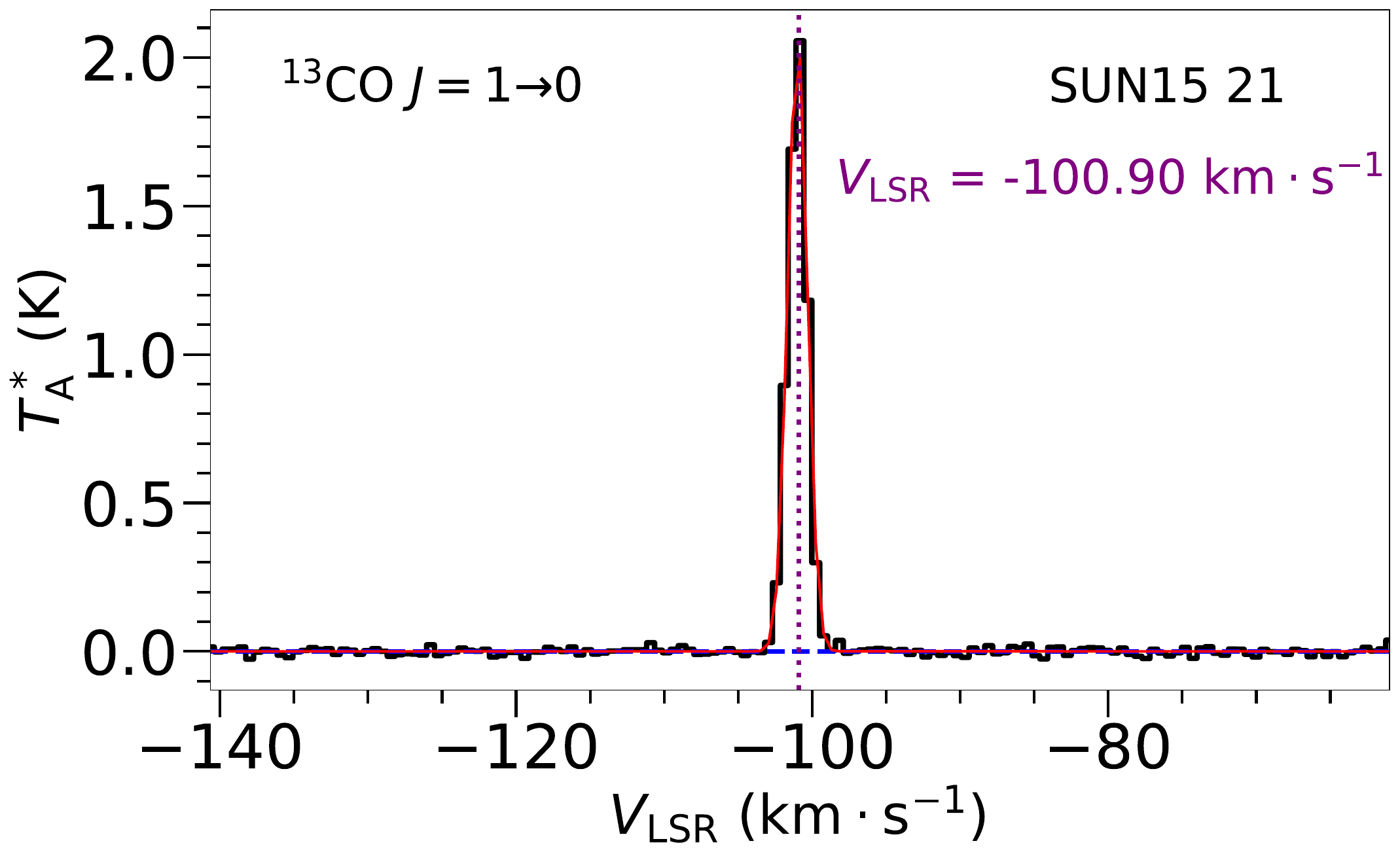}
                \includegraphics[scale=0.16]{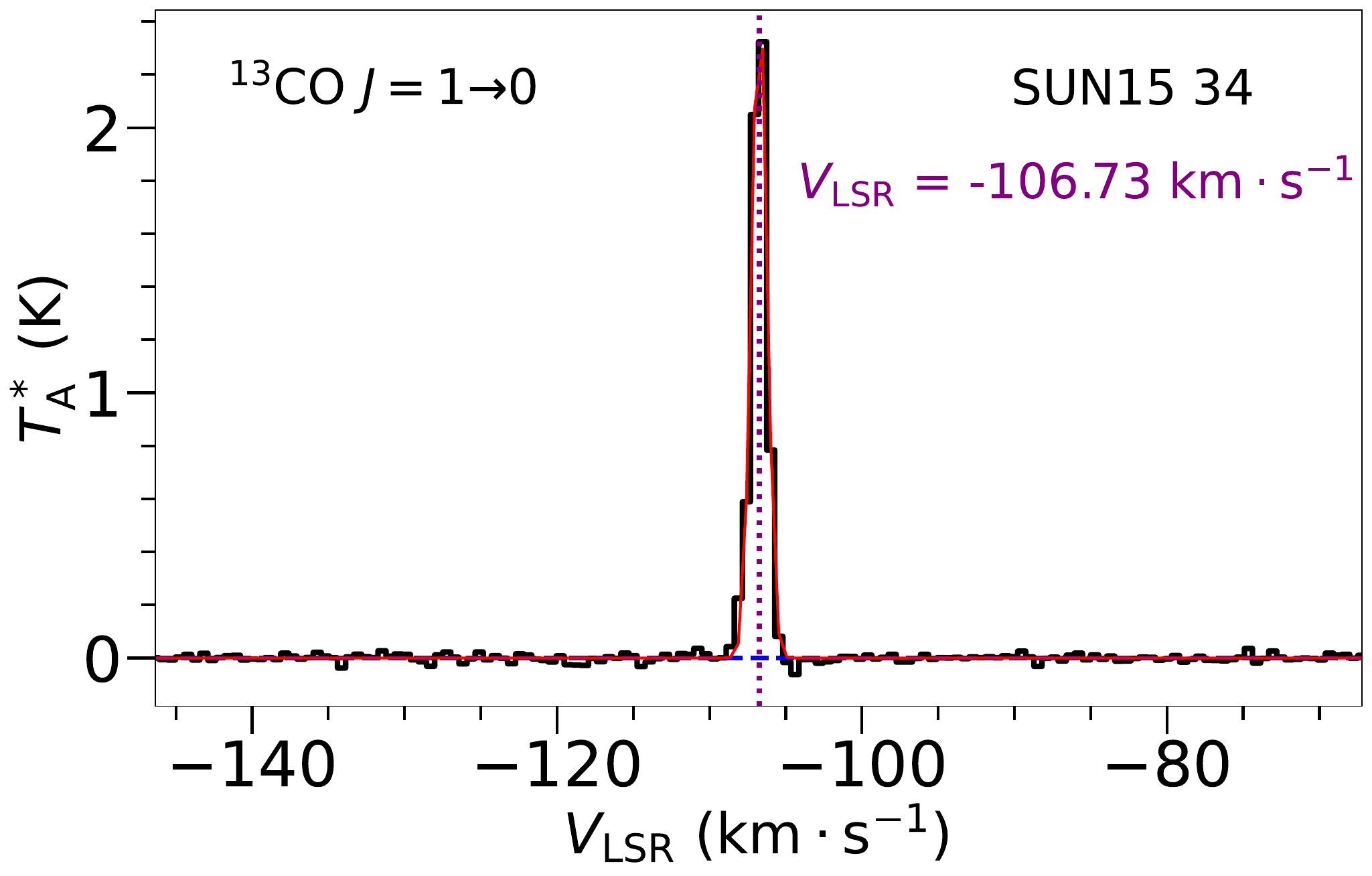}
                \includegraphics[scale=0.16]{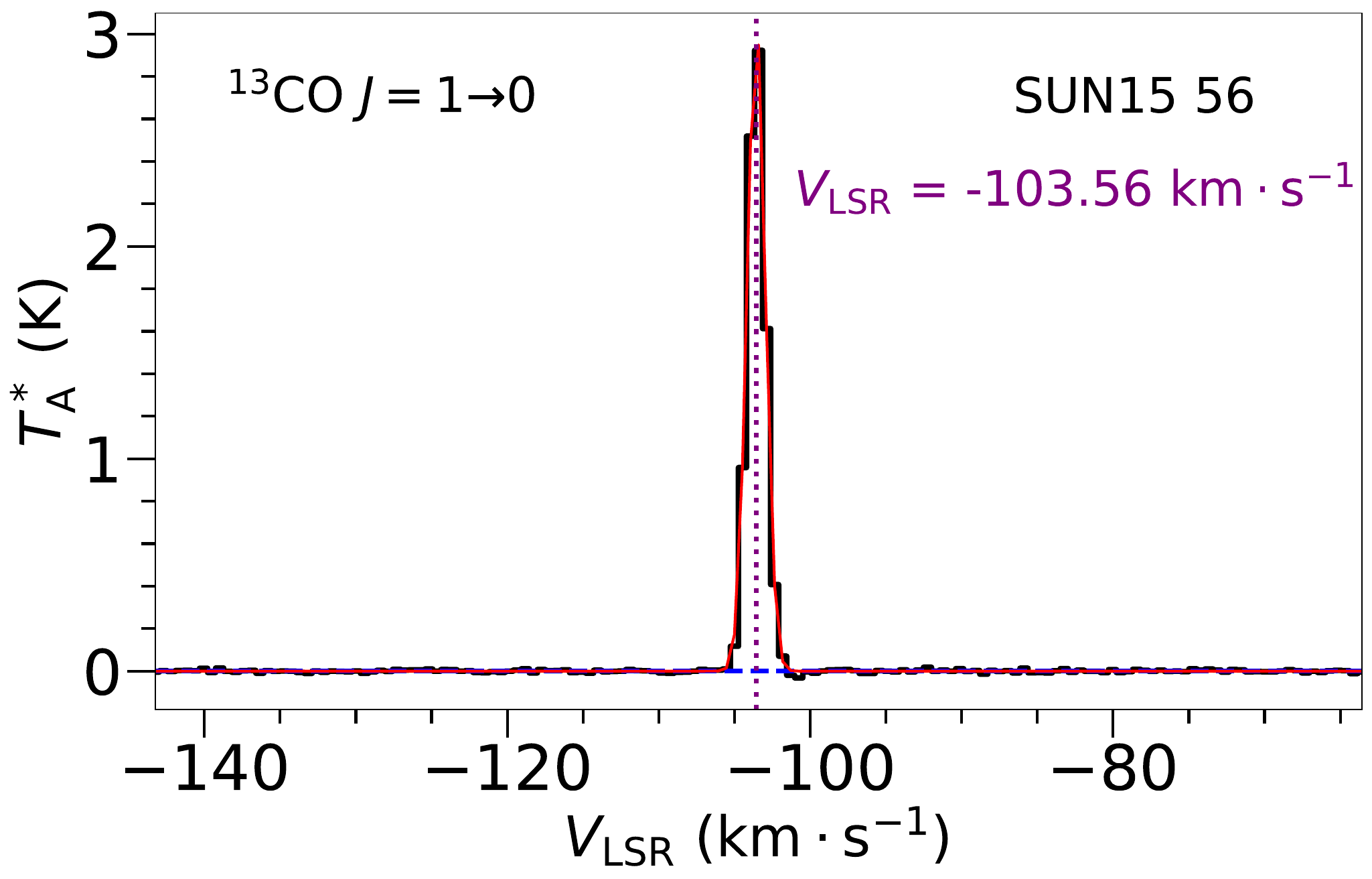}
		    \includegraphics[scale=0.16]{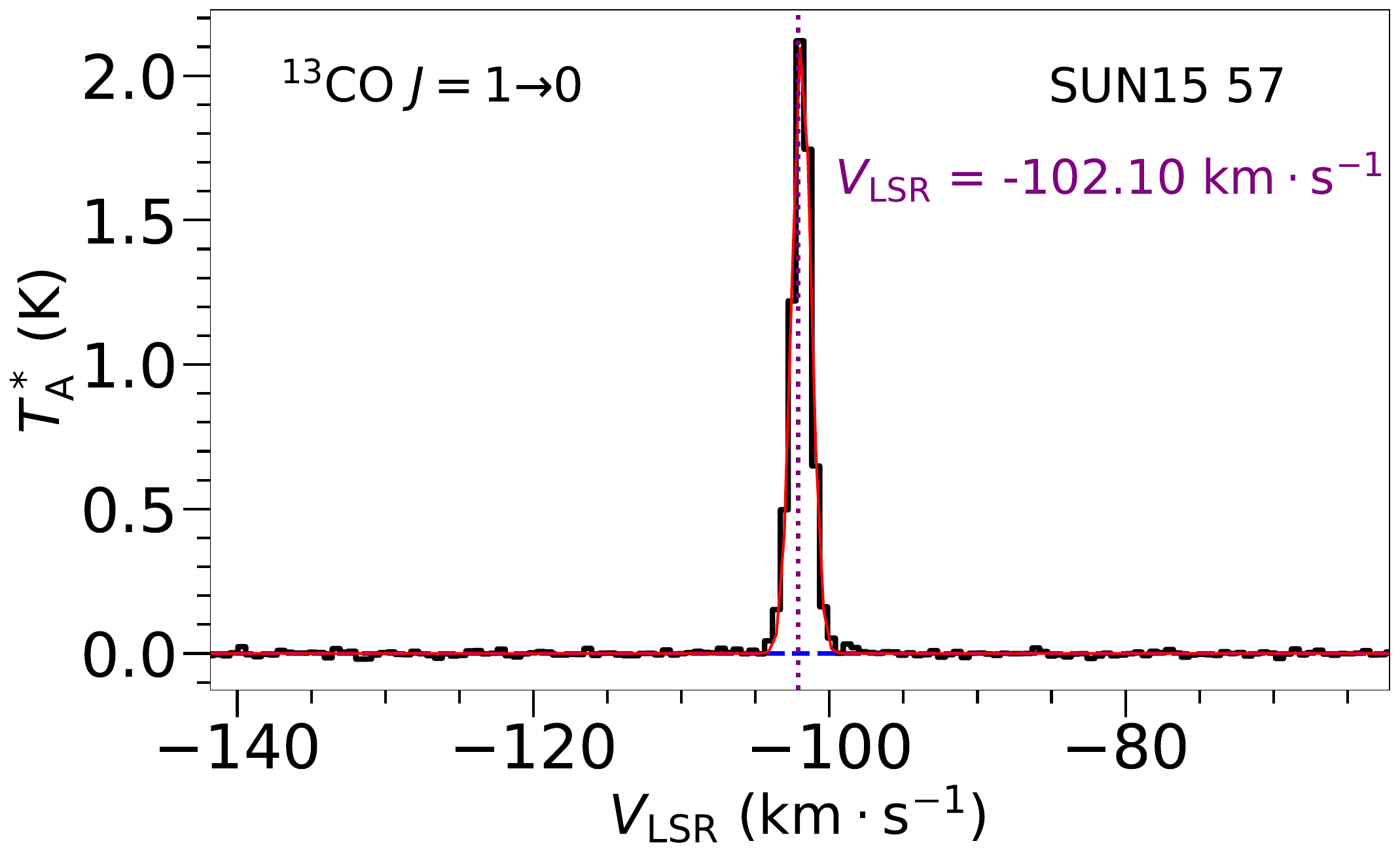}
		    \includegraphics[scale=0.16]{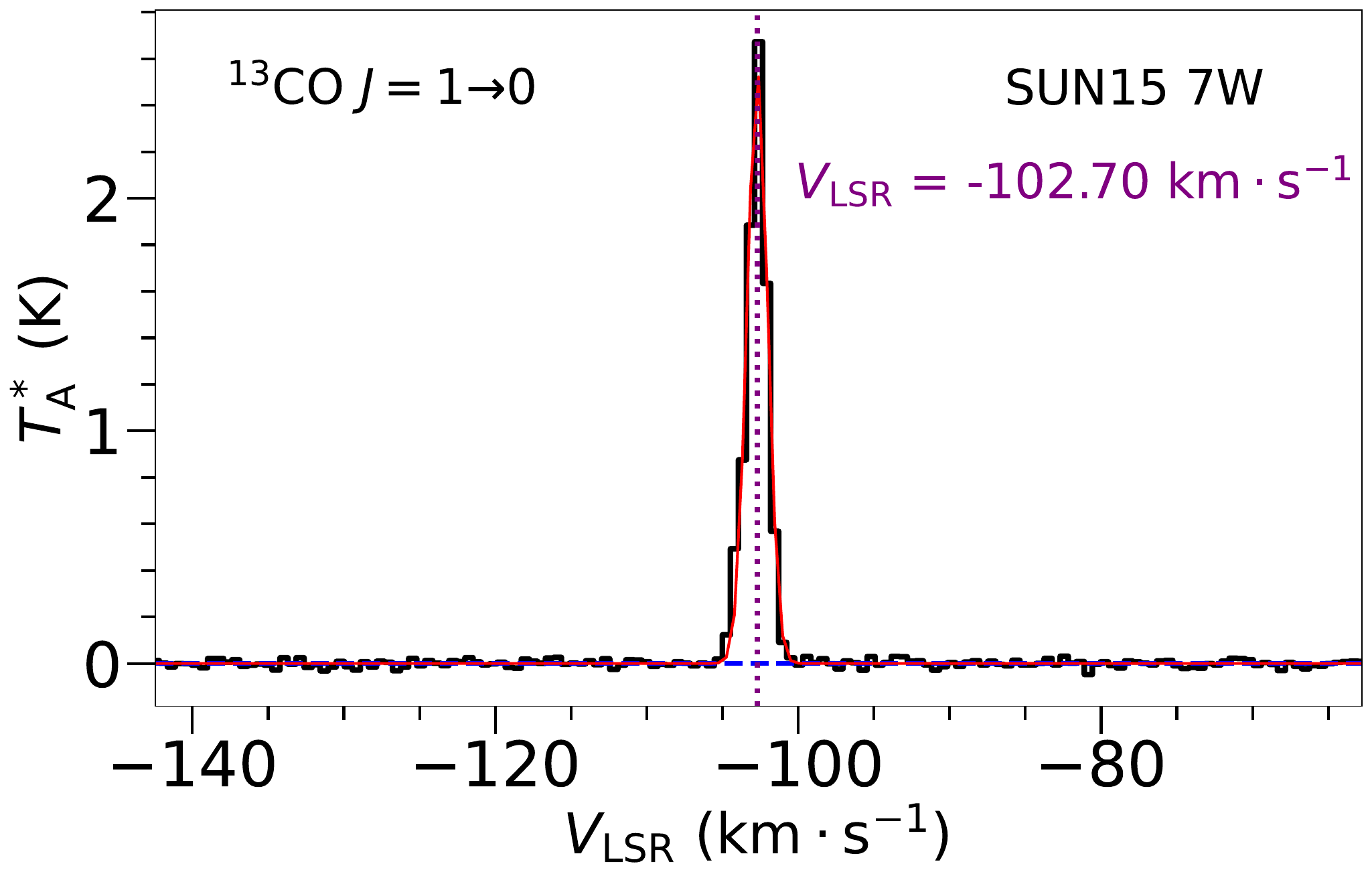}
                \includegraphics[scale=0.16]{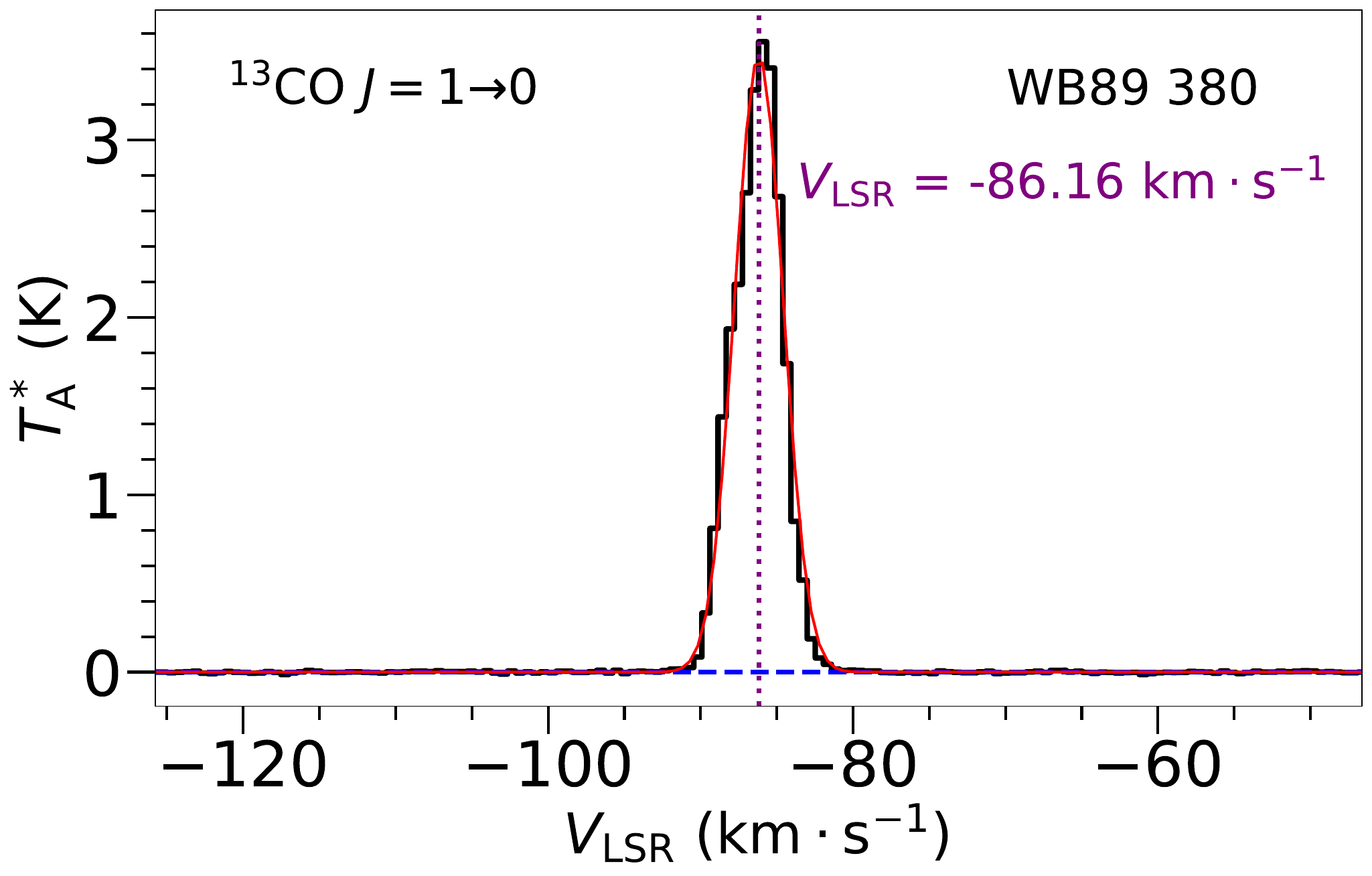}
                \includegraphics[scale=0.16]{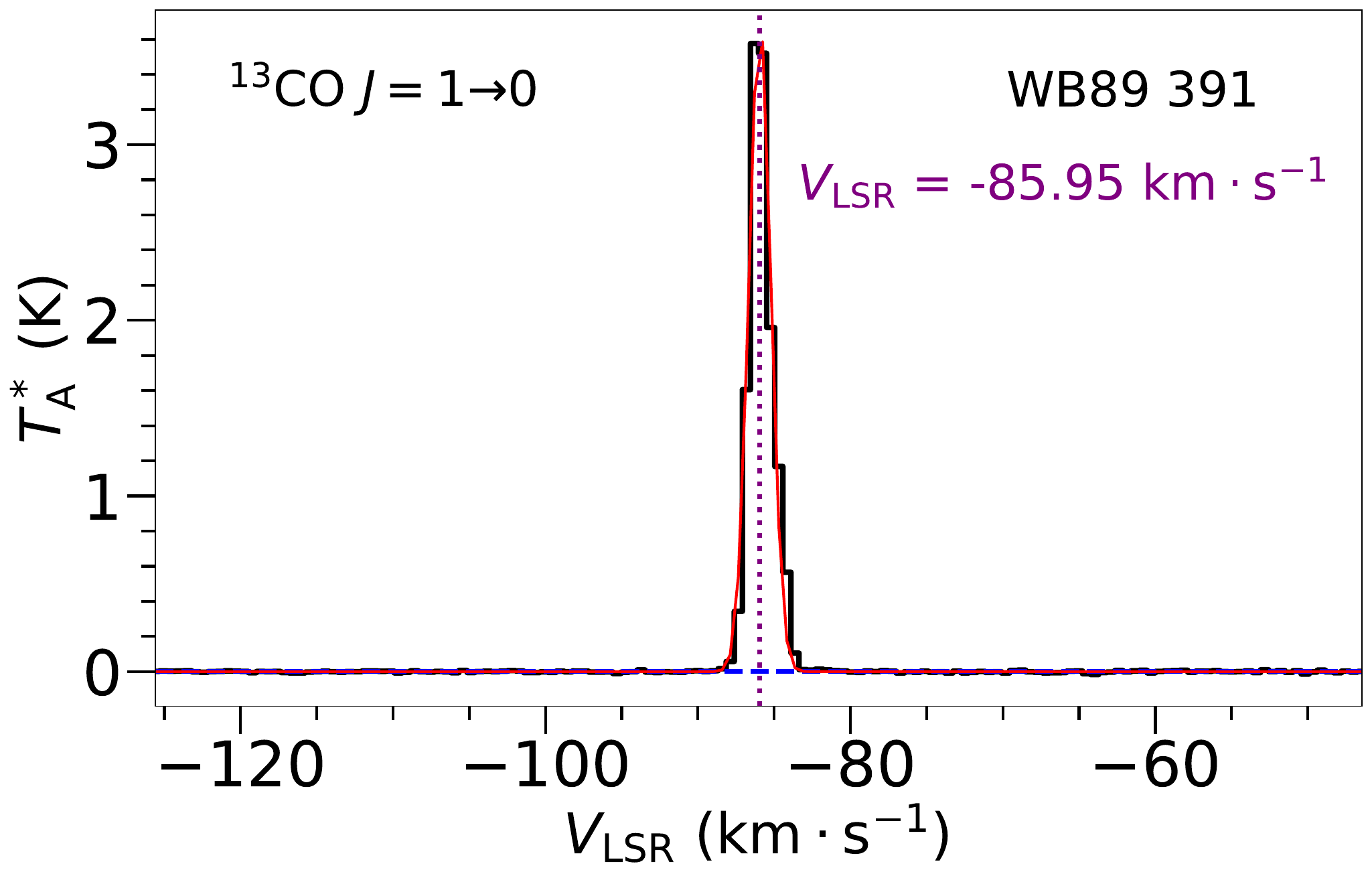}
                \includegraphics[scale=0.16]{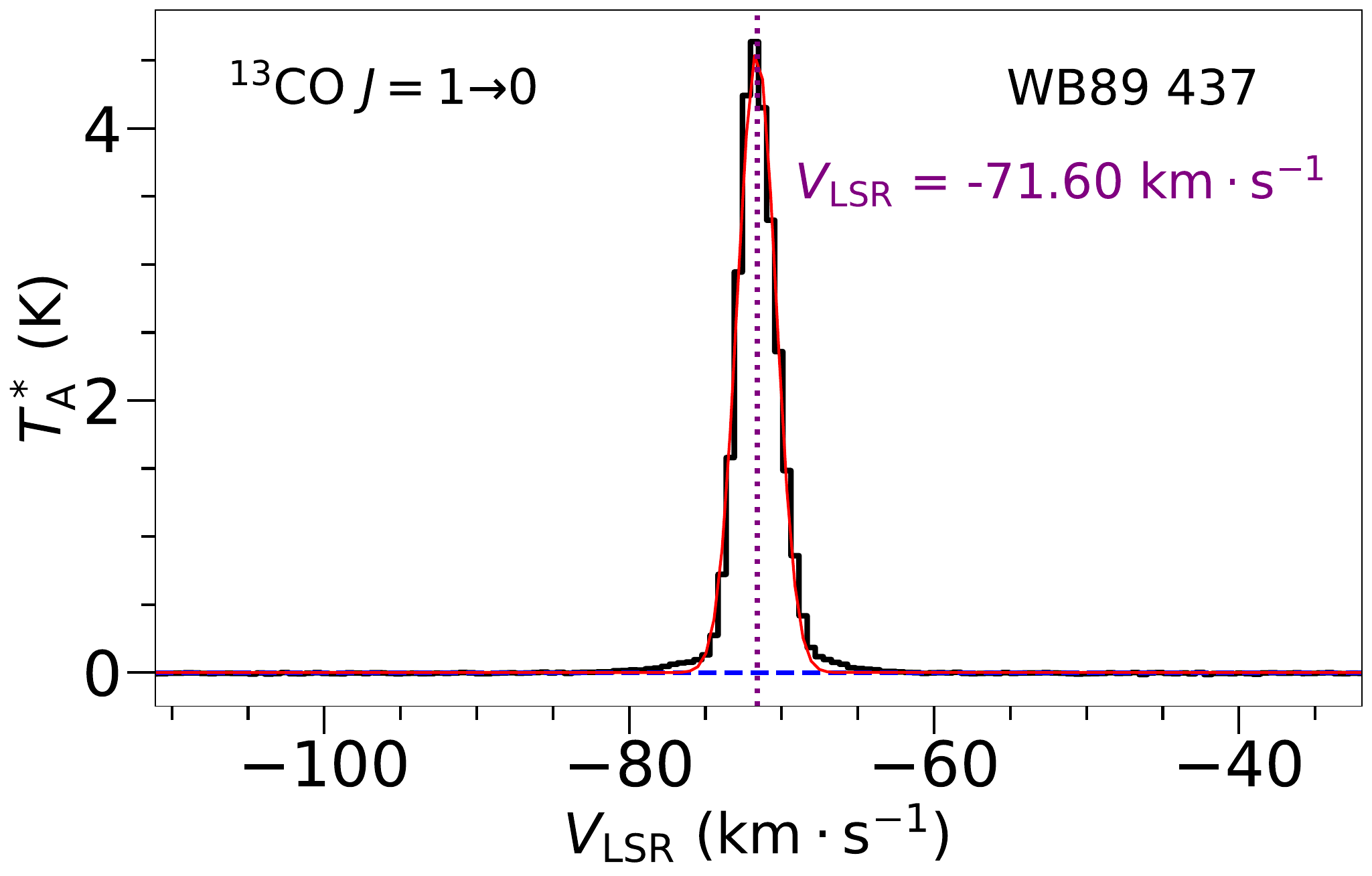}
                \includegraphics[scale=0.16]{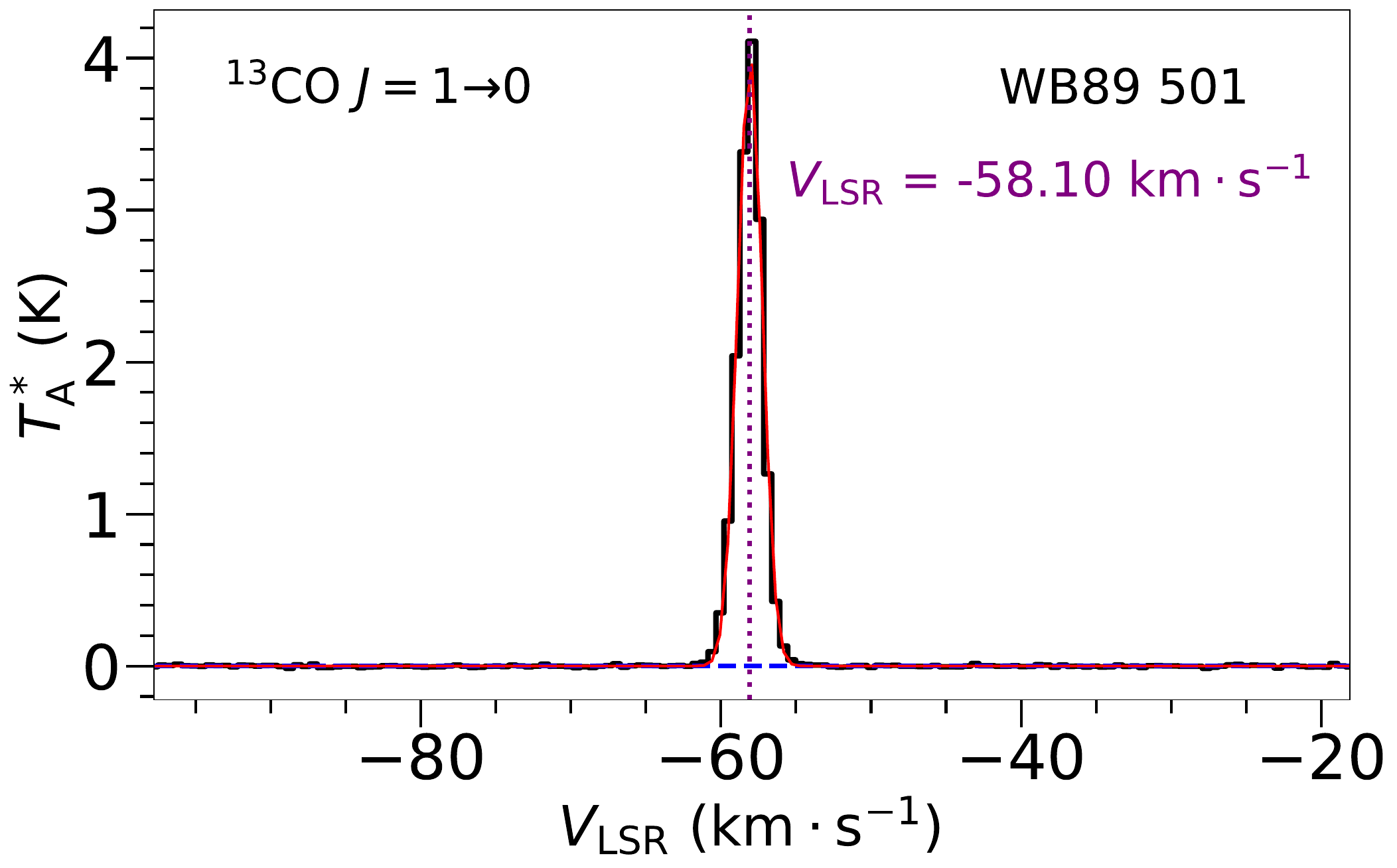}
        \caption{\label{fig_13CO_fitting_spectra}The $\rm ^{13}CO$ $J=1\to0$ spectra of our targets. The black steps show the original spectra and the red curves show the Gaussian fitting results. The blue dashed lines show the baseline. The vertical purple dotted lines show the velocity at the local standard rest.}
\end{figure*}

\begin{figure*}
        \centering
		        \includegraphics[scale=0.14]{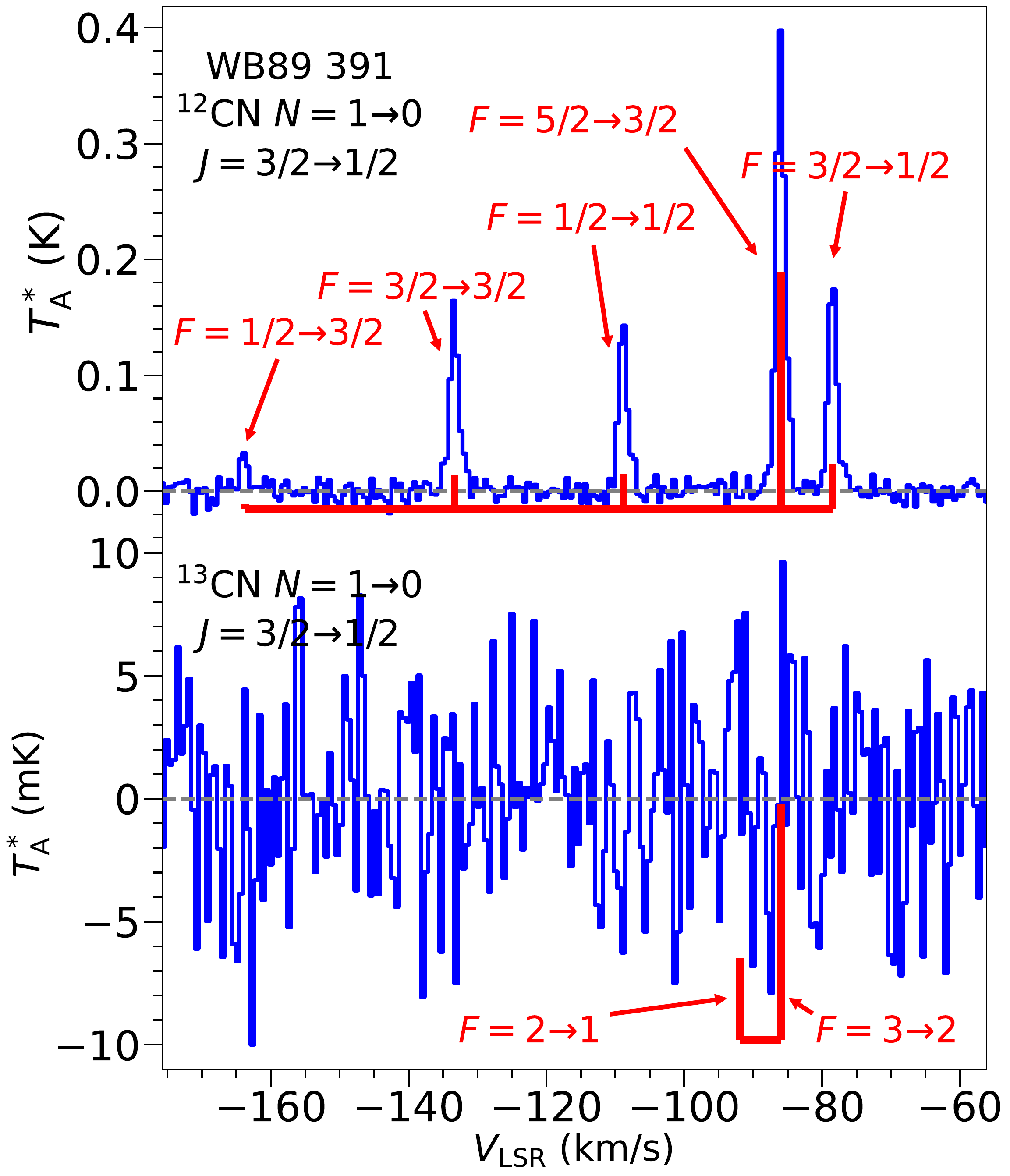}
                \includegraphics[scale=0.14]{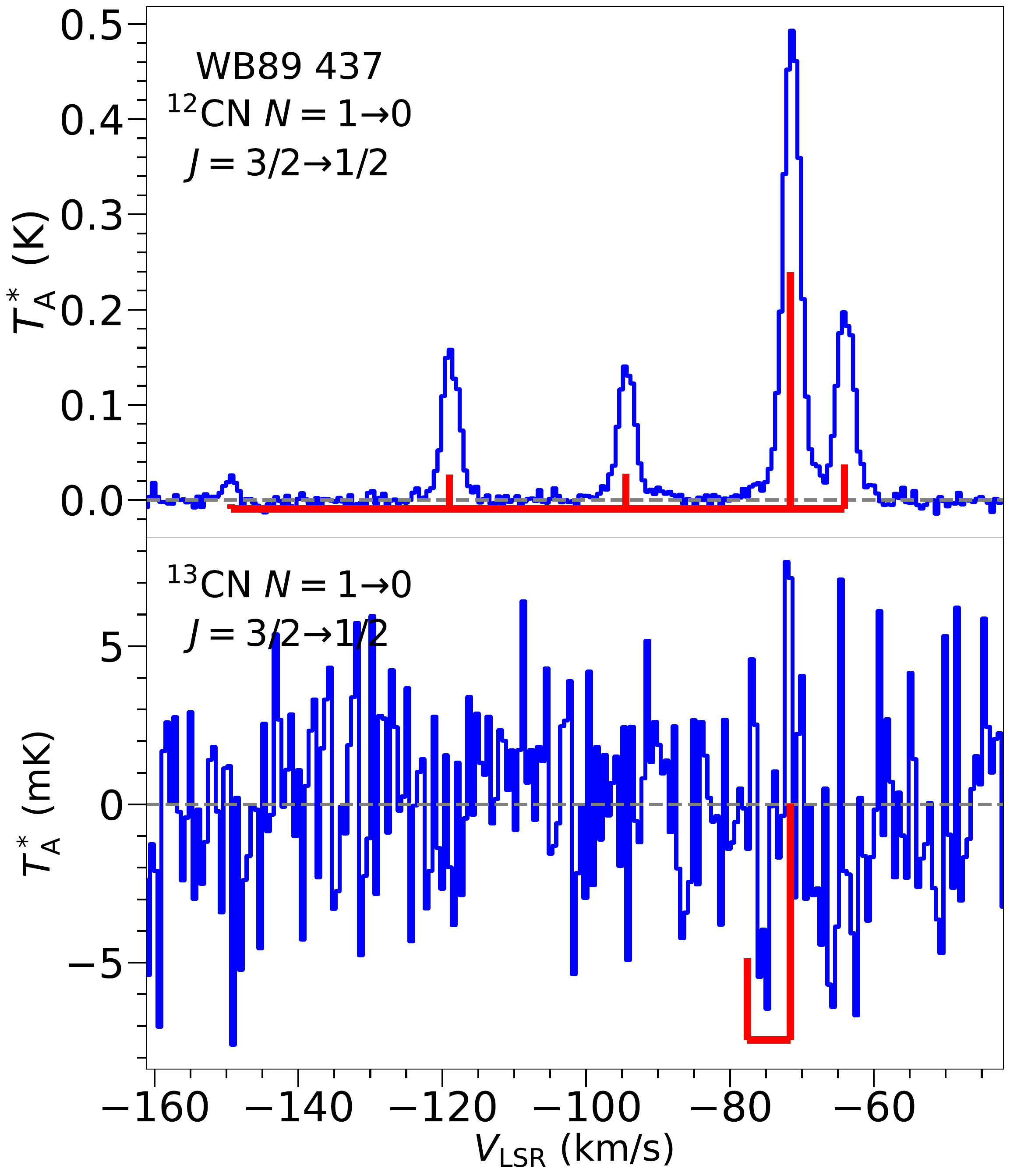}
                \includegraphics[scale=0.14]{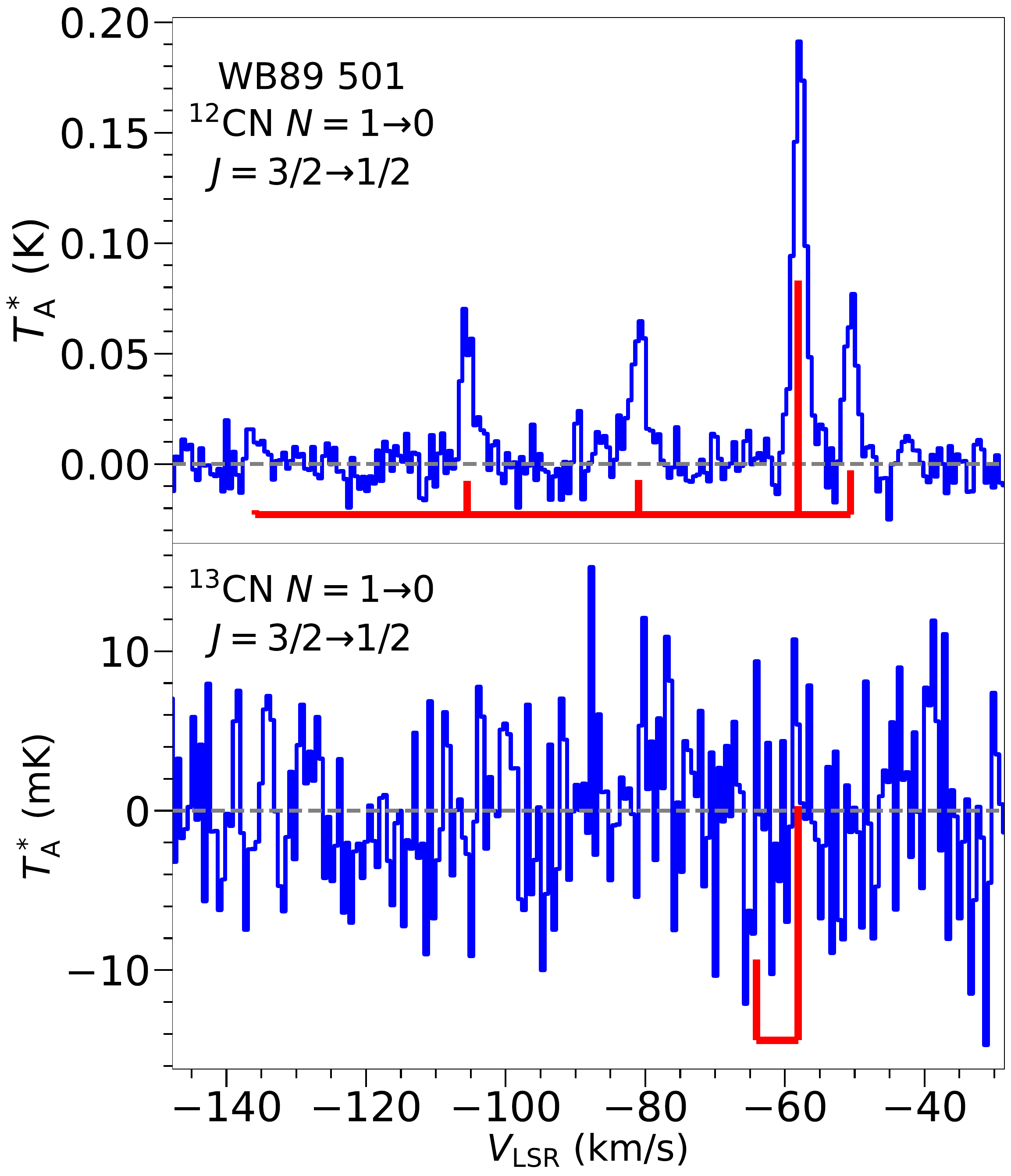}
                \includegraphics[scale=0.14]{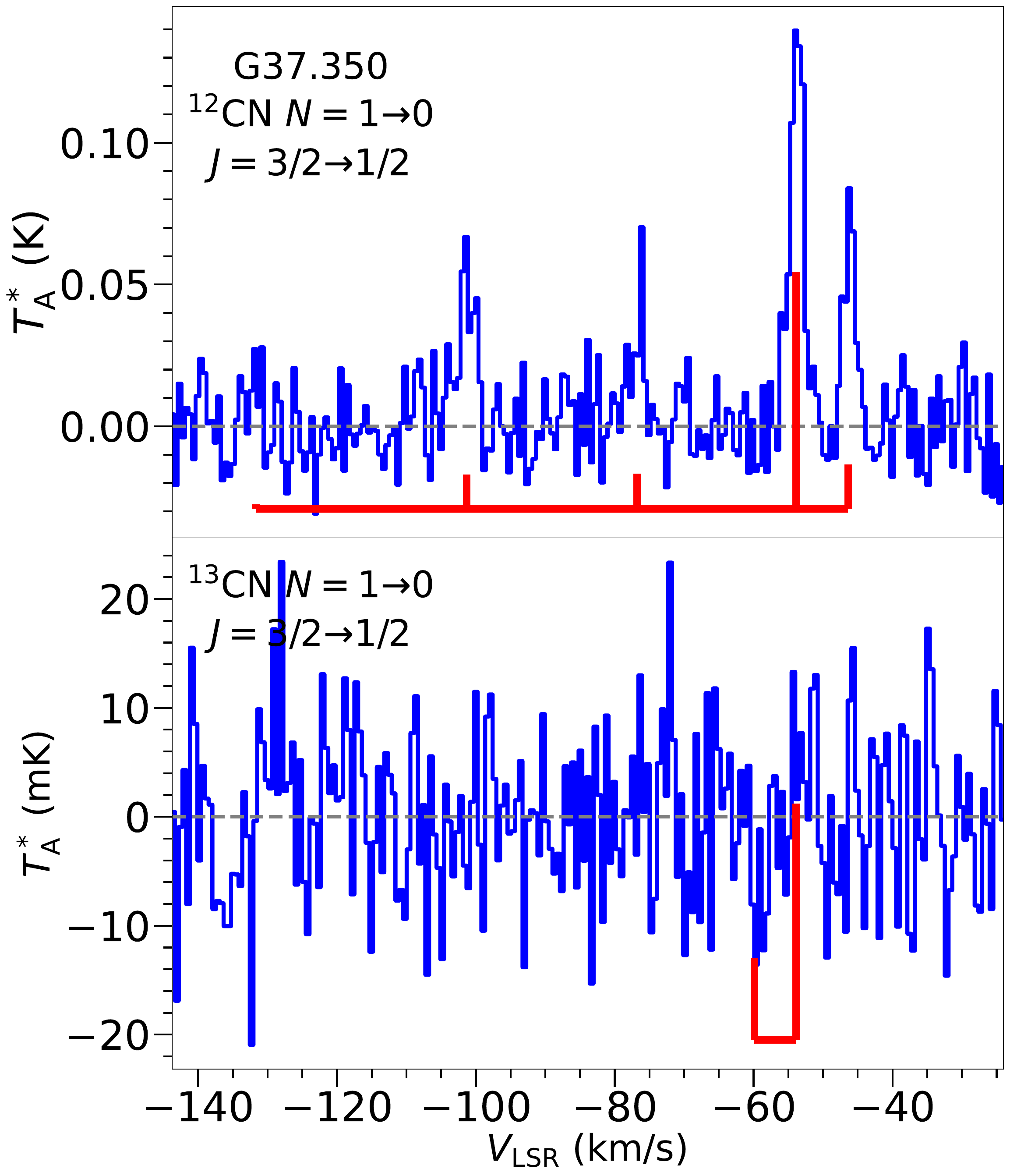}
                \includegraphics[scale=0.14]{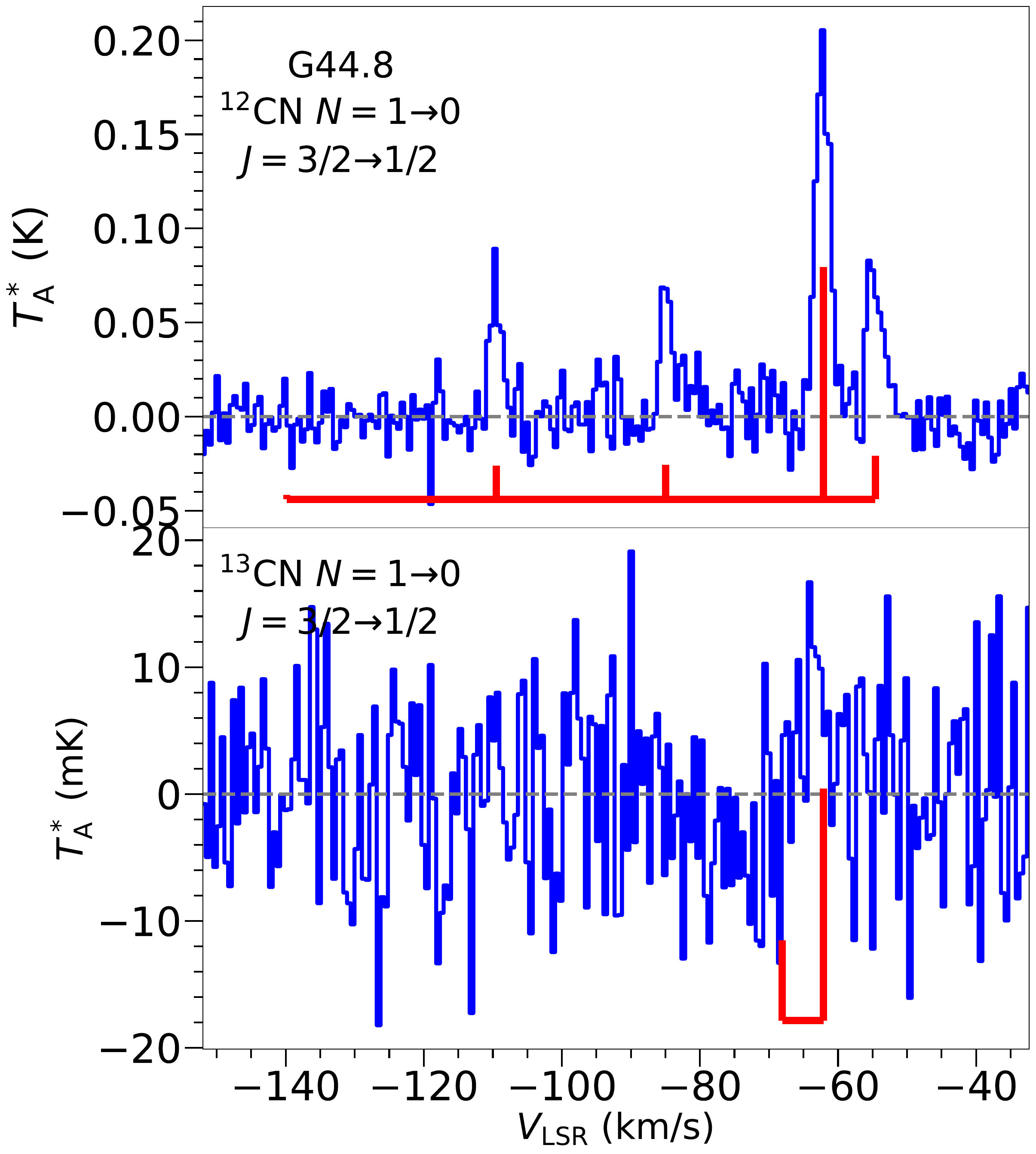}
                \includegraphics[scale=0.14]{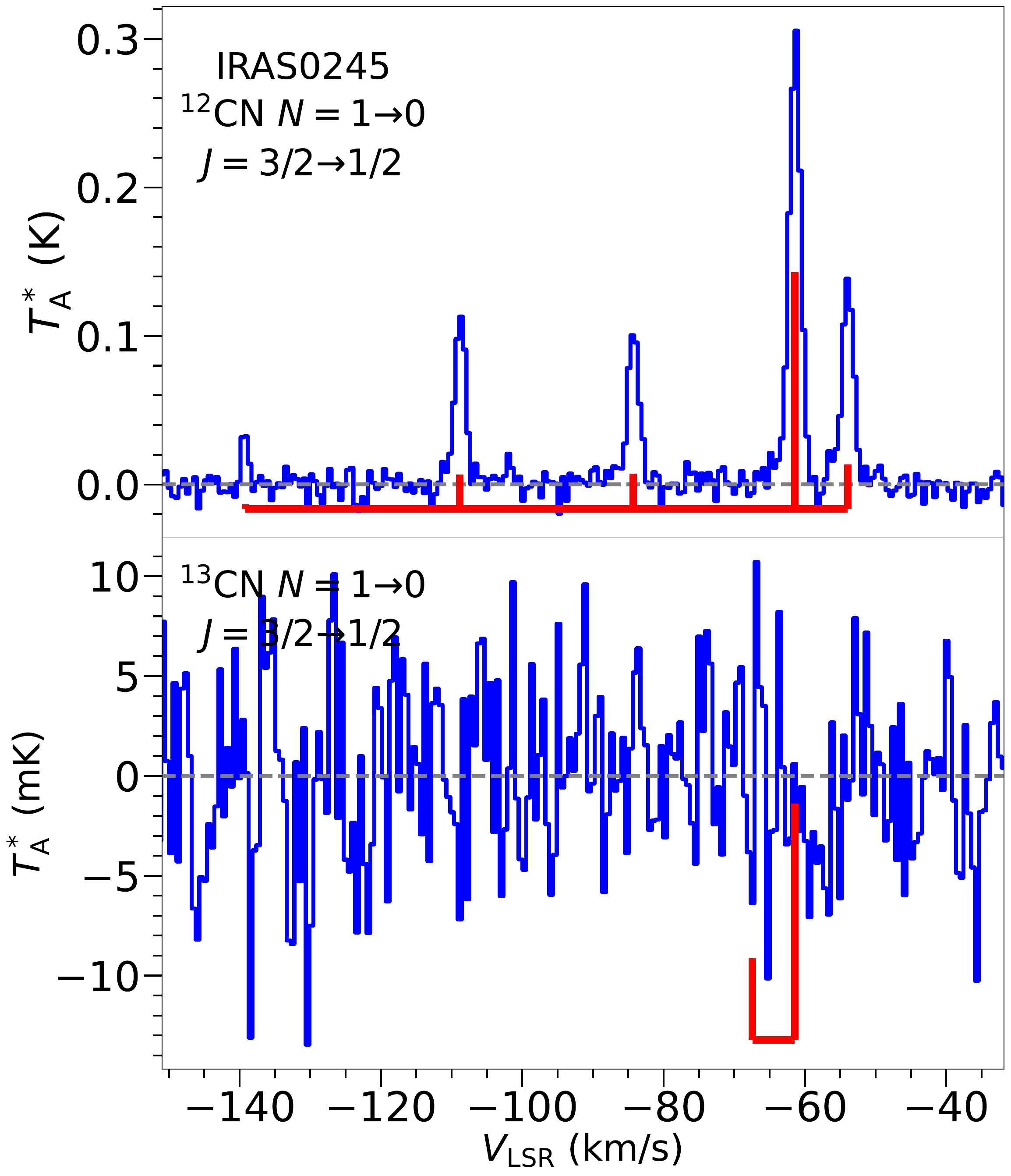}
                \includegraphics[scale=0.14]{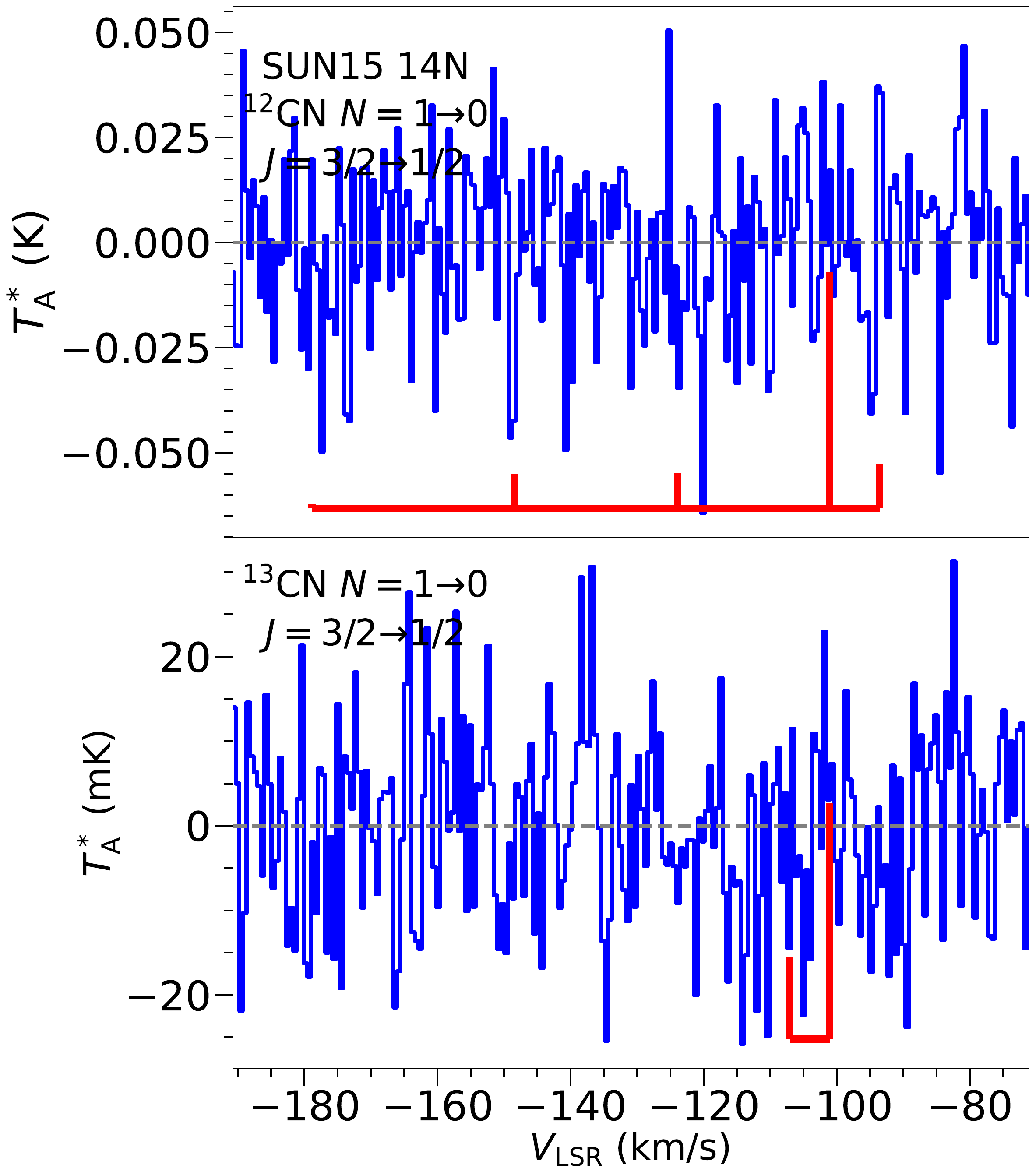}
                \includegraphics[scale=0.14]{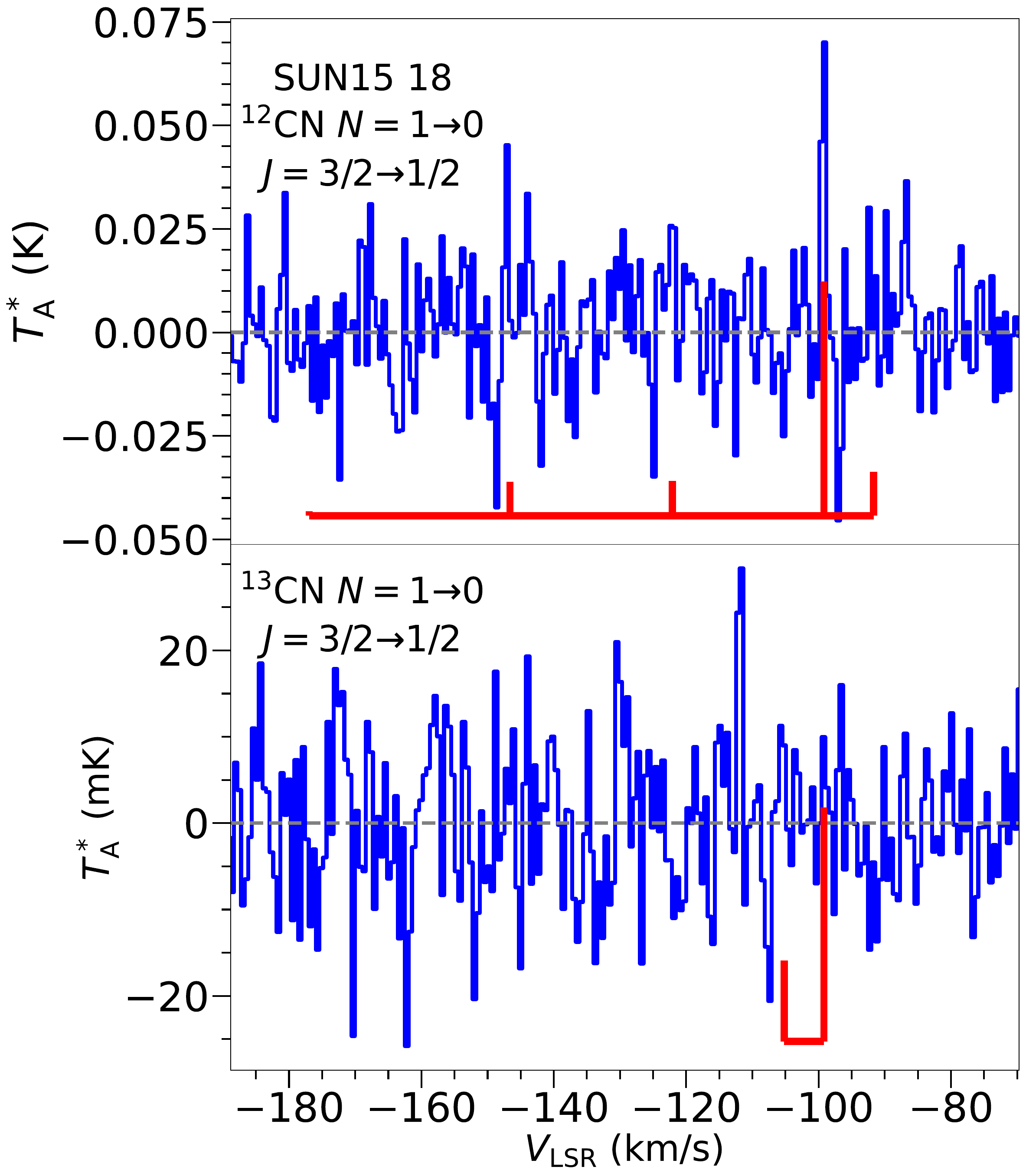}
                \includegraphics[scale=0.14]{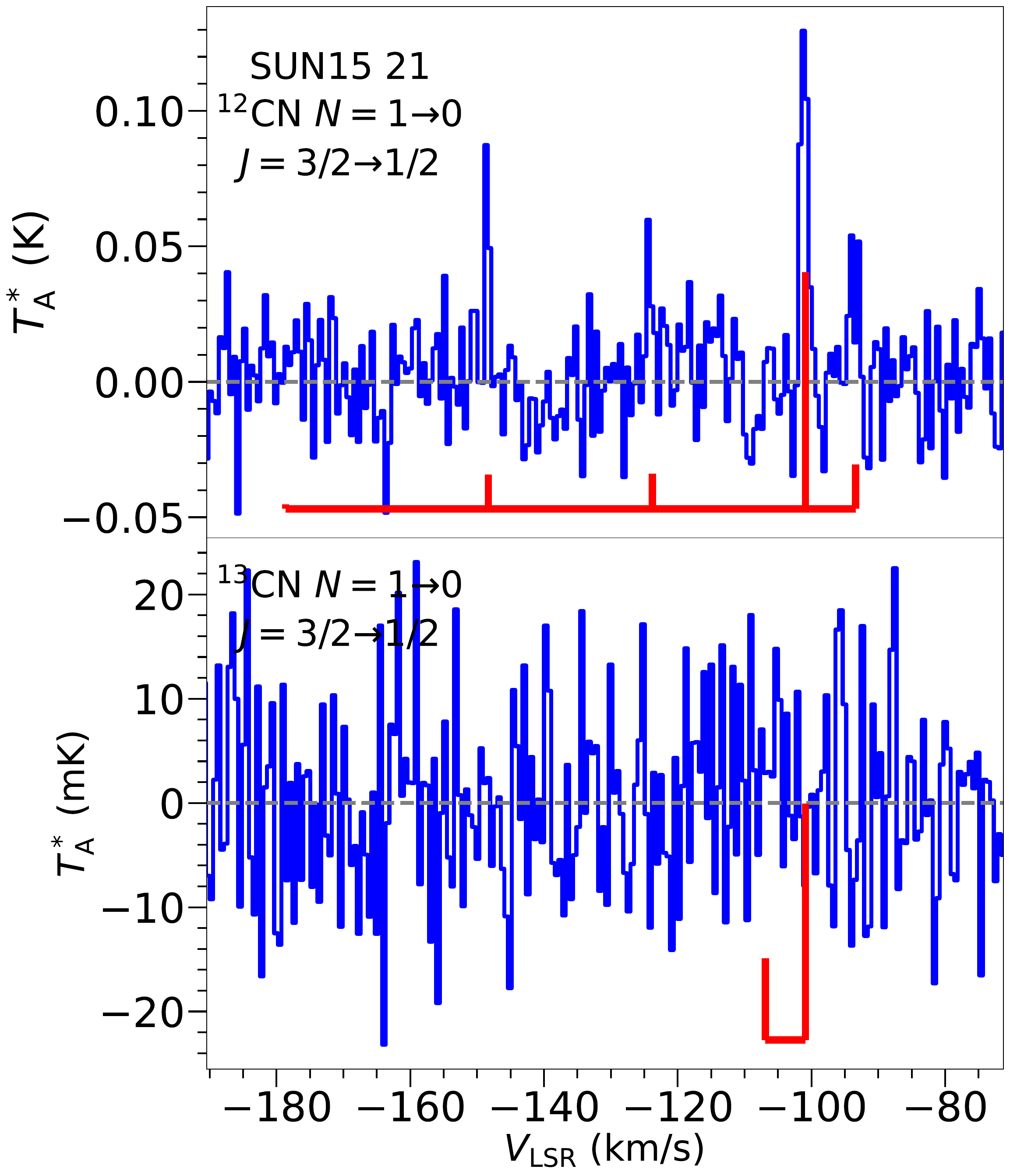}
        \caption{\label{fig:undetected_13CN_spectra} Spectra of \twCN\ and \thCN\ $N=1\to0$ for targets without \thCN\ (and \twCN\ for some cases)  detections.}
\end{figure*}

\begin{figure*}
        \centering
                \includegraphics[scale=0.14]{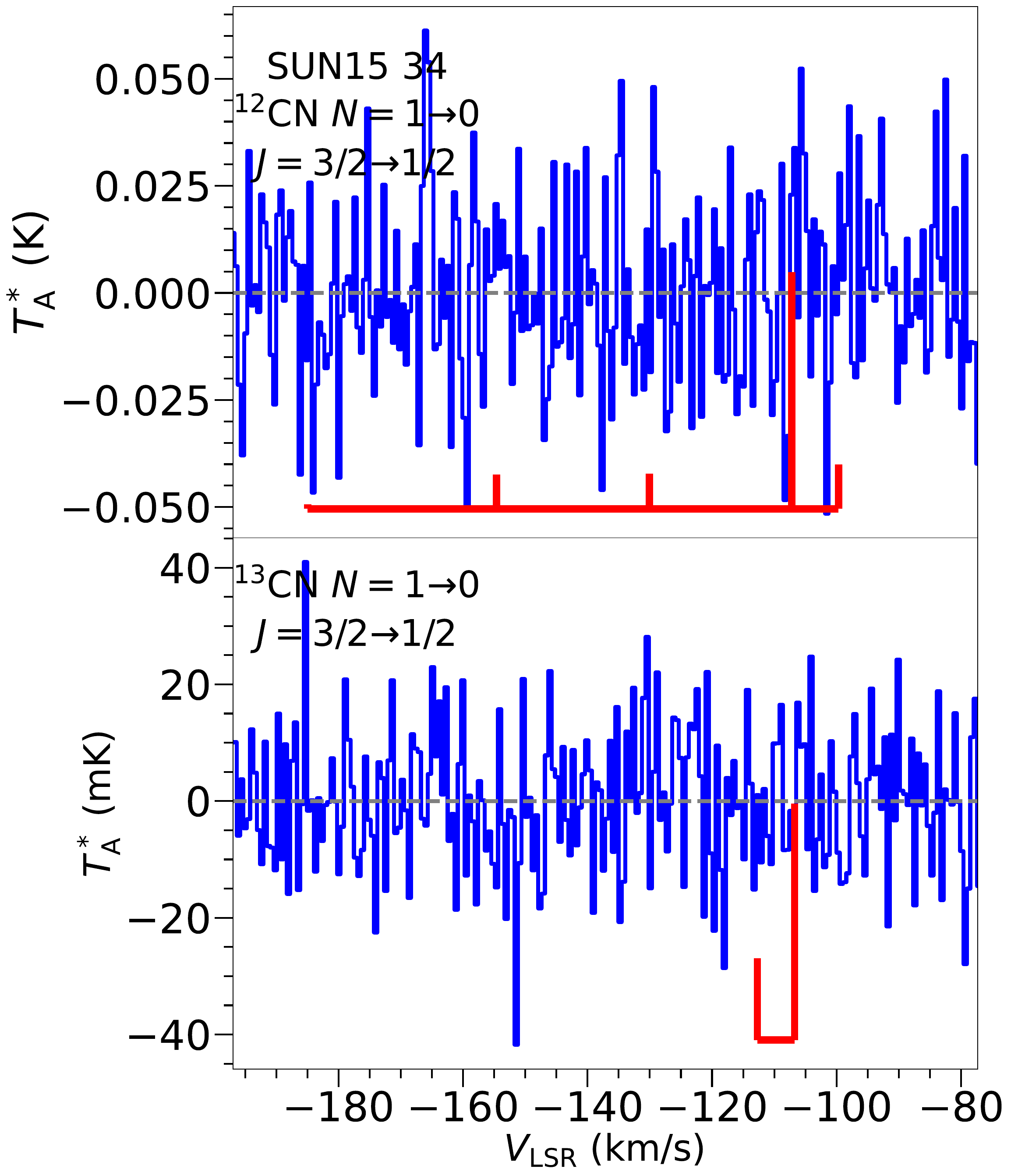}
		    \includegraphics[scale=0.14]{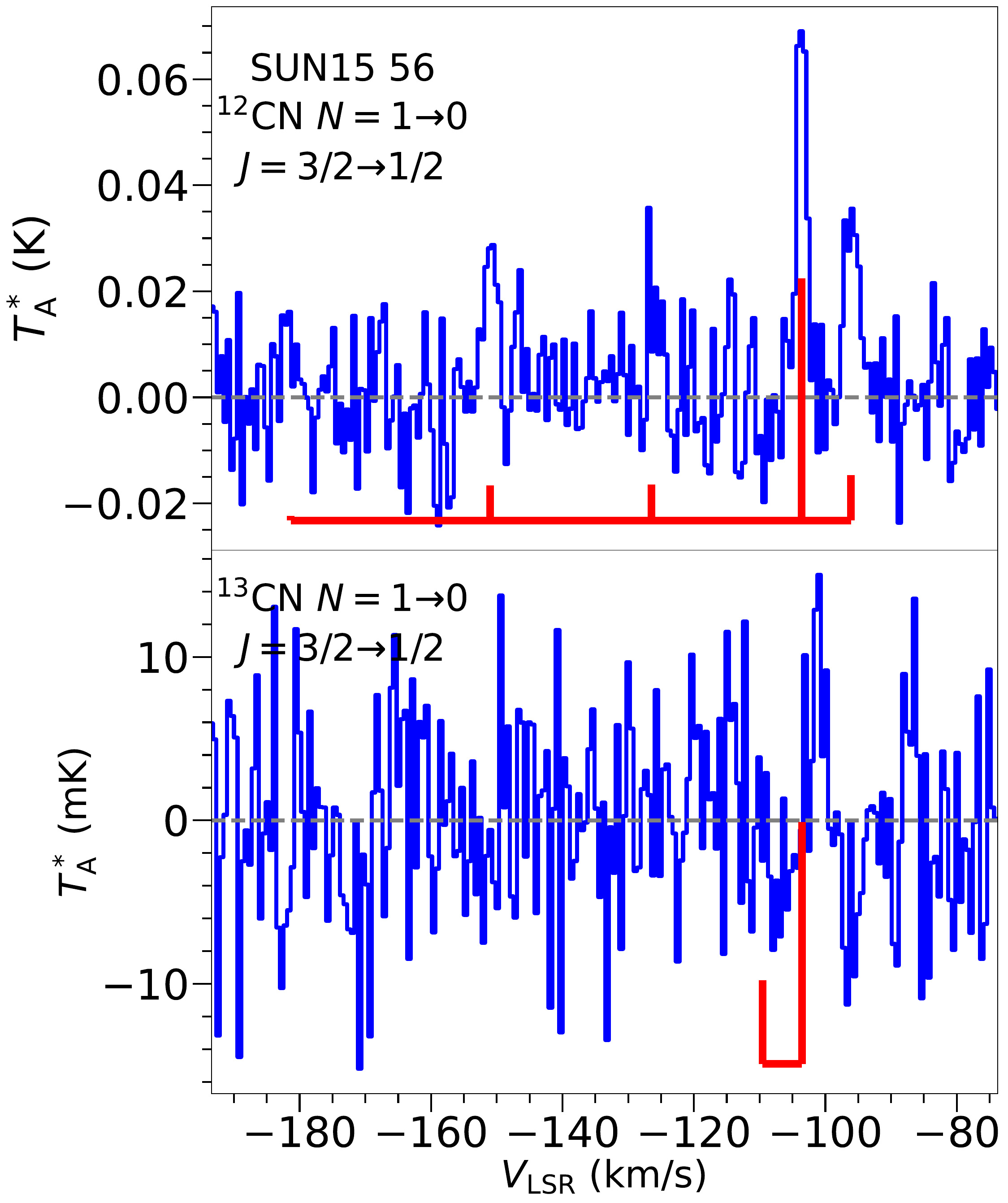}
                \includegraphics[scale=0.14]{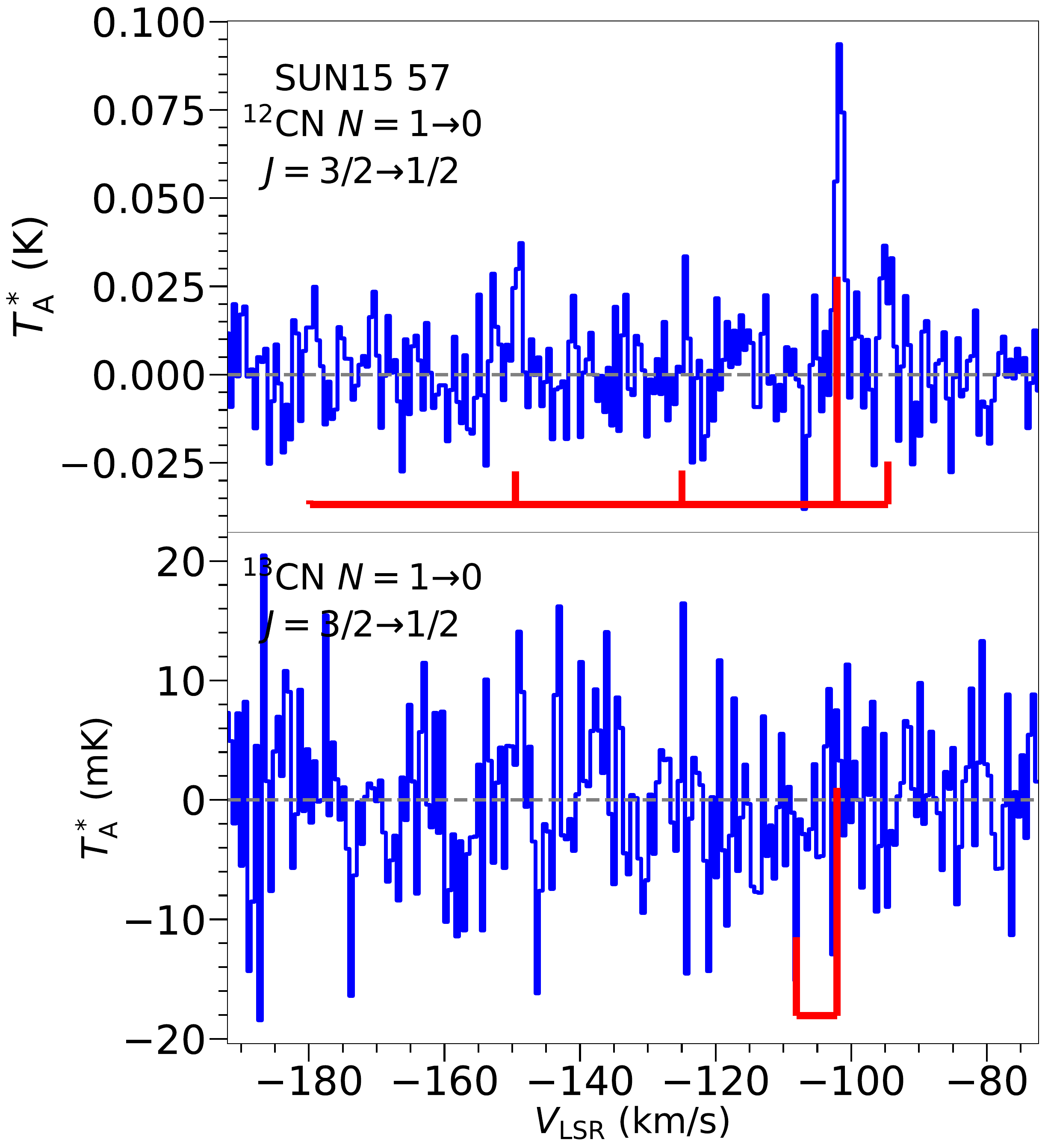}
                \includegraphics[scale=0.14]{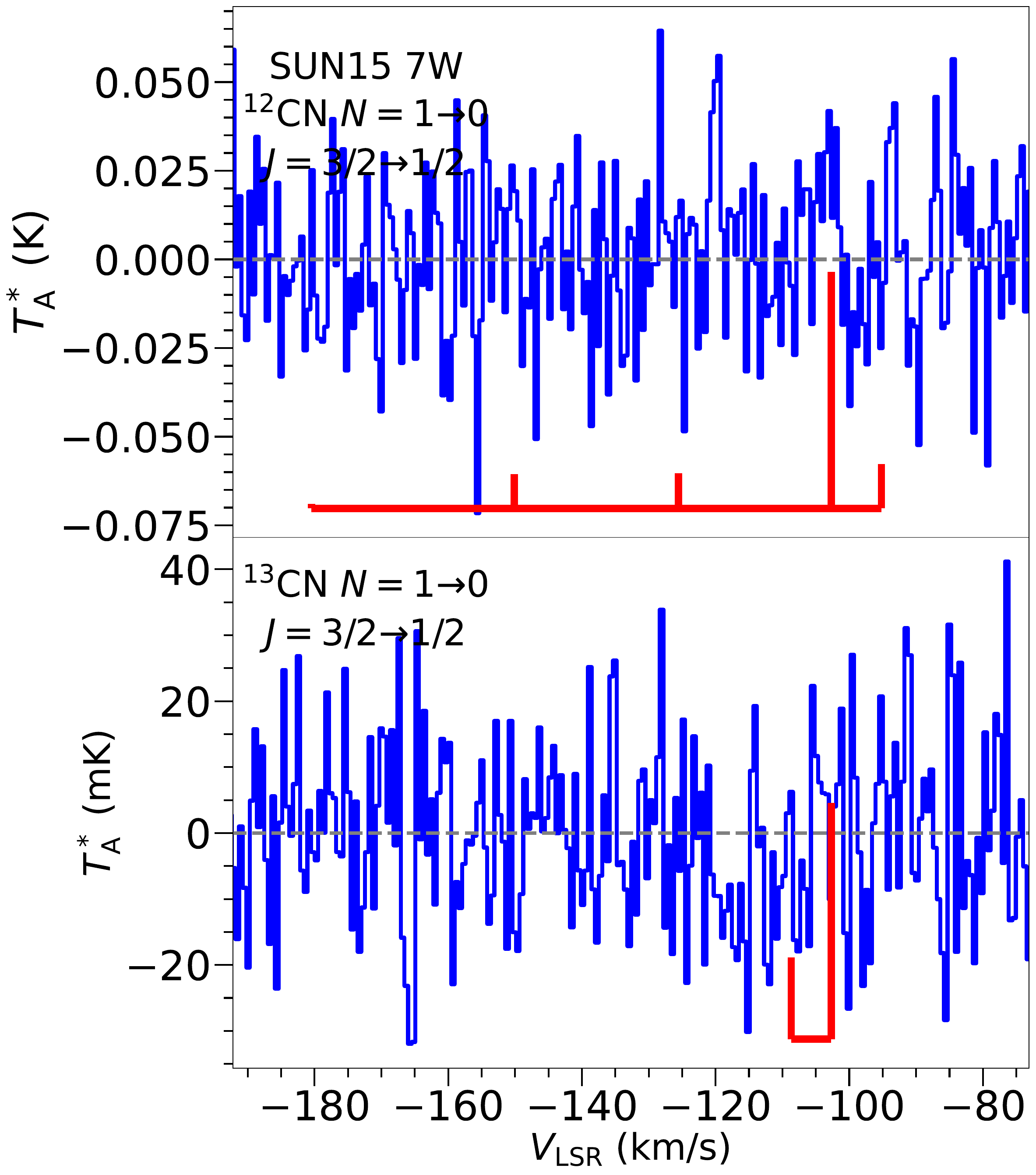}
        \caption{\label{fig:undetected_13CN_spectra_conti} Spectra of \twCN\ and \thCN\ $N=1\to0$ for targets without \thCN\ (and \twCN\ for some cases) detections. (Continued.).}
\end{figure*}

\begin{figure*}
        \centering
                \includegraphics[scale=0.14]{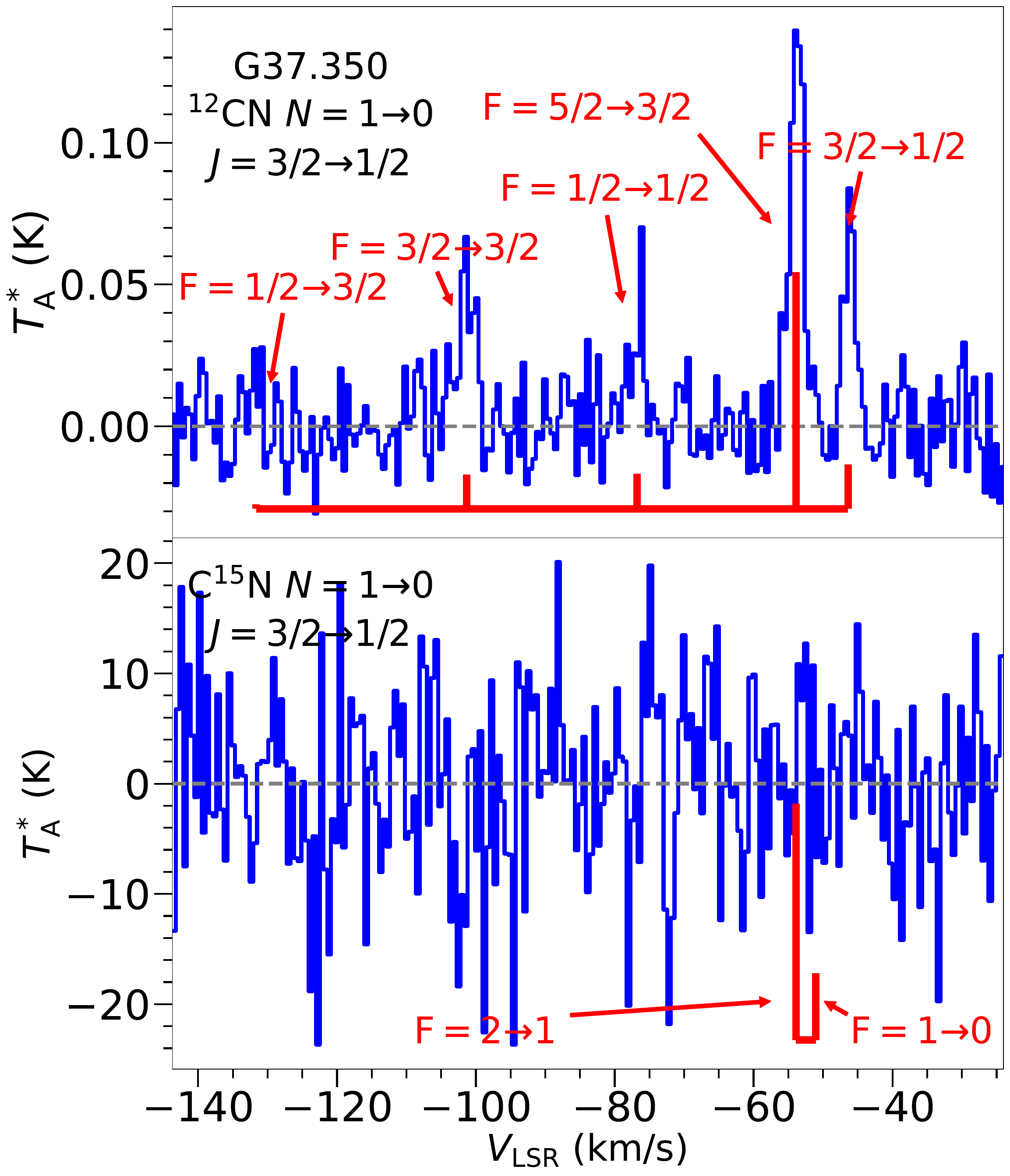}
                \includegraphics[scale=0.14]{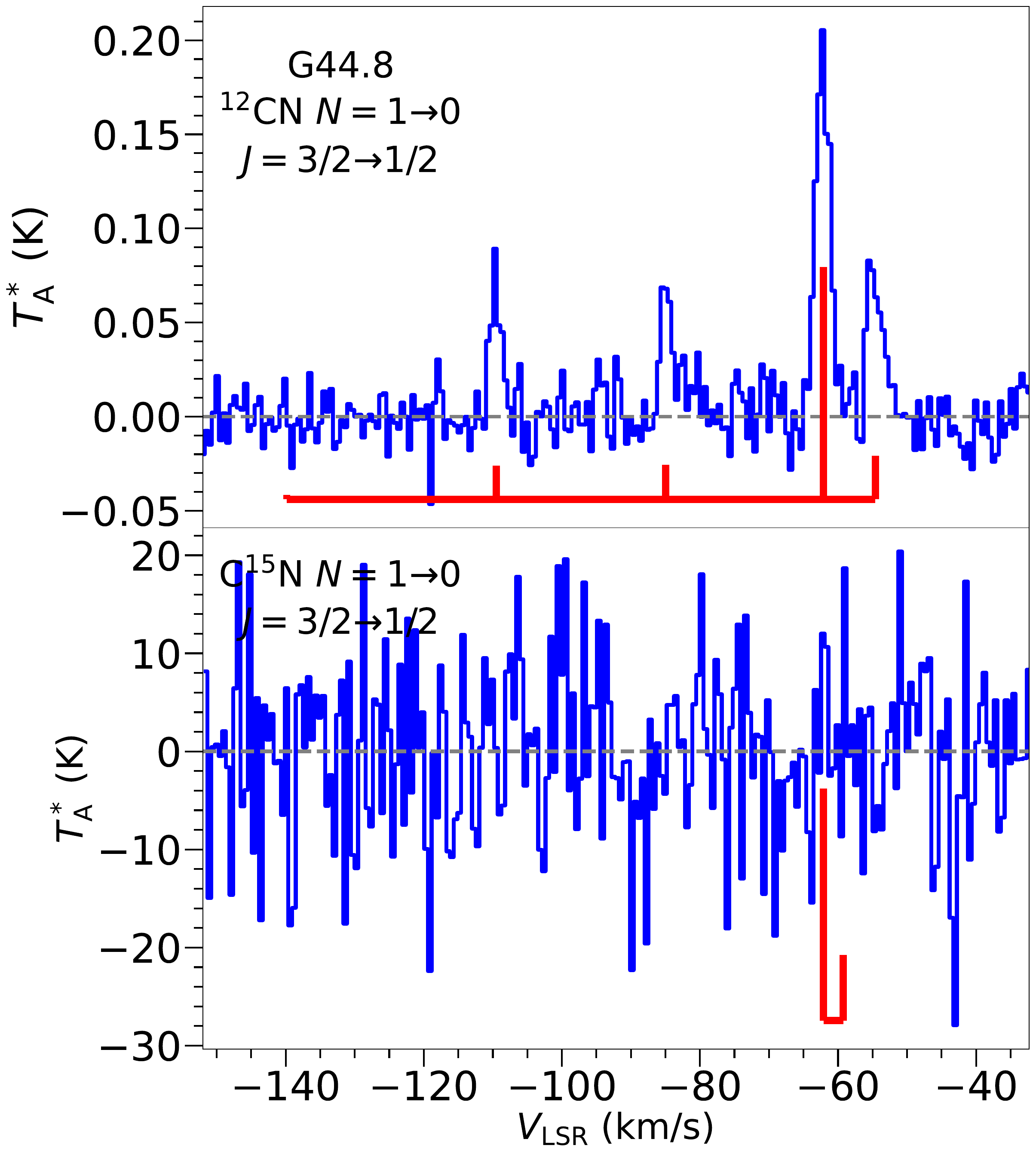}
                \includegraphics[scale=0.14]{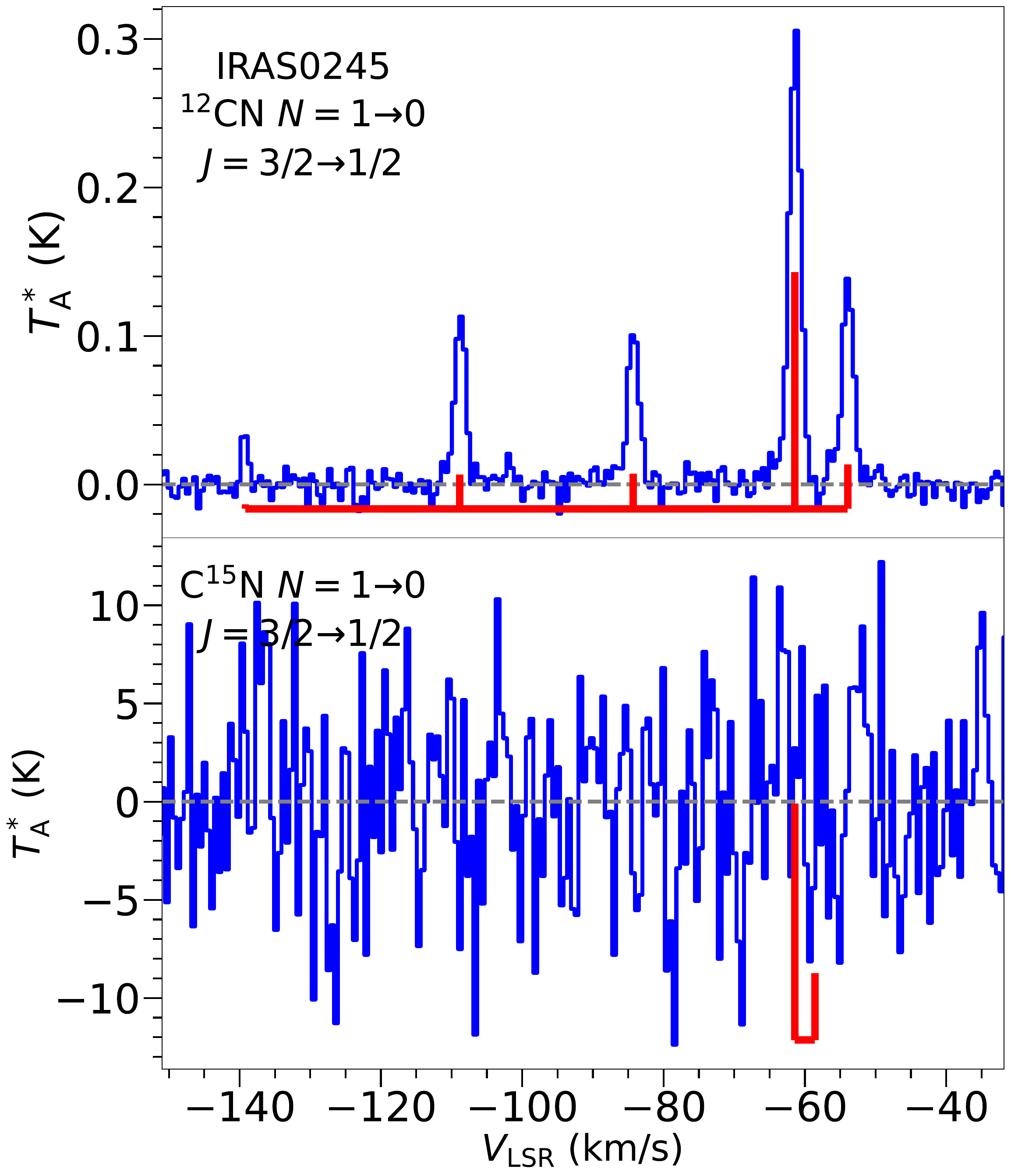}
                \includegraphics[scale=0.14]{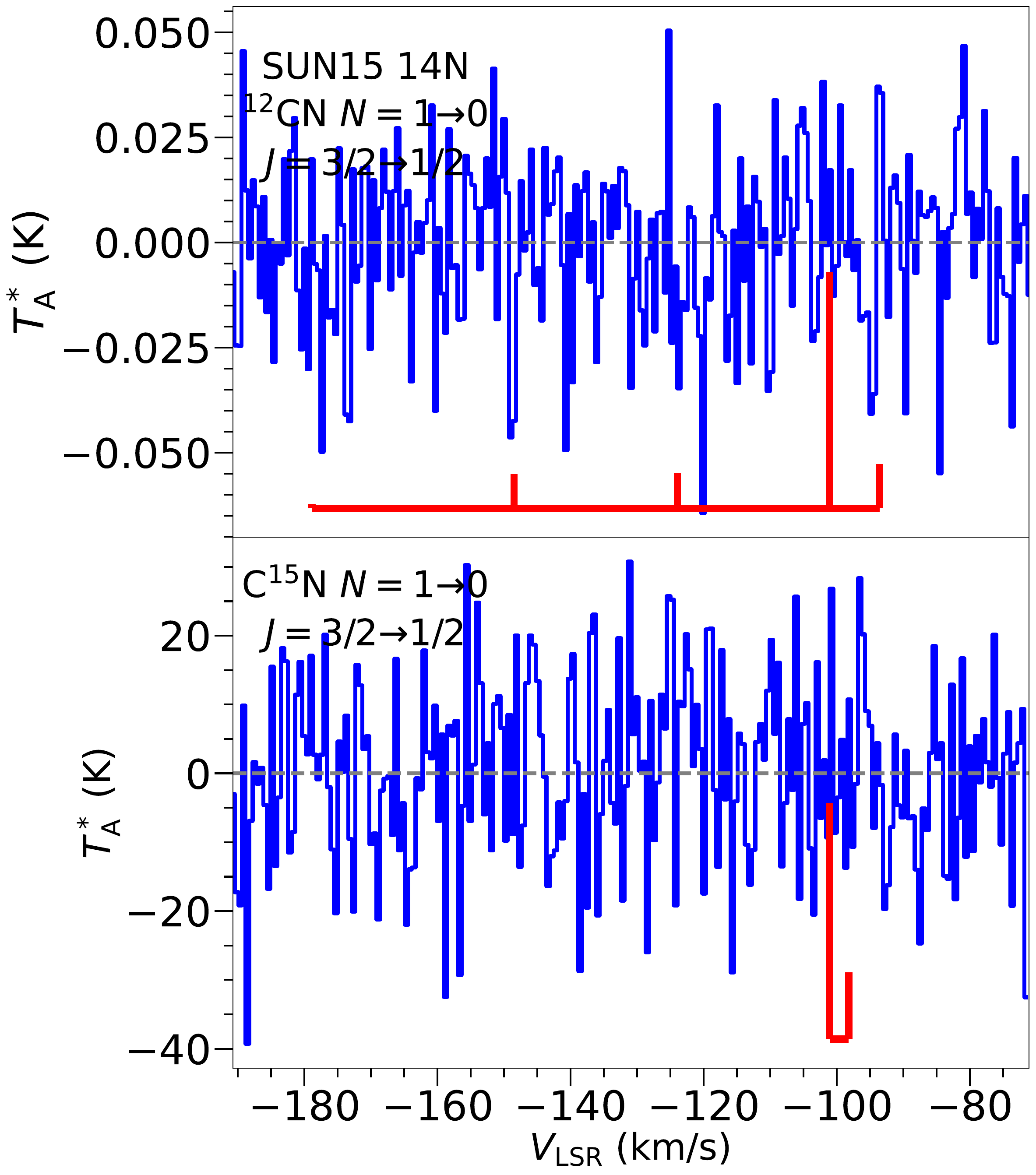}
                \includegraphics[scale=0.14]{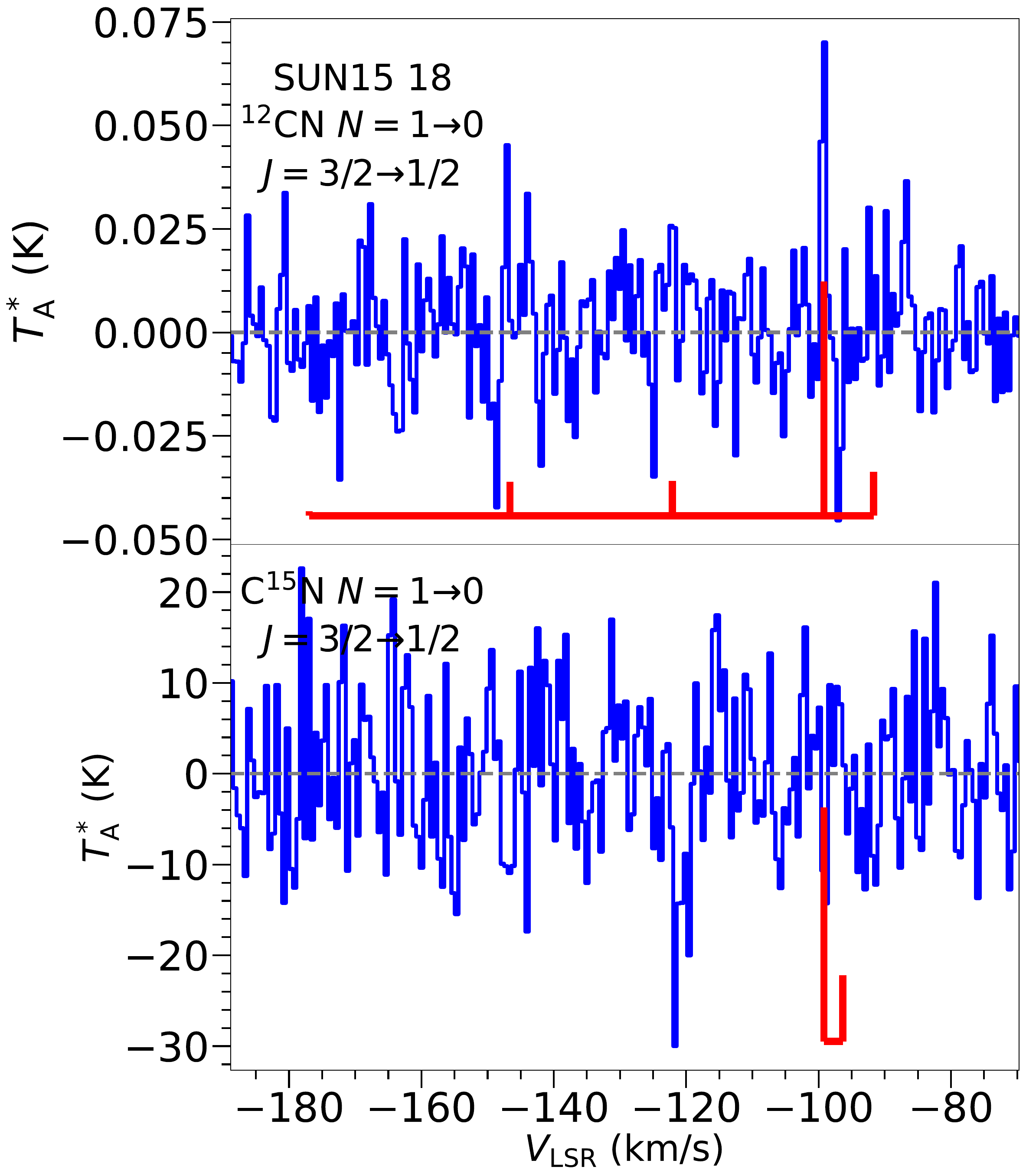}
		    \includegraphics[scale=0.14]{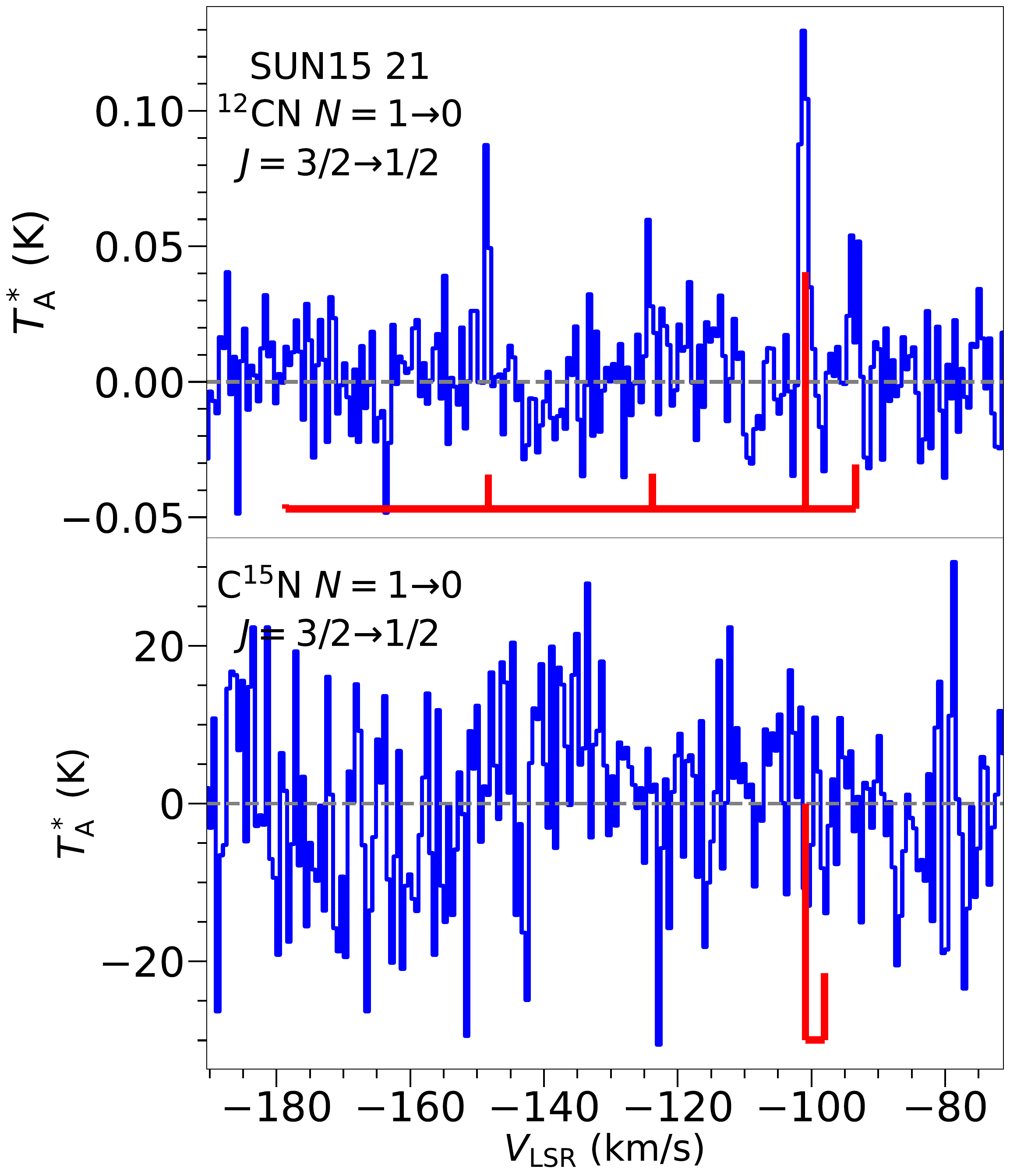}
                \includegraphics[scale=0.14]{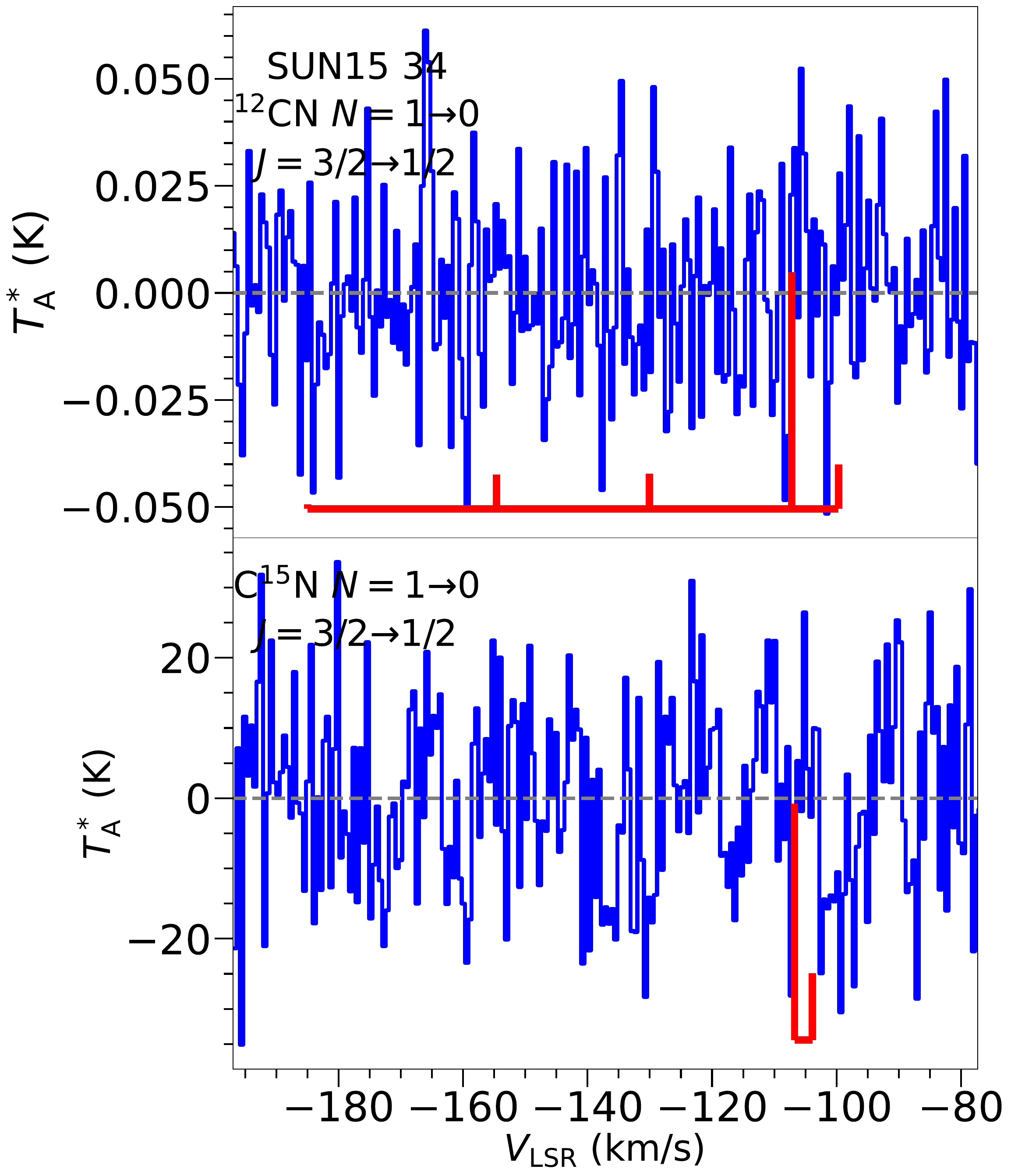}
                \includegraphics[scale=0.14]{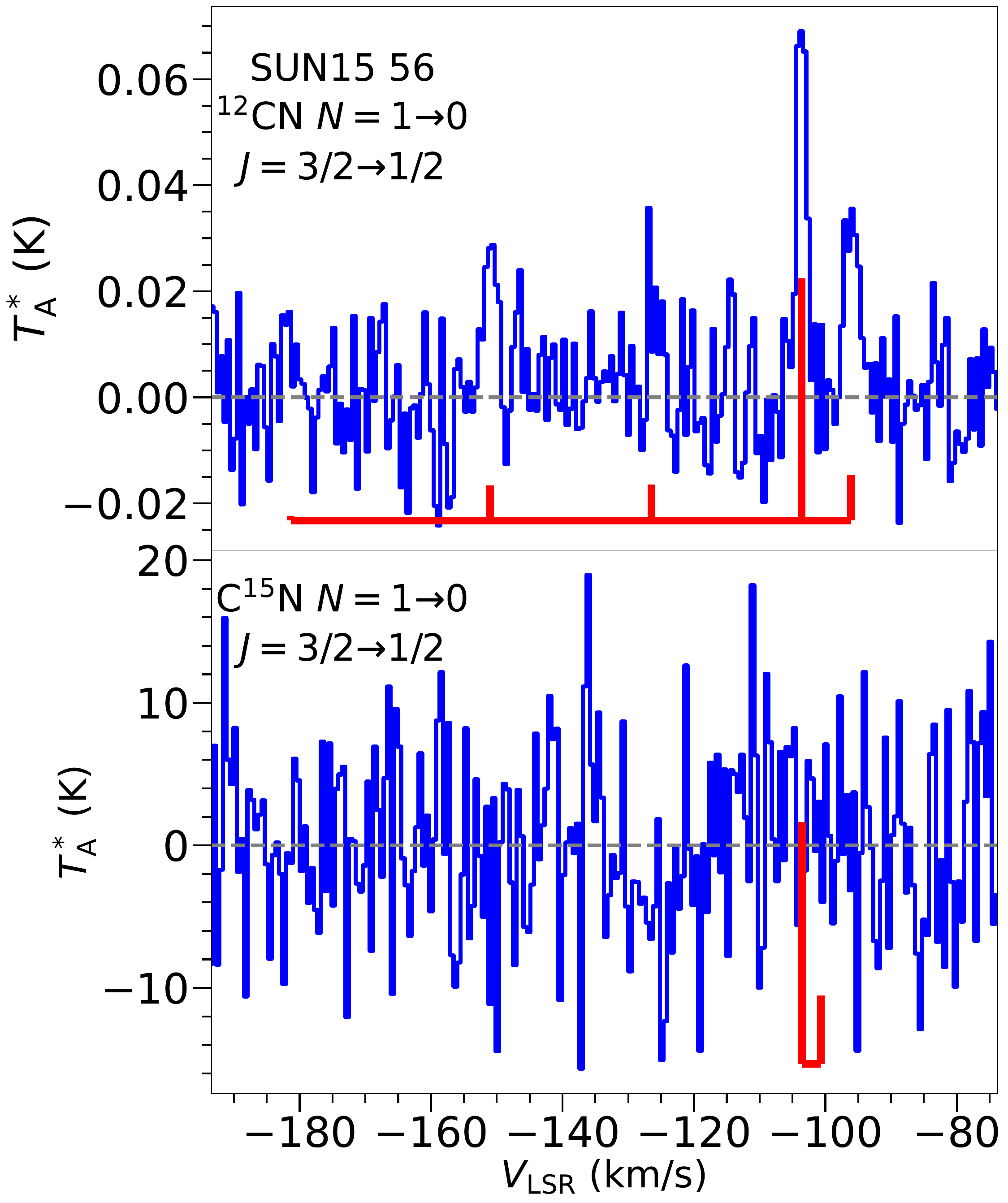}
                \includegraphics[scale=0.14]{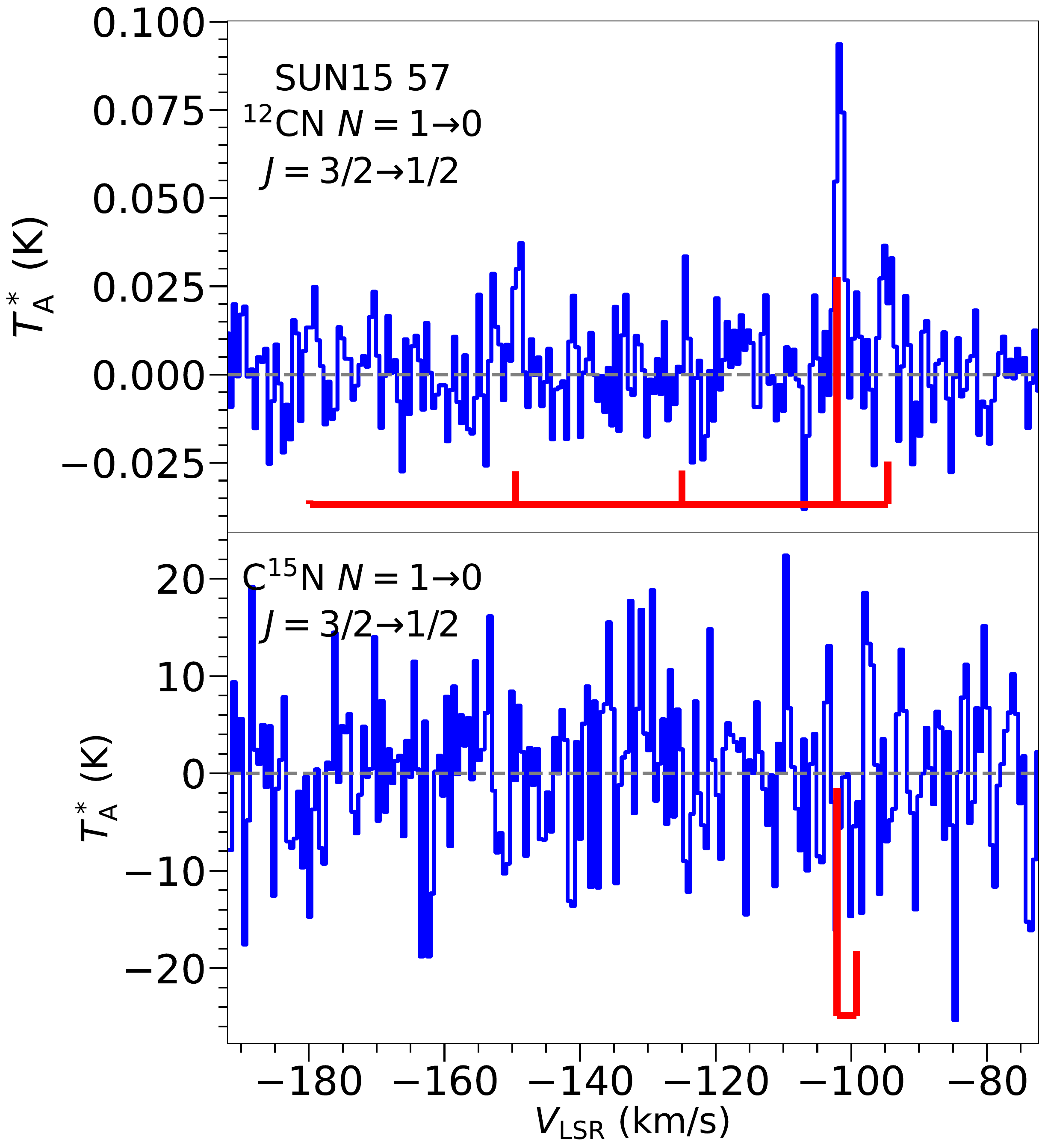}
        \caption{\label{fig:12CN_C15N_spectra} Spectra of \twCN\ and \CftN\ $N=1\to0$ for targets without \CftN\ detections.}
\end{figure*}

\begin{figure*}
        \centering
		    \includegraphics[scale=0.14]{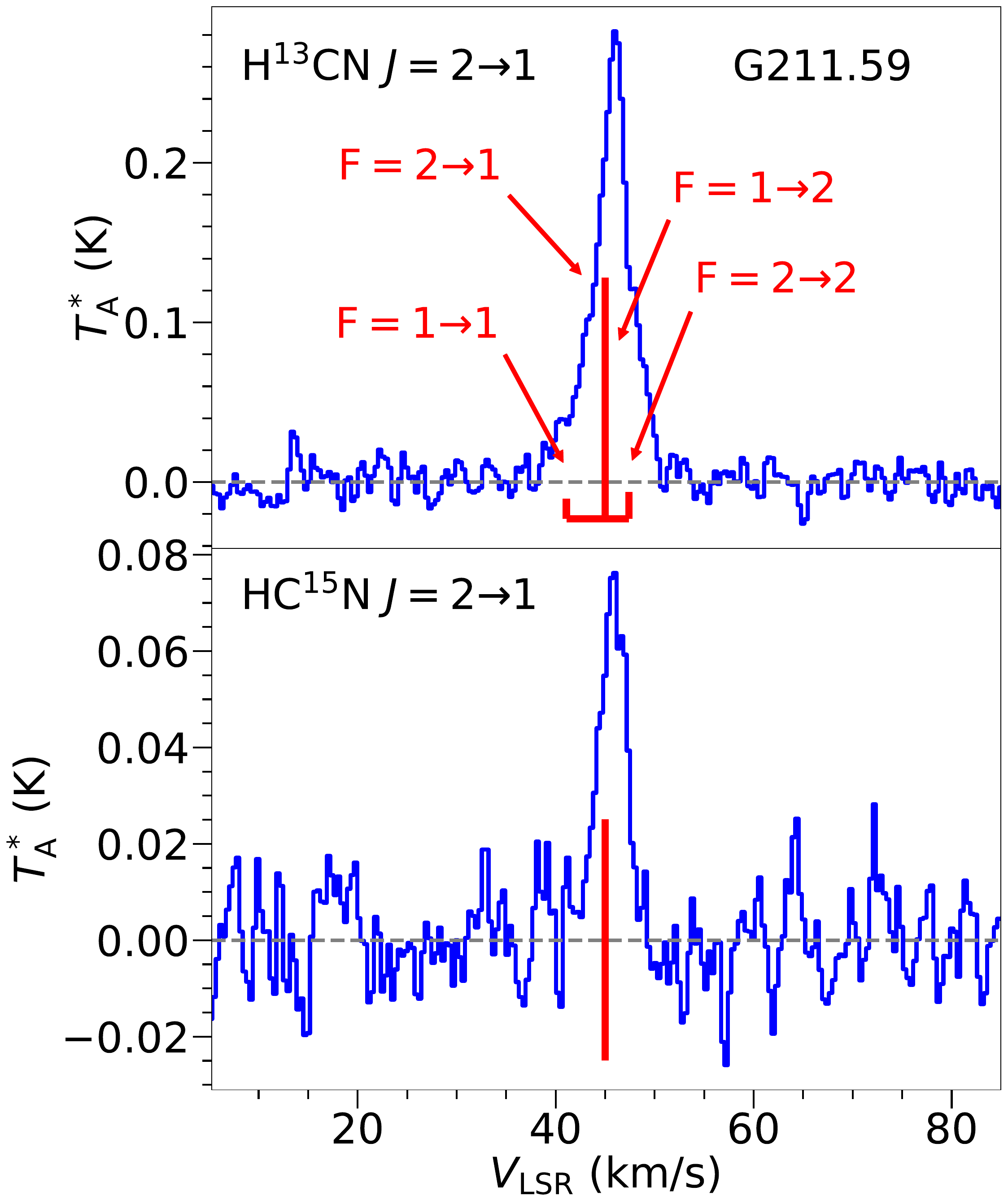}
                \includegraphics[scale=0.14]{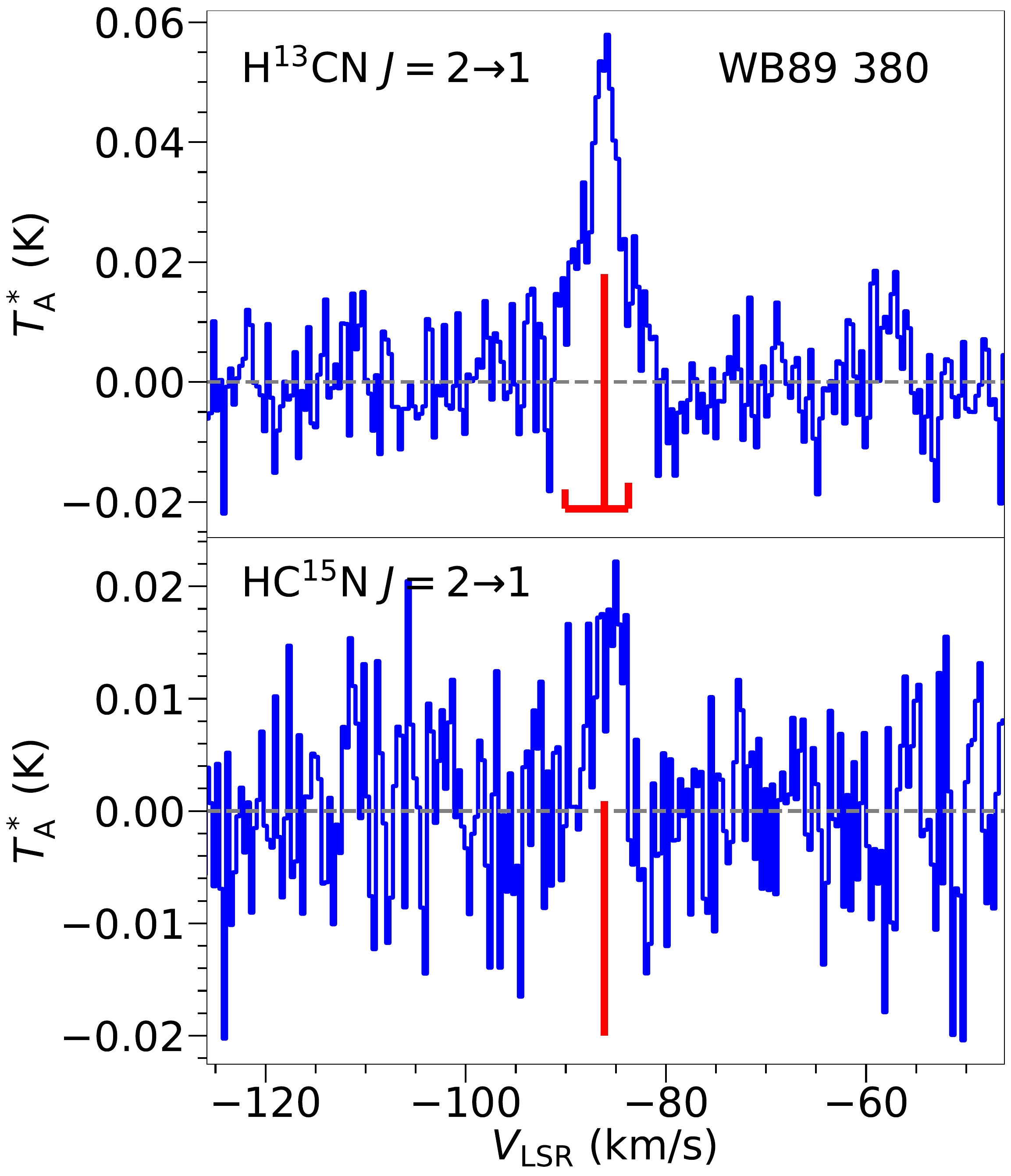}
                \includegraphics[scale=0.14]{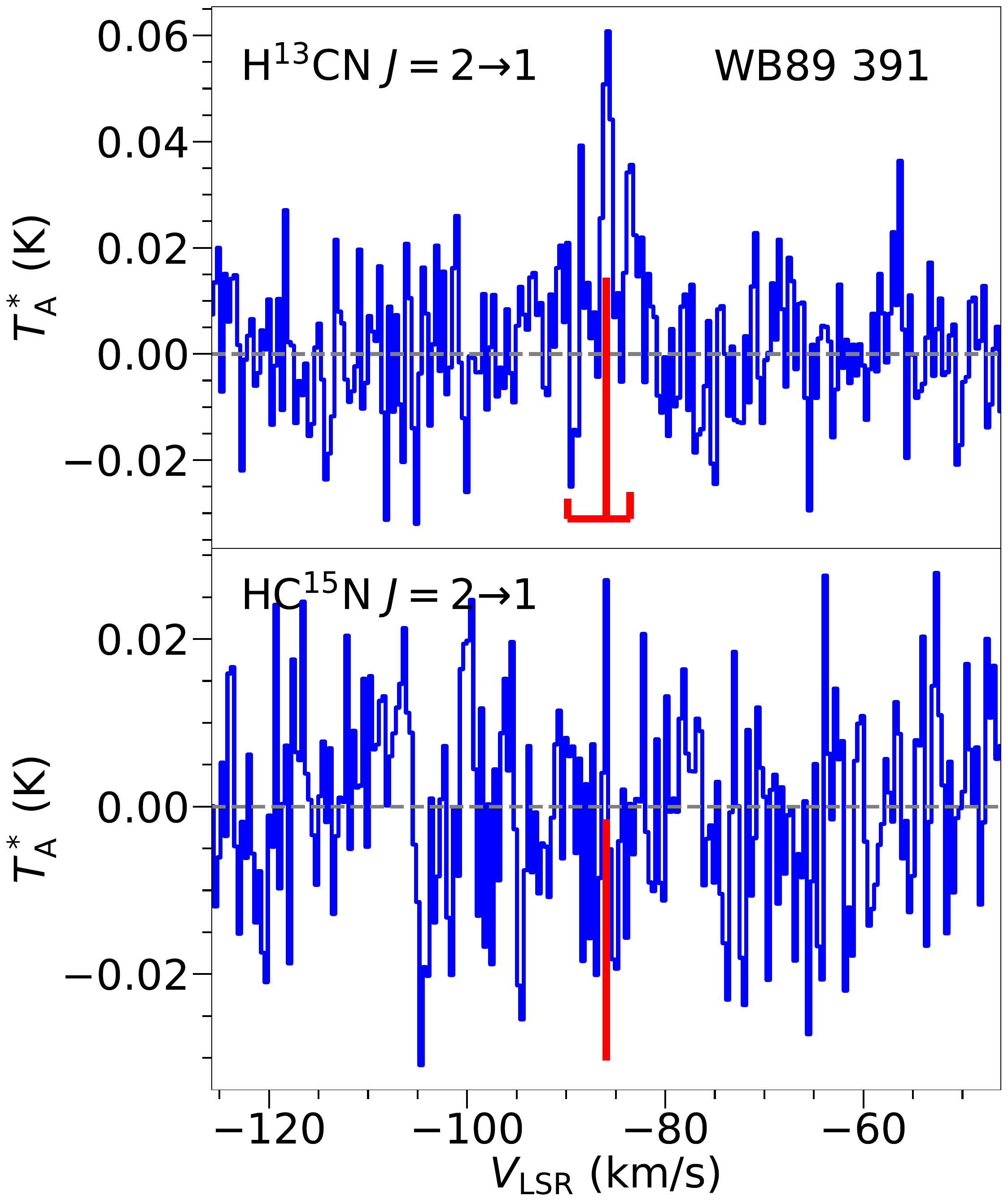}
                \includegraphics[scale=0.14]{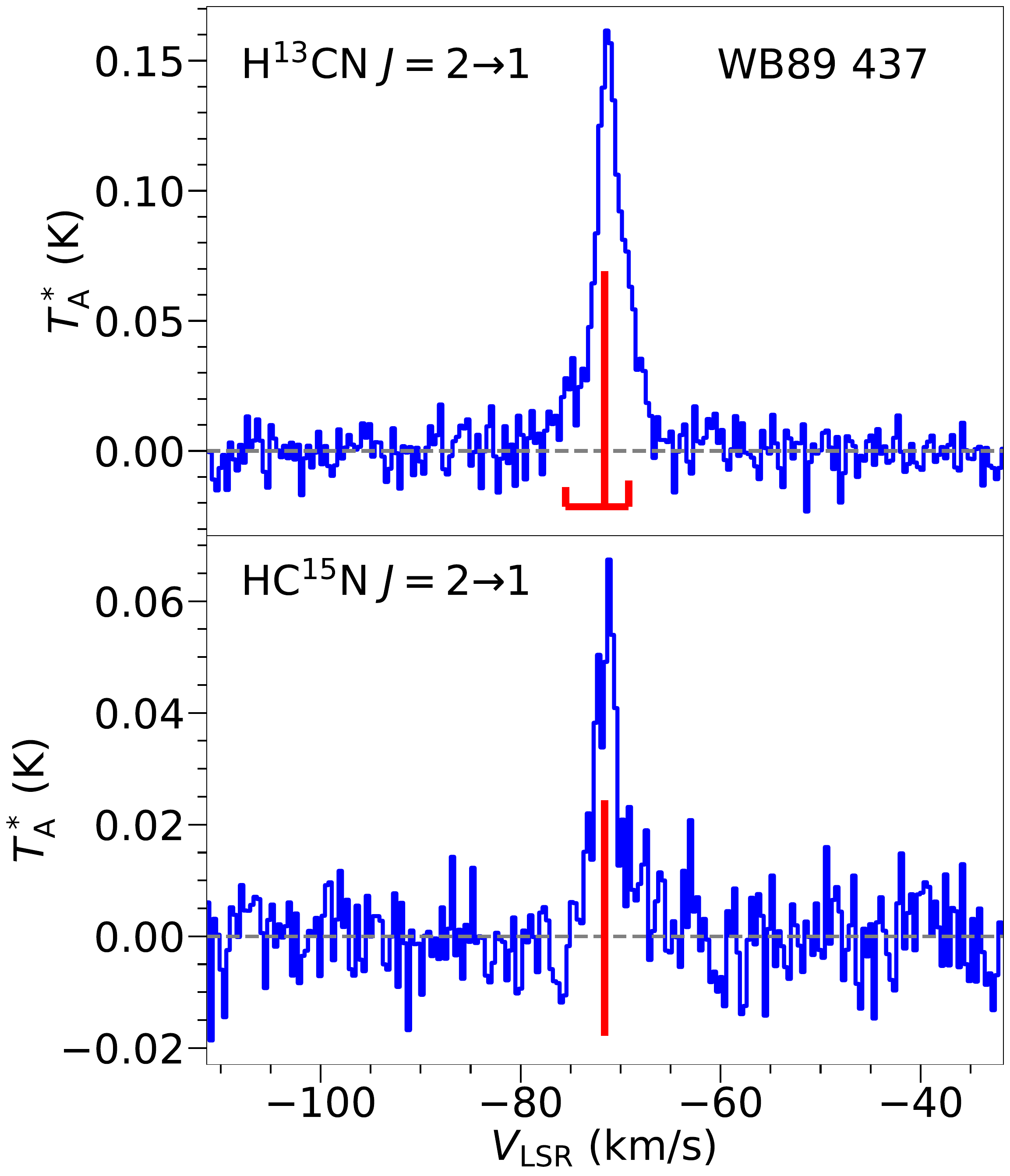}
                \includegraphics[scale=0.14]{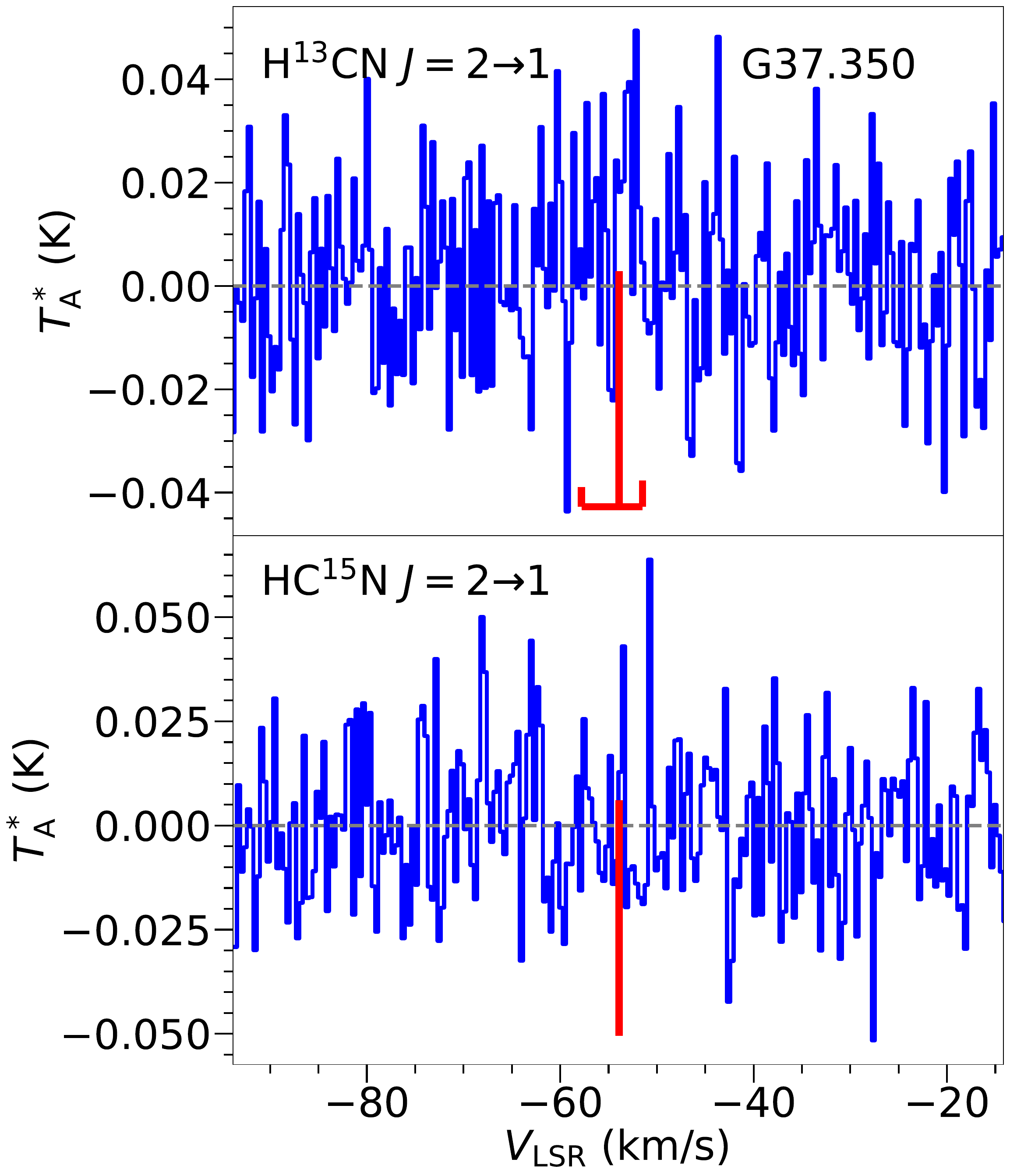}
		    \includegraphics[scale=0.14]{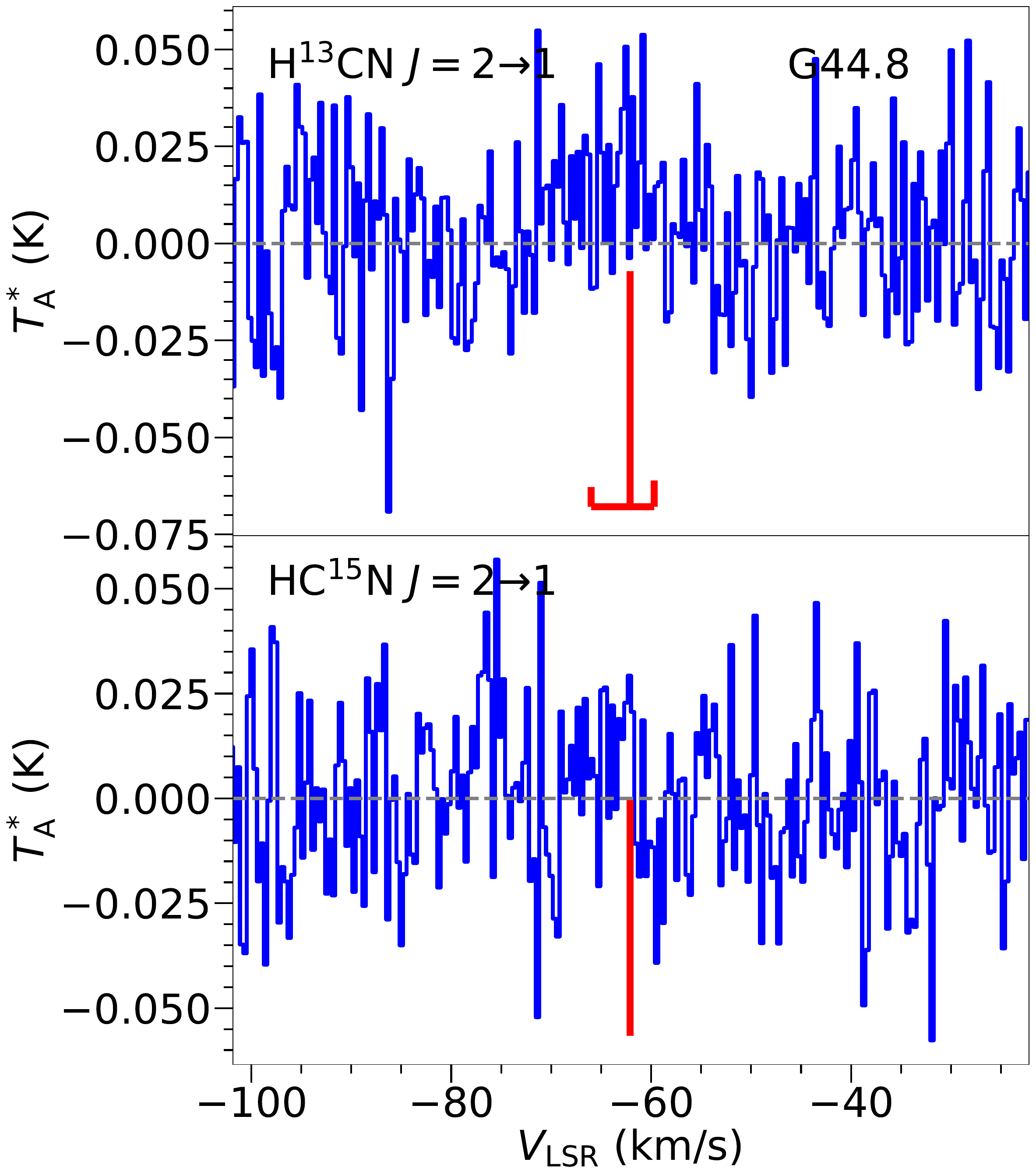}
                \includegraphics[scale=0.14]{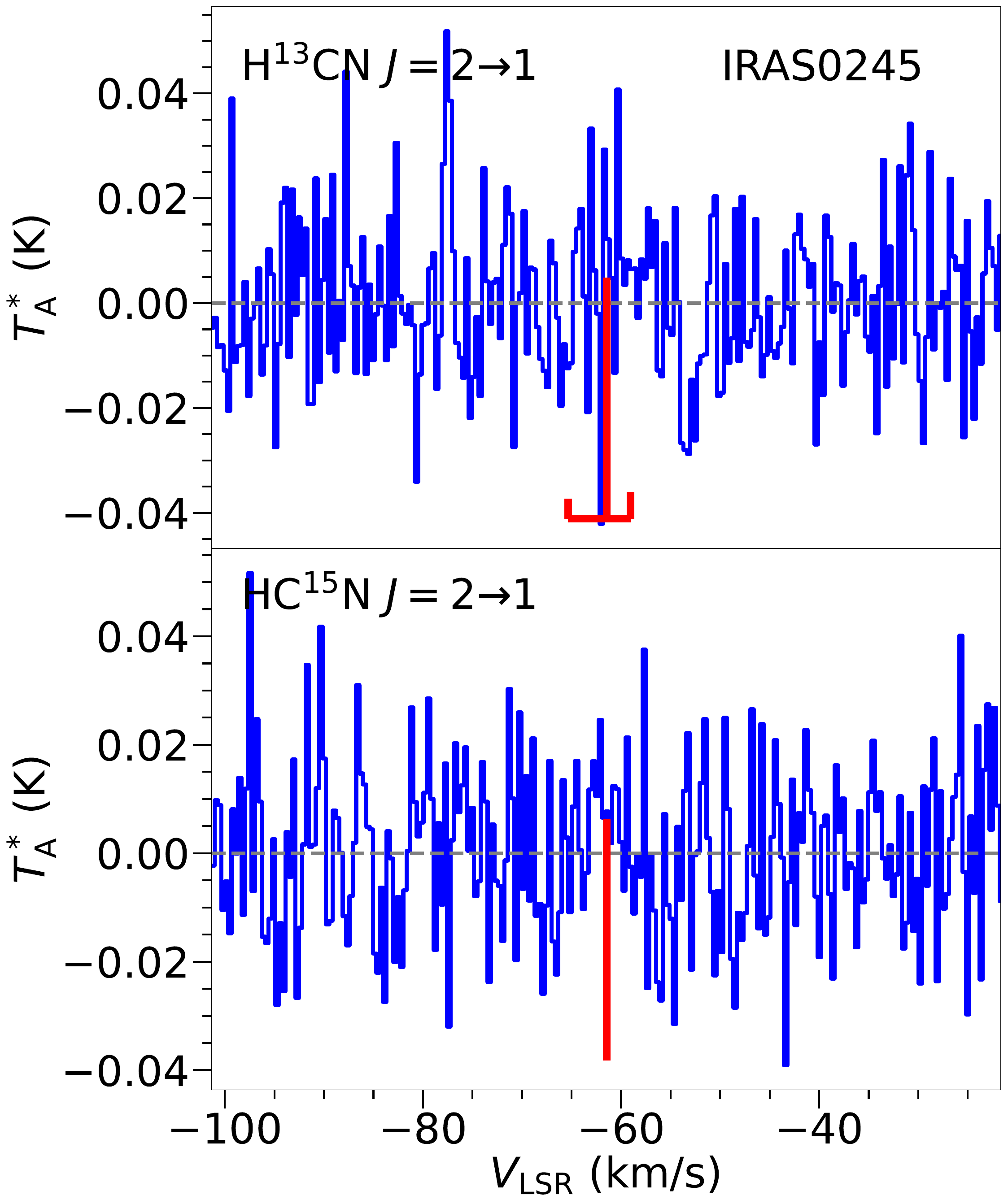}
                \includegraphics[scale=0.14]{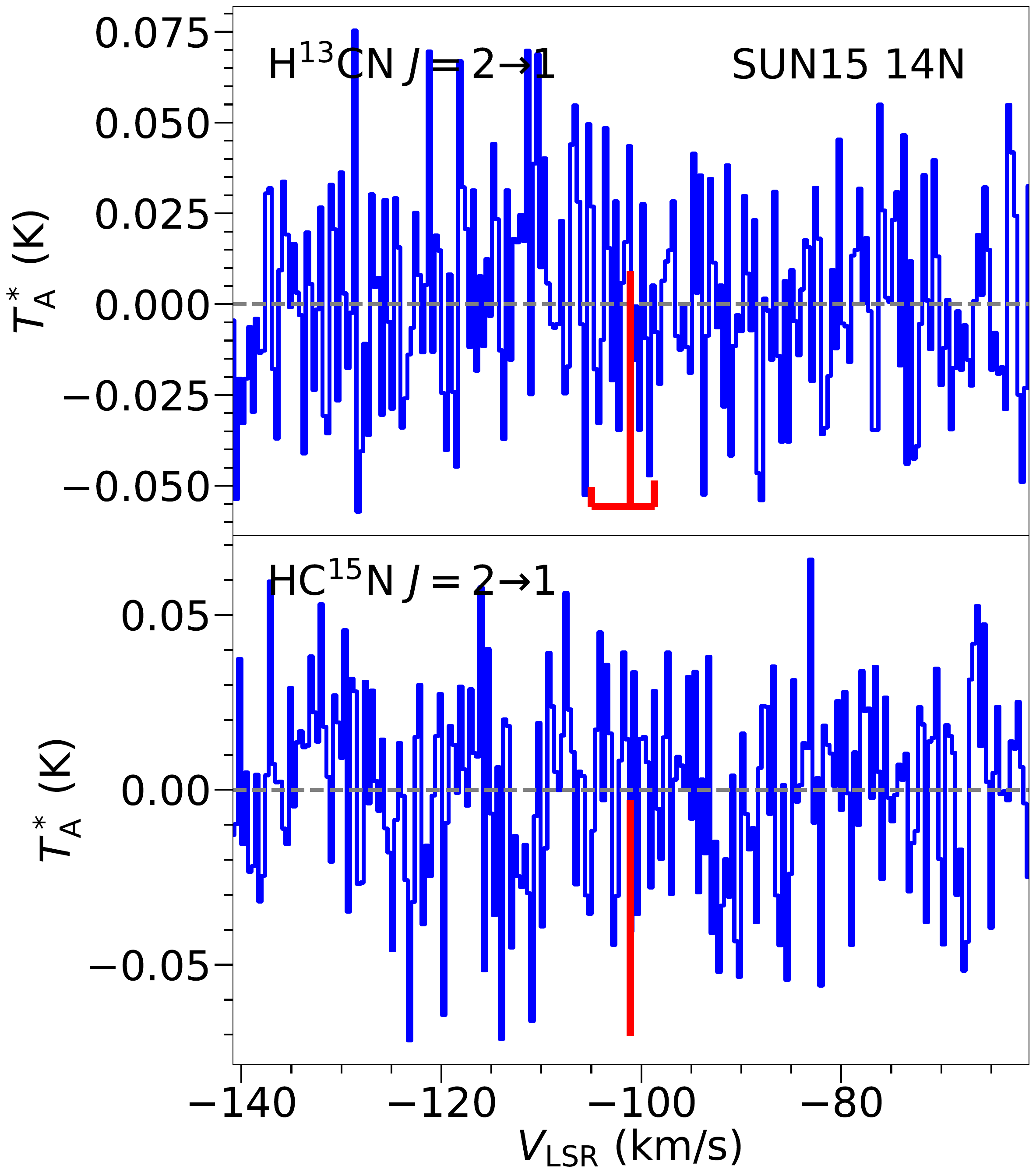}
                \includegraphics[scale=0.14]{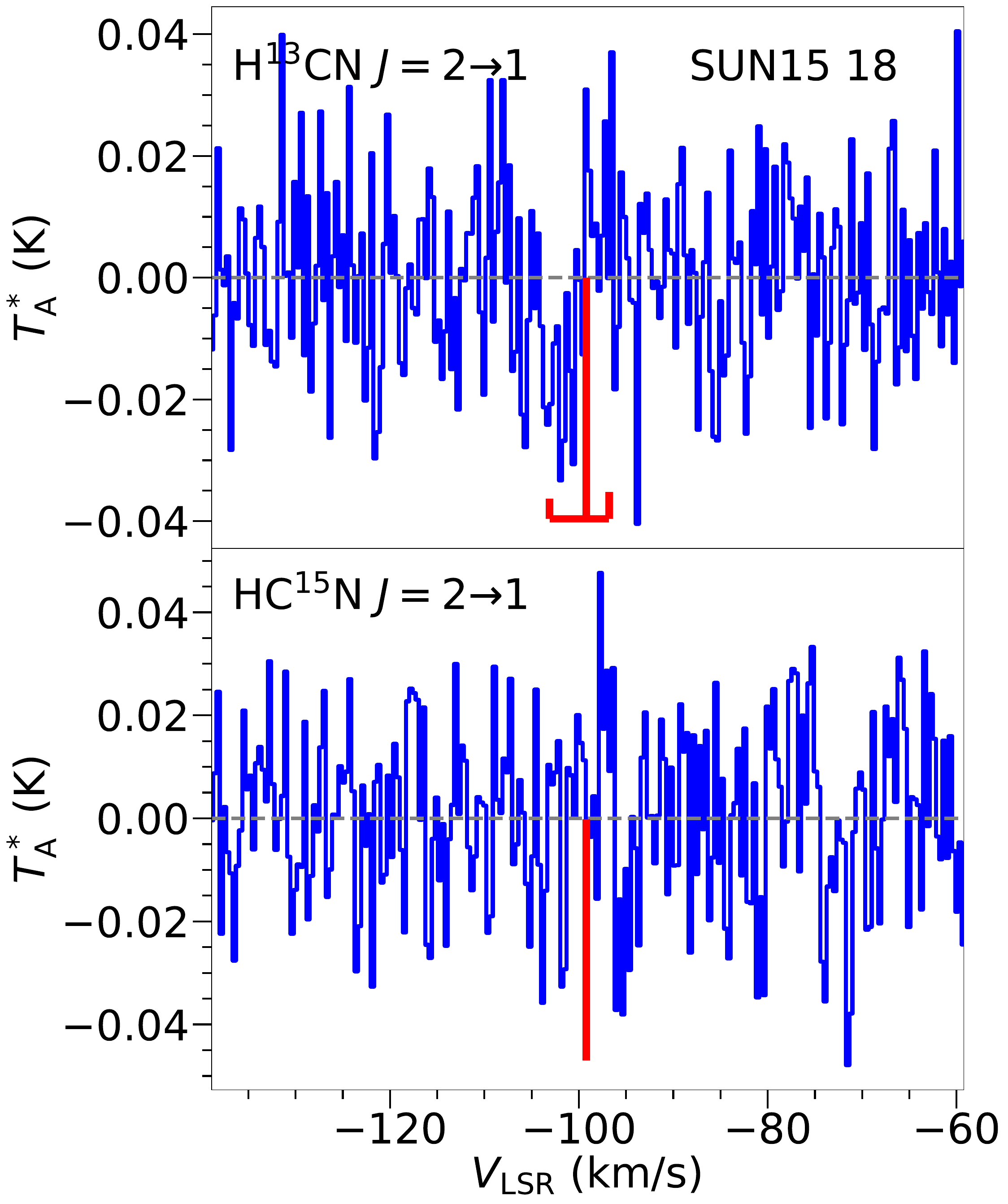}
        \caption{\label{fig:H13CN_HC15N_spectra} The spectra of $\rm H^{13}CN$ and $\rm HC^{15}N$ at $N=2\to1$ transition.}
\end{figure*}

\begin{figure*}
        \centering
		    \includegraphics[scale=0.14]{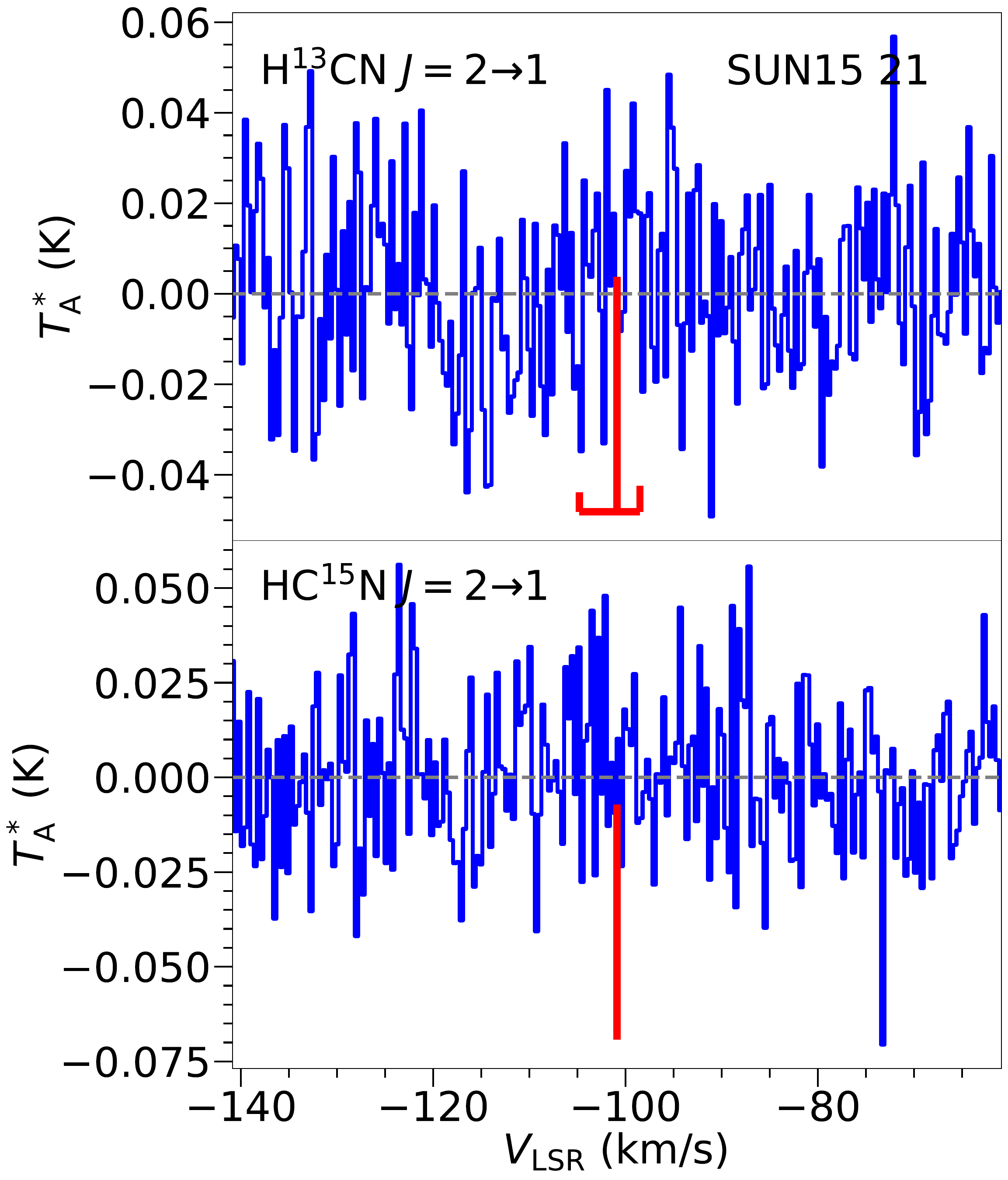}
                \includegraphics[scale=0.14]{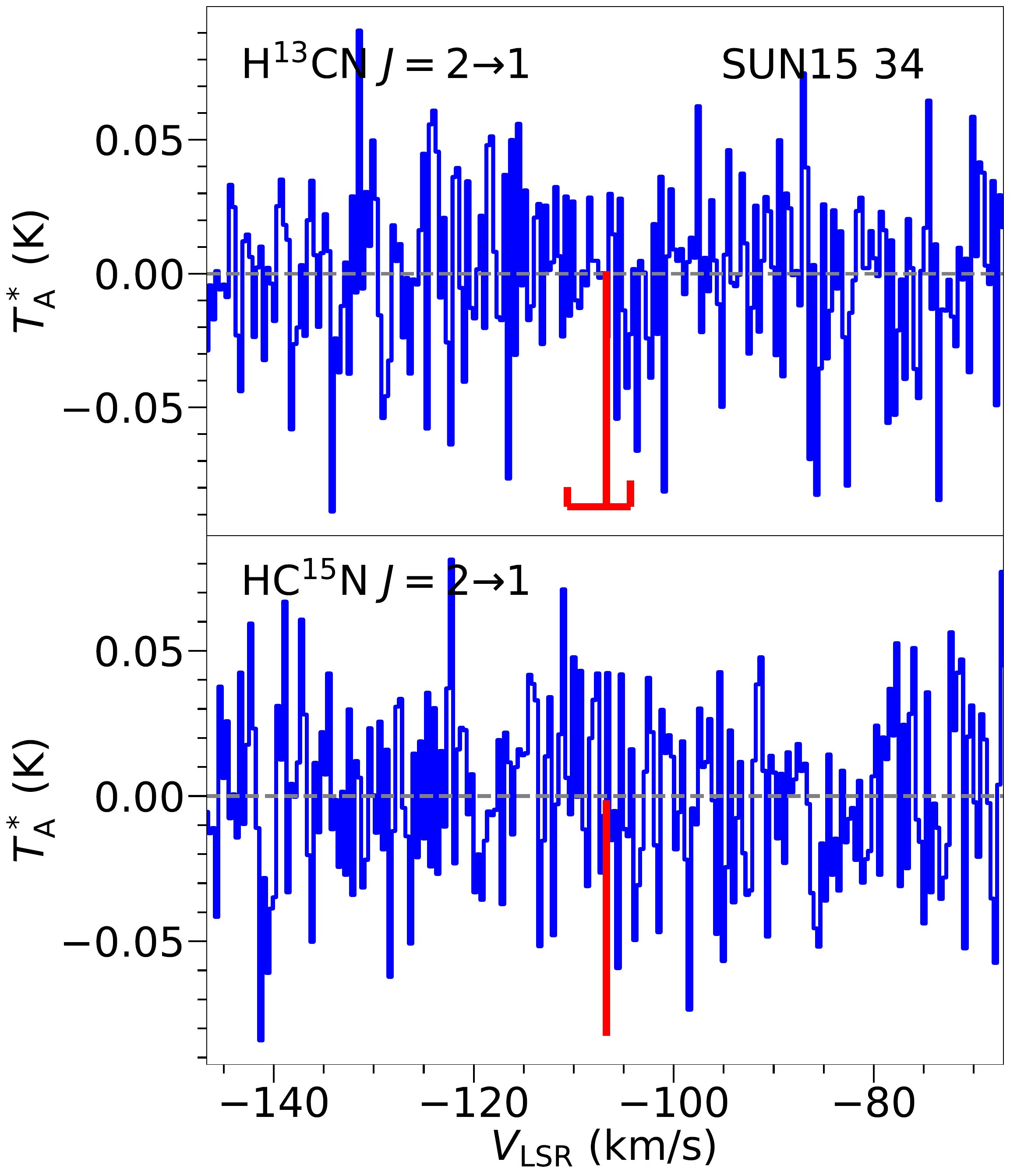}
                \includegraphics[scale=0.14]{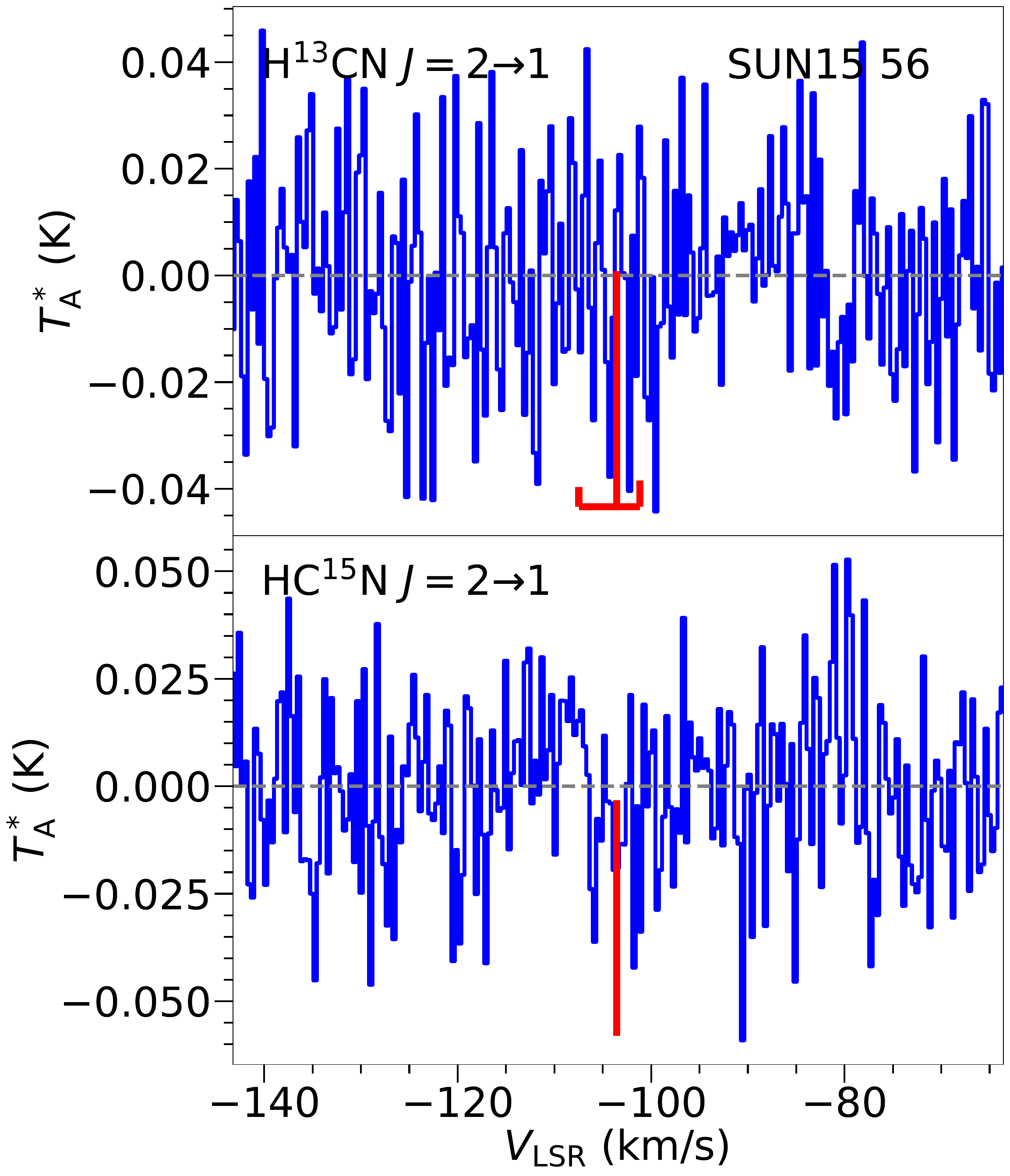}
                \includegraphics[scale=0.14]{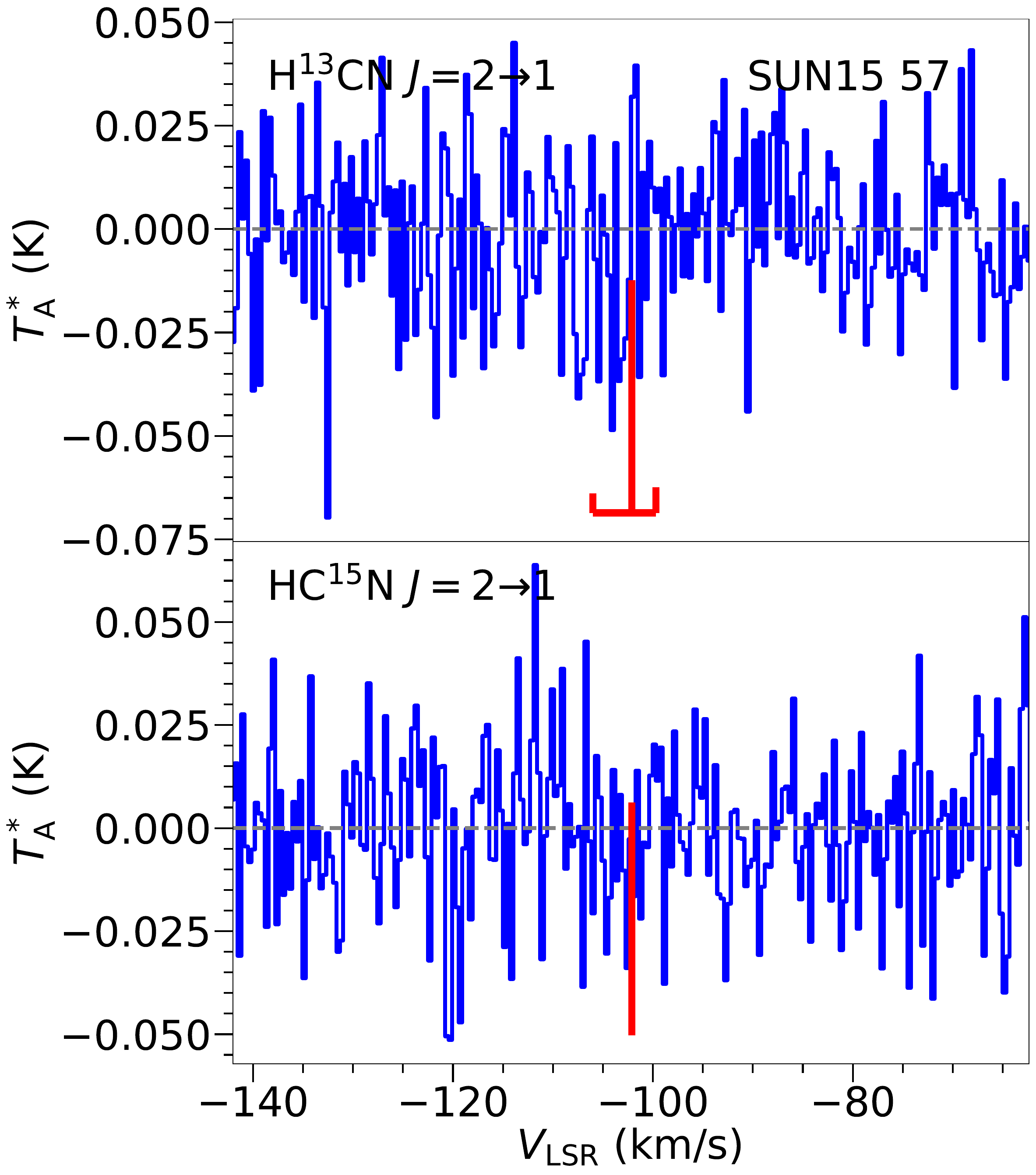}
                \includegraphics[scale=0.14]{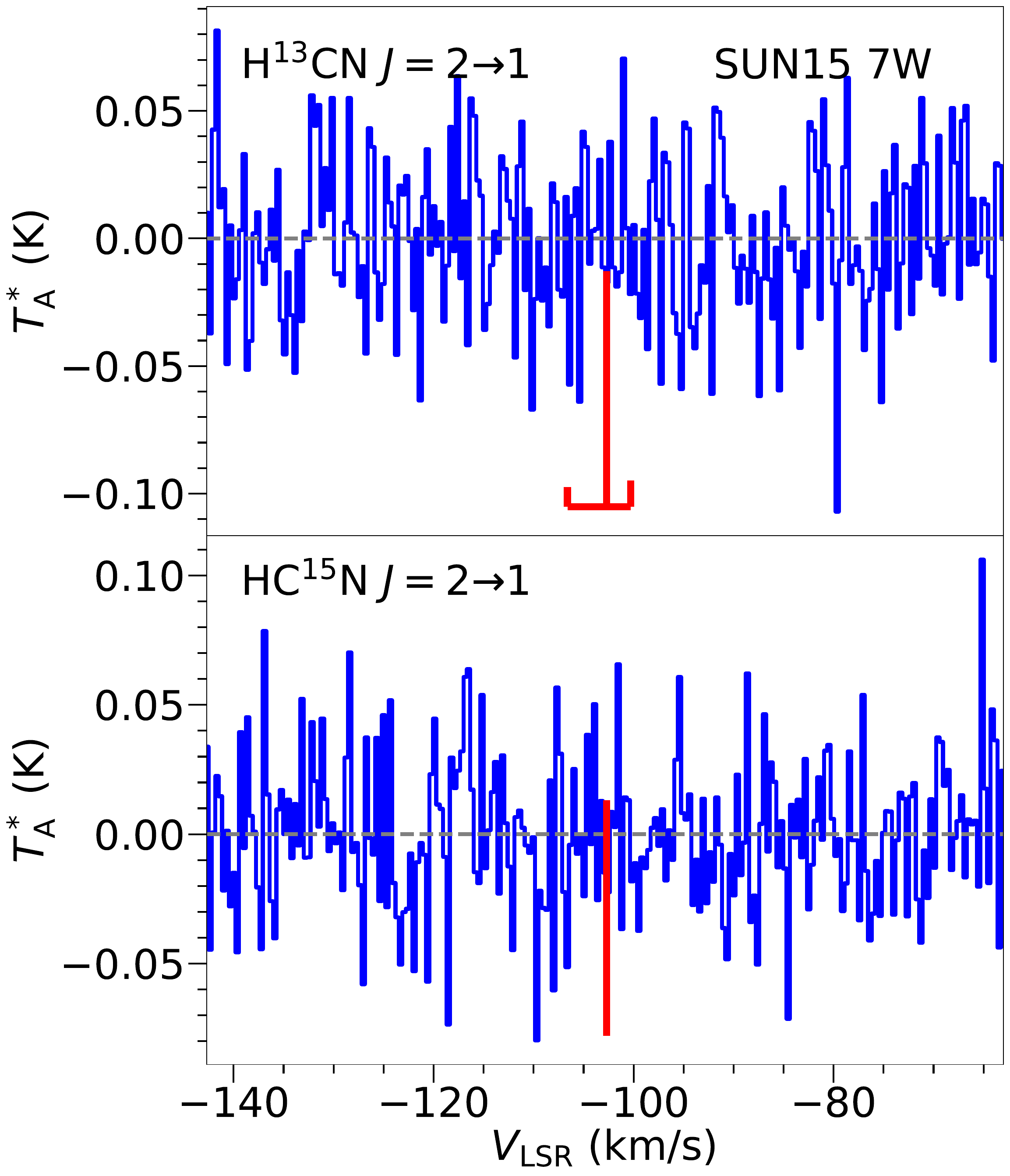}
		    \includegraphics[scale=0.14]{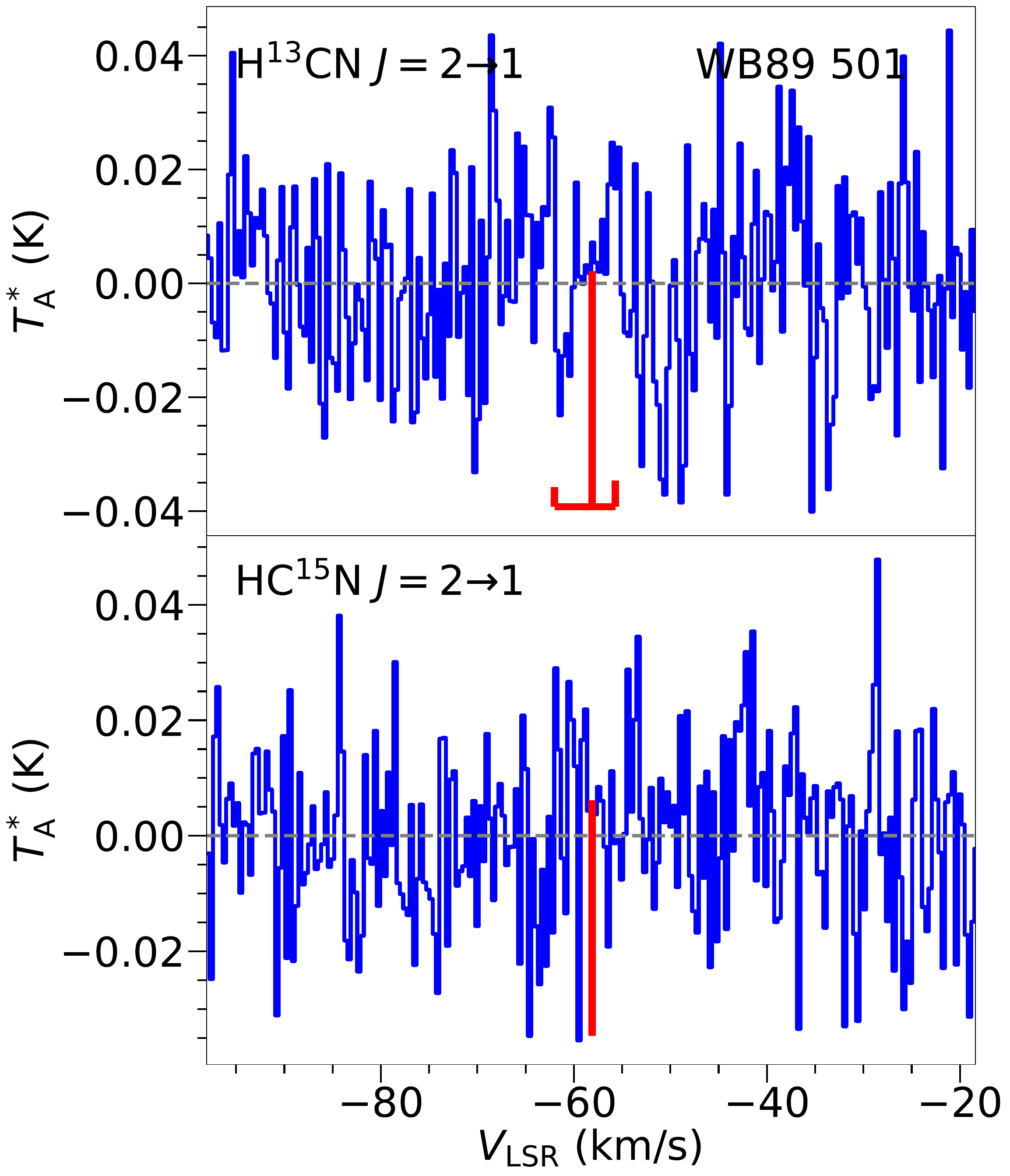}
        \caption{\label{fig:conti_H13CN_HC15N_spectra} The spectra of $\rm H^{13}CN$ and $\rm HC^{15}N$ at $N=2\to1$ transition (Continued.).}
\end{figure*}

\begin{figure*}
        \centering
		        \includegraphics[scale=0.4]{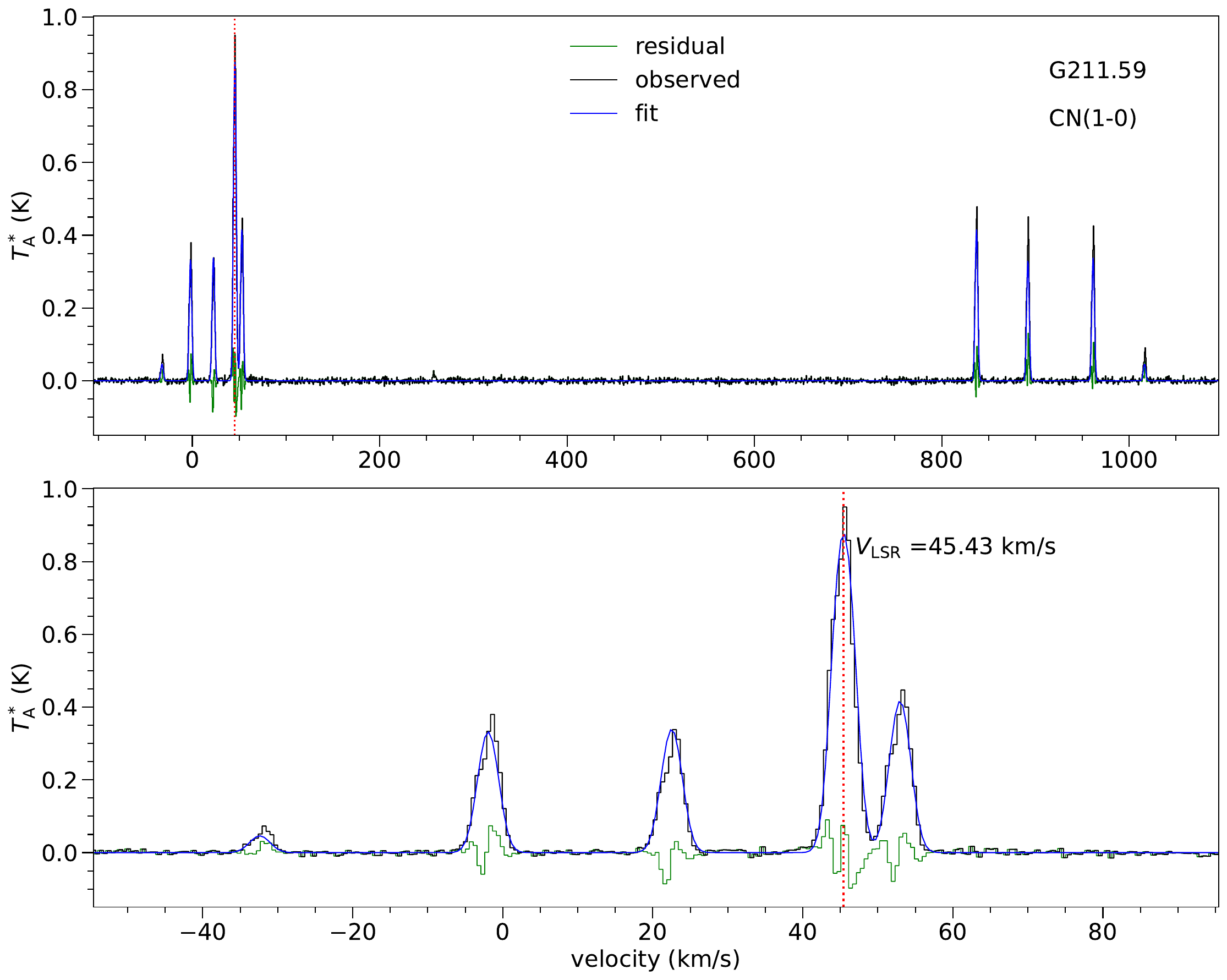}
        \caption{\label{fig:HfS_fitting_spectra}The HfS fitting spectra. The upper panels: the whole spectra of $\rm ^{12}CN$ $N=1\to0$ nine components used to do HfS fitting. The lower panels: the zoom-in spectra showing the five components of $\rm ^{12}CN$ $N=1\to0\ J=3/2\to1/2$. The original spectra, HfS fitting results, and residuals are shown in black, blue, and green, respectively. The red dashed lines locate the $V_{\rm LSR}$ of the main component (the HfS fitting results of the other targets are shown below).}
\end{figure*}

\begin{figure*}
        \centering
		    \includegraphics[scale=0.38]{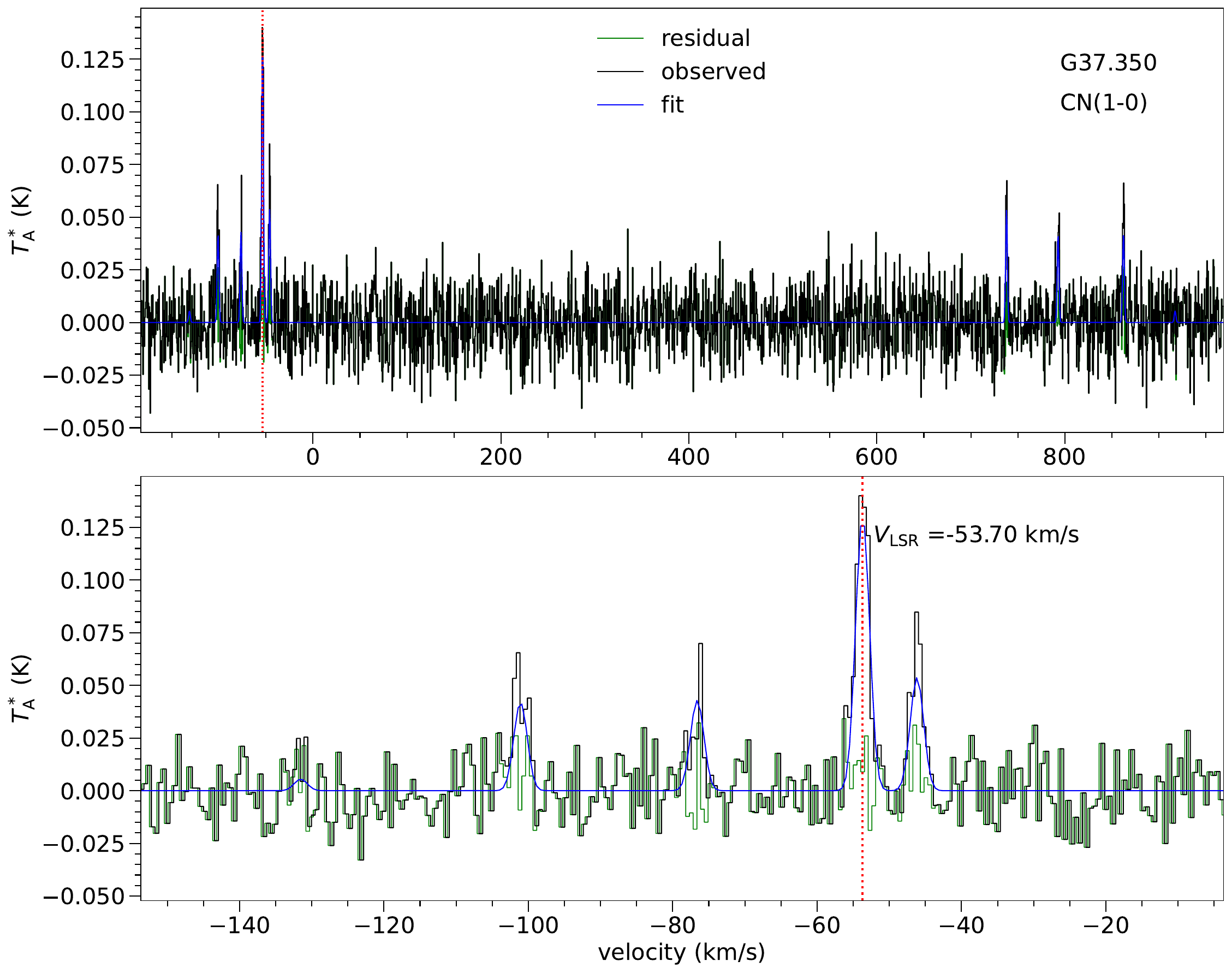}
                \includegraphics[scale=0.38]{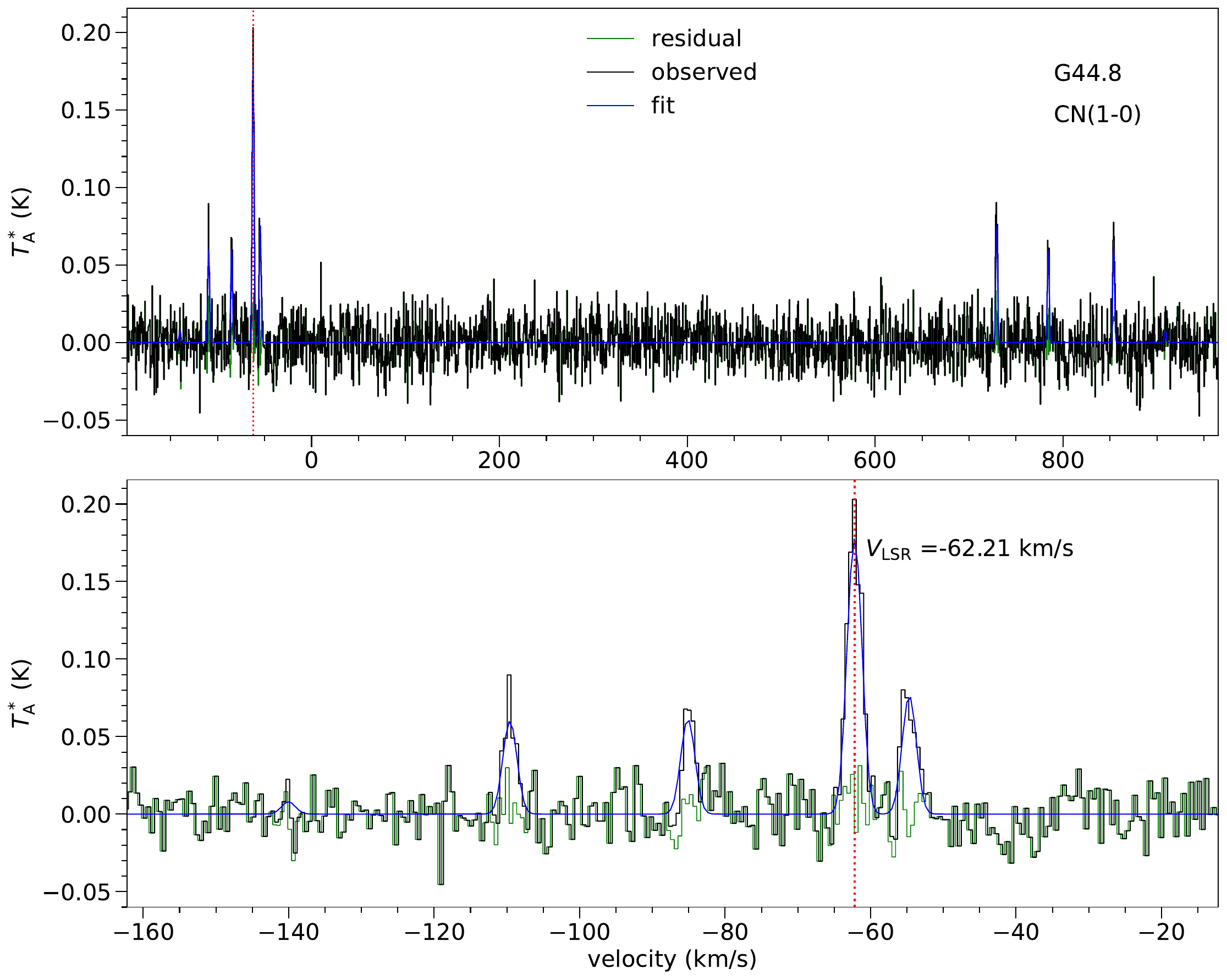}
        \caption{\label{fig:cont1_HfS_fitting_spectra}The HfS fitting spectra (Continued.)}
\end{figure*}

\begin{figure*}
        \centering
		    \includegraphics[scale=0.38]{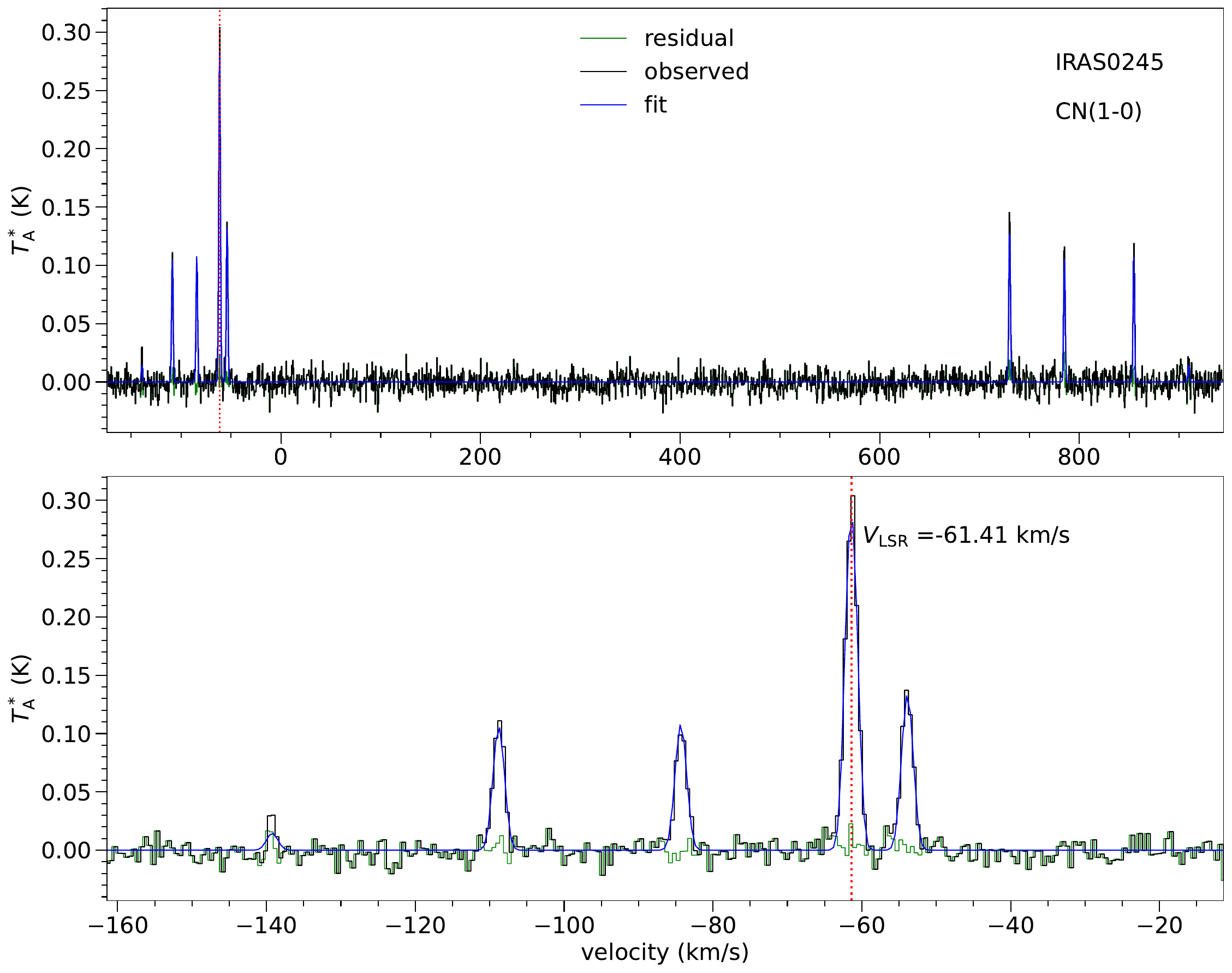}
                \includegraphics[scale=0.38]{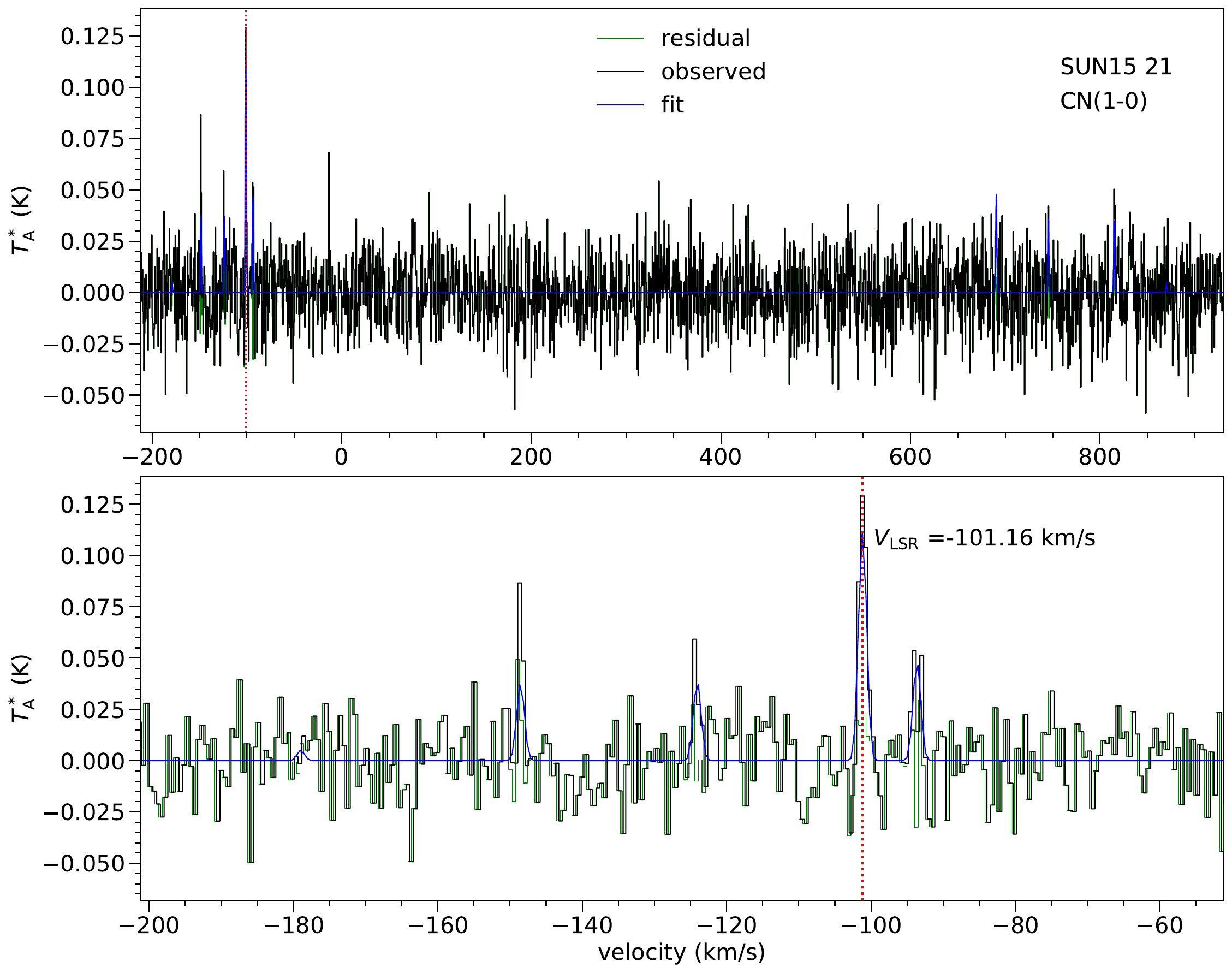}
        \caption{\label{fig:cont2_HfS_fitting_spectra}The HfS fitting spectra (Continued.)}
\end{figure*}

\begin{figure*}
        \centering
		    \includegraphics[scale=0.38]{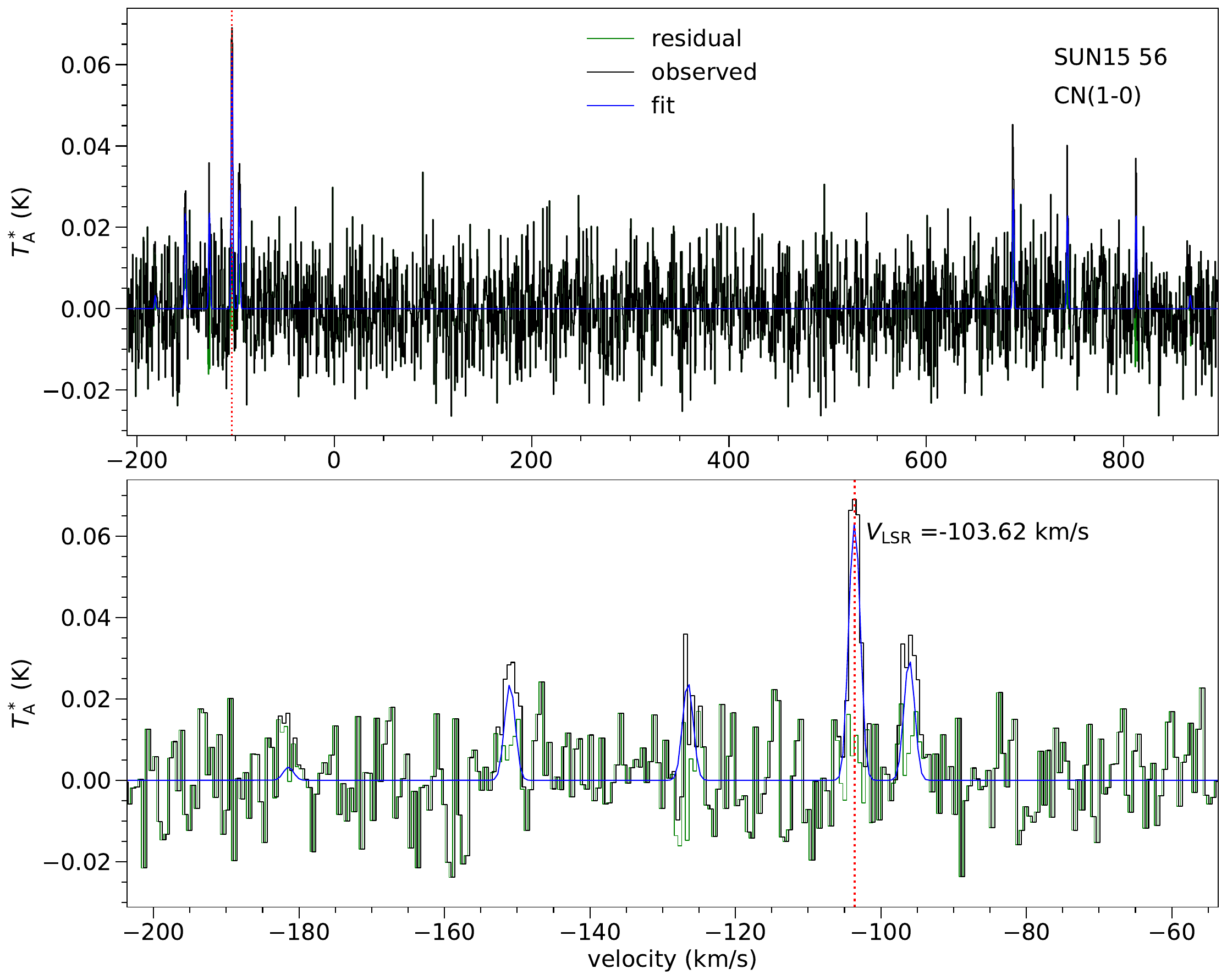}
                \includegraphics[scale=0.38]{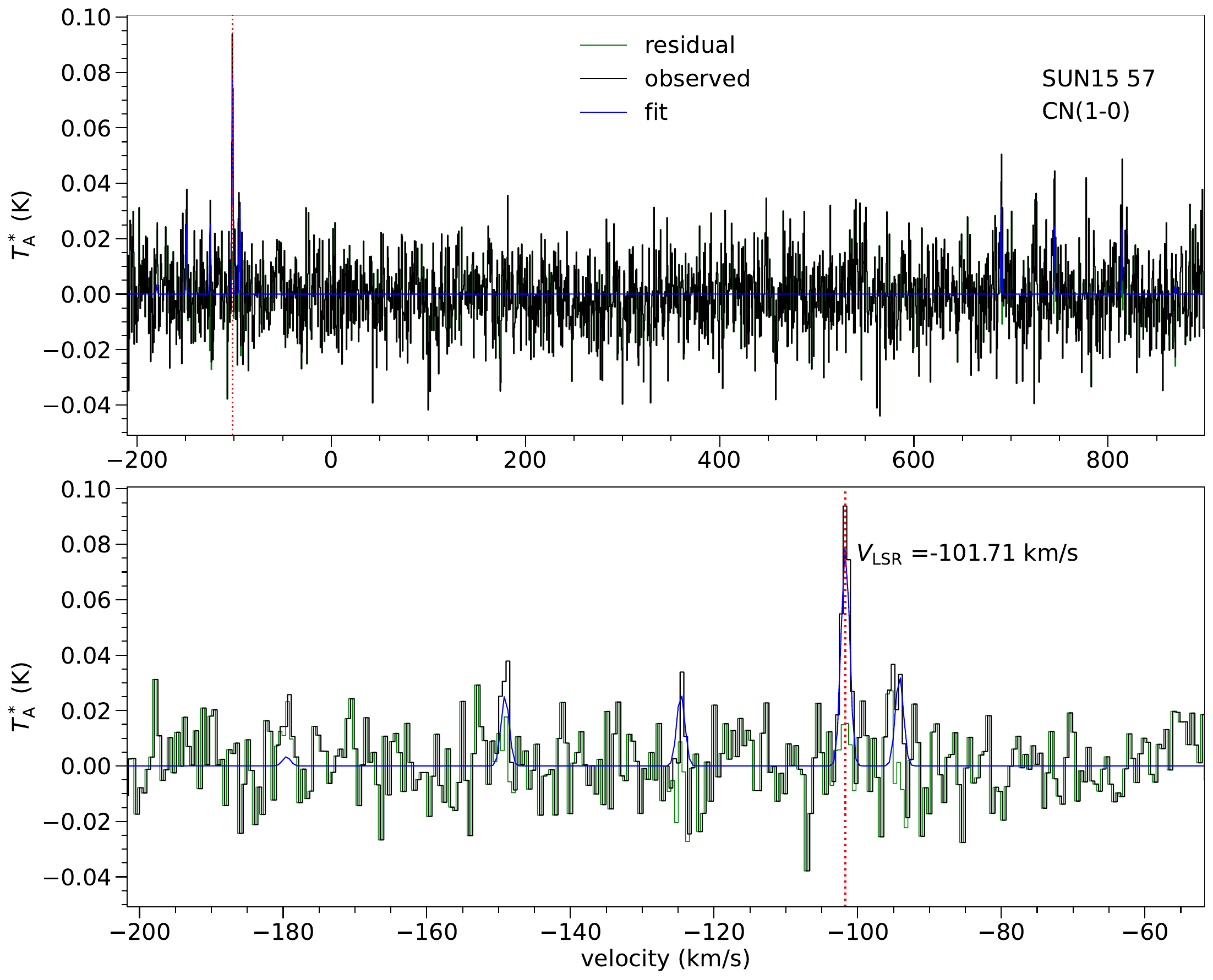}
        \caption{\label{fig:cont3_HfS_fitting_spectra}The HfS fitting spectra (Continued.)}
\end{figure*}

\begin{figure*}
        \centering
		    \includegraphics[scale=0.38]{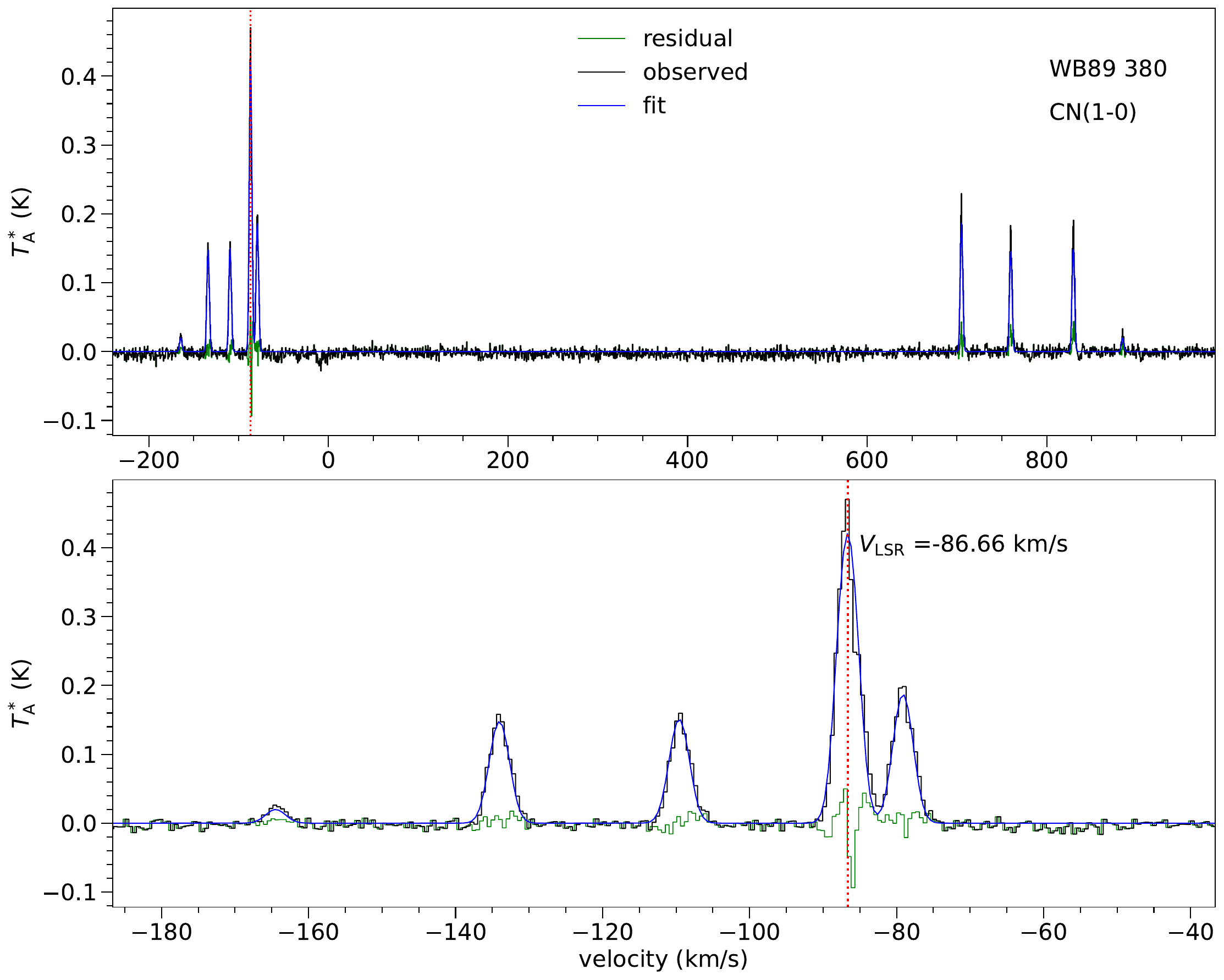}
                \includegraphics[scale=0.38]{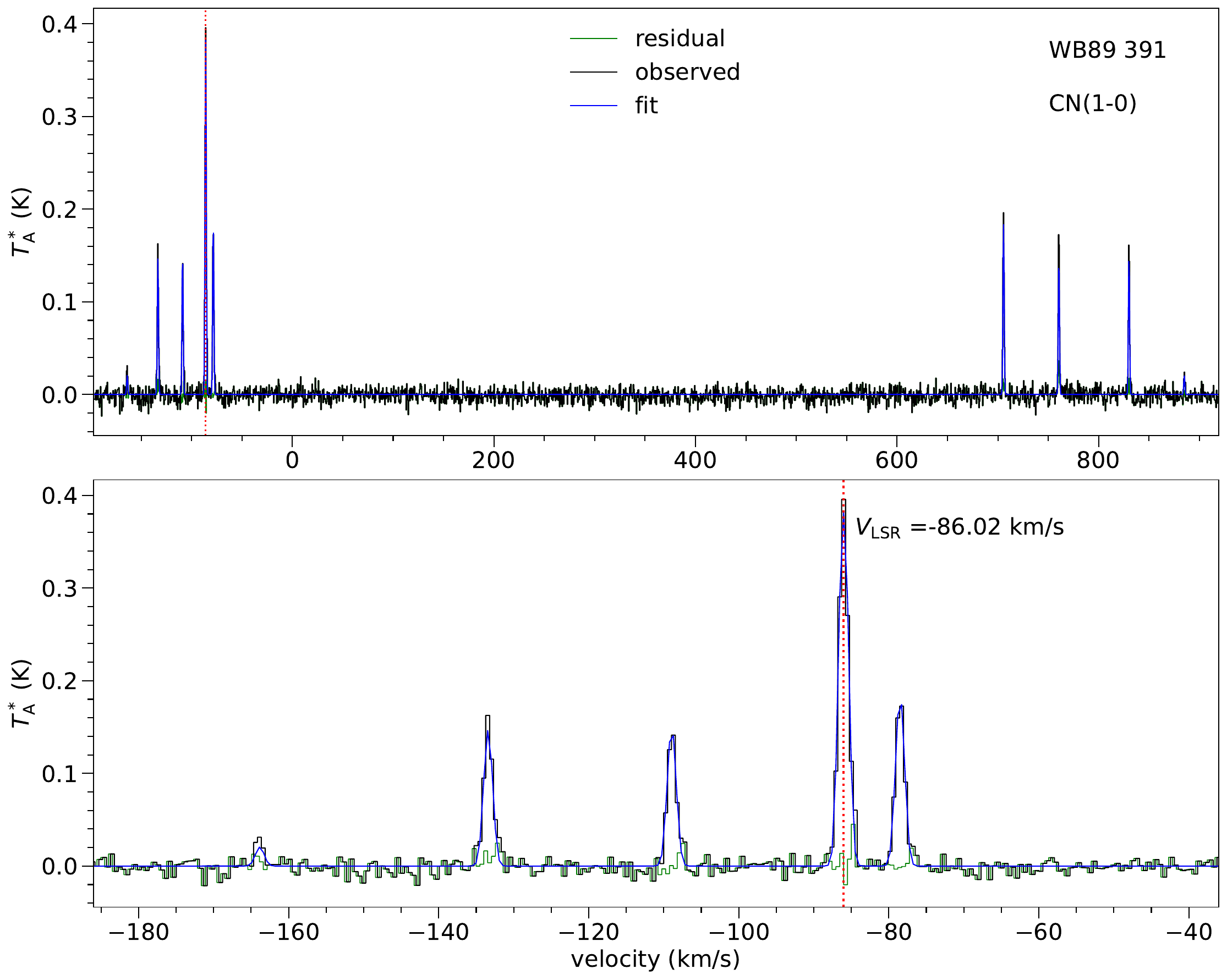}
        \caption{\label{fig:cont4_HfS_fitting_spectra}The HfS fitting spectra (Continued.)}
\end{figure*}

\begin{figure*}
        \centering
		    \includegraphics[scale=0.38]{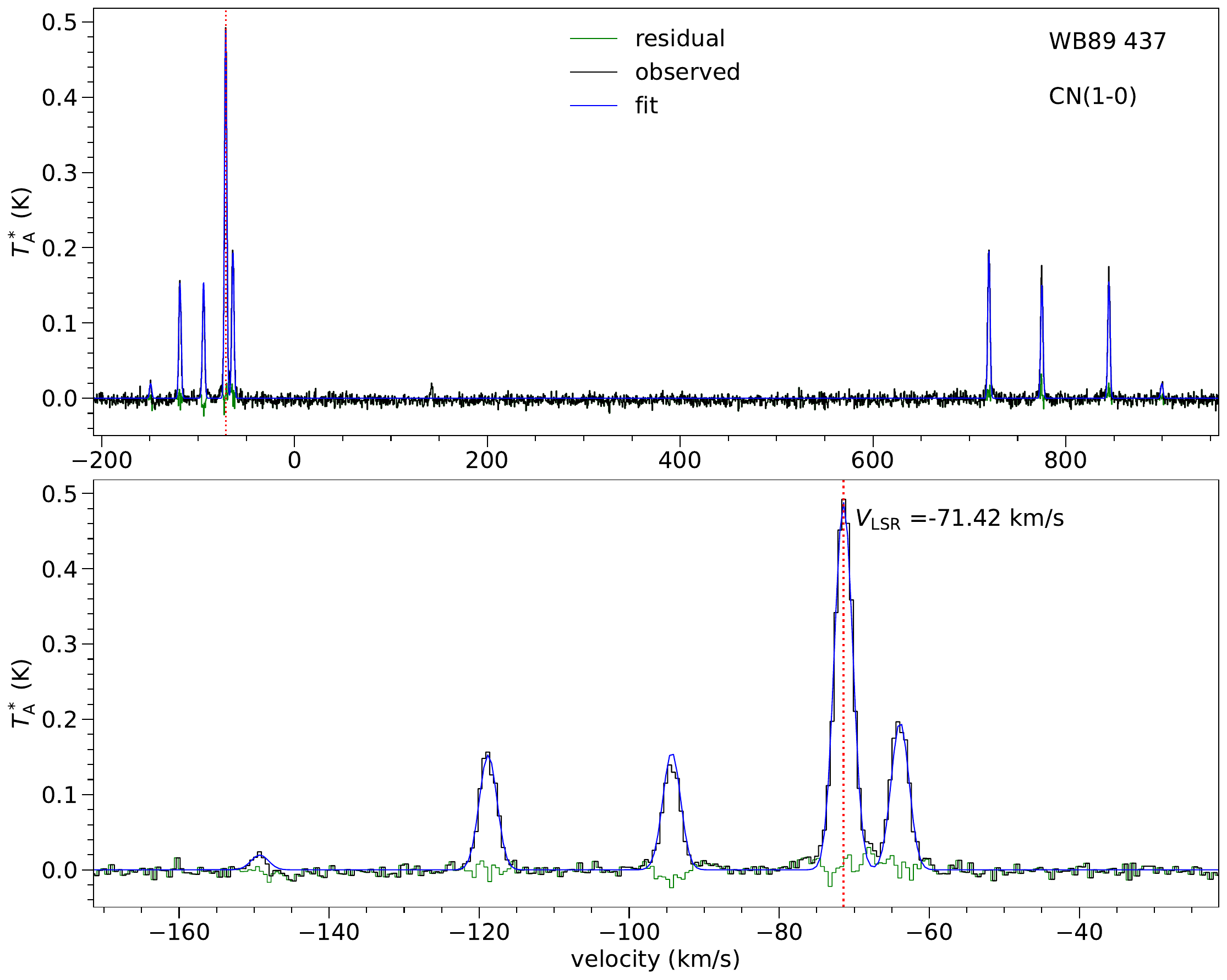}
                \includegraphics[scale=0.38]{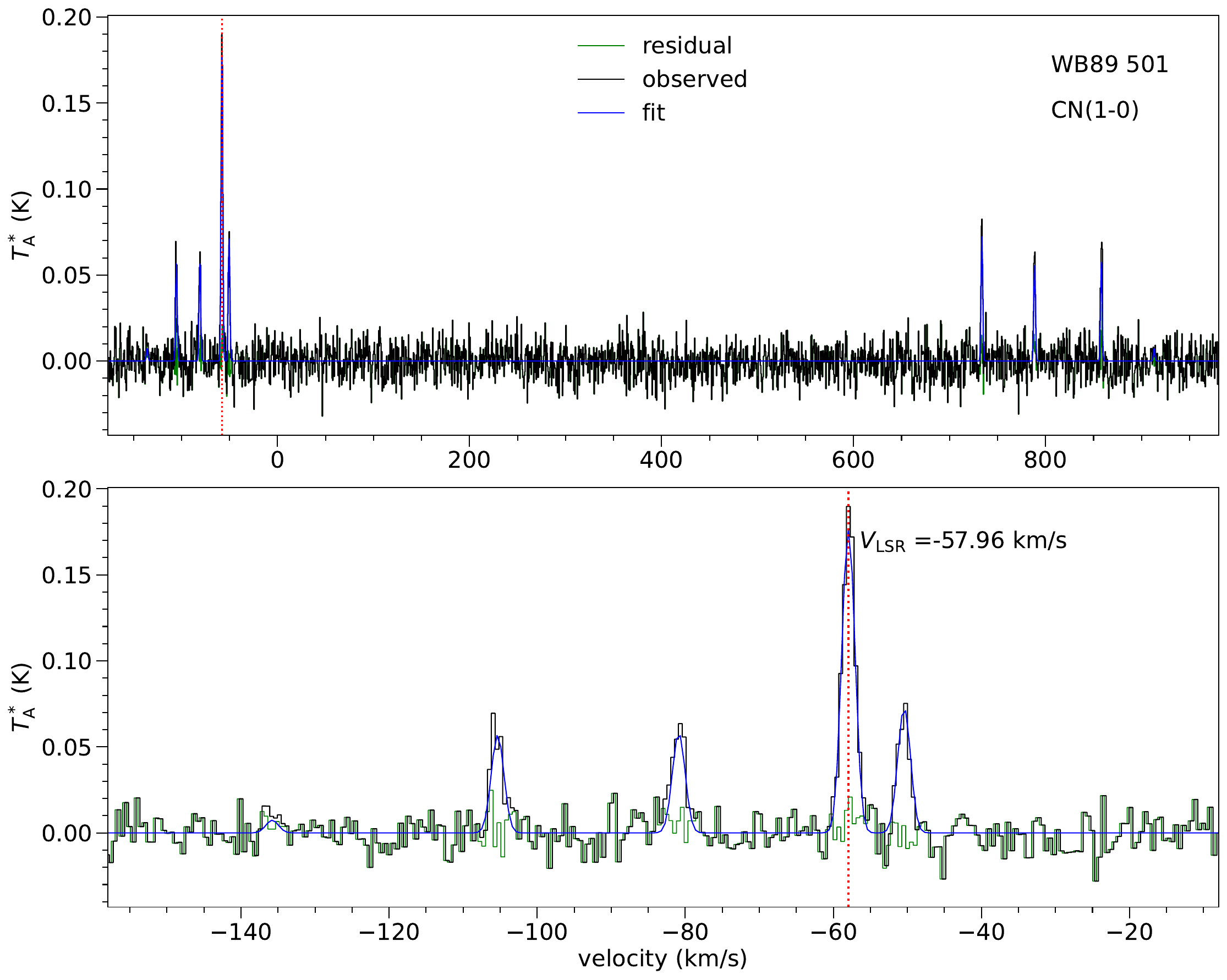}
        \caption{\label{fig:cont5_HfS_fitting_spectra}The HfS fitting spectra (Continued.)}
\end{figure*}

In this section, we show the spectra of the lines we used to derive \Rtwth\ and \Rftfif.  In Fig.~\ref{fig_13CO_fitting_spectra}, we show the $\rm ^{13}CO$ $J=1\to0$ lines and the Gaussian fitting to these lines which derives the $V_{\rm LSR}$ of our targets. In Figures ~\ref{fig:undetected_13CN_spectra} and \ref{fig:undetected_13CN_spectra_conti} we show the $\rm ^{12}CN$ and $\rm ^{13}CN$ $N=1\to0$ spectra without $\rm ^{13}CN$ detections.  We show the $\rm ^{12}CN$ and $\rm C^{15}N$ $N=1\to0$ spectra in Figures~\ref{fig:12CN_C15N_spectra}. The $\rm H^{13}CN$ and $\rm HC^{15}N$ $J=2\to1$ spectra are shown in Figures~\ref{fig:H13CN_HC15N_spectra} and \ref{fig:conti_H13CN_HC15N_spectra}. In Figures~\ref{fig:HfS_fitting_spectra},~\ref{fig:cont1_HfS_fitting_spectra},~\ref{fig:cont2_HfS_fitting_spectra},~\ref{fig:cont3_HfS_fitting_spectra},~\ref{fig:cont4_HfS_fitting_spectra} and~\ref{fig:cont5_HfS_fitting_spectra}, we show the HfS fitting results for the detected $\rm ^{12}CN$. 

\begin{figure*}
        \centering
                \includegraphics[scale=0.14]{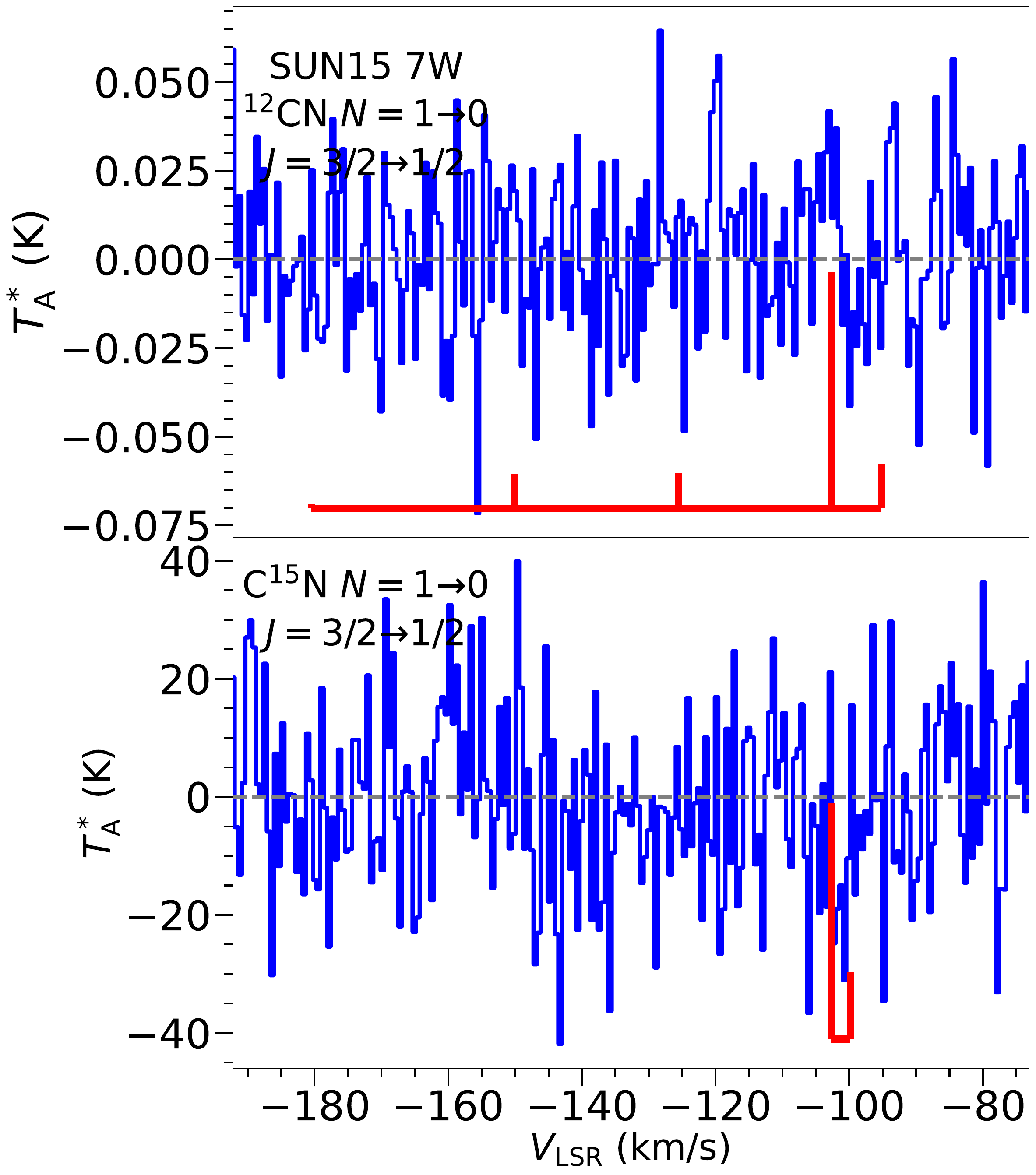}
                \includegraphics[scale=0.14]{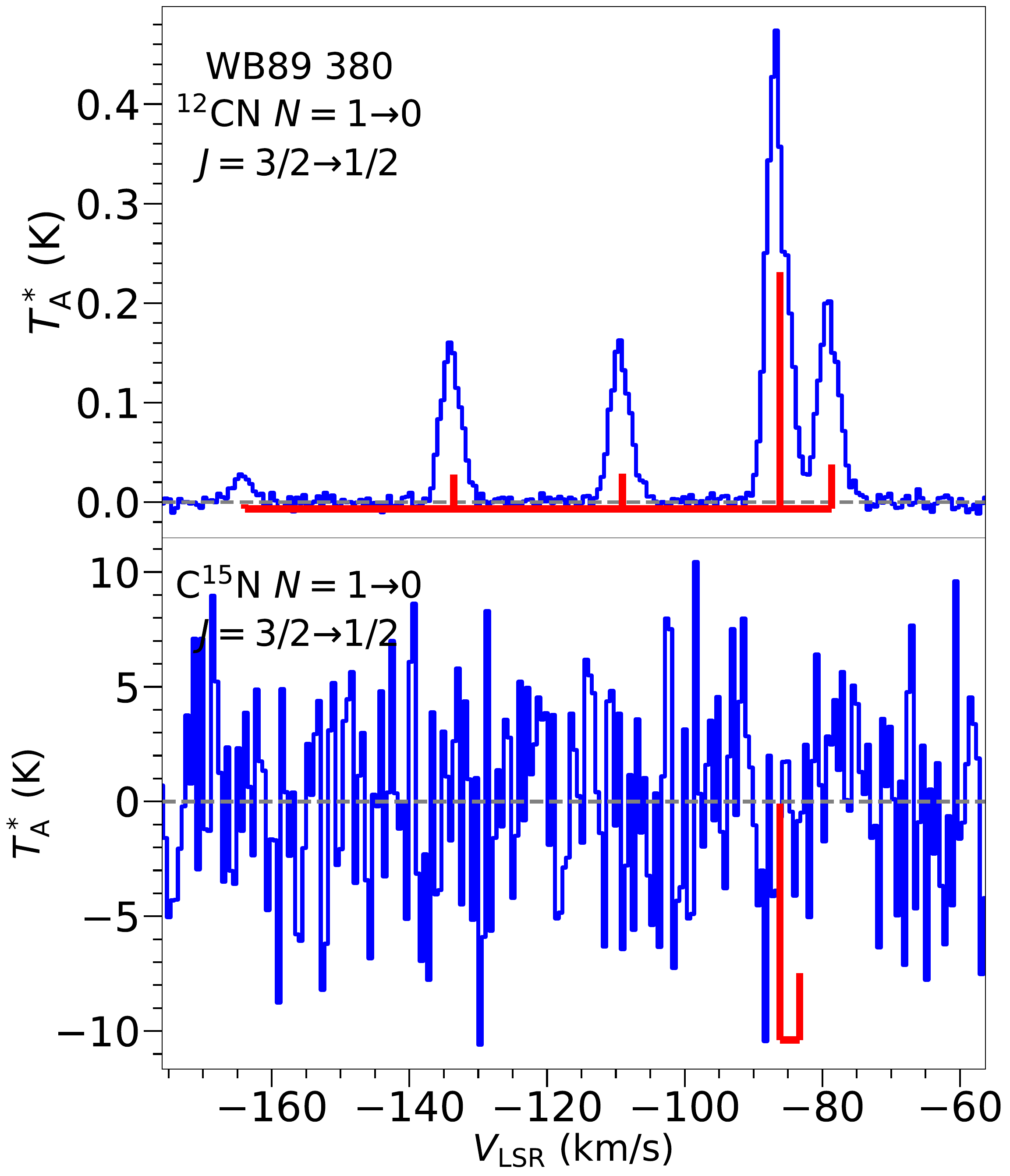}
                \includegraphics[scale=0.14]{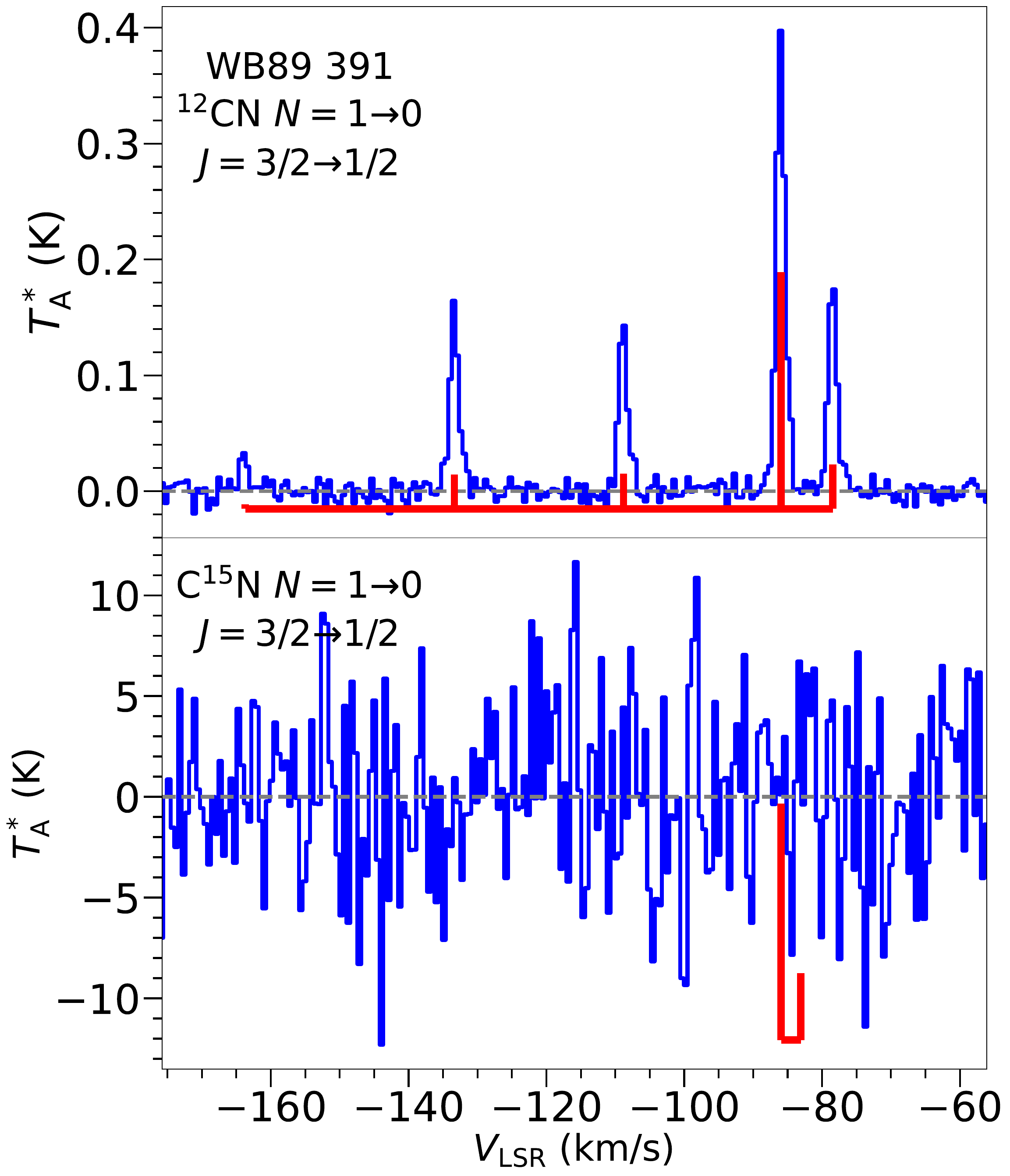}
                \includegraphics[scale=0.14]{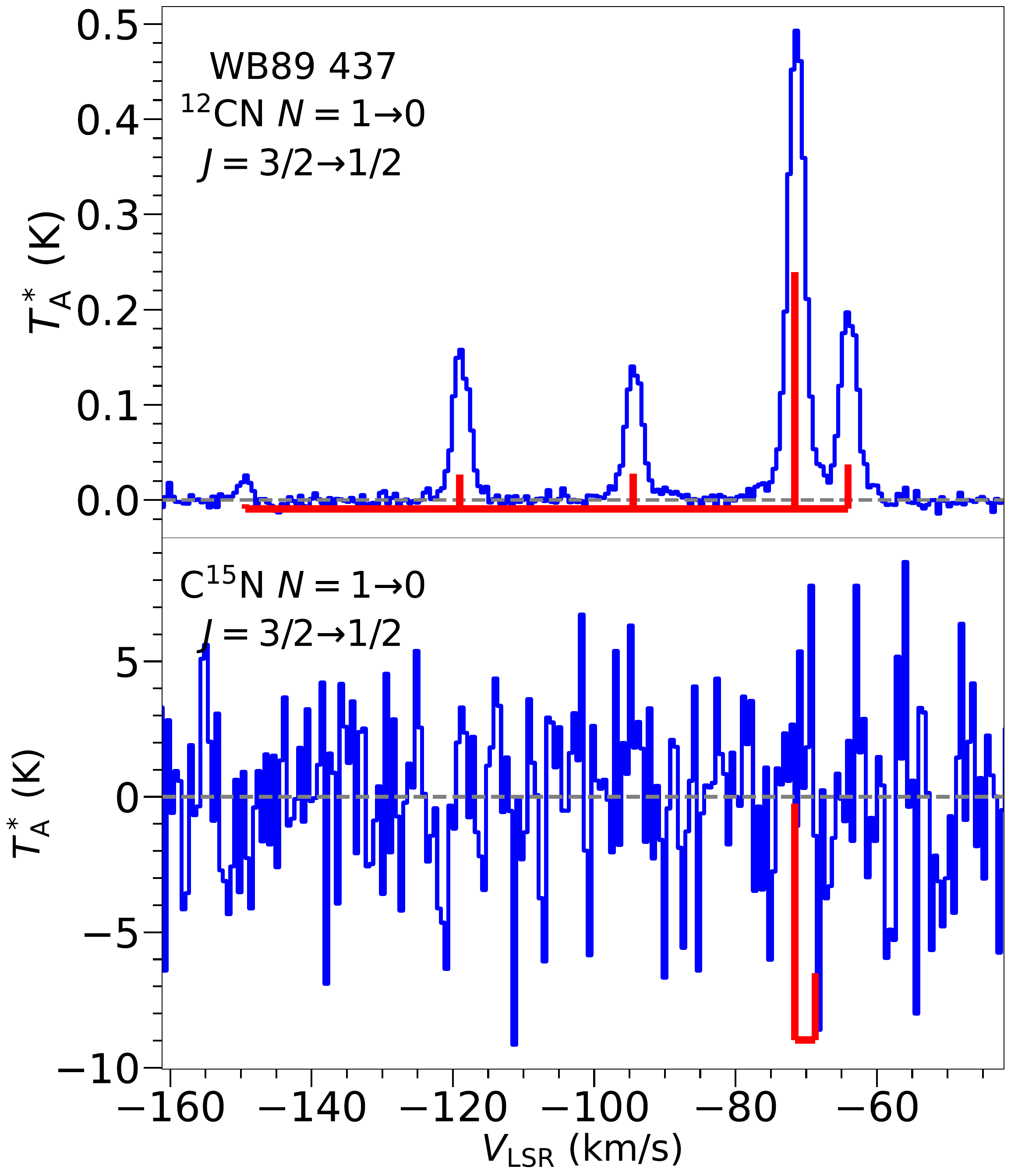}
                \includegraphics[scale=0.14]{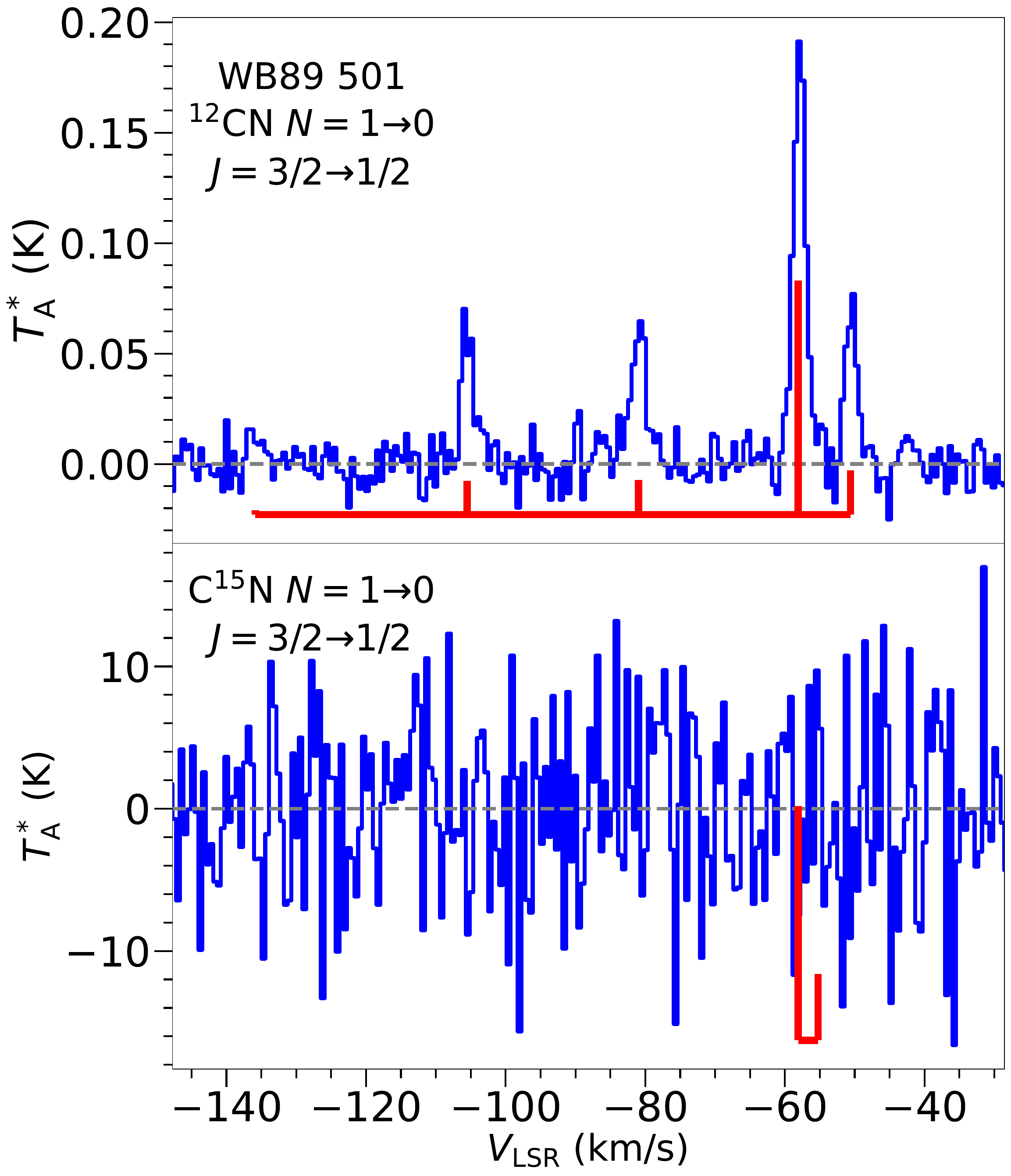}
        \caption{\label{fig:conti_12CN_C15N_spectra} Spectra of \twCN\ and \CftN\ $N=1\to0$ for targets without \CftN\ detections (Continued.).}
\end{figure*}

\section{\NftNfif\ derived from HCN isotopologues}
\label{Appendix:14N_15N_from_HCN}

For targets with detected \HthCN\ and \HCfifN\ $J=2\to1$ at 2-mm band, the \Rftfif\ can also be derived from their \Rtwth, as the following equation:

\begin{equation}
\label{eq:cal14N15N_withHCN}
     \frac{\rm{^{14}N}}{\rm{^{15}N}}=\frac{I_{\rm H^{13}CN}}{I_{\rm HC^{15}N}} \cdot R_{\rm 12C13C}
\end{equation}

Here $I_{\rm H^{13}CN}$ and $I_{\rm HC^{15}N}$ are the integrated intensities of the \HthCN\ and \HCfifN\ $J=2\to1$, respectively. Because the integrated intensity we measured is the integral of antenna temperature and the velocity, we revise the difference between the main beam efficiency at the frequency of \HthCN\ and \HCfifN\ $N=1\to0$. For IRAM 30-m, we have $\eta_{\rm mb, H^{13}CN} = 0.687$ and $\eta_{\rm mb, H^{13}CN} = 0.688$.

We have detected both \HthCN\ and \HCfifN\ $J=2\to1$ in three targets, two of them have measurements of \Rtwth\ to get \Rftfif. The line intensities and derived \Rftfif\ from \HthCN\ and \HCfifN\ $J=2\to1$ are shown in Table~\ref{tab:14N_15N_results_from_HCN}. The spectra of \HthCN\ and \HCfifN\ $J=2\to1$ are shown in Figures~\ref{fig:H13CN_HC15N_spectra} and \ref{fig:conti_H13CN_HC15N_spectra}. 

In Table \ref{tab:14N_15N_results_from_HCN} and Table \ref{tab:14N_15N_results_from_HCN_opthin}, we show the \Rftfif\ derived from multiplying \Rtwth\ from the $\tau$-correction method and multiplying \Rtwth\ from optically-thin satellite lines, respectively. Adopting \Rtwth\ from the optically-thin satellite line method, the \Rftfif\ value in G211.59 will be $\sim 100$ higher than the one adopting \Rtwth\ from the $\tau$-correction method. For other targets, the change is not significant.

\begin{table*}
\caption{$\rm{^{14}N/^{15}N}$ ratios derived from HCN isotopologues in our targets (The traditional HfS method).} \label{tab:14N_15N_results_from_HCN}
\begin{threeparttable}
\begin{tabular}{ccccc}
\hline
\hline
Sources            &  $I_{\rm H^{13}CN}$ & $I_{\rm HC^{15}N}$ & $\rm H^{13}CN/HC^{15}N$ & $\rm ^{14}N/^{15}N$  \\
                   &   ($\rm{10^{-2}}$ K $\cdot$ km s $\rm{^{-1}}$)  & ($\rm{10^{-2}}$ K $\cdot$ km s $\rm{^{-1}}$)      &      & \\
 \hline
G211.59            & 186.1 $\pm$  4.8   &    40.0 $\pm$ 4.1   &  4.7 $\pm$ 0.5 &  242 $\pm$ 40   \\  
G37.350            & 20.7  $\pm$  6.3   &    $\le$ 14       &  $\ge$ 1.4      &  $\ge$ 8.5            \\
G44.8              & 29.2  $\pm$  8.0   &    $\le$ 17        &  $\ge$ 1.7      &  $\ge$ 16   \\
IRAS0245           & $\le$  144        &    $\le$ 10       &        -        &  -   \\
SUN15 14N          & $\le$  241        &    $\le$ 15        &        -        &  -   \\
SUN15 18           & $\le$  123        &    $\le$ 9.0       &        -        &  -   \\
SUN15 21           & $\le$  191         &    $\le$ 12       &        -        &  -   \\
SUN15 34           & $\le$  276         &    $\le$ 16      &        -        &  -   \\
SUN15 56           & $\le$  172         &    $\le$ 11       &        -        &  -   \\
SUN15 57           & $\le$  174         &    $\le$ 11       &        -        &  -   \\
SUN15 7W           & $\le$  298         &    $\le$ 17        &        -        &  -   \\
WB89 380           & 34.6 $\pm$ 3.1     &    6.4 $\pm$ 2.5   &  5.4 $\pm$ 2.1  &  329 $\pm$ 171   \\
WB89 391           & 22.8 $\pm$ 4.1     &    $\le$  7.1      &  $\ge$ 3.2      &  $\ge$ 102  \\
WB89 437           & 89.0 $\pm$ 2.8     &    24.6 $\pm$ 2.0  &  3.6 $\pm$ 0.3  &  $\ge$ 223  \\
WB89 501           & $\le$  170         &    $\le$ 10       &        -        &  -   \\
\hline
\end{tabular}
\end{threeparttable}
\end{table*}

\begin{table*}
\caption{$\rm{^{14}N/^{15}N}$ ratios from HCN isotopologues with $\rm ^{12}C/^{13}C$ from $\rm ^{12}CN$ satellite lines in our targets.} \label{tab:14N_15N_results_from_HCN_opthin}
\begin{threeparttable}
\begin{tabular}{cccc}
\hline
\hline
Sources            & $\rm H^{13}CN/HC^{15}N$ & $\rm ^{12}C/^{13}C$   & $\rm ^{14}N/^{15}N$  \\
 \hline
G211.59            &    4.7 $\pm$ 0.5  &   71.7 $\pm$ 9.5    &  336 $\pm$ 57   \\
WB89 380           &    5.4 $\pm$ 2.1  &   57 $\pm$ 13       &  312 $\pm$ 141  \\
WB89 391           &    $\ge$ 3.2      &   $\ge$ 36          &  $\ge$  121     \\
WB89 437           &    3.6 $\pm$ 0.3  &   $\ge$ 59          &  $\ge$  211     \\
\hline
\end{tabular}
\end{threeparttable}
\end{table*}

\section{The linear fitting functions of Galactic \Rtwth\ and \Rftfif\ gradients}
\label{Appendix:linear_fitting_fuction}

We use the Markov Chain Monte Carlo  procedures\footnote{Use the Python procedure \texttt{emcee} \citep{Goodman2010,emcee}.} to fit linear functions for the measurements shown in Fig.~\ref{fig:12C_13C_grad_diff_tracer} (b) and Fig.~\ref{fig:14N_15N_grad} (b), respectively. The linear fitting function of the Galactic \Rtwth\ ratio is:

\begin{equation}
    {\rm ^{12}C/^{13}C} = 4.08\,(^{+0.86}_{-0.43}) R_{\rm gc} + 18.8\,(^{+2.6}_{-6.6})
\end{equation}
The linear fitting function of the Galactic \Rftfif\ gradient is:

\begin{equation}
    {\rm ^{14}N/^{15}N} = 10.6\,(^{+5.6}_{-8.8}) R_{\rm gc} + 96\,(^{+61}_{-44})
\end{equation}

Here, \Rgc\ is the Galactocentric distance in the unit of kpc.
However, it is highly risky to read a fitted value of \Rtwth\ or \Rftfif\ from these gradients. These gradients suffer large scatters, possibly because of the inhomogeneous mixing, the non-LTE conditions, and the astrochemistry of individual targets. The Galactic chemical evolution models predict that both \Rtwth\ and \Rftfif\ gradients are non-linear curves \citep[e.g., ][]{Romano2017,Romano2019}. For these reasons, we highly recommend using the direct measurements of \Rtwth\ and \Rftfif\ of individual targets, instead of values fitted from curves.

\section{The Figure of Galactic H$\;${\sc II} regions}
\label{appendix:HII_region_figure}

\begin{figure}
    \includegraphics[scale=0.35]{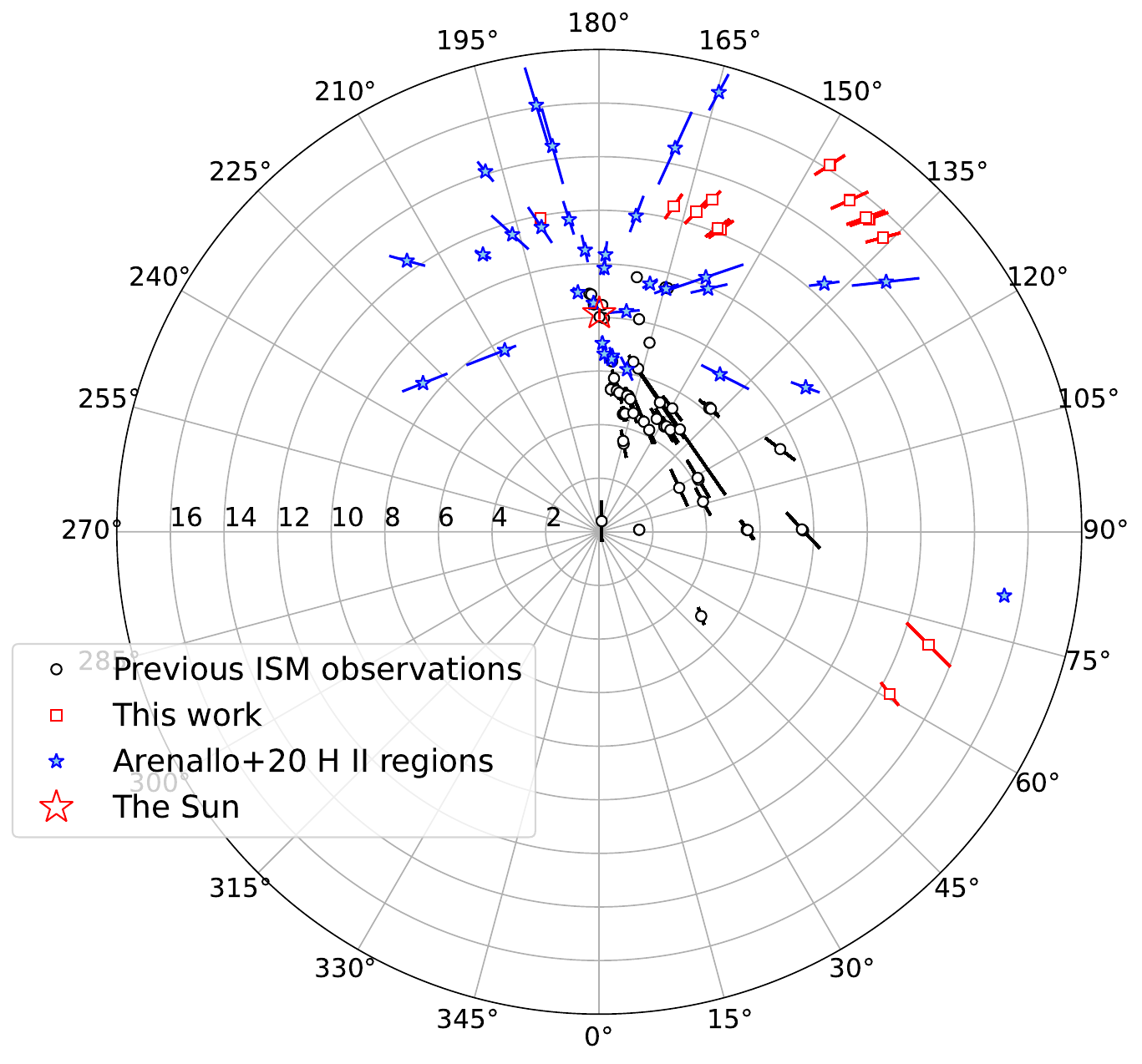}
\centering
	\caption{\label{fig:fig_posi_compared_with_HIIregion} The location of ISM targets in isotopic ratio works and H$\;${\sc II} regions in the Galactic plane. The pole is the Galactic center and the big red star shows the location of the Sun. The black dots represent the locations of targets in previous works \citep{Savage2002,Milam2005,Langer1990,Wouterloot1996,Giannetti2014,Jacob2020} and the red squares represent the targets in this work. The blue stars represent the location of H$\;${\sc II} regions in \citet{Arellano2020}.} 
\end{figure}

In Fig~\ref{fig:fig_posi_compared_with_HIIregion}, we compare the locations of H$\;${\sc II} regions in \citet{Arellano2020} and the locations of ISM targets for Galactic \Rtwth\ gradient on the Galactic plane. We find that most of the ISM targets are located in the same quadrants as those H$\;${\sc II} regions. It indicates the element abundance may be similarly well-mixed in the ISM targets compared to the mixed condition in those  H$\;${\sc II} regions. 

\bsp	
\label{lastpage}
\end{document}